\documentclass[12pt,a4paper]{article}
\usepackage[utf8]{inputenc}
\usepackage[T1]{fontenc}
\usepackage{amsmath,amssymb,amsthm}
\usepackage{mathtools}
\usepackage{booktabs}
\usepackage{geometry}
\usepackage{enumitem}
\usepackage{adjustbox}
\usepackage{comment}
\usepackage{hyperref}
\hypersetup{colorlinks=true,linkcolor=blue,citecolor=blue}
    \usepackage{rotating}
\newtheorem{theorem}{Theorem}[section]

\newtheorem{corollary}[theorem]{Corollary}
\newtheorem{proposition}[theorem]{Proposition}
\newtheorem{lemma}[theorem]{Lemma}
\newtheorem{definition}[theorem]{Definition}
\newtheorem{remark}{Remark}[section]
\newtheorem{assumption}{Assumption}[section]
\newcommand{\TK}{\widehat{\mathcal{T}}_{\widehat{K}}}

\def\*#1{\mathbf{#1}}
\def\+#1{\boldsymbol{#1}}
\usepackage{hyperref}
\hypersetup{
    colorlinks=true,
    linkcolor=blue,
    filecolor=magenta,
    urlcolor=cyan,
}

\usepackage{xcolor}

\usepackage{natbib}
\hypersetup{
    citecolor=blue,        
}
\setcitestyle{authoryear,open={(},close={)}}

\numberwithin{equation}{section}

\begin{document}
\title{\textbf{Estimating  and Testing Kinks in Panel Data Models} 
\Large{\emph{}}}

{\author{Yousef Kaddoura\thanks{\"Orebro University
 E-mail address: \texttt{yousef.kaddoura@oru.se}.}\\
{\small \"Orebro University}\\ }}

\maketitle
\begin{abstract}
Many economic and financial relationships may change gradually rather than abruptly. We study panel data models in which the coefficient vector is continuous and piecewise linear in calendar time, with a finite number of unknown kink dates at which its slope changes. We propose a penalised least squares estimator that applies adaptive weighted group penalties to the second differences of the coefficient path, and develop asymptotic theory showing that it recovers both the number and the locations of the kinks with probability approaching one. To our knowledge, this is the first panel framework to estimate an unknown number of common kink dates in a time-varying coefficient path under fixed effects. We establish that endpoint slopes converge at the usual cubic regime-length rate and interior slopes at rates determined by their own and adjacent regime lengths. We also develop a coefficient-by-coefficient extension allowing individual regressors to kink at different dates. Monte Carlo evidence supports the good finite sample properties, and we illustrate the method through an application in macro-finance, specifically the relationship between debt and growth.
\end{abstract}




\section{Introduction} \label{one}
Certain economic and financial relationships may not change overnight. When a 
central bank shifts its policy stance, when a trade agreement phases 
in, or when a tax reform gradually alters incentives, the response 
of economic and financial agents unfolds over time. The coefficients governing 
these relationships may not jump discontinuously, but they might 
``bend.'' Yet the dominant paradigm in the structural change 
literature focuses precisely on such jumps: abrupt, discrete breaks 
in parameters that partition a sample into distinct regimes with 
different levels \citep[see][among others]{Andrews1993, BaiPerron1998, BaiPerron2003,
HorvathHuskova2012, HorvathHuskovaRiceWang2017,
AntochHanousekHorvathHuskovaWang2019}. This paper argues that for 
a broad and economically important class of phenomena, the right 
model is not a break but a \emph{kink}, i.e.\ a change in the rate 
at which a coefficient evolves.

Kink models have a growing presence in the threshold literature \citep[see, for example,][]{HidalgoLeeSeo2019, ZhongWanZhang2022, SunWanZhangZhong2024, ZhangXieXiao2025}. 
\citet{Hansen2017} formalizes estimation and inference for the regression 
kink model in time series, where the effect of a threshold variable $x$ on 
the response is linear on each side of an unknown threshold $\gamma$ but 
continuous at $\gamma$ itself, producing a ``kink'' in the regression 
function at that point. \citet{ZhangWangJiang2017} extend this framework to 
panel data. These papers model nonlinearity as a function 
of \textbf{\emph{where agents are}}: the coefficient path bends when a 
covariate crosses some threshold level $\gamma$, and crucially, an agent can 
move back and forth across that threshold as the variable fluctuates. This is 
distinct again from the regression kink design of \citet{CardLeePeiWeber2015}, 
where the kink is located in a known running variable at a policy threshold 
and the change in slope identifies a causal effect. There the kink is a 
feature of the cross-sectional design.

The present paper takes a different and complementary perspective. We model 
nonlinearity as a function of \textbf{\emph{when agents are}}. The coefficient 
vector is modeled as piecewise linear in time, with multiple unknown kink dates 
at which the rate of change shifts permanently. This distinction matters 
economically. A central bank may not react because 
inflation crossed some level last quarter, but it may do so because of a persistent 
shift in its policy framework. \citet{ClaridaGaliGertler2000} document 
  a shift around the Volcker appointment in 1979, but such a 
transition need not be instantaneous  as 
credibility may need to be established. Similarly, \citet{Trefler2004} shows that the 
employment and productivity effects of the Canada--US Free Trade Agreement 
unfolded over several years after ratification. 
Returning to the growth literature,  \citet{Hansen2017} studies the relationship between 
public debt and growth studied by \citet{ReinhartRogoff2010}. The question 
is whether the effect of debt on growth changes once the debt-to-GDP ratio 
crosses some threshold.  In our empirical application we revisit a similar question and find that the bend is dateable and that the relationship was already evolving before that date. This complements his analysis and 
emphasizes that when agents are is an important complement 
to where agents are in obtaining a fuller picture of the 
underlying economic relationship. 

Another related literature allows coefficients to evolve smoothly over time. In panel data, however, this literature remains relatively sparse; see \citet{LiChenGao2011}, \citet{SuWangJin2019}, \citet{Chen2019}, \citet{DongGaoPeng2021}, and \citet{SuJin2012}. Besides the when agents are point, a natural question, then, is why one should prefer the kink model for this task over its two closest neighbours: the level break model on one side and the fully nonparametric time-varying parameter (TVP) model on the other. The break model, as in \citet{BaiPerron1998}, is well suited to changes that occur overnight, but it rules out, by assumption, the possibility that the relationship was already evolving before the break date, constraining the within-regime path to be flat and discarding the information about the {speed} at which structural change unfolds. The TVP model imposes no such restriction, since its goal is to recover the function itself rather than to identify a small number of structural change dates. The cost, however, is that it delivers neither a dateable event nor a within-regime slope with a clear economic interpretation. Our model lies between these two extremes. It identifies a dateable kink and estimates how fast the relationship was evolving on each side of that date.  The within-regime slope measures the speed of structural change in a way that neither alternative can provide, making the kink model the natural choice whenever both the timing and the pace of a shifting economic relationship are objects of interest.

The estimator suggested in this paper proceeds in two steps. We
first run an unpenalised regression of the first-period differenced
model, which eliminates the fixed effects while leaving the
time-varying structure of the coefficients \citep[see][for a discussion in related settings]{KaddouraWesterlund2023, Kaddoura2025}, to construct
adaptive weights in the spirit of \citet{Zou2006} that are small
where the coefficient path already appears to bend and large where
it is locally linear. We then minimise a penalised least squares
objective that penalises the second differences of the coefficient
path using a weighted group-$\ell_2$ norm \citep[see][for the group
lasso and its adaptive extension]{YuanLin2006, Zou2006}. Since a
nonzero second difference at period $t$ is equivalent to a kink at $t$,
the penalty directly targets slope changes while enforcing continuity
throughout, distinguishing our approach from level break estimators
that allow jumps. The adaptive weights recover the oracle property in the
sense of \citet{Zou2006}, where the estimated kink dates coincide with the
true ones with probability approaching one, recovering both the
correct number of kinks and their exact locations. Conditional on
correct recovery, a post-kink ordinary least squares regression on
the piecewise linear basis estimates the initial level and
within-regime slopes, delivering asymptotic normality.

The estimation strategy proposed in this paper is related to several
strands of the penalisation literature, though it differs from each
in important ways. At a conceptual level, our use of second differences is reminiscent of the Hodrick--Prescott filter \citep{HodrickPrescott1997}. Our method is not an HP filter, however, since we work in a panel regression setting with observed covariates and fixed effects, and use an adaptive group penalty on second differences to induce sparse kinks. This panel structure, where first-period differencing removes the incidental fixed effect before the penalty is applied, is what separates our setting from the time series filtering and trend filtering literature \citep{KimEtAl2009, TibshiraniRyan2014} and the generalised lasso
of \citet{TibshiraniTaylor2011}. The closest to our setting is
\citet{QianSuET2016} and \citet{QianSu2016} who apply an adaptive
group fused Lasso to detect level breaks by penalising \emph{first}
differences of the coefficient path. We penalise second differences instead, 
which induces piecewise linearity rather than jumps and so targets kinks rather 
than breaks. Because a slope keeps shifting the 
coefficient path in every later regime, it is identified over the whole 
subsequent span rather than within a single regime alone.

The contributions of this paper are therefore threefold. First, we are the first to 
introduce the kink panel model as a new class of structural change
specifications in which the coefficient path is continuous and
piecewise linear in time, filling the gap between the fully flexible
time-varying parameter literature and the level break literature. To the best of our knowledge, this is the first panel-data structural-change framework in which common unknown kink dates are estimated in a time-varying coefficient path.
Second, we establish selection consistency, exact recovery of the kink dates, and oracle asymptotic normality for the post-kink estimator. Unlike the level-break model of \citet{QianSu2016}, in which each regime coefficient is identified from observations within that regime, continuity in our kink model links adjacent segments and changes the convergence rates of the slopes. Endpoint slopes retain the classical within-regime trend rates, whereas interior slopes borrow strength from neighboring regimes through their shared boundary levels. To the best of our knowledge, this is the first explicit characterization of such adjacent-regime slope rates in a panel model with multiple unknown kinks.   Third, we introduce a
coefficient-by-coefficient extension, in the spirit of
\citet{Kaddoura2025}, in which each regressor is permitted to kink
at its own set of dates independently of the others, accommodating
richer forms of partial structural change in which some coefficients
bend while others remain globally linear. Together, these
contributions provide applied researchers with a toolkit for dating
and estimating gradual, continuous shifts in economic relationships.

The remainder of the paper is organized as follows. Section~\ref{sec:model}
sets out the kink panel model and its key representation. Section~\ref{sec:estimator}
introduces the penalised estimator and the post-kink estimation step.
Section~\ref{asymptotics} develops the asymptotic theory, including
selection consistency, recovery of the kink dates, asymptotic normality
of the post-kink estimator, and the coefficient-by-coefficient
extension. Section~\ref{sec:MC} reports the Monte Carlo evidence, and Section~\ref{sec:debtgrowth}
presents the empirical applications.

\section{Model}\label{sec:model}

Consider a scalar  variable $y_{i,t}\in\mathbb{R}$ and a vector of covariates $\*x_{i,t}:=[x_{i,t,1},\ldots,x_{i,t,p}]'\in\mathbb{R}^{p\times 1}$, observed for cross-sectional units $i=1,\ldots,N$ over time periods $t=1,\ldots,T$, where $p$ is fixed. Consider the following data-generating process (DGP):
\begin{align}
    y_{i,t} = \mu_i + \*x_{i,t}'\+\theta_t + \varepsilon_{i,t},
    \label{eq:DGP}
\end{align}
where $\+\theta_t\in\mathbb{R}^{p\times 1}$ is a conformable vector of unknown coefficients that may vary across time periods in a manner made precise below, $\mu_i\in\mathbb{R}$ is a cross-sectional fixed effect, and $\varepsilon_{i,t}\in\mathbb{R}$ is an idiosyncratic error term. Throughout this paper ``$a:=b$'' indicates that $a$ is defined by $b$.

The object of interest is to estimate $\+\theta_t$ and its  kink behaviour. To allow for kinks, we let $\+\theta_t$ evolve over $K+1$ distinct regimes. Let $1=:T_0<T_1<\cdots<T_K<T<T_{K+1}:=T+1$ and define the $\ell$th regime set as
\(
    \mathcal{K}_\ell := \{T_{\ell-1},\ldots,T_\ell-1\},
    \) for \( \ell=1,\ldots,K+1,
     \)
where $K\in\{0,\ldots,T-2\}$. For each regime $\ell$, let $\+\kappa_\ell\in\mathbb{R}^{p\times 1}$ denote the within-regime slope (in~$t$). We model $\+\theta_t$ as piecewise linear in $t$ via
\begin{align}
    \+\theta_t = \+\theta_{T_{\ell-1}} + (t-T_{\ell-1})\+\kappa_\ell
    \label{eq:Kink}
\end{align}
for $ t\in\mathcal{K}_\ell$ and $\ \ell=1,\ldots,K+1$.
Under~\eqref{eq:Kink}, $\+\theta_{T_{\ell-1}}$ is the level of the coefficient path at the start of regime~$\ell$, and $\+\kappa_\ell$ governs the linear evolution of $\+\theta_t$ within that regime. Kinks occur at periods $t\in\mathcal{T}_K:=\{T_1,\ldots,T_K\}$, where the slope changes across regimes. Since regime~$\ell$ ends at time $T_\ell-1$ and regime~$\ell+1$ starts at time $T_\ell$, we impose continuity at kink locations (no jumps) via the recursion
\begin{align}
    \+\theta_{T_\ell} = \+\theta_{T_{\ell-1}} + (T_\ell-T_{\ell-1})\+\kappa_\ell
    \label{eq:continuity}
\end{align}
for $\ell = 1,\hdots, K$. In particular,~\eqref{eq:continuity} implies $\+\theta_{T_\ell}-\+\theta_{T_\ell-1}=\+\kappa_\ell$,\footnote{%
To see this, note that $T_\ell-1\in\mathcal{K}_\ell$, so~\eqref{eq:Kink} gives $\+\theta_{T_\ell-1}=\+\theta_{T_{\ell-1}}+\bigl((T_\ell-1)-T_{\ell-1}\bigr)\+\kappa_\ell$, while~\eqref{eq:continuity} gives $\+\theta_{T_\ell}=\+\theta_{T_{\ell-1}}+(T_\ell-T_{\ell-1})\+\kappa_\ell$. Subtracting yields $\+\theta_{T_\ell}-\+\theta_{T_\ell-1}=\+\kappa_\ell$.}
so the coefficient path does not jump when moving from $T_\ell-1$ to $T_\ell$.

In fact, how we model the parameters is closely related to the literature in which local linear methods are used to approximate smoothly evolving coefficient functions (see, for example, \citealt{FanZhang1999}; \citealt{Cai2007}). The key difference is that, in our setting, piecewise linearity is imposed as a global structural approximation with a set of unknown kink dates, rather than serving as a local approximation to an otherwise smooth coefficient path.\footnote{In the special case where $K=1$, $p=1$,  $x_{i,t,1}=1$,
our model becomes close to the deterministic panel trend-break
model of \citet{Kim2011}.}

\begin{remark}[Unified representation]\label{rem:unified}
Equations~\eqref{eq:Kink}--\eqref{eq:continuity} can be condensed into a single representation. Iterating~\eqref{eq:continuity} yields
\(
    \+\theta_{T_{\ell-1}}
    =
    \+\theta_1
    +
    \sum_{j=1}^{\ell-1}(T_j-T_{j-1})\+\kappa_j \) for \(
     \ell=1,\ldots,K+1,\)
where the sum is empty (hence zero) when $\ell=1$. Substituting this into~\eqref{eq:Kink} gives, for $t\in\mathcal{K}_\ell$,
\begin{align}
    \+\theta_t
    =
    \+\theta_1
    +
    \sum_{j=1}^{\ell-1}(T_j-T_{j-1})\+\kappa_j
    +
    (t-T_{\ell-1})\+\kappa_\ell
    \label{eq:Kinkalt}
\end{align}
for $ \ell=1,\ldots,K+1$. Hence, the entire path $\{\+\theta_t\}_{t=1}^T$ is determined by the initial level $\+\theta_1$, the within-regime slopes $\+\kappa_1,\ldots,\+\kappa_{K+1}$, and the kink dates $\mathcal{T}_K$.
\end{remark}


An equivalent way of expressing the piecewise-linear restriction~\eqref{eq:Kink}--\eqref{eq:continuity} is through first differences. Define $\Delta\+\theta_t:=\+\theta_t-\+\theta_{t-1}$ for $t=2,\ldots,T$ and consider the following proposition. 

\begin{proposition}[Piecewise-constant first differences]\label{prop:FD}
Under~\eqref{eq:Kink}--\eqref{eq:continuity}, the first differences are piecewise constant:
\begin{align}
    \Delta\+\theta_t = \+\kappa_\ell
    \label{eq:FDconst}
\end{align}
for $ t\in\{T_{\ell-1}+1,\ldots,\min(T_\ell,\, T)\}$ and $\ \ell=1,\ldots,K+1$.
Conversely,~\eqref{eq:FDconst} implies~\eqref{eq:Kink} and~\eqref{eq:continuity}.
\end{proposition}

\begin{proof}
\emph{Forward direction.} Fix a regime $\ell\in\{1,\ldots,K+1\}$. For $t\in\{T_{\ell-1}+1,\ldots,T_\ell-1\}$, both $t$ and $t-1$ belong to
$\mathcal{K}_\ell$, so~\eqref{eq:Kink} applies to both and
\(
    \Delta\+\theta_t
    = \bigl[\+\theta_{T_{\ell-1}}+(t-T_{\ell-1})\+\kappa_\ell\bigr]
    -\bigl[\+\theta_{T_{\ell-1}}+(t-1-T_{\ell-1})\+\kappa_\ell\bigr]
    = \+\kappa_\ell.
\)
For $\ell\leq K$ and $t=T_\ell$, continuity~\eqref{eq:continuity} combined
with~\eqref{eq:Kink} at $t=T_\ell-1$ gives
$\+\theta_{T_\ell}-\+\theta_{T_\ell-1}=\+\kappa_\ell$, so
$\Delta\+\theta_{T_\ell}=\+\kappa_\ell$ as well. For the last regime
$\ell=K+1$, no boundary $t=T_{K+1}=T+1$ exists since $\Delta\+\theta_t$
is defined only for $t=2,\ldots,T$. This establishes~\eqref{eq:FDconst}.
\emph{Converse direction.} Assume~\eqref{eq:FDconst}. Fix $\ell$ and any
$t\in\mathcal{K}_\ell$. For every $s\in\{T_{\ell-1}+1,\ldots,t\}$, we have
$s\leq t\leq T_\ell-1\leq\min(T_\ell,T)$, with equality in the last bound
only in the final regime $\ell=K+1$, where $T_{K+1}=T+1$ and
$\min(T_\ell,T)=T$. In every case $s\leq\min(T_\ell,T)$, so~\eqref{eq:FDconst}
gives $\Delta\+\theta_s=\+\kappa_\ell$. Telescoping yields
\(
    \+\theta_t
    = \+\theta_{T_{\ell-1}}+\sum_{s=T_{\ell-1}+1}^{t}\Delta\+\theta_s
    = \+\theta_{T_{\ell-1}}+(t-T_{\ell-1})\+\kappa_\ell,
\)
which is~\eqref{eq:Kink}. For $\ell\leq K$, taking $t=T_\ell-1$ in this
identity gives
$\+\theta_{T_\ell-1}=\+\theta_{T_{\ell-1}}+(T_\ell-1-T_{\ell-1})\+\kappa_\ell$,
and combining with $\Delta\+\theta_{T_\ell}=\+\kappa_\ell$
from~\eqref{eq:FDconst} gives
$\+\theta_{T_\ell}=\+\theta_{T_{\ell-1}}+(T_\ell-T_{\ell-1})\+\kappa_\ell$,
which is~\eqref{eq:continuity}.
\end{proof}

\begin{figure}[h!]
    \centering    \includegraphics[width=0.8\linewidth]{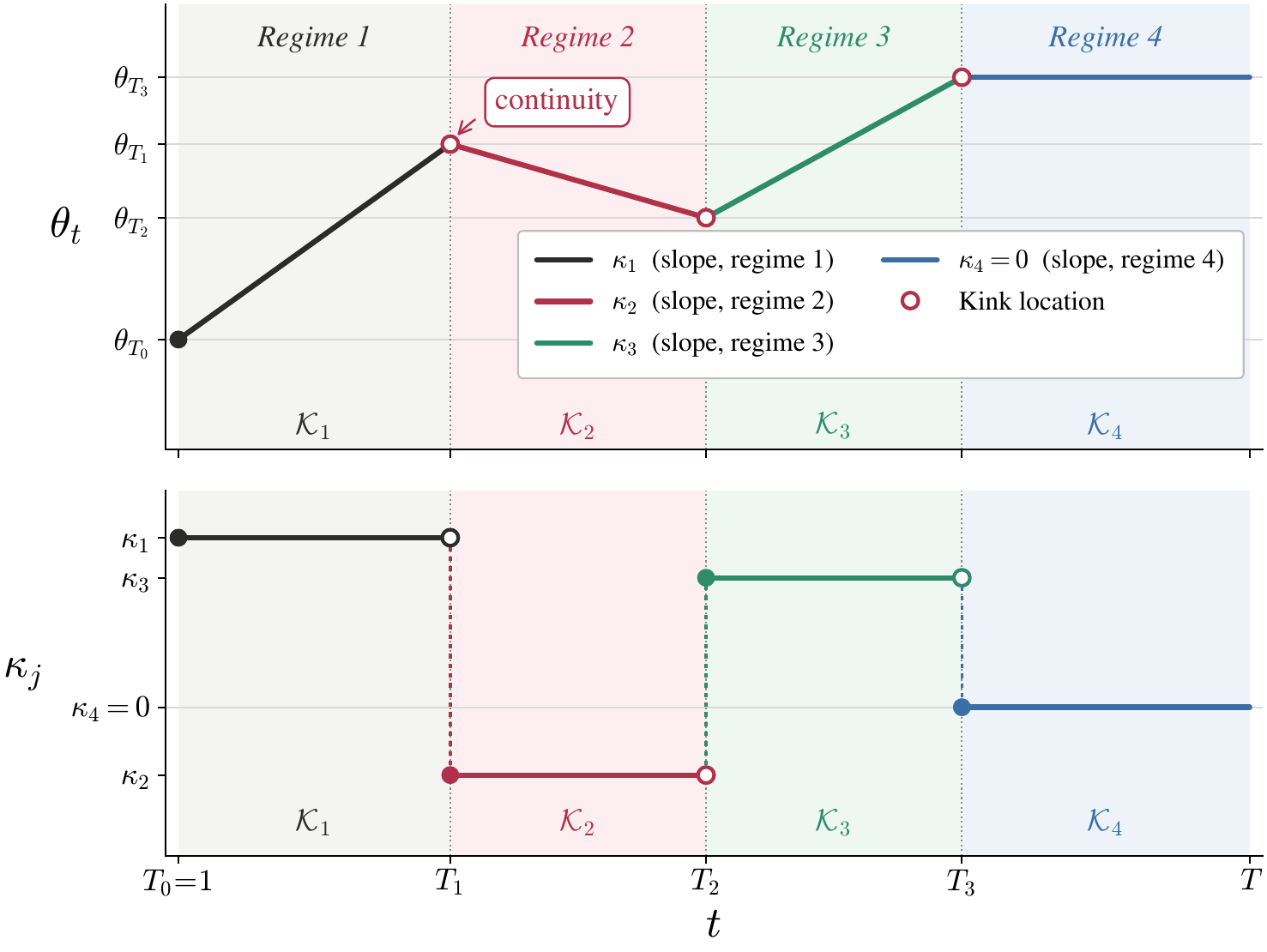}
    \caption{\footnotesize{Example of the piecewise-linear coefficient path. The upper panel shows $\+\theta_t$ across three regimes $\mathcal{K}_1,\mathcal{K}_2,\mathcal{K}_3$ and $\mathcal{K}_4$, with kinks at $T_1$, $T_2$ and $T_3$. The lower panel shows the corresponding piecewise-constant first differences $\Delta\+\theta_t$.}}
    \label{fig:kink}
\end{figure}

An immediate consequence of Proposition~\ref{prop:FD} is that a kink at $T_\ell$ corresponds to a change in slope:
\begin{align}
    \+\kappa_{\ell+1}\neq\+\kappa_\ell
    \Longleftrightarrow
    \|\Delta\+\theta_{T_\ell+1}-\Delta\+\theta_{T_\ell}\|
    = \|\+\kappa_{\ell+1}-\+\kappa_\ell\| > 0.
    \label{eq:kinkdetect}
\end{align}
Hence, estimating kink locations reduces to detecting the times at which the \emph{second differences} $\Delta^2\+\theta_t:=\Delta\+\theta_{t+1}-\Delta\+\theta_t$ are nonzero. This motivates a sparsity assumption on $\{\Delta^2\+\theta_t\}_{t=2}^{T-1}$: most second differences are zero (corresponding to the linear segments), and the nonzero ones identify the kink dates. 
Take Figure \ref{fig:kink} as an illustration. The upper panel shows a coefficient path $\+\theta_t$ across three regimes $\mathcal{K}_1,\mathcal{K}_2,\mathcal{K}_3$ and $\mathcal{K}_4$, while the lower panel shows the corresponding piecewise constant slope $\+\kappa_\ell$, with jumps at the two kink dates.

\bigskip

\bigskip

\begin{remark}[Second-derivative interpretation]\label{rem:secondderiv}
The second difference has a natural continuous time counterpart. For a twice differentiable function $b(\cdot)$, the standard finite-difference approximation gives $b''(t)\approx\bigl[b(t+1)-2b(t)+b(t-1)\bigr]$. In our discrete-time setting, $\Delta^2\+\theta_t=\+\theta_{t+1}-2\+\theta_t+\+\theta_{t-1}$ is the exact analog. Hence, penalising $\sum_{t=2}^{T-1}\|\Delta^2\+\theta_t\|$ can be interpreted as penalising the ``curvature'' of the coefficient path: the penalty is zero when $\+\theta_t$ is globally linear and nonzero only at kink points.
\end{remark}

It is also worth asking what happens if the truth contains a level break rather
than a kink. In fact, the kink model contains the break
model as a class of discrete-time observed sequences. Suppose the coefficient equals \(\+\theta_A\) through \(T^*-1\) and
\(\+\theta_A+\+\alpha\) from \(T^*\) onward, similar to  \citet{BaiPerron1998}.
Although continuity rules out a literal jump in the interpolated path, it does not rule out the same sequence at the observed integer dates. Suppose further that  \(K=2\) with
\(T_1=T^*-1\) and \(T_2=T^*\), \(\+\theta_1=\+\theta_A\),
\(\+\kappa_1=\+\kappa_3=\*0_{p\times1}\) and \(\+\kappa_2=\+\alpha\),
equations~\eqref{eq:Kink} and~\eqref{eq:continuity} reproduce the break path
exactly, with the jump carried by the middle slope. Thus, the observed level break is represented by two adjacent kinks, with slopes \(\*0_{p\times1}\), \(\+\alpha\), and \(\*0_{p\times1}\).
The reason is that the coefficient is only ever observed at integer dates.
Between \(T^*-1\) and \(T^*\) there are no observations, so the data cannot distinguish a literal jump from a steep but continuous movement across that interval. The kink model therefore connects the two observed coefficient levels by a straight line. A change occurring over several sampling periods is gradual, whereas one occurring over a single period is abrupt at the observed sampling frequency.

\section{Estimator}\label{sec:estimator}

By~\eqref{eq:Kink}, the behaviour of the coefficient path $\+\Theta_T:=[\+\theta_1',\ldots,\+\theta_T']'\in\mathbb{R}^{Tp\times 1}$ is governed by the within-regime slopes $\*K_K:=[\+\kappa_1',\ldots,\+\kappa_{K+1}']'\in\mathbb{R}^{(K+1)p\times 1}$, the initial level $\+\theta_1$, and the kink dates $\mathcal{T}_K$. Conversely, once $\+\theta_1$, $\*K_K$, and $\mathcal{T}_K$ are known, the entire path $\+\Theta_T$ can be recovered via~\eqref{eq:Kinkalt}. We use superscript ``$0$'' to denote true values: $K^0$ is the true number of kinks, $\mathcal{T}^0_{K^0}:=\{T_1^0,\ldots,T_{K^0}^0\}$ the true kink dates, $\*K^0_{K^0}:=[\+\kappa_1^{0\prime},\ldots,\+\kappa_{K^0+1}^{0\prime}]'\in\mathbb{R}^{(K^0+1)p\times 1}$ the true slopes, and $\+\Theta^0_T:=[\+\theta_1^{0\prime},\ldots,\+\theta_T^{0\prime}]'\in\mathbb{R}^{Tp\times 1}$ the true coefficient path.

Two transformations of the data are used to handle the fixed effects
$\mu_i$, one per estimation stage. The selection stage works with
deviations from the initial observation. For each $i$ and $t=2,\ldots,T$,
define $\check{a}_{i,t}:=a_{i,t}-a_{i,1}$, and in particular
$\check{y}_{i,t}:=y_{i,t}-y_{i,1}$ and
$\check{\varepsilon}_{i,t}:=\varepsilon_{i,t}-\varepsilon_{i,1}$. The
post-kink stage instead removes $\mu_i$ by the within transformation. For
each $i$ and $t=1,\ldots,T$, define
$\ddot{a}_{i,t}:=a_{i,t}-T^{-1}\sum_{s=1}^{T}a_{i,s}$, and in particular
$\ddot{y}_{i,t}$ and $\ddot{\varepsilon}_{i,t}$.

\subsection{Penalised estimator}\label{sec:initial}

As a preliminary step, we compute an unpenalised  OLS estimator from the first-period-differenced model:
\begin{align}
    \{\dot{\+\theta}_t\}_{t=1}^T
    := \underset{\{\+\theta_t\}_{t=1}^T}{\arg\min}\;
    \frac{1}{NT}\sum_{i=1}^N\sum_{t=2}^{T}
    \bigl( \check{y}_{i,t} - \*x_{i,t}'\+\theta_t + \*x_{i,1}'\+\theta_1\bigr)^2.
    \label{eq:initial}
\end{align}
Since~\eqref{eq:initial} imposes no piecewise-linear structure, the resulting estimator $\dot{\+\theta}_t$ is $\sqrt{N}$-consistent for $\+\theta_t^0$ under the assumptions stated in Section~\ref{sec:assumptions}, but does not exploit the kink structure.

The initial estimator is used to construct adaptive weights that steer the penalty in the penalised objective function defined below. Specifically, define for all $t=2,\hdots, T-1$
\(
    \dot{\omega}_t
    := \|\Delta\dot{\+\theta}_{t+1}-\Delta\dot{\+\theta}_t\|^{-\zeta_1},
\)
where $\zeta_1>0$ is a user-specified constant (typically $\zeta_1=2$).   This adaptive weighting is essential for the oracle property established later.\footnote{For numerical stability, one may replace 
$\|\Delta\dot{\+\theta}_{t+1}-\Delta\dot{\+\theta}_t\|^{-\zeta_1}$ 
with 
$\bigl(\|\Delta\dot{\+\theta}_{t+1}-\Delta\dot{\+\theta}_t\|+\delta_t\bigr)^{-\zeta_1}$, 
where $\delta_t>0$ is a small perturbation of order $O(N^{-1/2})$ satisfying 
$\delta_t\to 0$ as $N\to\infty$. This prevents near-division-by-zero in 
finite samples while leaving the asymptotic properties unchanged.}
Now, given these weights, we propose minimising the following objective function:
\begin{align}
    \mathcal{L}_{\vartheta_1}(\+\Theta_T)
    &:= \frac{1}{NT}\sum_{i=1}^N\sum_{t=2}^{T}
    \bigl( \check{y}_{i,t} - \*x_{i,t}'\+\theta_t + \*x_{i,1}'\+\theta_1\bigr)^2
    +  {\vartheta_1}\sum_{t=2}^{T-1}\dot{\omega}_t\,
    \|\Delta\+\theta_{t+1}-\Delta\+\theta_t\|,
    \label{eq:objective}
\end{align}
where $\vartheta_1=\vartheta_1(N,T)>0$ is a tuning parameter that can be chosen by minimizing an information criterion (see \citealt{QianSu2016} for details). We do this explicitly Sections \ref{sec:MC} and \ref{sec:debtgrowth}. Three features of~\eqref{eq:objective}  are worth commenting on.

\medskip
\noindent\textbf{($i$)~\emph{Loss function.}}\;
The first term is a least-squares loss computed on the first-period-differenced data. Taking first-period differences of~\eqref{eq:DGP} gives $ \check{y}_{i,t}=\*x_{i,t}'\+\theta_t-\*x_{i,1}'\+\theta_1+\check{\varepsilon}_{i,t}$, so the fixed effect $\mu_i$ cancels. The initial level $\+\theta_1$ enters through the term $\*x_{i,1}'\+\theta_1$.

\bigskip

\noindent\textbf{($ii$)~\emph{Penalty}.}\;
The second term penalises the second differences $\Delta^2\+\theta_t=\Delta\+\theta_{t+1}-\Delta\+\theta_t$ using the group-$\ell_2$ norm $\|\cdot\|$. By~\eqref{eq:kinkdetect}, $\|\Delta^2\+\theta_t\|>0$ if and only if there is a kink at~$t$. The group-$\ell_2$ structure encourages the \emph{entire} $p\times 1$ vector $\Delta^2\+\theta_t$ to be set to zero simultaneously, thereby inducing piecewise linearity in the coefficient path. The sum runs over $t=2,\ldots,T-1$ because $\Delta^2\+\theta_t=\+\theta_{t+1}-2\+\theta_t+\+\theta_{t-1}$ requires $t-1\geq 1$ and $t+1\leq T$.

\bigskip

\noindent\textbf{($iii$)~\emph{Adaptive weights}.}\;
The weights $\dot{\omega}_t$  make the penalty adaptive. At true kink dates, the penalty is weak (small $\dot{\omega}_t$), so the nonzero second differences survive; at non-kink dates, the penalty is strong (large $\dot{\omega}_t$), so $\Delta^2\+\theta_t$ is shrunk to $\*0_{p\times 1}$ exactly. This is the mechanism by which the estimator simultaneously detects kink locations and estimates the coefficient path.

\medskip

The penalised estimator of $\+\Theta_T^0$ is then given by
\begin{align}
    \widehat{\+\Theta}_T
    :=
    \begin{bmatrix}
        \widehat{\+\theta}_1 \\ \vdots \\ \widehat{\+\theta}_T
    \end{bmatrix}
    =
    \underset{\+\Theta_T}{\arg\min}\;\mathcal{L}_{\vartheta_1}(\+\Theta_T).
    \label{eq:penest}
\end{align}
From the penalised estimator $\widehat{\+\Theta}_T$, we extract the estimated kink dates and regime sets as follows.\footnote{Demeaning would also eliminate $\mu_i$ and is linear in
$\+\Theta_T$, so it could serve at this stage too. However, under the within
transformation the regressor attaching to $\+\theta_s$ at observation
$(i,t)$ is $\*x_{i,t}\*1\{s=t\}-T^{-1}\*x_{i,s}$, so every date enters
every equation which leads to a matrix of
order $Tp\times Tp$. Differencing against $t=1$ instead  reduces the  dimension to 
$2p\times2p$.}

\begin{definition}[Estimated kinks and regimes]\label{def:estkinks}
For a given $\widehat{\+\Theta}_T$, define the set of estimated kink locations as
\begin{align}
    \widehat{\mathcal{T}}_{\widehat{K}}
    := \left\{t\in\{2,\ldots,T-1\}:\|\Delta\widehat{\+\theta}_{t+1}-\Delta\widehat{\+\theta}_t\|>0\right\}
    = \left\{\widehat{T}_1,\ldots,\widehat{T}_{\widehat{K}}\right\},
    \label{eq:estkinks}
\end{align}
where $\widehat{T}_1<\cdots<\widehat{T}_{\widehat{K}}$ are the ordered elements and $\widehat{K}:=|\widehat{\mathcal{T}}_{\widehat{K}}|$ is the estimated number of kinks. By convention, set $\widehat{T}_0:=1$ and $\widehat{T}_{\widehat{K}+1}:=T+1$. These kink dates partition $\{1,\ldots,T\}$ into $\widehat{K}+1$ estimated regime sets:
\(
    \widehat{\mathcal{K}}_\ell
    :=  \left\{\widehat{T}_{\ell-1},\ldots,\widehat{T}_\ell-1\right\}
\) for $\ell=1,\ldots,\widehat{K}+1.$
If $\|\Delta\widehat{\+\theta}_{t+1}-\Delta\widehat{\+\theta}_t\|=0$ for all $t=2,\ldots,T-1$, then $\widehat{K}=0$ and $\widehat{\mathcal{T}}_0=\varnothing$, corresponding to a single linear regime spanning all periods.
\end{definition}

\subsection{Post kink estimator}\label{sec:postkink}
Recall from Definition~\ref{def:estkinks} that
$\TK=\{\widehat{T}_1,\ldots,\widehat{T}_{\widehat{K}}\}$, with
$\widehat{T}_0:=1$ and $\widehat{T}_{\widehat{K}+1}:=T+1$, and that the
induced regime sets are
$\widehat{\mathcal{K}}_\ell:=\{\widehat{T}_{\ell-1},\ldots,
\widehat{T}_\ell-1\}$ for $\ell=1,\ldots,\widehat{K}+1$. Holding $\TK$
fixed, the unified representation in Remark~\ref{rem:unified} gives for $ t\in\widehat{\mathcal{K}}_\ell$
\begin{align}
    \+\theta_t
    =
    \+\theta_1
    +\sum_{j=1}^{\ell-1}\bigl(\widehat{T}_j-\widehat{T}_{j-1}\bigr)\+\kappa_j
    +\bigl(t-\widehat{T}_{\ell-1}\bigr)\+\kappa_\ell,
    \label{eq:pk_rep}
\end{align}
where $\+\theta_1$ is the initial level and $\+\kappa_\ell$ the slope in
regime $\widehat{\mathcal{K}}_\ell$. The continuity
conditions~\eqref{eq:continuity} are already embedded
in~\eqref{eq:pk_rep}, so the entire $Tp$-dimensional path $\+\Theta_T$ is
determined by the $(\widehat{K}+2)p$-dimensional  parameter
vector
\(
    \+\psi_{\widehat{K}}
    := [\+\theta_1', \*K_{\widehat{K}}']' =
    \bigl[\+\theta_1',\,\+\kappa_1',\ldots,\+\kappa_{\widehat{K}+1}'\bigr]'.
\)
To express~\eqref{eq:pk_rep} as a function of $\+\psi_{\widehat{K}}$,
define for each $t\in\widehat{\mathcal{K}}_\ell$ and each
$j=1,\ldots,\widehat{K}+1$ the scalar \emph{basis loading}
\begin{align}
    d_{j,t}\big(\TK\big)
    :=
    \begin{cases}
        \widehat{T}_j-\widehat{T}_{j-1} & j < \ell, \\[4pt]
        t-\widehat{T}_{\ell-1}           & j = \ell, \\[4pt]
        0                                & j > \ell,
    \end{cases}
    \label{eq:djt}
\end{align}
where $\ell$ is the unique regime with $t\in\widehat{\mathcal{K}}_\ell$.
The loading measures how long the slope $\+\kappa_j$ has been active by
period $t$, i.e. the full length $\widehat{T}_j-\widehat{T}_{j-1}$ of
$\widehat{\mathcal{K}}_j$ once that regime has ended, the time elapsed
since the current regime began while $t$ lies within it, and zero before
it starts. With this notation~\eqref{eq:pk_rep} takes the compact form
\begin{align}
    \+\theta_t
    =
    \+\theta_1
    +\sum_{j=1}^{\widehat{K}+1}d_{j,t}\big(\TK\big)\,\+\kappa_j
    \label{eq:pk_loading}
\end{align}
for each $t=1,\hdots T$ and  is linear in $\+\psi_{\widehat{K}}$. Collecting the loadings into
the $(\widehat{K}+2)p\times 1$ regressor
\(
    \+\phi_{i,t}\big(\TK\big)
    :=
    \bigl[
        \*x_{i,t}',\,
        d_{1,t}\big(\TK\big)\,\*x_{i,t}',\ldots,
        d_{\widehat{K}+1,t}\big(\TK\big)\,\*x_{i,t}'
    \bigr]'
\)
gives $\*x_{i,t}'\+\theta_t=\+\phi_{i,t}\big(\TK\big)'\+\psi_{\widehat{K}}$
for every $t$, so that the DGP~\eqref{eq:DGP} reads for each $i=1,\ldots,N$ and  
$t=1,\ldots,T,$
\begin{align}
     y_{i,t}
    =
    \mu_i
    +\+\phi_{i,t}\big(\TK\big)'\+\psi_{\widehat{K}}
    +\varepsilon_{i,t},
    \label{eq:pk_expanded}
\end{align}
a standard linear panel regression in which the coefficient vector
$\+\psi_{\widehat{K}}$ is common to all $i$ and $t$ and the time
variation of the path is carried entirely by the regressor.

It remains to remove $\mu_i$, and here the choice of transformation
matters in a way it did not for selection. The penalised problem of
Section~\ref{sec:initial} differences against the initial period, which
is enough to recover $\TK$ but subtracts the single equation $t=1$ from
every other, returning the  error $\varepsilon_{i,1}$ in all
$T-1$ rows, where it accumulates with the time index. The post-kink
stage subtracts the unit mean instead, which spreads the transformation
evenly across periods and leaves a transformed error whose covariance
remains bounded in $T$. Averaging the
DGP~\eqref{eq:DGP} over $t$ for fixed $i$ and subtracting gives
\begin{align}
    \ddot y_{i,t}
    =
    \*x_{i,t}'\+\theta_t
    -\dfrac{1}{T}\sum_{s=1}^{T}\*x_{i,s}'\+\theta_s
    +\ddot\varepsilon_{i,t},
    \label{eq:pk_demeanraw}
\end{align}
and the subtracted term carries a different coefficient $\+\theta_s$ at
every date, so the equation for period $t$ involves the entire path
$\+\theta_1,\ldots,\+\theta_T$ rather than $\+\theta_t$ alone.  However, every coefficient in the sum is the same linear
function of $\+\psi_{\widehat{K}}$. Substituting
$\*x_{i,s}'\+\theta_s=\+\phi_{i,s}\big(\TK\big)'\+\psi_{\widehat{K}}$ at
every date $s$, the common vector factors out of the sum,
\begin{align}
    \dfrac1T\sum_{s=1}^{T}\*x_{i,s}'\+\theta_s
   & =
    \dfrac1T\sum_{s=1}^{T}\+\phi_{i,s}\big(\TK\big)'\+\psi_{\widehat{K}}
    \notag \\ & =
    \Bigl(\frac1T\sum_{s=1}^{T}\+\phi_{i,s}\big(\TK\big)\Bigr)'
    \+\psi_{\widehat{K}}
     \notag \\ &=
    \bar{\+\phi}_i\big(\TK\big)'\+\psi_{\widehat{K}},
    \label{eq:pk_factor}
\end{align}
so the path collapses to the single structural vector
and~\eqref{eq:pk_demeanraw} becomes
\begin{align}
    \ddot y_{i,t}
    =
    \ddot{\+\phi}_{i,t}\big(\TK\big)'\+\psi_{\widehat{K}}
    +\ddot\varepsilon_{i,t}
    \label{eq:pk_linear}
\end{align}
 for all $i=1,\ldots,N$ and $t=1,\ldots,T,$ a regression on the same parameter vector as~\eqref{eq:pk_expanded},
with no coefficients of other periods entering. The transformation acts
on the regressor alone.

In matrix form, stack the observations as
$\*y:=[y_{1,1},\ldots,y_{1,T},\ldots,y_{N,T}]'\in\mathbb{R}^{NT\times1}$,
let $\+\Phi\big(\TK\big)\in\mathbb{R}^{NT\times(\widehat{K}+2)p}$ collect
the rows $\+\phi_{i,t}\big(\TK\big)'$ in the same order, and let
$\+\varepsilon$ be conformable. The transformation
in~\eqref{eq:pk_linear} is then the centering matrix
\(
    \*C
    :=\*I_N\otimes\*C_T,
    \) with \(
    \*C_T:=\*I_T-\frac1T\+{\iota}_T\+{\iota}_T',
\)
which is symmetric and idempotent, demeans each unit's $T$ observations,
and satisfies $\*C(\*I_N\otimes\+{\iota}_T)=\*0$, so that the fixed effects are
annihilated. The post-kink estimator of $\+\psi_{\widehat{K}}$ is the OLS
solution in~\eqref{eq:pk_linear},
\begin{align}
\widetilde{\+\psi}_{\widehat{K}}
&:=
\underset{\+\psi_{\widehat{K}}\in\mathbb{R}^{(\widehat{K}+2)p}}{\arg\min}
\;\frac{1}{NT}\sum_{i=1}^{N}\sum_{t=1}^{T}
\bigl(\ddot y_{i,t}
-\ddot{\+\phi}_{i,t}\big(\TK\big)'\+\psi_{\widehat{K}}\bigr)^2
\notag\\
&=
\underset{\+\psi_{\widehat{K}}\in\mathbb{R}^{(\widehat{K}+2)p}}{\arg\min}
\;\frac{1}{NT}
\bigl(\*y-\+\Phi\big(\TK\big)\+\psi_{\widehat{K}}\bigr)'\*C
\bigl(\*y-\+\Phi\big(\TK\big)\+\psi_{\widehat{K}}\bigr)
\notag\\
&=
\bigl[\+\Phi\big(\TK\big)'\*C\,\+\Phi\big(\TK\big)\bigr]^{-1}
\+\Phi\big(\TK\big)'\*C\,\*y,
    \label{eq:postest}
\end{align}
where the second equality holds by symmetry and idempotency of $\*C$, and the third
holding under Assumption~\ref{ass:rank} below. Partitioning
$\widetilde{\+\psi}_{\widehat K}=[\widetilde{\+\theta}_1',\,
\widetilde{\+\kappa}_1',\ldots,\widetilde{\+\kappa}_{\widehat K+1}']'$
and defining the $p\times(\widehat K+2)p$ loading matrix
\(
    \*L_t\bigl(\TK\bigr)
    :=\bigl[\*I_p,\;
    d_{1,t}\bigl(\TK\bigr)\*I_p,\ldots,
    d_{\widehat K+1,t}\bigl(\TK\bigr)\*I_p\bigr],
\)
the post-kink coefficient path for each $t=1,\ldots,T$  follows from~\eqref{eq:pk_loading} as
\begin{align}
    \widetilde{\+\theta}_t
    :=\*L_t\bigl(\TK\bigr)\widetilde{\+\psi}_{\widehat K}
    =\widetilde{\+\theta}_1
    +\sum_{j=1}^{\widehat K+1}
    d_{j,t}\bigl(\TK\bigr)\,\widetilde{\+\kappa}_j
    \label{eq:pk_path}
\end{align}

\begin{remark}[The regressor matrix]
\label{rem:structure}
Sorting rows by regime, the raw design is block lower triangular,
\begin{align*}
\+\Phi\big(\TK\big)
\;=\;
\begin{bmatrix}
\*D_{1,\theta}   & \*D_{1,1}   & \*0         & \cdots & \*0 \\[4pt]
\*D_{2,\theta}   & \*D_{2,1}   & \*D_{2,2}   & \cdots & \*0 \\[4pt]
\vdots           & \vdots      & \vdots      & \ddots & \vdots \\[4pt]
\*D_{\widehat{K}+1,\theta} &
\*D_{\widehat{K}+1,1} &
\*D_{\widehat{K}+1,2} &
\cdots &
\*D_{\widehat{K}+1,\widehat{K}+1}
\end{bmatrix},
\end{align*}
with row block $\ell$ collecting the observations $(i,t)$,
$t\in\widehat{\mathcal{K}}_\ell$, and
\(
    \*D_{\ell,\theta}
    := \bigl[\*x_{i,t}'\bigr]_{i,\,t\in\widehat{\mathcal{K}}_\ell},
\)
\(
    \*D_{\ell,j}
    := \bigl[d_{j,t}\big(\TK\big)\,\*x_{i,t}'\bigr]_{i,\,
    t\in\widehat{\mathcal{K}}_\ell}
\)
for $j=1,\ldots,\ell$. The blocks above the diagonal vanish because
slopes from regimes not yet begun contribute nothing at $t$. The
diagonal block carries the loading $t-\widehat{T}_{\ell-1}$, whose
variation in $t$ separates $\+\kappa_\ell$ from the level $\+\theta_1$,
and the blocks below it carry the frozen loading
$d_{j,t}=\widehat{T}_j-\widehat{T}_{j-1}$, through which an earlier
slope keeps shifting the path in every later regime.
\end{remark}

\section{Asymptotic results}\label{asymptotics}

We first introduce some notation. If $\*A$ is a matrix,
$\mu_{\min}(\*A)$ and $\mu_{\max}(\*A)$ denote its smallest and largest
singular values (its smallest and largest eigenvalues when $\*A$ is
symmetric), $\mathrm{tr}\,\*A$ its trace,
$\|\*A\|:=\sqrt{\mathrm{tr}\,\*A'\*A}$ its Frobenius norm, and
$\|\*A\|_{op}:=\mu_{\max}(\*A)$ its operator norm, so that
$\|\*A\|_{op}\le\|\*A\|$. For two symmetric matrices $\*A$ and $\*B$, we
write $\*A\preceq\*B$ if $\*B-\*A$ is positive semidefinite. All limits
are taken as $N,T\to\infty$ jointly unless stated otherwise, and we
abbreviate $\lim_{N,T\to\infty}$ to $\lim$ where no confusion arises.
For a sequence of random variables or vectors $\{Z_N\}$, we write
$Z_N=O_p(a_N)$ if $Z_N/a_N$ is bounded in probability, i.e.\ for every
$\varepsilon>0$ there exists $M<\infty$ such that
$\sup_N\mathbb{P}(|Z_N/a_N|>M)<\varepsilon$, and $Z_N=o_p(a_N)$ if
$Z_N/a_N\to_p0$. For two positive sequences $\{a_N\}$ and $\{b_N\}$, we
write $a_N\asymp b_N$ if there exist finite constants $0<c\le C<\infty$
such that $c\le a_N/b_N\le C$ for all $N$ sufficiently large. For
positive scalars $a$ and $b$, we write $a\wedge b:=\min\{a,b\}$. The
terms w.p.1, w.p.a.1, $\to_d$, $\to_p$, $\mathcal{N}(\cdot,\cdot)$, and
$\*0_{r\times1}$ signify with probability one, with probability
approaching one, convergence in distribution, convergence in
probability, a normal distribution, and an $r\times1$ vector of zeros,
respectively.

\subsection{Assumptions for selection}\label{sec:assumptions}

Let $I_\ell^0:=T_\ell^0-T_{\ell-1}^0$ denote the length of regime
$\ell$ for $\ell=1,\ldots,K^0+1$, and let
$I_{\min}^0:=\min_{1\le\ell\le K^0+1}I_\ell^0$ be the length of the
shortest regime, with the conventions $I_0^0:=0$ and
$I_{K^0+2}^0:=0$, which will be convenient at the sample endpoints.
When $K^0\ge1$, the relevant signal for selection is
\(
    J_{\min}
    :=\min_{1\le\ell\le K^0}
    \bigl\|\+\kappa^0_{\ell+1}-\+\kappa^0_\ell\bigr\|,
\)
the smallest change in slope at a true kink date or, equivalently, the smallest nonzero second difference of
the true coefficient path. This is the quantity that the adaptive
penalty must distinguish from the estimation error in the unrestricted
path. Because selection works with deviations from the initial
observation, the transformed error is
$\check\varepsilon_{i,t}=\varepsilon_{i,t}-\varepsilon_{i,1}$ for
$t=2,\ldots,T$, and the following moment conditions are stated in terms
of it.

\begin{assumption}[Selection: errors and regressors]\label{ass:errors}
\leavevmode
\begin{enumerate}[label=(\alph*)]
    \item We have that $\mathbb{E}(\varepsilon_{i,t}\*x_{i,s})
    =\*0_{p\times1}$ for all $i$, $t$ and $s$;
    \item We have that for any $N$ and $T$,
\(
    \sup_{2\le t\le T}\frac1N\sum_{i=1}^N\sum_{j=1}^N
    \bigl|\mathbb{E}(\check{\varepsilon}_{i,t}\*x_{i,t}'
    \*x_{j,t}\check{\varepsilon}_{j,t})\bigr|\le C
\)
and
\(
    \sup_{2\le t\le T}\frac1N\sum_{i=1}^N\sum_{j=1}^N
    \bigl|\mathbb{E}(\check{\varepsilon}_{i,t}\*x_{i,1}'
    \*x_{j,1}\check{\varepsilon}_{j,t})\bigr|\le C
\)
for some finite constant $C$;
  \item We have that for any $N$ and $T$,
    \(
        \frac1N\sum_{i=1}^N\sum_{j=1}^N
        \mathbb{E}\!\left(\*x_{i,1}'\*x_{j,1}
        \Bigl(\sum_{t=2}^T\check{\varepsilon}_{i,t}\Bigr)
        \!\Bigl(\sum_{s=2}^T\check{\varepsilon}_{j,s}\Bigr)\right)
        =O(T^2).
    \)
\end{enumerate}
\end{assumption}

Assumption~\ref{ass:errors} controls the score of the
first-period-differenced least squares criterion. Part (a)
imposes strict exogeneity, and parts (b) and (c) permit at most weak
cross-sectional dependence together with heteroskedasticity of
unrestricted form (see \citealt{KaddouraWesterlund2023};
\citealt{Kaddoura2025}). Part (c) is specific to first-period
differencing. The reference error $\varepsilon_{i,1}$ is repeated in
all $T-1$ transformed equations of unit $i$, so its cumulative score
contribution may grow with $T$. The strict
exogeneity in part (a) can be relaxed by replacing the penalised least
squares objective with a GMM-type analog as in \citet{QianSu2016}. 
The preliminary estimator of \eqref{eq:initial} must also identify
$\+\theta_1^0$ and each $\+\theta_t^0$ well enough for the adaptive
weights to be informative. For $t=2,\ldots,T$, let
$\*z_{i,t}:=[\*x_{i,1}',(\*x_{i,t}-\*x_{i,1})']'$ and
$\*G_t:=N^{-1}\sum_{i=1}^N\*z_{i,t}\*z_{i,t}'$, the joint second moment
matrix of the initial-period regressor and its deviation at date $t$.

\begin{assumption}[Selection: identification and rank]\label{ass:rank}
There exist constants $0<\underline c\le\overline c<\infty$ such that,
w.p.1,
\(
    \underline c
    \le\inf_{2\le t\le T}\mu_{\min}(\*G_t)
    \le\sup_{2\le t\le T}\mu_{\max}(\*G_t)
    \le\overline c.
\)
\end{assumption}

The lower bound in Assumption~\ref{ass:rank} ensures that $\+\theta_1$
and $\+\theta_t$ are jointly identified in every differenced
cross-section, which is what the unrestricted regression
\eqref{eq:initial} estimates, while the upper bound prevents the design
from becoming arbitrarily ill scaled. The condition concerns the
selection stage only. The tuning parameter has two tasks. It must be small enough at the true
kink dates that a genuine change in slope survives the shrinkage, and
large enough at the remaining dates that a second difference generated
only by sampling noise is set exactly to zero.

\begin{assumption}[Selection: signal strength and tuning parameter]
\label{ass:signal}
\leavevmode
\begin{enumerate}[label=(\alph*)]
    \item $\sqrt{NTK^0}\,\vartheta_1 J_{\min}^{-\zeta_1}\to_p
    c_1\in[0,\infty)$ and $\sqrt{N/K^0}\,J_{\min}\to\infty$ as
    $N,T\to\infty$;
    \item $\dfrac{N^{(\zeta_1+1)/2}\,\vartheta_1}{T^{1+\zeta_1/2}}
    \to_p\infty$ as $N,T\to\infty$.
\end{enumerate}
\end{assumption}

Assumption~\ref{ass:signal} is similar in spirit to the corresponding
condition in \citet{QianSu2016}, to which we refer for further
discussion. The first clause of part (a) and part (b) squeeze
$\vartheta_1$ from both sides, small enough that the penalty vanishes
at the true kinks and large enough that it diverges at the non-kinks.
The power $T^{1+\zeta_1/2}$ in part (b) exceeds the corresponding
requirement in \citet{QianSu2016} because a slope loads on
$d_{j,t}=t-T_{j-1}^0$ with $\sum_t d_{j,t}^2$ of order up to $T^3$,
rather than the order $T$ of the indicator loading in the flat break
case; at fixed $T$, part (b) reduces to their Assumption A.2(iii). The
second clause of part (a) keeps the adaptive weights bounded at every
true kink, and permits $J_{\min}$ to be fixed or to shrink at any rate
slower than $\sqrt{K^0/N}$. No separate detection condition is needed:
combining parts (a) and (b) yields $\sqrt{N/T}\,J_{\min}\to\infty$
whenever a tuning parameter satisfying both parts exists, so any kink
small enough to escape detection is also too small for the tuning
window to accommodate.\footnote{By part (a), w.p.a.1
$\vartheta_1\le(c_1+1)J_{\min}^{\zeta_1}(NTK^0)^{-1/2}$, which is
compatible with part (b) only if
$J_{\min}^{\zeta_1}N^{\zeta_1/2}T^{-(3+\zeta_1)/2}(K^0)^{-1/2}
\to\infty$. Multiplying by $T^{3/2}\sqrt{K^0}\ge1$ gives the claim.}
Parts (a) and (b) are compatible whenever
$(\sqrt N\,J_{\min})^{\zeta_1}\gg T^{(3+\zeta_1)/2}\sqrt{K^0}$. At
$\zeta_1=2$ with $K^0$ fixed and $J_{\min}$ bounded away from zero,
this reduces to $N/T^{5/2}\to\infty$. The constant
$\zeta_1>0$ enters through the adaptive weights and is set to
$\zeta_1=2$ in practice (see, e.g., \citealt{QianSu2016};
\citealt{Kaddoura2025}).\footnote{When $K^0=0$, the coefficient path is
globally linear and $J_{\min}$ is not needed; part (a) is replaced by
the single requirement $\vartheta_1\to_p0$.}

\subsection{Selection consistency}
\label{sec:selectionasymptotics}

The selection argument proceeds in three steps. We first control the
penalised estimator of the unrestricted coefficient path, which bounds
the error in its second differences. We then show that the adaptive
penalty sets every second difference outside the true kink set exactly
to zero. The minimum-signal condition finally guarantees that no true
kink is lost, and exact recovery follows.

\begin{theorem}[Consistency of the penalised estimator]
\label{thm:consistency}
Suppose that Assumptions~\ref{ass:errors}, \ref{ass:rank}
and~\ref{ass:signal}(a) hold. Then, as $N,T\to\infty$,
\begin{enumerate}[label=\textnormal{(\alph*)}]
\item $\widehat{\+\theta}_1-\+\theta_1^0=O_p\!\left(N^{-1/2}\right)$;
\item $\dfrac1T\displaystyle\sum_{t=1}^T
      \bigl\|\widehat{\+\theta}_t-\+\theta_t^0\bigr\|^2
      =O_p\!\left(N^{-1}\right)$;
\item $\displaystyle\max_{1\le t\le T}
      \bigl\|\widehat{\+\theta}_t-\+\theta_t^0\bigr\|
      =O_p\!\left(\sqrt{T/N}\right)$.
\end{enumerate}
\end{theorem}

Theorem~\ref{thm:consistency} establishes three rates for the
penalised estimator. Part (a) gives the $\sqrt N$ rate for the initial
level, part (b) the $N^{-1}$ rate for the path on average over $t$, and
part (c) the uniform rate $\sqrt{T/N}$. Part (c) bounds the largest
deviation along the whole path and should not be read as a pointwise
$\sqrt N$ rate at each date. The reason is the second difference
penalty: a slope shift at one date moves the path in every later
period, so the penalty loads cumulatively and the dates cannot be
separated before the kink structure is known. These are rates for the
unrestricted $Tp$-dimensional selection problem; the faster rates
available after refitting the selected model are the subject of
Section~\ref{sec:postselection}.\footnote{These rates can be improved upon.   
Under cross-sectional independence and sub-Gaussian errors, a maximal
inequality should give $O_p(\sqrt{\log T/N})$ instead, weakening the
requirement at $\zeta_1=2$ from $N/T^{5/2}\to\infty$ to
$N/(T^{3/2}\log T)\to\infty$.} The theorem says that the estimated
path is steered towards the truth, but it does not by itself establish
that the estimated kink dates coincide with the true ones, which is the
content of the next result.

\begin{theorem}[Elimination of false kinks]
\label{thm:sign}
Suppose that Assumptions~\ref{ass:errors}, \ref{ass:rank}
and~\ref{ass:signal} hold. Then, as $N,T\to\infty$,
\begin{align*}
    \mathbb{P}\!\left(\left\{
      \left\|\Delta\widehat{\+\theta}_{t+1}
        -\Delta\widehat{\+\theta}_t\right\|=0
        \;\;\forall\;t\in\mathcal{T}^{0c}_{K^0}
   \right\}\right)\to1,
\end{align*}
where
$\mathcal{T}^{0c}_{K^0}:=\{2,\ldots,T-1\}\setminus\mathcal{T}^0_{K^0}$.
\end{theorem}

Theorem~\ref{thm:sign} establishes one half of the recovery argument,
namely that $\widehat{\mathcal T}_{\widehat K}\subseteq
\mathcal T^0_{K^0}$ w.p.a.1. The other half follows from
Theorem~\ref{thm:consistency} and the triangle inequality. At a true
kink date $t\in\mathcal T^0_{K^0}$,
\(
    \bigl\|\Delta^2\widehat{\+\theta}_t\bigr\|
    \ge\bigl\|\Delta^2\+\theta_t^0\bigr\|
    -\bigl\|\Delta^2\bigl(\widehat{\+\theta}_t-\+\theta_t^0\bigr)\bigr\|
    \ge J_{\min}
    -4\max_{1\le s\le T}
    \bigl\|\widehat{\+\theta}_s-\+\theta_s^0\bigr\|.
\)
The maximum is $O_p(\sqrt{T/N})$ by Theorem~\ref{thm:consistency}(c),
and $\sqrt{N/T}\,J_{\min}\to\infty$ under Assumption~\ref{ass:signal},
so the right side is strictly positive
w.p.a.1 and every true kink date survives. Combining the two halves
gives the following corollary.

\begin{corollary}[Exact recovery of the kink structure]
\label{cor:kinks}
Suppose that the assumptions of Theorem~\ref{thm:sign} hold. Then, as
$N,T\to\infty$,
\begin{enumerate}
    \item $\mathbb{P}\!\left(\left\{\widehat K=K^0\right\}\right)\to1$;
    \item $\mathbb{P}\!\left(\left\{\widehat T_1=T_1^0,\ldots,
    \widehat T_{K^0}=T^0_{K^0}
    \;\Big|\;\widehat K=K^0\right\}\right)\to1$.
\end{enumerate}
\end{corollary}

Corollary~\ref{cor:kinks} is the oracle property of the estimator. It
recovers both the correct number of kinks and their exact locations
w.p.a.1. Consequently, for all $N$ and $T$ large enough, the post-kink
regression~\eqref{eq:postest} is run on the correct design matrix
$\+\Phi(\mathcal{T}^0_{K^0})$, and we may now study its distribution in the next subsection.

\subsection{Post-selection asymptotics}
\label{sec:postselection}

Now that exact recovery has been established, we turn to the post-kink
estimator of Section~\ref{sec:postkink}. On the event
$\{\widehat{\mathcal T}_{\widehat K}=\mathcal T^0_{K^0}\}$, whose
probability tends to one by Corollary~\ref{cor:kinks}, the feasible
estimator coincides with the oracle estimator that treats the kink
dates as known. The distribution theory can therefore be developed for the
oracle regression, which is ordinary least squares in a
$(K^0+2)p$-dimensional panel regression, and transferred to the
feasible estimator at the end. 

The estimator is computed in the cumulative coordinates
\(\+\psi_{K^0}^0
=[\+\theta_1^{0\prime},\+\kappa_1^{0\prime},\ldots,
\+\kappa_{K^0+1}^{0\prime}]'\),
but these are not the coordinates in which the rates are read. For
\(t\in\mathcal K_\ell^0\), the coefficient path satisfies
\(\+\theta_t^0
=\+\theta_{T_{\ell-1}^0}^0+(t-T_{\ell-1}^0)\+\kappa_\ell^0\),
so regime \(\ell\) is directly informative about its starting level
\(\+\theta_{T_{\ell-1}^0}^0
=\+\theta_1^0+\sum_{j=1}^{\ell-1}I_j^0\+\kappa_j^0\)
and its own slope, but not about the individual terms of the sum,
which are separated only by the early regimes themselves. The
precision of a cumulative coordinate therefore depends on the full
configuration of regime lengths. We therefore reparameterise the
model in terms of the boundary levels, which are exactly what the regimes identify.
Each is informative only through its two adjacent regimes, so each
carries its own rate, and the within-regime slopes are recovered as
scaled differences of consecutive boundary levels.

To that end, we consider the following reparameterisation. For
$j=0,\ldots,K^0+1$, define the boundary levels
\begin{align}
    \+\eta_j^0
    :=\+\theta_1^0+\sum_{m=1}^{j}I_m^0\+\kappa_m^0,
    \label{eq:knotdef}
\end{align}
where the sum is empty when $j=0$, so that $\+\eta_0^0=\+\theta_1^0$.
Each interior $\+\eta_j^0$ is the level of the coefficient path at the
$j$th kink date. Indeed, $T_0^0=1$, so the continuity
recursion~\eqref{eq:continuity} at $\ell=1$ gives
\(
    \+\theta_{T_1^0}^0
    =\+\theta_1^0+\bigl(T_1^0-T_0^0\bigr)\+\kappa_1^0
    =\+\theta_1^0+I_1^0\+\kappa_1^0
    =\+\eta_1^0,\)
and if $\+\theta_{T_{j-1}^0}^0=\+\eta_{j-1}^0$ for some $j\le K^0$,
then~\eqref{eq:continuity} at $\ell=j$ gives
\begin{align}
    \+\theta_{T_j^0}^0
    =\+\theta_{T_{j-1}^0}^0+I_j^0\+\kappa_j^0
    =\+\eta_{j-1}^0+I_j^0\+\kappa_j^0
    =\+\eta_j^0,
    \label{eq:etainduction}
\end{align}
so that $\+\eta_j^0=\+\theta_{T_j^0}^0$ for $j=1,\ldots,K^0$ by
induction. The final coordinate is the endpoint obtained by extending
the last linear segment one period beyond the sample. To see this, note
that $T\in\mathcal K_{K^0+1}^0$, so~\eqref{eq:Kink} and
$I_{K^0+1}^0=T_{K^0+1}^0-T_{K^0}^0=T+1-T_{K^0}^0$ give
\begin{align}
    \+\eta_{K^0+1}^0-\+\theta_T^0
    &=\bigl[\+\eta_{K^0}^0+I_{K^0+1}^0\+\kappa_{K^0+1}^0\bigr]
    -\bigl[\+\eta_{K^0}^0
    +\bigl(T-T_{K^0}^0\bigr)\+\kappa_{K^0+1}^0\bigr]
    \notag\\
    &=\+\kappa_{K^0+1}^0.
    \label{eq:etaendpoint}
\end{align}
It is not a new parameter, since $T_{K^0}^0<T$ implies
$I_{K^0+1}^0\ge2$, so the last slope, and with it the endpoint, is
identified from the observed part of the final regime. Including it
makes every regime an interval between two boundary levels.

Stacking $\+\eta^0
:=[\+\eta_0^{0\prime},\ldots,\+\eta_{K^0+1}^{0\prime}]'$,
equation~\eqref{eq:knotdef} reads $\+\eta^0=\*M_{K^0}\+\psi_{K^0}^0$
with
\begin{align}
    \*M_{K^0}
    :=\begin{bmatrix}
    \*I_p & \*0 & \*0 & \cdots & \*0\\
    \*I_p & I_1^0\*I_p & \*0 & \cdots & \*0\\
    \*I_p & I_1^0\*I_p & I_2^0\*I_p & \cdots & \*0\\
    \vdots & \vdots & \vdots & \ddots & \vdots\\
    \*I_p & I_1^0\*I_p & I_2^0\*I_p & \cdots & I_{K^0+1}^0\*I_p
    \end{bmatrix},
    \label{eq:Mmatrix}
\end{align}
which is block lower triangular with diagonal blocks
$\*I_p,I_1^0\*I_p,\ldots,I_{K^0+1}^0\*I_p$ and hence nonsingular for
every $N$ and $T$, every regime length being positive. Its inverse is
the successive difference operator. Subtracting consecutive terms
in~\eqref{eq:knotdef},
\begin{align}
    \+\eta_j^0-\+\eta_{j-1}^0
    =\Bigl(\+\theta_1^0+\sum_{m=1}^{j}I_m^0\+\kappa_m^0\Bigr)
    -\Bigl(\+\theta_1^0+\sum_{m=1}^{j-1}I_m^0\+\kappa_m^0\Bigr)
    =I_j^0\+\kappa_j^0,
    \label{eq:etadiff}
\end{align}
so that $\+\theta_1^0=\+\eta_0^0$ and
\begin{align}
    \+\kappa_j^0=\dfrac{\+\eta_j^0-\+\eta_{j-1}^0}{I_j^0}  =  \dfrac{
        \+\theta_{T_j^0}^0-\+\theta_{T_{j-1}^0}^0
    }{
        T_j^0-T_{j-1}^0
    }
    \label{eq:slopefromknots}
\end{align}
for $j=1,\ldots,K^0+1$, or,  In matrix
form, we have $\+\psi_{K^0}^0=\*M_{K^0}^{-1}\+\eta^0$, where
\begin{align}
    \*M_{K^0}^{-1}
    =\begin{bmatrix}
    \*I_p & \*0 & \*0 & \cdots & \*0\\
    -(I_1^0)^{-1}\*I_p & (I_1^0)^{-1}\*I_p & \*0 & \cdots & \*0\\
    \*0 & -(I_2^0)^{-1}\*I_p & (I_2^0)^{-1}\*I_p & \cdots & \*0\\
    \vdots & \vdots & \ddots & \ddots & \vdots\\
    \*0 & \*0 & \cdots & -(I_{K^0+1}^0)^{-1}\*I_p
    & (I_{K^0+1}^0)^{-1}\*I_p
    \end{bmatrix}.
    \label{eq:Minverse}
\end{align}
Equation~\eqref{eq:slopefromknots} is the map by which the slopes are
reported once the boundary levels have been estimated, and it is exact
and linear, so no approximation is involved at any point. The
boundary levels are therefore an exact nonsingular reparameterisation
of the post-kink model, changing the coordinates of the theory but
neither the estimator, its fitted values, nor the column space of the
design. It shows that the rate of \(\widetilde{\+\kappa}_j\) depends locally on the rates of the two boundary-level estimators enclosing regime \(j\), rather than on the full span of subsequent regimes.

The representation of the path in the new coordinates is an
interpolation. Fix a regime $\ell$ and a date
$t\in\mathcal K_\ell^0$, and let
$u_{\ell,t}:=(t-T_{\ell-1}^0)/I_\ell^0$ denote the fraction of regime
$\ell$ elapsed by date $t$, so that $u_{\ell,t}\in[0,1)$ at observed
dates. Substituting $\+\theta_{T_{\ell-1}^0}^0=\+\eta_{\ell-1}^0$
and~\eqref{eq:slopefromknots} into~\eqref{eq:Kink} gives
\begin{align}
    \+\theta_t^0
    =(1-u_{\ell,t})\+\eta_{\ell-1}^0+u_{\ell,t}\+\eta_\ell^0,
    \label{eq:knotrep}
\end{align}
so that every date is a convex combination of the two boundary levels
enclosing its regime. The identity is exact, being nothing but the
model's continuity and piecewise linearity, and it makes the source of
the convergence rates transparent, where the information about $\+\theta_t^0$
is the information about the two boundary levels adjacent to date $t$.

The implied loading of date $t$ on $\+\eta_j^0$ is
\begin{align}
    h_{j,t}
    :=
    \begin{cases}
    \displaystyle
    u_{j,t}
    =
    \frac{t-T_{j-1}^{0}}
         {T_{j}^{0}-T_{j-1}^{0}},
    & t\in\mathcal{K}_{j}^{0},
    \\[8pt]
    \displaystyle
    1-u_{j+1,t}
    =
    \frac{T_{j+1}^{0}-t}
         {T_{j+1}^{0}-T_{j}^{0}},
    & t\in\mathcal{K}_{j+1}^{0},
    \\[8pt]
    0,
    & \text{otherwise},
    \end{cases}
    \label{eq:hdef}
\end{align}
for $j=0,\ldots,K^0+1$, with the conventions
$\mathcal K_0^0=\mathcal K_{K^0+2}^0:=\varnothing$. The loading is
supported only on the two regimes adjacent to boundary $j$. It rises
linearly from zero over regime $j$, falls back linearly over regime
$j+1$, and vanishes elsewhere. At any date exactly two loadings are
nonzero and they sum to one, so \eqref{eq:knotrep} reads
$\+\theta_t^0=\sum_{j=0}^{K^0+1}h_{j,t}\+\eta_j^0$.

\begin{remark}[Linear B-spline representation]\label{rem:bspline}
A useful and perhaps surprising implication of this representation is that
the coefficient path admits an exact B-spline form. In particular, the path
$t\mapsto\+\theta_t^0$ is a spline of degree one with knots at the kink
dates, and $h_{0,t},\ldots,h_{K^0+1,t}$ are precisely its first-degree
B-spline basis functions, restricted to the observed dates
\citep[]{deBoor2001}. The associated B-spline coefficients are the
boundary levels and, because first-degree B-splines interpolate at the
knots, they coincide with the values of the coefficient path at those dates.
Unlike in nonparametric spline regression, however, the spline is not an
approximation as it is the exact data-generating coefficient path with finitely
many unknown knots and therefore introduces no approximation bias.
\end{remark}

Collecting the loadings into the regressor
$\+\phi_{i,t}^\eta:=[h_{0,t}\*x_{i,t}',\ldots,h_{K^0+1,t}\*x_{i,t}']'$,
 DGP~\eqref{eq:DGP} becomes
$y_{i,t}=\mu_i+\+\phi_{i,t}^{\eta\prime}\+\eta^0+\varepsilon_{i,t}$, a
linear panel regression in which every observation loads on at most two
parameter blocks. As noted earlier, this representation is equivalent to a varying-coefficient regression in which the time variation in the coefficient on $\*x_{i,t}$ is captured by a B-spline basis \citep{HastieTibshirani1993,HuangWuZhou2002}. This connection is noteworthy in the sense that although the estimator is derived from a structural kink model rather than from a conventional spline specification, the post-kink regression admits an exact first-degree B-spline representation.
Now, let $\+\Phi^\eta(\mathcal T^0_{K^0})$ stack the rows
$\+\phi_{i,t}^{\eta\prime}$ conformably with~\eqref{eq:pk_expanded}.
The two designs are related through the partial sums of the
loadings. Fix $t\in\mathcal K_\ell^0$. By~\eqref{eq:hdef}, the only potentially 
nonzero loadings at date $t$ are $h_{\ell-1,t}=1-u_{\ell,t}$ and
$h_{\ell,t}=u_{\ell,t}$, so, for $m=0,\ldots,K^0+1$,
\begin{align}
    \sum_{j=m}^{K^0+1}h_{j,t}
    =\mathbf 1\{m<\ell\}+u_{\ell,t}\,\mathbf 1\{m=\ell\},
    \label{eq:partialsums}
\end{align}
and multiplying by $I_m^0$ and using
$I_\ell^0u_{\ell,t}=t-T_{\ell-1}^0$,
\begin{align}
    I_m^0\sum_{j=m}^{K^0+1}h_{j,t}
    &=I_m^0\,\mathbf 1\{m<\ell\}
    +\bigl(t-T_{\ell-1}^0\bigr)\mathbf 1\{m=\ell\}
    \notag \\
    &=d_{m,t}\bigl(\mathcal T^0_{K^0}\bigr),
    \label{eq:partialsumd}
\end{align}
where the second equality follows from~\eqref{eq:djt} for
$m=1,\ldots,K^0+1$ and from the convention of setting
$d_{0,t}(\mathcal T_{K^0}^0):=0$ for $m=0$. Now consider the product
$\+\Phi^\eta(\mathcal T^0_{K^0})\*M_{K^0}$. By~\eqref{eq:Mmatrix},
block column $0$ of $\*M_{K^0}$ carries $\*I_p$ in every block row,
and block column $m\ge1$ carries $I_m^0\*I_p$ in block rows
$j\ge m$ and $\*0_{p\times p}$ above, so the row belonging to
observation $(i,t)$ has $m$th block entry
\begin{align}
    \bigl[\+\phi_{i,t}^{\eta\prime}\*M_{K^0}\bigr]_m
    &=I_m^0\,\Bigl(\sum_{j=m}^{K^0+1}h_{j,t}\Bigr)\*x_{i,t}'\,
    \mathbf 1\{m\ge1\}
    +\Bigl(\sum_{j=0}^{K^0+1}h_{j,t}\Bigr)\*x_{i,t}'\,
    \mathbf 1\{m=0\}
    \notag \\
    &=d_{m,t}\bigl(\mathcal T^0_{K^0}\bigr)\*x_{i,t}'\,
    \mathbf 1\{m\ge1\}
    +\*x_{i,t}'\,\mathbf 1\{m=0\}
    \notag \\
    &=\bigl[\+\phi_{i,t}
    \bigl(\mathcal T^0_{K^0}\bigr)'\bigr]_m,
    \label{eq:rowident}
\end{align}
where the second equality holds by~\eqref{eq:partialsumd} and by the case $m=0$
of~\eqref{eq:partialsums}, which is the partition of unity
$\sum_{j=0}^{K^0+1}h_{j,t}=1$, and the third by the definition of
$\+\phi_{i,t}(\mathcal T^0_{K^0})$. Since~\eqref{eq:rowident} holds
at every observation, stacking the rows gives
\begin{align}
    \+\Phi(\mathcal T^0_{K^0})
    =\+\Phi^\eta(\mathcal T^0_{K^0})\,\*M_{K^0},
    \label{eq:designequiv}
\end{align}
so the two designs share the same column space and identical fitted
values, and the OLS estimators satisfy
$\widetilde{\+\eta}=\*M_{K^0}\widetilde{\+\psi}$ exactly in finite
samples.\footnote{The identity~\eqref{eq:designequiv} is what permits the
distribution theory to be developed on
$\+\Phi^\eta(\mathcal T^0_{K^0})$ while the estimator is computed
in the $\+\psi$ coordinates of~\eqref{eq:postest}. The two designs
have identical fitted values and residuals, so
$\widetilde{\+\eta}=\*M_{K^0}\widetilde{\+\psi}$ holds exactly in
finite samples, covariance matrices transform by the fixed
nonsingular map $\*M_{K^0}$.}

\begin{remark}[Identification as a degree-one B-spline basis]
To identify the basis precisely, let
\(\+\tau:=(T_0^0,T_0^0,T_1^0,\ldots,T_{K^0}^0,
T_{K^0+1}^0,T_{K^0+1}^0)\)
be the open knot vector.  Starting from the degree-zero interval indicators
\(B_{j,0}(t):=\mathbf{1}\{\tau_j\leq t<\tau_{j+1}\}\), the Cox--de Boor
recursion gives
\(\displaystyle
B_{j,1}(t)
=
\frac{t-\tau_j}{\tau_{j+1}-\tau_j}B_{j,0}(t)
+
\frac{\tau_{j+2}-t}{\tau_{j+2}-\tau_{j+1}}B_{j+1,0}(t)
\),
where a term with zero denominator is defined as zero. On the interval
to the left of \(T_j^0\), this recursion yields \(u_{j,t}\), while on
the interval to its right it yields \(1-u_{j+1,t}\). Hence
\(h_{j,t}=B_{j,1}(t)\) for \(j=0,\ldots,K^0+1\).
\end{remark}

Different boundary levels are estimated at different rates, and the
right scale for $\+\eta_j^0$ is the norm of its own loading sequence,
\(
    r_j^2:=\sum_{t=1}^{T}h_{j,t}^2
\)
for $j=0,\ldots,K^0+1$. Dividing the $j$th block of the design by
$r_j$ produces a time loading of unit length, so $r_j$ removes the
deterministic magnitude of the basis before the regressors and the
within transformation are brought in. The norms are available in
closed form. By~\eqref{eq:hdef}, $h_{j,t}$ is nonzero only on
$\mathcal K_j^0\cup\mathcal K_{j+1}^0$, rising linearly over regime
$j$ and falling linearly over regime $j+1$, so the sum splits into
two ramp contributions,
\(
    r_j^2
    =\sum_{t\in\mathcal K_j^0}u_{j,t}^2
    +\sum_{t\in\mathcal K_{j+1}^0}(1-u_{j+1,t})^2.
\)
Substituting $q=t-T_{j-1}^0$ in the first sum and $q=t-T_j^0$ in the
second, so that $u_{j,t}=q/I_j^0$ and $u_{j+1,t}=q/I_{j+1}^0$ with
$q$ running over the regime, gives $r_j^2=A(I_j^0)+B(I_{j+1}^0)$,
where for an integer $L\ge1$,
\(
    A(L):=\sum_{q=0}^{L-1}(q/L)^2=(L-1)(2L-1)/(6L)
\)
and
\(
    B(L):=\sum_{q=0}^{L-1}(1-q/L)^2=(L+1)(2L+1)/(6L),
\)
with $A(0):=B(0):=0$ at the endpoints, so that $r_0^2=B(I_1^0)$ and
$r_{K^0+1}^2=A(I_{K^0+1}^0)$. A ramp over a regime of length $L$
thus contributes $L/3$ per regime up to a bounded term, and
$r_j^2\asymp I_j^0+I_{j+1}^0$, so $r_j^2$ counts the periods in the
two regimes adjacent to boundary $j$, which are the only ones
informative about $\+\eta_j^0$. The following lemma records the
exact bounds.

\begin{lemma}[The rate of convergence]\label{lem:basis}
Let $\*H^0:=[h_{j,t}]_{1\le t\le T,\,0\le j\le K^0+1}$ and
$\*D^0:=\mathrm{diag}(r_0,\ldots,r_{K^0+1})$. Then, uniformly over all
admissible kink configurations,
\(
    \frac18\bigl(I_j^0+I_{j+1}^0\bigr)
    \le r_j^2
    \le I_j^0+I_{j+1}^0
\)
for $j=0,\ldots,K^0+1$, and
\(
    \frac12\*I_{K^0+2}
    \preceq(\*D^0)^{-1}\*H^{0\prime}\*H^0(\*D^0)^{-1}
    \preceq\frac32\*I_{K^0+2}.
\)
\end{lemma}
The normalised column by column is then  well
conditioned, whatever the number of kinks and however the sample is
divided across regimes. Non-adjacent loadings have disjoint support,
while adjacent ones overlap on a single regime, and the overlap
carries less than half of the mass of either column. This is the
first-degree case of the stability property of B-splines, whose
condition number is bounded independently of the knot sequence
\citep{deBoor1972,SchererShadrin1999}. In particular, since the
 bounds in Lemma \ref{lem:basis} imply
\(
    \frac23 r_j^{-2}
    \le
    \bigl[(\*H^{0\prime}\*H^0)^{-1}\bigr]_{jj}
    \le
    2 r_j^{-2},
\)
the diagonal of the inverse  satisfies
\([(\*H^{0\prime}\*H^0)^{-1}]_{jj}\asymp r_j^{-2}\),
which is precisely what gives us the rate of
\(\widetilde{\+\eta}_j\) below. Further, collect the scales in
$\mathbb D_{K^0+1}:=\mathrm{diag}(r_0,\ldots,r_{K^0+1})\otimes\*I_p$
and let $\*C$ be the centering matrix.
Evaluating the design at the true kink dates, define the normalised  matrix
\(
    \widehat{\*Q}_N
    :=\mathbb D_{K^0+1}^{-1}\,\frac1N\,
    \+\Phi^\eta(\mathcal T^0_{K^0})'\*C\,
    \+\Phi^\eta(\mathcal T^0_{K^0})\,\mathbb D_{K^0+1}^{-1}
\)
and 
\(
    \+\Psi_{NT}
    :=\frac1{\sqrt N}\+\Phi^\eta(\mathcal T^0_{K^0})'\*C\+\varepsilon.
\)
 They are
linked to the estimator by the exact factorisation
\(
    \+\Phi^\eta(\mathcal T^0_{K^0})'\*C\,
    \+\Phi^\eta(\mathcal T^0_{K^0})
    =N\,\mathbb D_{K^0+1}\widehat{\*Q}_N\mathbb D_{K^0+1}.
\)

\begin{assumption}[Post-selection: design and rank]\label{ass:postrank}
We have $\|\widehat{\*Q}_N-\*Q_0\|_{op}=o_p(1)$, where $\*Q_0$ is
finite and positive definite with
$0<c\le\mu_{\min}(\*Q_0)\le\mu_{\max}(\*Q_0)\le C<\infty$ for some
constants $c$ and $C$.
\end{assumption}

The lower eigenvalue bound excludes the possibility that the centering
or the regressors asymptotically eliminate a normalised direction of
the design, and the upper bound prevents any normalised direction from
accumulating information at an order larger than its scale. By
Lemma~\ref{lem:basis}, the normalised time basis is uniformly well
conditioned, so positive definiteness of $\*Q_0$ is a restriction on
the regressors and not a restriction on the kink configuration.

\begin{assumption}[Post-selection: score and central limit theorem]
\label{ass:postclt}
Let $\{\*H_{NT}\}$ be any  sequence of
$l\times p(K^0+2)$ selection matrices with $l$ fixed and
$\limsup_{N,T\to\infty}\|\*H_{NT}\|_{op}<\infty$.
\begin{enumerate}[label=(\alph*)]
    \item The limit
    \(
        \+\Sigma_0
        :=\lim\mathrm{Var}\!\left(
        \mathbb D_{K^0+1}^{-1}\+\Psi_{NT}\right)
    \)
    exists and is finite and positive definite;
    \item whenever
    \(
        \+\Upsilon
        :=\lim\*H_{NT}\*Q_0^{-1}\+\Sigma_0\*Q_0^{-1}\*H_{NT}'
    \)
    exists and is positive definite, we have
    \(
        \*H_{NT}\*Q_0^{-1}\mathbb D_{K^0+1}^{-1}\+\Psi_{NT}
        \to_d\mathcal N\!\left(\*0_{l\times1},\+\Upsilon\right).
    \)
\end{enumerate}
\end{assumption}

Assumption~\ref{ass:postclt} is a high-level central limit condition
for fixed-dimensional contrasts of the normalised score, in the spirit
of \citet{QianSu2016}. It holds under arbitrary within-unit serial
correlation and heteroskedasticity with bounded spectrum, and weak
cross-sectional dependence can be maintained directly. The matrices
$\*H_{NT}$ allow the same result to deliver the individual boundary
levels, the slopes, and the path values below, and $\+\Upsilon$ is the
usual sandwich variance of the corresponding OLS contrast. We stress
again that Assumptions~\ref{ass:postrank} and~\ref{ass:postclt}
supplement, and do not replace,
Assumptions~\ref{ass:errors}--\ref{ass:signal}.

Let $\widetilde{\+\eta}^{\,o}$ denote the OLS estimator of $\+\eta^0$
from the centered boundary-level regression evaluated at the true kink
dates, so that $\widetilde{\+\eta}^{\,o}=\*M_{K^0}\widetilde{\+\psi}^{\,o}$
with $\widetilde{\+\psi}^{\,o}$ the oracle version
of~\eqref{eq:postest}. Since $\*C$ is symmetric and idempotent,
\begin{align}
    \widetilde{\+\eta}^{\,o}-\+\eta^0
    =\bigl[\+\Phi^\eta(\mathcal T^0_{K^0})'\*C
    \+\Phi^\eta(\mathcal T^0_{K^0})\bigr]^{-1}
    \+\Phi^\eta(\mathcal T^0_{K^0})'\*C\+\varepsilon,
    \label{eq:oracleerr}
\end{align}
and inserting the factorisation~\eqref{eq:gramfact} together with
$\+\Phi^\eta(\mathcal T^0_{K^0})'\*C\+\varepsilon=\sqrt N\+\Psi_{NT}$
yields the exact identity
\(
    \sqrt N\,\mathbb D_{K^0+1}
    \bigl(\widetilde{\+\eta}^{\,o}-\+\eta^0\bigr)
    =\widehat{\*Q}_N^{-1}\mathbb D_{K^0+1}^{-1}\+\Psi_{NT}.
\)
Because $\widehat{\*Q}_N^{-1}=\*Q_0^{-1}+o_p(1)$ in operator norm under
Assumption~\ref{ass:postrank}, combining the identity  with
Assumption~\ref{ass:postclt} and Slutsky's theorem gives the oracle
limit.

\begin{theorem}[Oracle post-kink asymptotic normality]
\label{thm:oracle}
Suppose that $K^0$ is fixed and that Assumptions~\ref{ass:postrank} and~\ref{ass:postclt} hold.
Then, as $N,T\to\infty$,
\begin{align*}
    \sqrt N\,\*H_{NT}\,\mathbb D_{K^0+1}
    \bigl(\widetilde{\+\eta}^{\,o}-\+\eta^0\bigr)
    \to_d
    \mathcal N\!\left(\*0_{l\times1},\+\Upsilon\right).
\end{align*}
\end{theorem}

The feasible estimator uses $\widehat{\mathcal T}_{\widehat K}$. Let
$\widetilde{\+\eta}_{\widehat K}
:=\*M_{\widehat K}\widetilde{\+\psi}_{\widehat K}$ be its boundary-level
representation. On the event
$\{\widehat{\mathcal T}_{\widehat K}=\mathcal T^0_{K^0}\}$ the feasible
and oracle regressions share the design, the response, and the OLS
formula, so $\widetilde{\+\eta}_{\widehat K}=\widetilde{\+\eta}^{\,o}$
exactly, and Corollary~\ref{cor:kinks} shows that the probability of
this event tends to one. The feasible estimator therefore inherits the
oracle limit.

\begin{theorem}[Feasible post-kink asymptotic normality]
\label{thm:normality}
Suppose that $K^0$ is fixed and that the assumptions of Corollary~\ref{cor:kinks} and
Assumptions~\ref{ass:postrank}--\ref{ass:postclt} hold. Then, as
$N,T\to\infty$,
\begin{enumerate}[label=(\roman*)]
\item $    \sqrt N\,\*H_{NT}\,\mathbb D_{K^0+1}
    \bigl(\widetilde{\+\eta}_{\widehat K}-\+\eta^0\bigr)
    \to_d
    \mathcal N\!\left(\*0_{l\times1},\+\Upsilon\right);$
    \item  $\widetilde{\+\eta}_j-\+\eta_j^0 
=O_p\!\left(\left[N\left(I_j^0+I_{j+1}^0\right)\right]^{-1/2}\right)$ for each for each $j=0,\hdots, K^0+1$.
\end{enumerate}
\end{theorem}

The selection assumptions enter Theorem~\ref{thm:normality} only
through exact recovery. Conditional on the correct kink set, the limit
is the ordinary OLS distribution of the centered post-kink regression.
The theorem contains the rates of every object of interest. Let
$\*P_j:=\*e_{j+1}'\otimes\*I_p$ select the $j$th boundary block, so
that $\*P_j\mathbb D_{K^0+1}=r_j\*P_j$. Choosing $\*H_{NT}=\*P_j$,
whose limit variance is the $(j+1)$th diagonal $p\times p$ block of
$\*Q_0^{-1}\+\Sigma_0\*Q_0^{-1}$ and hence positive definite, gives a
nondegenerate limit for
$\sqrt N\,r_j(\widetilde{\+\eta}_j-\+\eta_j^0)$, and
Lemma~\ref{lem:basis} gives us the following rate of convergence
\begin{align}
    \widetilde{\+\eta}_j-\+\eta_j^0
    =O_p\!\left(\left[N\left(I_j^0+I_{j+1}^0\right)\right]^{-1/2}\right)
    \label{eq:etarate}
\end{align}
for $j=0,\ldots,K^0+1$. This says that a boundary level is estimated at the rate of
the combined length of the two regimes that inform it.

Consider next the within-regime slopes. By~\eqref{eq:slopefromknots},
the estimation error of a slope is the successive difference of two
boundary-level errors divided by the regime length, and this identity
already contains the rate. For the limiting distribution, define the
slope scale
\begin{align}
    s_j:=I_j^0\left(r_{j-1}^{-2}+r_j^{-2}\right)^{-1/2}
    \label{eq:sj}
\end{align}
and the contrast
$\*H_{j,NT}^{\kappa}
:=s_j(I_j^0)^{-1}\bigl(r_j^{-1}\*P_j-r_{j-1}^{-1}\*P_{j-1}\bigr)$,
which produces $\sqrt N\,s_j(\widetilde{\+\kappa}_j-\+\kappa_j^0)$
exactly when applied to the left side of Theorem~\ref{thm:normality}
and satisfies $\*H_{j,NT}^{\kappa}\*H_{j,NT}^{\kappa\prime}=\*I_p$
because the two selectors act on disjoint blocks. Let
\(
    \+\Upsilon_j^{\kappa}
    :=\lim\*H_{j,NT}^{\kappa}\*Q_0^{-1}\+\Sigma_0\*Q_0^{-1}
    \*H_{j,NT}^{\kappa\prime},
\)
whenever the limit exists.

\begin{corollary}[Distribution and convergence rates of the slopes]
\label{cor:kappa}
Under the assumptions of Theorem~\ref{thm:normality}, for each
$j=1,\ldots,K^0+1$, suppose that $\+\Upsilon_j^{\kappa}$ exists and
is positive definite. Then, as $N,T\to\infty$,
\begin{enumerate}[label=(\roman*)]
    \item
    \(
    \sqrt N\,s_j
    \bigl(\widetilde{\+\kappa}_j-\+\kappa_j^0\bigr)
    \to_d
    \mathcal N\!\left(\*0_{p\times1},\+\Upsilon_j^{\kappa}\right);
    \)
    \item
    \(
    \widetilde{\+\kappa}_j-\+\kappa_j^0
    =O_p\!\left(
    \left[N\left(I_j^0\right)^2
    \left\{\left(I_{j-1}^0+I_j^0\right)\wedge\left(I_j^0+I_{j+1}^0\right)\right\}
    \right]^{-1/2}
    \right).
    \)
\end{enumerate}
\end{corollary}

The order of $s_j^2$ follows from Lemma~\ref{lem:basis} because
$s_j^2/(I_j^0)^2$ is the harmonic mean of $r_{j-1}^2$ and $r_j^2$,
which lies within a factor of two of their minimum.\footnote{To see this, let \(H(a,b):=2ab/(a+b)\) denote the harmonic mean of two positive scalars \(a\) and \(b\). With \(a=r_{j-1}^2\) and \(b=r_j^2\), the definition of \(s_j\) in~\eqref{eq:sj} gives \(s_j^2/(I_j^0)^2 = (r_{j-1}^{-2}+r_j^{-2})^{-1} = r_{j-1}^2 r_j^2/(r_{j-1}^2+r_j^2) = \frac12 H(r_{j-1}^2,r_j^2)\). The  inequality \(\frac12(a\wedge b)\le ab/(a+b)\le a\wedge b\) bounds this quantity between half the minimum and the minimum itself, so \(s_j^2/(I_j^0)^2\) and \(r_{j-1}^2\wedge r_j^2\) differ by at most a factor of two.} The rate has a simple reading. By~\eqref{eq:slopefromknots}, a slope
is the difference of the two boundary levels enclosing its regime,
divided by the regime length $I_j^0$, so its precision is
$(I_j^0)^{-1}$ times the precision of the worse of its two
endpoints. The first boundary level has
no regime to its left, so it is identified from regime~$1$ alone at
rate $(NI_1^0)^{-1/2}$, and dividing by $I_1^0$ gives
$[N(I_1^0)^3]^{-1/2}$ for the first slope and  the same argument applied at the right end gives
$[N(I_{K^0+1}^0)^3]^{-1/2}$ for the last. These are exactly the rates of a trend regression run on the
regime alone. An interior slope inherits the precision of its two endpoints. Each
boundary level enclosing regime $j$ converges at the
rate $[N(I_j^0+I_{j+1}^0)]^{-1/2}$ of its own pair of adjacent regimes, the
slope is their difference divided by $I_j^0$, and so it converges at
the worse of the two endpoint rates divided by $I_j^0$, which is
Corollary~\ref{cor:kappa}.

\begin{remark}[Equivalent representations of the slope rates]\label{rem:rateinterpretation}  Since \(\kappa_j^0=(\eta_j^0-\eta_{j-1}^0)/I_j^0\), and since \(r_{j-1}^2\asymp I_{j-1}^0+I_j^0\) and \(r_j^2\asymp I_j^0+I_{j+1}^0\), the exact slope scale satisfies \(s_j^2=(I_j^0)^2(r_{j-1}^{-2}+r_j^{-2})^{-1}\asymp (I_j^0)^2\frac{(I_{j-1}^0+I_j^0)(I_j^0+I_{j+1}^0)}{I_{j-1}^0+2I_j^0+I_{j+1}^0}\). Moreover, \((I_{j-1}^0+I_j^0)\wedge(I_j^0+I_{j+1}^0)=I_j^0+(I_{j-1}^0\wedge I_{j+1}^0)\), so an interior slope has rate \(O_p([N(I_j^0)^2\{I_j^0+(I_{j-1}^0\wedge I_{j+1}^0)\}]^{-1/2})\). At the endpoints, the conventions \(I_0^0=I_{K^0+2}^0=0\) imply \((I_0^0+I_1^0)\wedge(I_1^0+I_2^0)=I_1^0\) and \((I_{K^0}^0+I_{K^0+1}^0)\wedge(I_{K^0+1}^0+I_{K^0+2}^0)=I_{K^0+1}^0\), yielding the usual rates \(O_p([N(I_1^0)^3]^{-1/2})\) and \(O_p([N(I_{K^0+1}^0)^3]^{-1/2})\) for the first and last slopes, respectively. \end{remark}

\begin{remark}[Balanced regimes]\label{rem:balanced}
If $I_\ell^0\asymp T$ for all $\ell$, then
$r_j^2\asymp T$ and $s_j^2\asymp T^3$ for every $j$, and the slope rate
reduces to $(NT^3)^{-1/2}$, that is $T^{-3/2}$ in the time dimension.
The rate in Corollary~\ref{cor:kappa} makes precise which
departures from balance are costly: a short first or last regime is
estimated at the cube of its own length, while a short interior regime
borrows strength from its neighbours through the shared boundary
levels.
\end{remark}

\begin{remark}[Connection to classical trend regression]
\label{rem:trend}
The rate in Corollary~\ref{cor:kappa} has a natural connection to OLS
estimation of a slope in a deterministic trend regression. In the
special case $K^0=0$, there is a single regime with $I_1^0=T$, so
$s_1^2\asymp T^3$ and the rate reduces to $O_p((NT^3)^{-1/2})$. This is
the panel analog of the classical $T^{-3/2}$ rate for OLS in
$y_t=\alpha+\beta t+\varepsilon_t$, where
$\mathrm{Var}(\hat\beta)\propto
\bigl(\sum_{t=1}^T(t-\overline t)^2\bigr)^{-1}\asymp T^{-3}$ with
$\overline t:=(T+1)/2$; see \citet[][Ch.~16]{Hamilton1994}.
\end{remark}

We finally consider the coefficient path. Fix a regime index $\ell$ and
a date $t\in\mathcal K_\ell^0$. By~\eqref{eq:knotrep}, the path error
is the same convex combination of the two adjacent boundary-level
errors, and the natural scale is
\(
    w_t
    :=\bigl[(1-u_{\ell,t})^2r_{\ell-1}^{-2}
    +u_{\ell,t}^2r_\ell^{-2}\bigr]^{-1/2},
\)
with contrast
$\*H_{t,NT}^{\theta}
:=w_t\bigl[(1-u_{\ell,t})r_{\ell-1}^{-1}\*P_{\ell-1}
+u_{\ell,t}r_\ell^{-1}\*P_\ell\bigr]$, again satisfying
$\*H_{t,NT}^{\theta}\*H_{t,NT}^{\theta\prime}=\*I_p$. Let
\(
    \+\Upsilon_t^{\theta}
    :=\lim\*H_{t,NT}^{\theta}\*Q_0^{-1}\+\Sigma_0\*Q_0^{-1}
    \*H_{t,NT}^{\theta\prime},
\)
whenever the limit exists.

\begin{corollary}[Pointwise distribution of the coefficient path]
\label{cor:thetat}
Suppose that the assumptions of Theorem~\ref{thm:normality} hold and
fix a regime index $\ell\in\{1,\ldots,K^0+1\}$. Then, as
$N,T\to\infty$,
\begin{enumerate}[label=(\roman*)]
    \item for any deterministic sequence of dates
    $t\in\mathcal K_\ell^0$ such that $\+\Upsilon_t^{\theta}$ exists
    and is positive definite,
    \(
    \sqrt N\,w_t
    \bigl(\widetilde{\+\theta}_t-\+\theta_t^0\bigr)
    \to_d
    \mathcal N\!\left(\*0_{p\times1},\+\Upsilon_t^{\theta}\right);
    \)
    \item uniformly over $t\in\mathcal K_\ell^0$,
    \(
    \widetilde{\+\theta}_t-\+\theta_t^0
    =O_p\!\left(
    \left[N\left\{\left(I_{\ell-1}^0+I_\ell^0\right)
    \wedge\left(I_\ell^0+I_{\ell+1}^0\right)\right\}\right]^{-1/2}
    \right).
    \)
\end{enumerate}
\end{corollary}

The second claim follows from the convex-combination representation
\(
\widetilde{\+\theta}_t-\+\theta_t^0
=(1-u_{\ell,t})(\widetilde{\+\eta}_{\ell-1}-\+\eta_{\ell-1}^0)
+u_{\ell,t}(\widetilde{\+\eta}_{\ell}-\+\eta_{\ell}^0),
\)
which implies, uniformly over \(t\in\mathcal K_\ell^0\),
\(
\|\widetilde{\+\theta}_t-\+\theta_t^0\|
\le
\max\{\|\widetilde{\+\eta}_{\ell-1}-\+\eta_{\ell-1}^0\|,
\|\widetilde{\+\eta}_{\ell}-\+\eta_{\ell}^0\|\}.
\)
Moreover, \((1-u_{\ell,t})^2+u_{\ell,t}^2\le1\) implies
\(w_t^2\ge r_{\ell-1}^2\wedge r_\ell^2\), showing that the pointwise rate is never slower than the stated uniform rate. The weight $w_t$ traces how precision moves
along the path. At the kink dates the rate is that of the corresponding
boundary level, and in between the two adjacent loadings share the
load. Precision at a date is therefore a local property, governed by
the lengths of the regimes adjacent to the nearby boundaries, and not
by the amount of sample preceding or following the date.

\begin{remark}[Relation to spline regression asymptotics]
\label{rem:splinerates}
The local character of the rates is familiar from least squares spline
regression, where the pointwise variance of the fit is governed by the
sample size times the local knot spacing, so that precision at a point
depends on the mesh near that point alone
\citep{AgarwalStudden1980,ZhouShenWolfe1998,Huang2003}.  
\end{remark}

\begin{remark}[Initial level]\label{rem:initlevel}
The initial level is the case $t=1$, $\ell=1$, and $u_{1,1}=0$ of
Corollary~\ref{cor:thetat}, in which $\*H_{1,NT}^{\theta}=\*P_0$ and
$w_1=r_0$ with $r_0^2\asymp I_1^0$. Hence
$\widetilde{\+\theta}_1-\+\theta_1^0=O_p((NI_1^0)^{-1/2})$, which
reduces to the familiar $(NT)^{-1/2}$ rate under balanced regimes. The
initial level is identified from the first regime alone, whose
observations are the only ones in which its loading is active.
\end{remark}

\begin{remark}[Fixed $T$]\label{rem:fixedT}
The results above extend to the fixed-$T$ case; see
\citet{KaddouraWesterlund2023} for a related discussion. When $T\ge3$
is fixed as $N\to\infty$, the normalisation $\mathbb D_{K^0+1}$ and the
transformation $\*M_{K^0}$ reduce to constant matrices, the selection
matrices may be taken fixed, and Assumption~\ref{ass:signal} simplifies
to $\sqrt N\vartheta_1J_{\min}^{-\zeta_1}\to_pc_1\in[0,\infty)$,
$\sqrt N\,J_{\min}\to\infty$, and
$N^{(\zeta_1+1)/2}\vartheta_1\to_p\infty$. All theorems continue to
hold, with every post-kink parameter converging at the usual
$N^{-1/2}$ rate.
\end{remark}

Theorem~\ref{thm:normality} implies that inference on $\+\eta^0$
requires consistent estimation of
\(
    \mathrm{Avar}(\widetilde{\+\eta}_{\widehat K})
    :=N^{-1}\mathbb D_{K^0+1}^{-1}\*Q_0^{-1}
    \+\Sigma_0\*Q_0^{-1}\mathbb D_{K^0+1}^{-1},
\)
and covariance matrices and Wald contrasts transform by the same fixed
nonsingular map $\*M_{K^0}$, so the estimator may be computed directly
in the original $\+\psi$ coordinates of Section~\ref{sec:postkink}. Let
$\*y_i:=[y_{i,1},\ldots,y_{i,T}]'$ and let
$\widehat{\*u}_i
:=\*y_i-\+\Phi_i\bigl(\widehat{\mathcal T}_{\widehat K}\bigr)
\widetilde{\+\psi}_{\widehat K}$ denote the post-kink residual vector
of unit $i$, so that $\*C_T\widehat{\*u}_i$ collects the fitted
residuals of~\eqref{eq:pk_linear}. A natural cluster-robust variance
estimator is
\(
    \widehat{\+\Omega}
    :=\frac1N\widehat{\*A}^{-1}\widehat{\*G}\,\widehat{\*A}^{-1},
\)
where
\(
    \widehat{\*A}
    :=\frac1N\+\Phi\bigl(\widehat{\mathcal T}_{\widehat K}\bigr)'\*C
    \+\Phi\bigl(\widehat{\mathcal T}_{\widehat K}\bigr)
\)
and
\(
    \widehat{\*G}
    :=\frac1N\sum_{i=1}^N\hat{\*g}_i\hat{\*g}_i'
\)
with
\(
    \hat{\*g}_i
    :=\+\Phi_i\bigl(\widehat{\mathcal T}_{\widehat K}\bigr)'
    \*C_T\widehat{\*u}_i.
\)
For a linear hypothesis $H_0:\*R\+\psi^0_{K^0}=\*r$, with $\*R$ of full
row rank, the Wald statistic is
\(
    \mathcal{W}
    :=\bigl(\*R\widetilde{\+\psi}_{\widehat K}-\*r\bigr)'
    \bigl(\*R\widehat{\+\Omega}\*R'\bigr)^{-1}
    \bigl(\*R\widetilde{\+\psi}_{\widehat K}-\*r\bigr).
\)
For the path, since $\widetilde{\+\theta}_t
=\*L_t(\widehat{\mathcal T}_{\widehat K})
\widetilde{\+\psi}_{\widehat K}$ by~\eqref{eq:pk_path}, an asymptotic
pointwise $95\%$ confidence interval for the $m$th component of
$\+\theta_t^0$ is
\(
    \widetilde\theta_{m,t}\pm1.96\,
    \bigl[\widehat{\+\Omega}_{\theta,t}\bigr]_{mm}^{1/2},
\)
where
\(
    \widehat{\+\Omega}_{\theta,t}
    :=\*L_t(\widehat{\mathcal T}_{\widehat K})\widehat{\+\Omega}
    \*L_t(\widehat{\mathcal T}_{\widehat K})'.
\)\footnote{On the recovery event of Corollary~\ref{cor:kinks},
$\widehat{\+\Omega}$ coincides with the oracle cluster-robust
estimator at the true design, so consistency follows by the usual
argument under cross-sectional independence of
$\{(\*x_i,\+\varepsilon_i)\}$ together with standard moment and
cluster-leverage conditions, and transfers to the feasible estimator
because that event has probability tending to one. Under weak
cross-sectional dependence a cross-sectionally robust estimator would
be needed.}
\subsection{Level breaks at the sampling frequency}\label{sec:breaks}

A level break and a kink are different structural objects. In the classical
break model the coefficient may jump between regimes
\citep{BaiPerron1998,BaiPerron2003}, whereas a kink is continuous and changes
the slope of the regression function. Hence the continuity
restriction in~\eqref{eq:continuity} means that our model does not nest a
literal discontinuity. It can, however, generate exactly the same coefficient
sequence at the dates observed by a discrete panel.

\begin{figure}[htbp]
    \centering    \includegraphics[width=0.78\linewidth]{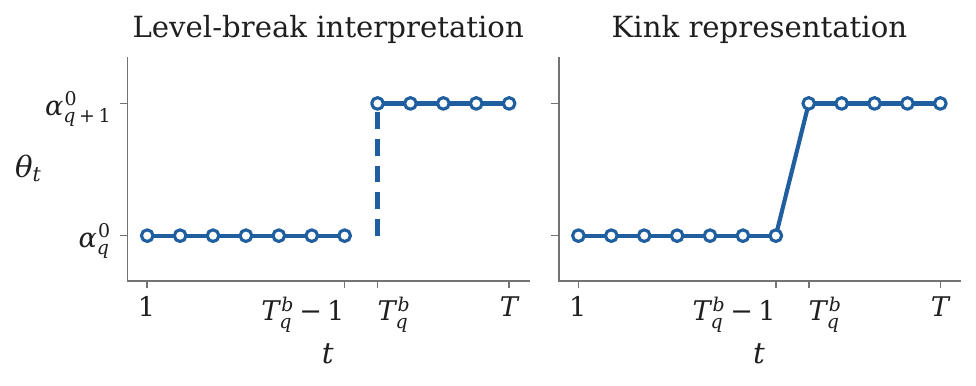}
    \caption{\footnotesize{The same sampled coefficient sequence under a level
    break and a continuous kink representation. The two readings differ only
    between $T_q^b-1$ and $T_q^b$, where the panel contains no observation.}}
    \label{fig:breakint}
\end{figure}

Following \citet{BaiPerron1998}, suppose that the coefficient sequence is
piecewise constant with breaks at $T_1^b<\cdots<T_m^b$, with $T_0^b:=1$ and
$T_{m+1}^b:=T+1$, break-model quantities carrying the superscript $b$ and
being understood as true values. In regime $q$, the coefficient is
$\+\alpha_q^0$, so
$\+\theta_t^0=\+\alpha_q^0$ for $t=T_{q-1}^b,\ldots,T_q^b-1$. Let
$I_q^b:=T_q^b-T_{q-1}^b\ge2$ denote the regime length\footnote{The spacing
condition keeps the two-kink pairs associated with distinct breaks from
overlapping. If a break regime contained only one observed date, consecutive
breaks could share a second difference and the simple $K^0=2m$ mapping below
would no longer apply.} and let
$\+\delta_q^0:=\+\alpha_{q+1}^0-\+\alpha_q^0$ denote the magnitude of break
$q$.
Around $T_q^b$ the panel therefore observes $\+\alpha_q^0$ at $T_q^b-1$ and
$\+\alpha_{q+1}^0$ at $T_q^b$, with no observation between them. The same sampled values can be represented by a continuous kink path. Place
one kink at $T_q^b-1$ and another at $T_q^b$, keep the path flat on either
side, and join the two observed levels over the intervening unit interval. The
slope of that short segment is
\begin{align}
    \+\kappa_{2q}^0
    &=\frac{\+\alpha_{q+1}^0-\+\alpha_q^0}
    {T_q^b-(T_q^b-1)}
    =\+\delta_q^0.
    \label{eq:breakslope}
\end{align}
Thus a sampled break appears to the kink estimator as the local slope pattern
$\*0_{p\times1}\rightarrow\+\delta_q^0\rightarrow\*0_{p\times1}$. 
Here, $\+\delta_q^0$ is a change in level, whereas $\+\kappa_{2q}^0$ is the rate of
the continuous interpolation between two sampled levels. A literal step has
no finite slope at the jump. If the same level change were spread over $h$
periods, the corresponding slope would instead be $\+\delta_q^0/h$. The two
models are therefore different between observations, but observationally
identical on the observed time grid. Figure~\ref{fig:breakint} illustrates
this, the two readings agreeing at every sampled date and parting only over
the empty interval between them.

Econometrically, this last distinction is largely immaterial at the sampling
frequency. Both readings imply the same pre- and post-break levels, the same
magnitude $\+\delta_q^0$, and the same date $T_q^b$ at which the new level is
first observed. The kink estimator can therefore date the observed change and
estimate its economically relevant size even when the event is interpreted as
a level break. What the data cannot determine is the path inside that empty
sampling interval: a literal jump and a sufficiently rapid crossing are
observationally equivalent there. That distinction matters only when the
economic mechanism concerns adjustment within the sampling interval, in which
case higher-frequency observations or additional restrictions are required.
Otherwise, the timing and magnitude of the observed structural change are the
same under either reading. 

\begin{remark}[Reporting breaks from the kink representation]\label{rem:breakindex}
For completeness, the break dates and break parameters can be reported
directly from the kink information. Break $q$ corresponds to the adjacent
pair $T_{2q-1}^0=T_q^b-1$ and $T_{2q}^0=T_q^b$, so $m$ breaks induce
$K^0=2m$ kinks. The intervening regime has length one and slope
$\+\kappa_{2q}^0=\+\delta_q^0$, while the surrounding regimes are flat. The
two boundary levels associated with a flat break regime coincide; in
particular, $\+\eta_{2q-2}^0=\+\eta_{2q-1}^0=\+\alpha_q^0$. Hence one may
report the break date as $\widehat T_q^b:=\widehat T_{2q}$, the regime level
as $\widetilde{\+\alpha}_q:=\widetilde{\+\eta}_{2q-2}$, and the break
magnitude as
$\widetilde{\+\delta}_q:=\widetilde{\+\eta}_{2q}-\widetilde{\+\eta}_{2q-1}$.
The last quantity equals $\widetilde{\+\kappa}_{2q}$ because the interval
between the two levels has unit length.
\end{remark}

\begin{proposition}[Rates under a level-break configuration]
\label{prop:breaks}
Suppose that the level-break sequence above has fixed $m$, with
$\+\delta_q^0\neq\*0_{p\times1}$ and $I_q^b\ge2$ for every
$q=1,\ldots,m+1$, and that the assumptions of
Theorem~\ref{thm:normality} hold for the induced configuration. Then the
adjacent kink pairs, and hence the break dates, are recovered exactly
w.p.a.1, with $J_{\min}=\min_{1\le q\le m}\|\+\delta_q^0\|$, and
\begin{enumerate}[label=\textnormal{\emph{(\roman*)}}]
\item $\widetilde{\+\alpha}_q-\+\alpha_q^0
=O_p\bigl([NI_q^b]^{-1/2}\bigr)$ for $q=1,\ldots,m+1$, with the same
order holding uniformly for $\widetilde{\+\theta}_t-\+\alpha_q^0$ over
the dates $t=T_{q-1}^b,\ldots,T_q^b-1$ of break regime $q$;
\item $\widetilde{\+\delta}_q-\+\delta_q^0
=O_p\bigl([N\{I_q^b\wedge I_{q+1}^b\}]^{-1/2}\bigr)$ for
$q=1,\ldots,m$;
\item $\widetilde{\+\kappa}_{2q+1}-\+\kappa_{2q+1}^0
=O_p\bigl([N(I_{q+1}^b)^3]^{-1/2}\bigr)$ for $q=0,\ldots,m$, where
$\+\kappa_{2q+1}^0=\*0_{p\times1}$ are the slopes of the flat regimes
surrounding the breaks.
\end{enumerate}
\end{proposition}

Part~\emph{(i)} is the usual regime-level rate. Part~\emph{(ii)} is governed
by the shorter neighbouring regime because $\+\delta_q^0$ is a contrast of the
two adjacent levels. Thus the two-kink representation changes the
parameterisation, but not the first-order rates of the break levels and
magnitudes. Part~\emph{(iii)} concerns the slopes the break specification
does not carry, which are zero under it and are estimated at the trend rate of
Remark~\ref{rem:trend} on their own regime. They are what the generality buys,
since $H_0:\+\kappa_{2q+1}^0=\*0_{p\times1}$ is a Wald test of the level-break
specification against a coefficient that was already evolving before the
break, computed from $\widehat{\+\Omega}$ as in
Section~\ref{sec:postselection}.
\subsection{Component-wise kinks}\label{sec:component}

The baseline estimator imposes a common set of kink dates on the
entire coefficient vector. This is appropriate when the relationship
shifts jointly across regressors, but restrictive when only some
coefficients change slope at a given date. Following the
coefficient-by-coefficient logic of \citet{Kaddoura2025}, we now let
each coefficient carry its own kink structure. The analysis mirrors
the baseline case. Selection first recovers the coordinate-specific
kink sets, and post-selection estimation then runs the
within-transformed regression on the selected structure, analysed one
coordinate at a time in boundary-level coordinates.

For each $j=1,\ldots,p$, let coefficient $j$ have $K_j$ kinks at dates
$\mathcal T_{j,K_j}:=\{T_{j,1},\ldots,T_{j,K_j}\}$, where
$1=:T_{j,0}<T_{j,1}<\cdots<T_{j,K_j}<T<T_{j,K_j+1}:=T+1$, with regimes
$\mathcal K_{j,\ell}:=\{T_{j,\ell-1},\ldots,T_{j,\ell}-1\}$ for
$\ell=1,\ldots,K_j+1$. Within regime $\ell$ the coefficient evolves as
\begin{align}
    \theta_{j,t}
    =\theta_{j,T_{j,\ell-1}}+(t-T_{j,\ell-1})\kappa_{j,\ell}
    \label{eq:cwpath}
\end{align}
for $t\in\mathcal K_{j,\ell}$, where $\kappa_{j,\ell}\in\mathbb R$ is
the within-regime slope, and continuity at the coordinate-specific
kink dates is imposed by
$\theta_{j,T_{j,\ell}}
=\theta_{j,T_{j,\ell-1}}+(T_{j,\ell}-T_{j,\ell-1})\kappa_{j,\ell}$ for
$\ell=1,\ldots,K_j$. Equivalently, the scalar second difference
$\Delta^2\theta_{j,t}:=\theta_{j,t+1}-2\theta_{j,t}+\theta_{j,t-1}$ is
nonzero exactly when coefficient $j$ changes slope at date $t$.  
To ease the exposition, let \(a^0\) and \(\widehat a\) denote the true
value and estimator, respectively, of any quantity \(a\). In particular,
for each \(j=1,\ldots,p\), let
\(\mathcal T_{j,K_j^0}^0=\{T_{j,1}^0,\ldots,T_{j,K_j^0}^0\}\) and
\(\widehat{\mathcal T}_{j,\widehat K_j}
=\{\widehat T_{j,1},\ldots,\widehat T_{j,\widehat K_j}\}\)
denote the true and estimated coordinate-specific kink sets, with
\(T_{j,0}^0=\widehat T_{j,0}:=1\) and
\(T_{j,K_j^0+1}^0=\widehat T_{j,\widehat K_j+1}:=T+1\).

The preliminary estimator of~\eqref{eq:initial} yields the adaptive
weights
$\dot\omega^{\mathrm{cw}}_{j,t}
:=|\Delta^2\dot\theta_{j,t}|^{-\zeta_2}$ for $j=1,\ldots,p$ and
$t=2,\ldots,T-1$, where $\zeta_2>0$, and the component-wise estimator
minimises
\begin{align}
    \mathcal L^{\mathrm{cw}}_{\vartheta_2}(\+\Theta_T)
    :=\frac1{NT}\sum_{i=1}^N\sum_{t=2}^{T}
    \bigl(\check y_{i,t}-\*x_{i,t}'\+\theta_t
    +\*x_{i,1}'\+\theta_1\bigr)^2
    +\vartheta_2\sum_{j=1}^{p}\sum_{t=2}^{T-1}
    \dot\omega^{\mathrm{cw}}_{j,t}\,
    \bigl|\Delta^2\theta_{j,t}\bigr|,
    \label{eq:objective_componentwise}
\end{align}
where $\vartheta_2>0$ is a tuning parameter. In contrast to the
group-$\ell_2$ penalty of~\eqref{eq:objective}, the penalty acts on
every scalar second difference separately, so coefficient $j$ may kink
at a date at which the remaining coefficients stay locally linear.
The relevant count of active constraints is no longer $K^0$ but the
total number of scalar kinks, $M^0:=\sum_{j=1}^pK_j^0$, and, when
$M^0\ge1$, the relevant signal is
$J^{\mathrm{cw}}_{\min}
:=\min_{j,\ell}|\kappa^0_{j,\ell+1}-\kappa^0_{j,\ell}|$, the minimum
taken over $j=1,\ldots,p$ and $\ell=1,\ldots,K_j^0$.

\begin{proposition}[Component-wise recovery of the kink structure]
\label{thm:cbc}
Suppose that Assumptions~\ref{ass:errors}, \ref{ass:rank}
and~\ref{ass:signal} hold, the latter with $J_{\min}$, $\vartheta_1$,
$\zeta_1$, and $K^0$ replaced by $J^{\mathrm{cw}}_{\min}$,
$\vartheta_2$, $\zeta_2$, and $M^0$, respectively, and with part (a)
replaced by $\vartheta_2\to_p0$ when $M^0=0$. Then, as $N,T\to\infty$,
\begin{enumerate}[label=(\roman*)]
    \item
    \(
    \mathbb{P}\!\left(\left\{
    \left|\Delta^2\widehat\theta^{\mathrm{cw}}_{j,t}\right|=0
    \;\;\forall\;t\in\mathcal{T}_{j,K_j^0}^{0c},
    \;\forall\;j=1,\ldots,p
    \right\}\right)\to1;
    \)
    \item
    \(
    \mathbb{P}\!\left(\left\{
    \widehat K_j=K_j^0
    \;\;\forall\;j=1,\ldots,p
    \right\}\right)\to1;
    \)
    \item
    \(
    \mathbb{P}\!\left(\left\{
    \widehat T_{j,\ell}=T_{j,\ell}^0,
    \;\ell=1,\ldots,K_j^0,
    \;\forall\;j
    \;\Big|\;
    \widehat K_j=K_j^0\;\forall\;j
   \right\}\right)\to1.
    \)
\end{enumerate}
\end{proposition}

Proposition~\ref{thm:cbc} is the component-wise analog of
Theorem~\ref{thm:sign} and Corollary~\ref{cor:kinks}. Selection
operates separately for each coefficient, so the estimator recovers
not only whether a kink is present, but also which coefficient kinks
and at which date.

Post-selection estimation again runs on the within transform and is
analysed in boundary-level coordinates, one coordinate at a time. For
coordinate $j$, the true kink dates induce regimes of lengths
$I_{j,\ell}^0:=T_{j,\ell}^0-T_{j,\ell-1}^0$, with the conventions
$I_{j,0}^0:=0$, $I_{j,K_j^0+2}^0:=0$, and
$\mathcal K_{j,0}^0=\mathcal K_{j,K_j^0+2}^0:=\varnothing$, and
boundary levels
\begin{align}
    \eta_{j,m}^0
    :=\theta_{j,1}^0
    +\sum_{\ell=1}^{m}I_{j,\ell}^0\kappa_{j,\ell}^0
    \label{eq:cwknotdef}
\end{align}
for $m=0,\ldots,K_j^0+1$, so that
$\kappa_{j,\ell}^0=(\eta_{j,\ell}^0-\eta_{j,\ell-1}^0)/I_{j,\ell}^0$.
With $u_{j,\ell,t}:=(t-T_{j,\ell-1}^0)/I_{j,\ell}^0$ the fraction of
regime $\ell$ of coordinate $j$ elapsed by date $t$, the loading of
date $t$ on $\eta_{j,m}^0$ is
\begin{align}
    h_{j,m,t}
    :=
    \begin{cases}
        u_{j,m,t}, & t\in\mathcal K_{j,m}^0,\\[2pt]
        1-u_{j,m+1,t}, & t\in\mathcal K_{j,m+1}^0,\\[2pt]
        0, & \text{otherwise},
    \end{cases}
    \label{eq:cwhdef}
\end{align}
the per-coordinate analog of~\eqref{eq:hdef}, and the interpolation
identity~\eqref{eq:knotrep} holds coordinate by coordinate,
$\theta_{j,t}^0
=(1-u_{j,\ell,t})\eta_{j,\ell-1}^0+u_{j,\ell,t}\eta_{j,\ell}^0$ for
$t\in\mathcal K_{j,\ell}^0$. Each coefficient is therefore a
free-knot linear spline on its own knot sequence, and the post-kink
model is a varying-coefficient regression in which the $p$
coordinates carry separate first-degree B-spline bases. Setting
$r_{j,m}^2:=\sum_{t=1}^Th_{j,m,t}^2$, Lemmas~\ref{lem:rjorder}
and~\ref{lem:basis} apply to each coordinate with the same absolute
constants, so that $r_{j,m}^2\asymp I_{j,m}^0+I_{j,m+1}^0$.

Substituting the interpolation identity into the DGP~\eqref{eq:DGP},
\begin{align}
    \*x_{i,t}'\+\theta_t^0
    =\sum_{j=1}^p\sum_{m=0}^{K_j^0+1}
    h_{j,m,t}\,x_{i,t,j}\,\eta_{j,m}^0,
    \label{eq:cwdesign}
\end{align}
which is linear in the boundary levels with known coefficients once
the kink dates are given. To write~\eqref{eq:cwdesign} as a
regression, stack the boundary levels coordinate by coordinate, the
$K_1^0+2$ boundary levels of coordinate $1$ first, then those of
coordinate $2$, and so on,
\(
    \+\eta^{0,\mathrm{cw}}
    :=[\eta_{1,0}^0,\ldots,\eta_{1,K_1^0+1}^0,\ldots,
    \eta_{p,0}^0,\ldots,\eta_{p,K_p^0+1}^0]'
    \in\mathbb R^{q^0\times1},
\)
where $q^0:=\sum_{j=1}^p(K_j^0+2)$ is fixed since every $K_j^0$ is
fixed, and define the conformable regressor
\begin{align}
    \+\phi^{\eta,\mathrm{cw}}_{i,t}
    :=\bigl[h_{1,0,t}\,x_{i,t,1},\ldots,h_{1,K_1^0+1,t}\,x_{i,t,1},
    \ldots,
    h_{p,0,t}\,x_{i,t,p},\ldots,h_{p,K_p^0+1,t}\,x_{i,t,p}\bigr]'
    \in\mathbb R^{q^0\times1},
    \label{eq:cwphidef}
\end{align}
so that~\eqref{eq:cwdesign} reads
$\*x_{i,t}'\+\theta_t^0
=\+\phi^{\eta,\mathrm{cw}\prime}_{i,t}\+\eta^{0,\mathrm{cw}}$ and the
DGP becomes
$y_{i,t}=\mu_i+\+\phi^{\eta,\mathrm{cw}\prime}_{i,t}
\+\eta^{0,\mathrm{cw}}+\varepsilon_{i,t}$. Let
$\+\Phi^{\eta,\mathrm{cw}}\in\mathbb R^{NT\times q^0}$ stack the rows
$\+\phi^{\eta,\mathrm{cw}\prime}_{i,t}$ and
$\+\varepsilon\in\mathbb R^{NT\times1}$ the errors
$\varepsilon_{i,t}$  conformably with $\*y$
in~\eqref{eq:pk_expanded}. 

Collecting the rates in
\(
    \mathbb D^{\mathrm{cw}}
    :=\mathrm{diag}\bigl(r_{1,0},\ldots,r_{1,K_1^0+1},\ldots,
    r_{p,0},\ldots,r_{p,K_p^0+1}\bigr)
    \in\mathbb R^{q^0\times q^0},
\)
which is nonsingular because every $r_{j,m}$ is positive by
Lemma~\ref{lem:rjorder} applied per coordinate, the last kink date of
every coordinate lying strictly before $T$. Further, define
\(
    \widehat{\*Q}_N^{\mathrm{cw}}
    :=\bigl(\mathbb D^{\mathrm{cw}}\bigr)^{-1}\,\frac1N\,
    \+\Phi^{\eta,\mathrm{cw}\prime}\*C\,\+\Phi^{\eta,\mathrm{cw}}\,
    \bigl(\mathbb D^{\mathrm{cw}}\bigr)^{-1}
\)
and
\(
    \+\Psi_{NT}^{\mathrm{cw}}
    :=\frac1{\sqrt N}\,
    \+\Phi^{\eta,\mathrm{cw}\prime}\*C\,\+\varepsilon,
\)
with $\*C=\*I_N\otimes\*C_T$ the centering matrix.

The structure of these objects is read off from the columns of the
design. The column of $\+\Phi^{\eta,\mathrm{cw}}$ belonging to the
boundary level $\+\eta_{j,m}^0$ has $(i,t)$ entry
\(
    \chi_{i,t;j,m}:=h_{j,m,t}\,x_{i,t,j},
\)
the $j$th regressor switched on with weight $h_{j,m,t}$ over the two
regimes of coordinate $j$ adjacent to boundary $m$, and
premultiplication by $\*C$ demeans each column unit by unit,
\(
    \ddot\chi_{i,t;j,m}
    :=\chi_{i,t;j,m}-\frac{1}{T}\sum_{s=1}^{T}\chi_{i,s;j,m}.
\)
Hence the $\bigl((j,m),(j',m')\bigr)$ entry of
$N\mathbb D^{\mathrm{cw}}\widehat{\*Q}_N^{\mathrm{cw}}
\mathbb D^{\mathrm{cw}}
=\+\Phi^{\eta,\mathrm{cw}\prime}\*C\,\+\Phi^{\eta,\mathrm{cw}}$ is,
using $\*C'\*C=\*C$,
\begin{align}
    \left[
    N\mathbb D^{\mathrm{cw}}\widehat{\*Q}_N^{\mathrm{cw}}
    \mathbb D^{\mathrm{cw}}
    \right]_{(j,m),(j',m')}
    &=\sum_{i=1}^{N}\sum_{t=1}^{T}
    \ddot\chi_{i,t;j,m}\,\ddot\chi_{i,t;j',m'}
    \notag\\
    &=\sum_{i=1}^{N}\sum_{t=1}^{T}
    h_{j,m,t}h_{j',m',t}\,x_{i,t,j}x_{i,t,j'}
    \notag\\
    &\qquad
    -\frac1T\sum_{i=1}^{N}
    \left(\sum_{s=1}^{T}\chi_{i,s;j,m}\right)
    \left(\sum_{r=1}^{T}\chi_{i,r;j',m'}\right),
    \label{eq:cwgramentry}
\end{align}
the second equality by expanding the demeaned products. The first
term is the one that carries the structure, since it is nonzero only
at dates where both hat functions are active. Within a coordinate
($j'=j$), the hat functions are built on the same kink set,
$h_{j,m,t}$ is supported on
$\mathcal K_{j,m}^0\cup\mathcal K_{j,m+1}^0$ and vanishes at
$t=T_{j,m-1}^0$ and $t=T_{j,m+1}^0$, so for $|m-m'|\ge2$ the supports
of $h_{j,m,\cdot}$ and $h_{j,m',\cdot}$ are disjoint and
\(
    \sum_{i=1}^{N}\sum_{t=1}^{T}
    h_{j,m,t}h_{j,m',t}\,x_{i,t,j}^2=0,
    \) for \( |m-m'|\ge2.
\)
The leading term of each diagonal block is therefore tridiagonal in
$m$, the deterministic hat overlaps of Lemma~\ref{lem:basis} weighted
by $x_{i,t,j}^2$. Across coordinates ($j'\neq j$), the hat functions
are built on different kink sets, so no zero pattern is available,
and the leading term is
\(    \sum_{i=1}^{N}\;\sum_{t\in\mathcal S_{j,m;j',m'}}
    h_{j,m,t}h_{j',m',t}\,x_{i,t,j}x_{i,t,j'},
    \) where
    \( \mathcal S_{j,m;j',m'}
    :=\bigl\{t:h_{j,m,t}h_{j',m',t}>0\bigr\}
\)
a hat-weighted sample cross moment of $x_{i,t,j}$ and $x_{i,t,j'}$
over the window on which both hats are active. In both cases the
demeaning term in~\eqref{eq:cwgramentry} is a within-unit mean
correction that does not respect the banding, and its control is part
of the rank condition below. In sum,
$\widehat{\*Q}_N^{\mathrm{cw}}$ couples boundary levels of the same
coordinate only when they are adjacent, and boundary levels of
different coordinates only through the comovement of their regressors
on overlapping regimes. The post-kink estimator
$\widetilde{\+\eta}^{\mathrm{cw}}$ is the OLS estimator of
$\+\eta^{0,\mathrm{cw}}$ from the centered regression evaluated at
the estimated kink sets,
\begin{align}
\widetilde{\+\eta}^{\mathrm{cw}}
    :=\bigl[\+\Phi^{\eta,\mathrm{cw}\prime}\*C\,
    \+\Phi^{\eta,\mathrm{cw}}\bigr]^{-1}
    \+\Phi^{\eta,\mathrm{cw}\prime}\*C\,\*y,
    \label{eq:cwols}
\end{align}
with $\+\Phi^{\eta,\mathrm{cw}}$ built from
$\widehat{\mathcal T}_{j,\widehat K_j}$, $j=1,\ldots,p$, the inverse
existing w.p.a.1 under Assumption~\ref{ass:cwrank}(a). On the
recovery event of Proposition~\ref{thm:cbc} it coincides with the
oracle estimator evaluated at the true coordinate-specific dates.
\begin{assumption}[Component-wise post-selection design and central
limit theorem]\label{ass:cwrank}
Every $K_j^0$, $j=1,\ldots,p$, is fixed, and
$q^0=\sum_{j=1}^p(K_j^0+2)$.
\begin{enumerate}[label=(\alph*)]
    \item We have
    $\|\widehat{\*Q}_N^{\mathrm{cw}}-\*Q_0^{\mathrm{cw}}\|_{op}
    =o_p(1)$, where $\*Q_0^{\mathrm{cw}}$ is finite and positive
    definite with
    $0<c\le\mu_{\min}(\*Q_0^{\mathrm{cw}})
    \le\mu_{\max}(\*Q_0^{\mathrm{cw}})\le C<\infty$;
    \item the limit
    \(
        \+\Sigma_0^{\mathrm{cw}}
        :=\lim\operatorname{Var}\!\left(
        (\mathbb D^{\mathrm{cw}})^{-1}
        \+\Psi_{NT}^{\mathrm{cw}}\right)
    \)
    exists and is finite and positive definite, and for any
    deterministic sequence of $l\times q^0$ matrices
    $\{\*H_{NT}^{\mathrm{cw}}\}$ with $l$ fixed and
    $\limsup_{N,T\to\infty}\|\*H_{NT}^{\mathrm{cw}}\|_{op}<\infty$,
    whenever
    \(
        \+\Upsilon^{\mathrm{cw}}
        :=\lim\*H_{NT}^{\mathrm{cw}}(\*Q_0^{\mathrm{cw}})^{-1}
        \+\Sigma_0^{\mathrm{cw}}(\*Q_0^{\mathrm{cw}})^{-1}
        \*H_{NT}^{\mathrm{cw}\prime}
    \)
    exists and is positive definite, we have
    \(
        \*H_{NT}^{\mathrm{cw}}(\*Q_0^{\mathrm{cw}})^{-1}
        (\mathbb D^{\mathrm{cw}})^{-1}
        \+\Psi_{NT}^{\mathrm{cw}}
        \to_d\mathcal N(\*0_{l\times1},\+\Upsilon^{\mathrm{cw}}).
    \)
\end{enumerate}
\end{assumption}

Assumption~\ref{ass:cwrank} is the component-wise analog of
Assumptions~\ref{ass:postrank} and~\ref{ass:postclt}, stated on the
true design so that every object has the fixed dimension $q^0$ and
the scaling matrix is deterministic. The division of labour is the
same as in the baseline case, with one addition. Within a coordinate,
Lemma~\ref{lem:basis} controls the time basis, so the lower
eigenvalue bound in part (a) is not a hidden restriction on any
coordinate's kink configuration. Across coordinates, however, the
loadings of different coefficients may overlap arbitrarily in time,
and their conditioning is governed by the regressors, so the positive
definiteness of $\*Q_0^{\mathrm{cw}}$ is where the cross-coordinate
regressor design enters.

\begin{proposition}[Asymptotic normality of the component-wise
post-kink estimator]\label{prop:cwnorm}
Suppose that the assumptions of Proposition~\ref{thm:cbc} and
Assumption~\ref{ass:cwrank} hold, and define
\(
    s_{j,\ell}
    :=I_{j,\ell}^0
    \bigl(r_{j,\ell-1}^{-2}+r_{j,\ell}^{-2}\bigr)^{-1/2}
\)
for $j=1,\ldots,p$ and $\ell=1,\ldots,K_j^0+1$. Then, as
$N,T\to\infty$,
\begin{enumerate}[label=(\roman*)]
    \item for any deterministic sequence
    $\{\*H_{NT}^{\mathrm{cw}}\}$ as in
    Assumption~\ref{ass:cwrank}(b),
    \(
    \sqrt N\,\*H_{NT}^{\mathrm{cw}}\,\mathbb D^{\mathrm{cw}}
    \bigl(\widetilde{\+\eta}^{\mathrm{cw}}
    -\+\eta^{0,\mathrm{cw}}\bigr)
    \to_d
    \mathcal N\bigl(\*0_{l\times1},\+\Upsilon^{\mathrm{cw}}\bigr);
    \)
    \item
    \(
    \widetilde\kappa^{\mathrm{cw}}_{j,\ell}-\kappa^0_{j,\ell}
    =O_p\!\left(\bigl[Ns_{j,\ell}^2\bigr]^{-1/2}\right)
    =O_p\!\left(\Bigl[N\bigl(I_{j,\ell}^0\bigr)^2
    \bigl\{(I_{j,\ell-1}^0+I_{j,\ell}^0)
    \wedge(I_{j,\ell}^0+I_{j,\ell+1}^0)\bigr\}\Bigr]^{-1/2}\right)
    \)
    for $j=1,\ldots,p$ and $\ell=1,\ldots,K_j^0+1$;
    \item
    \(
    \widetilde\theta^{\mathrm{cw}}_{j,t}-\theta^0_{j,t}
    =O_p\!\left(\Bigl[N
    \bigl\{(I_{j,\ell-1}^0+I_{j,\ell}^0)
    \wedge(I_{j,\ell}^0+I_{j,\ell+1}^0)\bigr\}\Bigr]^{-1/2}\right)
    \)
    uniformly over $t\in\mathcal K^0_{j,\ell}$, for
    $j=1,\ldots,p$ and $\ell=1,\ldots,K_j^0+1$, and in particular
    $\widetilde\theta^{\mathrm{cw}}_{j,1}-\theta^0_{j,1}
    =O_p\bigl((NI_{j,1}^0)^{-1/2}\bigr)$.
\end{enumerate}
\end{proposition}

The component-wise rates therefore inherit the adjacent-regime
structure of the baseline model. Each coordinate-specific slope is the
contrast of the two boundary levels enclosing its regime, and each
boundary level is informed by the two regimes of that coordinate
adjacent to it, so the precision of coefficient $j$ is governed
entirely by its own regime configuration, independently of where the
remaining coefficients kink.

\section{Monte Carlo simulations}\label{sec:MC}
 
\subsection{Design}\label{sec:MCdesign}
 
We study the finite sample behaviour of the estimator in a restricted
version of the DGP in~\eqref{eq:DGP}. The regressors are generated as
\begin{align*}
    \*x_{i,t}=0.2\,\left(\*1_{p\times1}\cdot\xi_i\right)+\*z_{i,t},
\end{align*}
where $\xi_i\sim\mathcal N(0,1)$ and
$\*z_{i,t}\sim\mathcal N(\*0_{p\times1},\*I_p)$ are independent across
$i$ and $t$, so that $0.2\,\xi_i$ induces a common component in the
regressors within a unit. Following \citet{bai2002determining}, the
error carries serial dependence, cross-sectional dependence, and
heteroskedasticity through
\begin{align*}
    \varepsilon_{i,t}
    =\varpi_1\varepsilon_{i,t-1}+\varsigma_{i,t}
    +\varpi_2\sum_{r=1}^{R}
    \bigl(\varsigma_{i-r,t}+\varsigma_{i+r,t}\bigr),
\end{align*}
with $\varepsilon_{i,0}=0$ and $\varsigma_{i,t}\sim\mathcal N(0,1)$,
where $\varpi_1$ governs serial dependence, $\varpi_2$ cross-sectional
dependence, and $R$ its range, with baseline
$(\varpi_1,\varpi_2,R)=(0.2,0.1,5)$. The fixed effects are
$\mu_i\sim\mathcal N(0,1)$, independent of all else. The initial level
is $\+\theta_1^0=[1,-1]'$ when $p=2$ and $\+\theta_1^0=[1,-1,1]'$ when
$p=3$, and the path is built from the within-regime slopes
through~\eqref{eq:Kinkalt}. The sample sizes cover all nine
combinations of $N\in\{50,100,200\}$ and $T\in\{20,40,80\}$, and all
results are based on $1000$ replications.
 
The eight designs fall into three groups. The first group holds the
error dependence at its baseline and varies the kink configuration,
so that the number of kinks and the way identification is spread
across regimes are the only things that change. The second group
perturbs that configuration in the two directions the theory
identifies as costly, namely unbalanced regime lengths and stronger
error dependence. The third group breaks the common-kink restriction
altogether and lets each coordinate carry its own kink structure.
DGPs~1--3 and~5--7 have $p=2$ and use the group estimator of
Section~\ref{sec:initial}; DGP~4 has $p=2$ and DGP~8 has $p=3$, and
both use the component-wise estimator of Section~\ref{sec:component}.
 
\bigskip
\noindent\textbf{DGP~1 ($K^0=0$).}\;
A single regime with $\+\kappa_1^0=[0,0]'$, so the path is flat. The
estimator should return $\widehat K=0$, and the design measures
over-selection when there is nothing to find.
 
\bigskip
\noindent\textbf{DGP~2 ($K^0=1$).}\;
A kink at $T_1^0=\lfloor T/2\rfloor$ with $\+\kappa_1^0=[0.3,-0.2]'$
and $\+\kappa_2^0=[-0.2,0.3]'$. Each coordinate rises and then
reverses, so the kink is a genuine change of direction. Both regimes
are of length $I^0_\ell\asymp T/2$, so both slopes are endpoint
slopes in the sense of Corollary~\ref{cor:kappa}.
 
\bigskip
\noindent\textbf{DGP~3 ($K^0=2$).}\;
Kinks at $\lfloor T/3\rfloor$ and $\lfloor2T/3\rfloor$ with slopes
$[0,0]'$, $[0.5,-0.4]'$, and $[0,0]'$. The path is flat in the outer
regimes and linear in the middle one, so all identification of the two
kinks comes from the middle regime. This is the most demanding
selection design among those with balanced regimes.
 
\bigskip
\noindent\textbf{DGP~4 (coordinate-specific kinks).}\;
Coordinate one kinks once at $\lfloor T/3\rfloor$ with regime slopes
$(0.4,-0.3)$ and coordinate two kinks once at $\lfloor2T/3\rfloor$
with regime slopes $(-0.3,0.4)$, so that $K_1^0=K_2^0=1$ but the kink
dates differ. No common kink set describes the path, and the
component-wise estimator should recover the two coordinate-specific
kink sets.
 
\bigskip
\noindent\textbf{DGP~5 (uneven regimes, $K^0=2$).}\;
Kinks at $\lfloor T/5\rfloor$ and $\lfloor 4T/5\rfloor$ with slopes
$[0,0]'$, $[0.5,-0.4]'$, and $[0,0]'$, so the middle regime is long
and the two outer regimes short. The design isolates the role of the
regime configuration in Corollary~\ref{cor:kappa}: $\+\kappa_2^0$ is
an interior slope flanked by a long own regime, while $\+\kappa_1^0$
and $\+\kappa_3^0$ are endpoint slopes identified from a short own
regime alone.
 
\bigskip
\noindent\textbf{DGP~6 (DGP~2 under stronger dependence).}\;
The kink configuration of DGP~2, with the error dependence raised to
$(\varpi_1,\varpi_2)=(0.5,0.2)$. The design measures the sensitivity
of selection and estimation to serial and cross-sectional dependence
in the errors when identification is otherwise strong.
 
\bigskip
\noindent\textbf{DGP~7 (DGP~3 under stronger dependence).}\;
The kink configuration of DGP~3, the most demanding selection design,
under the same stronger dependence $(\varpi_1,\varpi_2)=(0.5,0.2)$.
This is the joint stress of weak identification and strong error
dependence.
 
\bigskip
\noindent\textbf{DGP~8 (heterogeneous kink counts, $p=3$).}\;
Three coordinates with different numbers of kinks: coordinate one is
flat ($K_1^0=0$), coordinate two kinks once at $\lfloor T/2\rfloor$
with slopes $(0.4,-0.3)$, and coordinate three kinks twice at
$\lfloor T/3\rfloor$ and $\lfloor2T/3\rfloor$ with slopes
$(0,0.5,0)$. No common kink set describes the path, and the
component-wise estimator must recover a different number of kinks in
each coordinate, including none in the first.
 
\subsection{Implementation}\label{sec:MCimpl}
 
The tuning parameter $\vartheta_1$ is chosen by minimising
\begin{align}
    \mathrm{IC}(\vartheta_1)
    :=\log\Biggl\{\frac{1}{NT}\sum_{i=1}^N\sum_{t=1}^T
    \Bigl(\ddot y_{i,t}
    -\ddot{\+\phi}_{i,t}
    \bigl(\widehat{\mathcal T}_{\widehat K_{\vartheta_1}}\bigr)'
    \widetilde{\+\psi}_{\widehat K_{\vartheta_1}}\Bigr)^2\Biggr\}
    +\varrho_{N,T}\,p\,\bigl(\widehat K_{\vartheta_1}+1\bigr),
    \label{eq:IC}
\end{align}
where $\varrho_{N,T}:=\log(NT)/NT$ is a BIC-type rate in the spirit
of \citet{QianSu2016}. The residuals are within transformed and
post-kinks. The convex objective~\eqref{eq:objective} is
solved along the grid $\vartheta_1=c\,(NT)^{-1/2}$ with $c$ on a
logarithmic grid. The kink set is built by stepwise selection on the
criterion, adding the candidate whose inclusion lowers it most and
dropping any date whose removal lowers it further, and each selected
date is then moved within a window of three periods to the position
minimising the post-kink sum of squared residuals, with a final
backward pass under the criterion. The adaptive weights use the
perturbed form
$\dot\omega_t=(\|\Delta\dot{\+\theta}_{t+1}
-\Delta\dot{\+\theta}_t\|+\delta_t)^{-\zeta_1}$ with
$\delta_t=N^{-1/2}$ and $\zeta_1=2$.
 
DGPs~4 and~8 use the separable
analogue~\eqref{eq:objective_componentwise} with $\zeta_2=2$. The
candidate pairs are the active scalar second differences accumulated
along the $\vartheta_2$ grid, and the coordinate-specific kink sets
are built by the same stepwise procedure on~\eqref{eq:IC}, with
$p(\widehat K+1)$ replaced by $\sum_{j=1}^p(\widehat K_j+1)$, adding
or dropping the pair, in whichever coordinate, that improves the
criterion most, followed by the same local refinement and final
backward pass applied to the selected pairs.

\subsection{Results}\label{sec:MCresults}

Selection is measured by three statistics, each averaged over the
replications. Correct selection is the frequency, in percent, with
which the estimated kink dates coincide exactly with the true ones,
which for DGP~1 is the frequency of $\widehat K=0$. The kink-count
bias $\mathbb E[\widehat K-K^0]$ separates missed kinks from spurious
ones. Misclassification is the percentage of dates assigned to the
wrong regime, the regime label of date $t$ implied by a kink set
$\mathcal T$ being
$r_t(\mathcal T):=\sum_{T_k\in\mathcal T}\*1\{T_k\le t\}$.
Estimation is measured by the bias and root mean squared error of the
within-regime slopes, regime by regime and pooled across coordinates,
conditional on exact recovery of the kink set, and of the coefficient
path, averaged over dates and coordinates, unconditional.
Table~\ref{tab:mcsel} reports selection for the common-kink designs,
Tables~\ref{tab:mcselcw} and~\ref{tab:mcselcw8} the component-wise
results, and Tables~\ref{tab:mcest}, \ref{tab:mcestrob},
\ref{tab:mccw} and~\ref{tab:mccw8} the estimation results.

\begin{center}
\textbf{[Tables 1--7 about here]}
\end{center}

Estimation improves in both dimensions and at the predicted rates.
Every RMSE column falls monotonically in $N$ and in $T$. Moreover, selection improves in $N$. In DGP~1 there are no kinks and the
estimator finds none, with correct selection never below $99.3$
percent. DGP~2, with one well-separated kink, runs from $97.8$ to
$99.8$ percent. DGP~3 is harder, since the outer regimes are flat and
both kinks are identified from the middle regime alone, and there
correct selection rises from $80.0$ percent at $N=50$ to $98.7$
percent at $N=200$ when $T=20$, and from $73.0$ to $96.2$ percent in
DGP~7 when $T=80$. Comparing DGP~6 with DGP~2 and DGP~7 with DGP~3
shows that error dependence costs a few percentage points.

The span matters less. Correct selection is roughly flat in $T$,
moving from $80.0$ to $83.9$ to $82.0$ percent in DGP~3 at $N=50$ and
from $70.9$ to $79.8$ to $73.0$ percent in DGP~7. What rises with $T$
is over-selection: in DGP~7 at $N=50$ the kink-count bias runs from
$0.05$ to $0.28$ and misclassification from $3.8$ to $12.5$ percent,
so the errors are spurious kinks rather than missed ones. With
$J_{\min}$ fixed, a longer span widens the $\sqrt{T/N}$ band of
Theorem~\ref{thm:consistency}(c) within which noise survives the
shrinkage, while also lengthening the regimes from which the true
slope changes are estimated; the two roughly offset in the
exact-recovery rate. The effect is confined to $N=50$, the kink-count
bias being flat in $T$ at $N=100$ and $N=200$ in every design. Precision depends on the configuration, not only the sample size.
DGP~5 has a long middle regime and short outer ones, so
$\+\kappa_2^0$ is an interior slope with two well-informed boundary
levels while $\+\kappa_1^0$ and $\+\kappa_3^0$ are endpoint slopes
converging at the cube of their own short regime. In all nine cells
the interior slope is estimated more precisely than the endpoint
slopes. Against DGP~3, which differs only in where the kinks sit, the
middle slope improves from $0.019$ to $0.010$ at $(N,T)=(50,20)$
while the first deteriorates from $0.030$ to $0.051$. 

The component-wise estimator behaves the same way coordinate by
coordinate. In DGP~4 both coordinates are recovered with comparable
accuracy, from about $90$ percent at $N=50$ to about $97$ percent at
$N=200$. In DGP~8 the three coordinates carry $0$, $1$ and $2$ kinks
and each performs according to its own configuration. The globally
linear first coordinate is correctly identified as such in $95.0$ to
$98.3$ percent of replications despite its neighbours kinking, the
second behaves like DGP~2 and reaches $97.7$ percent, and the third
has the configuration of DGP~3 and runs from $61.2$ to $93.4$
percent. This is what Proposition~\ref{prop:cwnorm} asserts. The cross-section is therefore the dimension that buys correct
dating, and a researcher with a short panel and a long span should
expect over-selection rather than missed kinks. That error is mild
for the fitted path. Even in DGP~7 the path RMSE falls with $T$ at
every $N$, from $0.100$ to $0.051$ at $N=50$. This implies that a spurious kink adds a basis function to a
design that already spans the truth, so the fitted values are barely
affected.

\section{Dating the Debt--Growth Relationship}\label{sec:debtgrowth}
 
The debt--growth literature asks whether the association changes once
public debt crosses a particular level. \citet{ReinhartRogoff2010} report
markedly lower growth above 90 percent of GDP.
\citet{HerndonAshPollin2014} replicate that exercise and show the finding
turns on data construction and weighting. The work that followed finds the
estimated nonlinearity fragile. It weakens once endogeneity and
cross-country heterogeneity are taken seriously, and it does not survive as
a universal cutoff \citep{PanizzaPresbitero2014, PescatoriSandriSimon2014,
EberhardtPresbitero2015, ChudikMohaddesPesaranRaissi2017}. Collecting 816
estimates from 47 studies, \citet{Heimberger2023} cannot reject a zero
average effect once publication selection is accounted for, and finds
threshold estimates to be sensitive to data and specification choices.
 
We ask when the association changed rather than at what level of debt it
differs. The two questions are distinct. A threshold model lets the
coefficient depend on where an economy sits in the debt distribution, and
an economy crosses back and forth as its debt ratio moves. Our model lets
the coefficient bend at calendar dates common to all economies, and asks
how fast it was moving on either side of each bend. The sample covers the
global financial crisis and the pandemic, two occasions on which debt and
output moved sharply for very different reasons, and the estimator is built
to separate a movement that reverses from one that does not. The exercise
is descriptive throughout. What it delivers is dates, annual rates of
change, and a distinction between a temporary excursion and a permanent
reorientation, not a causal effect of debt on growth.
 
\subsection{A debt-only benchmark}\label{sec:dg_benchmark}
 
Let $g_{i,t}$ denote annual real GDP growth in percentage points and
$d_{i,t}$ general government gross debt as a percentage of GDP. The
benchmark specification is
\begin{align}
    g_{i,t}
    &= \mu_i+\theta_t d_{i,t}+\varepsilon_{i,t},
    \label{eq:dg_model}\\[-2pt]
    \theta_t
    &= \theta_1+\kappa_1(t-1)
       +\sum_{m=1}^{K}\gamma_m(t-T_m)_+,
    \qquad
    \gamma_m:=\kappa_{m+1}-\kappa_m,
    \label{eq:dg_theta}
\end{align}
where $(a)_+:=\max\{a,0\}$. Within regime $\ell$ the slope $\kappa_\ell$ is
the annual change in the debt coefficient and $T_\ell$ is a date at which
that rate changes. Thus $\theta_t=-0.03$ means that a ten point higher debt
ratio is associated with 0.3 percentage points lower contemporaneous
growth, and $\kappa_\ell=0.004$ means that this association becomes 0.04
percentage points less negative each year for the same ten point
difference in debt.

\begin{figure}[htbp]
    \centering
    \includegraphics[width=0.74\linewidth]{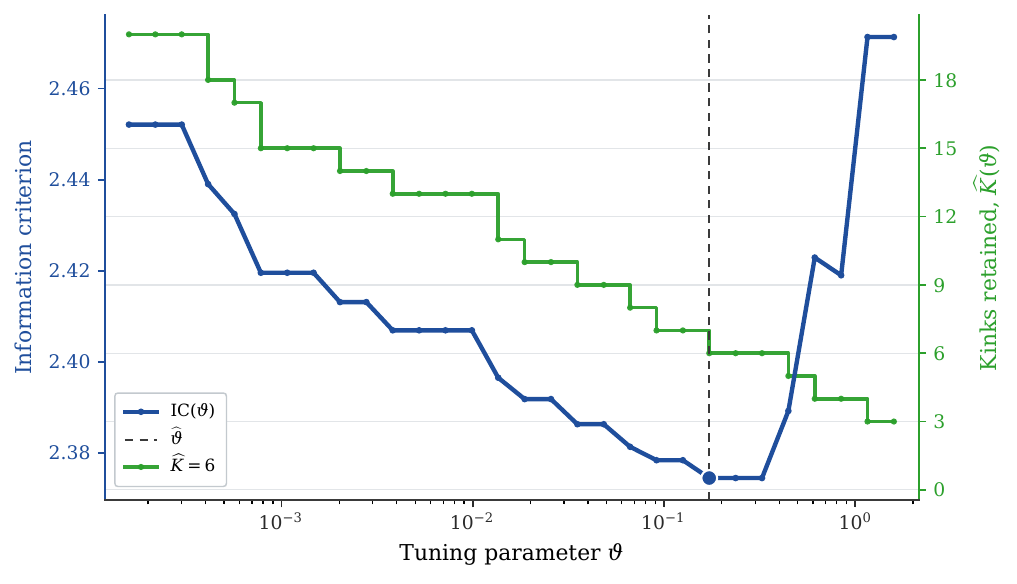}
    \caption{\footnotesize{Selection of the tuning parameter in the
    debt-only benchmark. The solid line is the post-kink information
    criterion along the penalised path, on the left axis. The step function
    is the number of kinks retained, on the right axis. The dashed vertical
    line marks $\widehat\vartheta_1$.}}
    \label{fig:dg_ic}
\end{figure}

The data are the April 2025 vintage of the IMF \emph{World Economic
Outlook} database \citep{IMFWEO2025}, from which we take
\texttt{NGDP\_RPCH} for real GDP growth and \texttt{GGXWDG\_NGDP} for
general government gross debt. The sample runs from 2000 to 2024. We drop
observations missing either series or with debt outside $(0,400)$ percent
of GDP and retain economies observed in every year, which leaves a balanced
panel of $N=158$ economies over $T=25$ years. Growth is winsorised at the
second and ninety-eighth percentiles, affecting four percent of the
observations. The final years should be read as a vintage result, since the
WEO carries IMF staff estimates where complete historical information is
unavailable and the April 2025 reference forecast was built on information
available as of 4 April 2025.

We set $\zeta_1=2$ and solve the penalised problem along a logarithmic grid
in $\vartheta_1$. The tuning parameter minimises the post-kink information
criterion, whose penalty
$\varrho_{N,T}=0.05\log(NT)/\sqrt{NT}$ is of the BIC type used by
\citet{QianSu2016}. Each selected date is then moved
within a three-year window to the date minimising the post-kink residual
sum of squares, and standard errors are clustered by economy.
Figure~\ref{fig:dg_ic} reports the selection path. The criterion is
minimised at $\widehat\vartheta_1=0.17$ and returns $\widehat K=6$, with
kinks at 2007, 2009, 2010, 2019, 2020 and 2021.

\begin{figure}[htbp]
    \centering
    \includegraphics[width=0.80\linewidth]{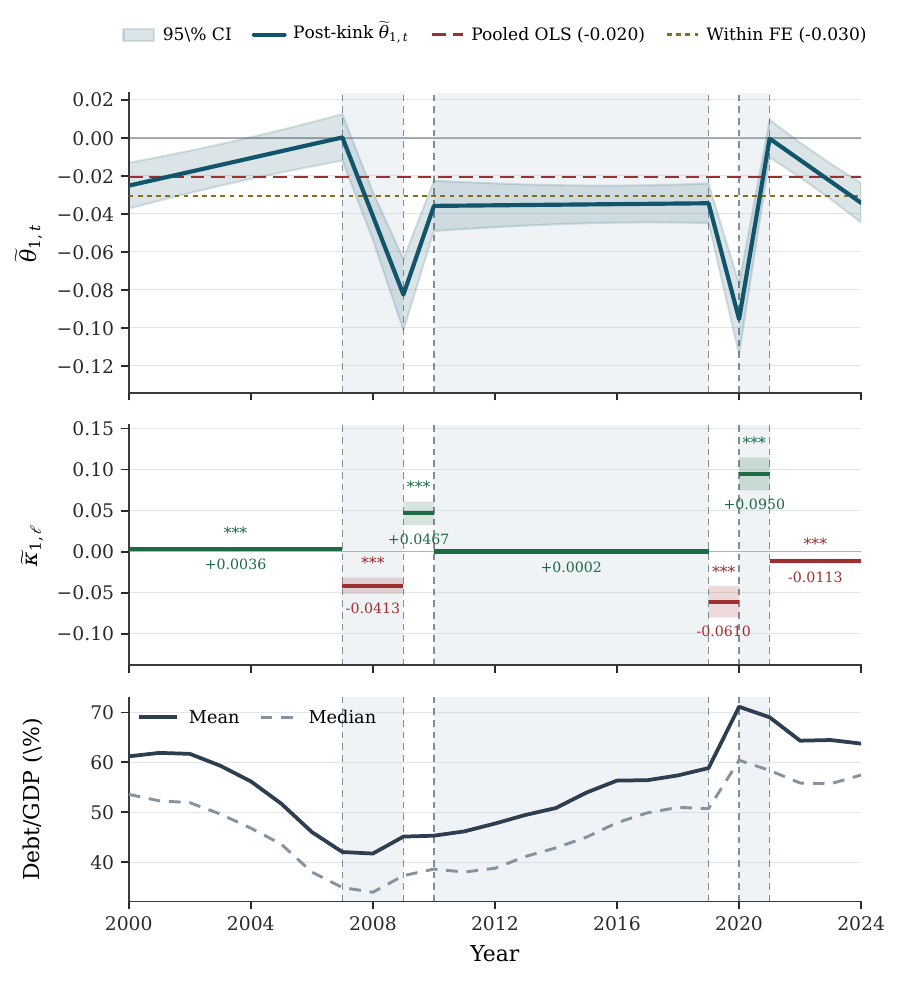}
    \caption{\footnotesize{The debt--growth association, 2000--2024. Upper
    panel: the post-kink debt coefficient with its pointwise 95 percent
    economy-clustered confidence band and the pooled and within
    benchmarks. Middle panel: the within-regime slopes with confidence
    intervals and printed estimates, starred at the one, five and ten
    percent levels. Lower panel: the cross-country mean and median debt
    ratio. Shaded regions are the estimated regimes.}}
    \label{fig:dg_main}
\end{figure}

Figure~\ref{fig:dg_main} reports the estimated path, the within-regime
slopes and the cross-country debt ratio. The coefficient begins at
$-0.025$ and climbs at $0.0036$ per year until it reaches zero in 2007. It
then falls to $-0.083$ by 2009 and returns to $-0.036$ in 2010, after
which it is flat for nine years, the 2010 to 2019 slope being $0.0002$
with a standard error of $0.0007$. The pandemic produces the same fall and
rebound in compressed form. The coefficient drops to $-0.095$ in 2020,
returns to zero in 2021, and then declines at $-0.0113$ per year to
$-0.034$ in 2024. Pooled OLS and the within estimator give $-0.020$ and
$-0.030$, values that run through the middle of the path and conceal both
movements.

Two clusters of dates therefore carry the crisis and the pandemic, and the
estimator distinguishes them by what happens after the rebound rather than
by the size of the fall. Both reach a similar trough. Only the pandemic
movement closes, with the coefficient back at zero within a year, whereas
after 2010 the coefficient settles near $-0.035$ and stays there for the
rest of the decade. A model that assigns a level to each regime can date
the two movements but cannot make this distinction, because it has no
within-regime rate to compare.
 
That flat decade is itself the more surprising feature. It runs from 2010
to 2019, years over which the cross-country mean debt ratio rises from 45
to 59 percent of GDP, and the coefficient does not become more negative
anywhere along the way. This is not a test of a threshold, since the model
dates common changes in calendar time rather than crossings by individual
economies. It does show that a large and sustained rise in average
indebtedness need not be accompanied by a progressively more negative
association, which sits closer to the literature emphasising debt
trajectories and heterogeneous environments than to a stable universal
cutoff \citep{PescatoriSandriSimon2014, EberhardtPresbitero2015,
ChudikMohaddesPesaranRaissi2017}. Over much of the same period the decline
in safe interest rates documented by \citet{Blanchard2019} also made the
economic meaning of any given debt ratio more dependent on the surrounding
environment.
 
\subsection{Coefficient-specific kinks}\label{sec:dg_cw}
 
A regression carrying one time-varying coefficient must assign to it
whatever common movement in the data happens to correlate with its
regressor. The dates in Figure~\ref{fig:dg_main} are therefore dates at
which \emph{something} in the growth process turned, and debt is only the
variable through which the turn is measured. The component-wise estimator
of Section~\ref{sec:component} lets each coefficient carry its own kink
dates, so the question of which variable turned when becomes an estimable
one.
 
\begin{figure}[ht]
    \centering
    \includegraphics[width=0.78\linewidth]{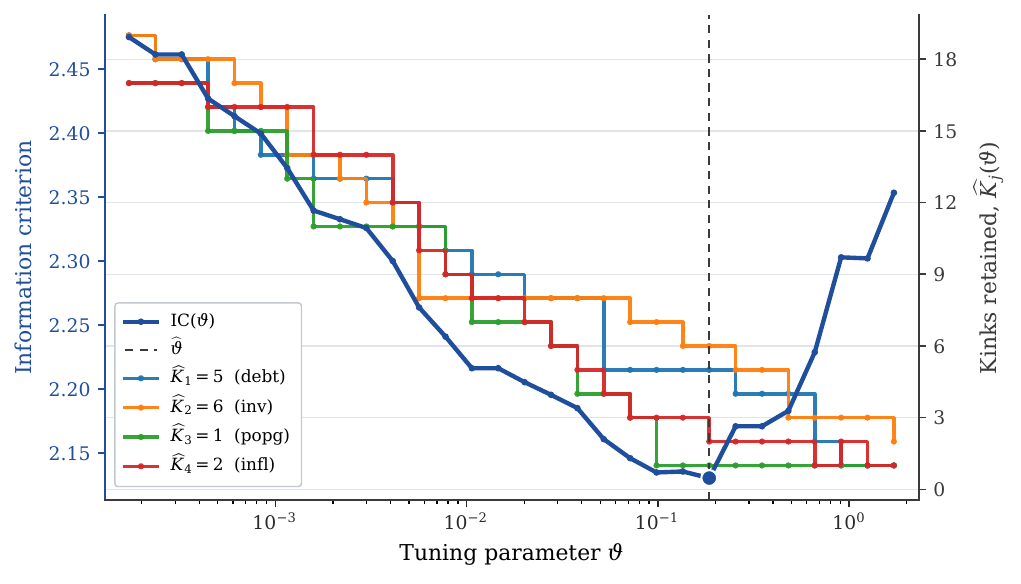}
    \caption{\footnotesize{Selection of the tuning parameter in the
    four-regressor model. Axes as in Figure~\ref{fig:dg_ic}, with one step
    function per coefficient.}}
    \label{fig:dg_cw_ic}
\end{figure}

We add total investment, population growth and inflation. Let
$\*x_{i,t}=[d_{i,t},q_{i,t},n_{i,t},\pi_{i,t}]'$, where $q_{i,t}$ is total
investment as a share of GDP, $n_{i,t}$ is population growth and
$\pi_{i,t}$ is average consumer price inflation, and estimate
\begin{align}
    g_{i,t}
    &=\mu_i+\sum_{j=1}^{4}\theta_{j,t}x_{i,t,j}+\varepsilon_{i,t},
    \label{eq:dg_cw_model}\\[-2pt]
    \theta_{j,t}
    &=\theta_{j,1}+\kappa_{j,1}(t-1)
      +\sum_{m=1}^{K_j}\gamma_{j,m}(t-T_{j,m})_+,
    \qquad
    \gamma_{j,m}:=\kappa_{j,m+1}-\kappa_{j,m},
    \label{eq:dg_cw_theta}
\end{align}
so that coefficient $j$ has its own number of kinks $K_j$, its own dates
and its own regime slopes. Estimation is by the component-wise objective
\eqref{eq:objective_componentwise} with $\zeta_2=2$, everything else as
above. Requiring investment and inflation in every year leaves $N=136$
economies, and regressors are standardised before the penalty is applied.
Figure~\ref{fig:dg_cw_ic} reports the selection, which retains fourteen
kinks: five in debt, at 2006, 2008, 2019, 2020 and 2021, six in
investment, at 2007, 2009, 2010, 2019, 2020 and 2021, two in inflation, at
2008 and 2015, and one in population growth, at 2009.
 
Investment's six dates are exactly the six the scalar model assigned to
debt. The sharp reversal of 2007 to 2010, and the pandemic movement that
follows it, belong to the investment coefficient once investment is allowed
its own clock, and the debt-only regression was measuring them through the
one regressor it had. Debt keeps neither of the two crisis dates. It bends
in 2006 instead, two years before the crisis, and its 2008 kink opens a
recovery rather than a fall.
 
\begin{figure}[p]
    \centering
    \begin{minipage}[t]{0.48\linewidth}\centering
        \includegraphics[width=\linewidth]{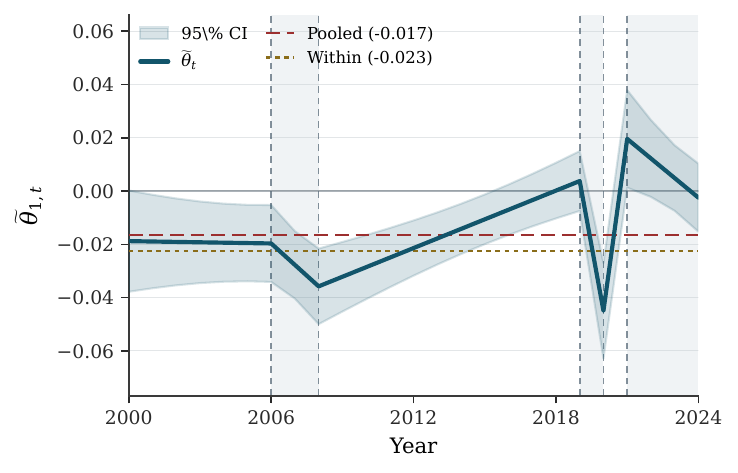}
    \end{minipage}\hfill
    \begin{minipage}[t]{0.48\linewidth}\centering
        \includegraphics[width=\linewidth]{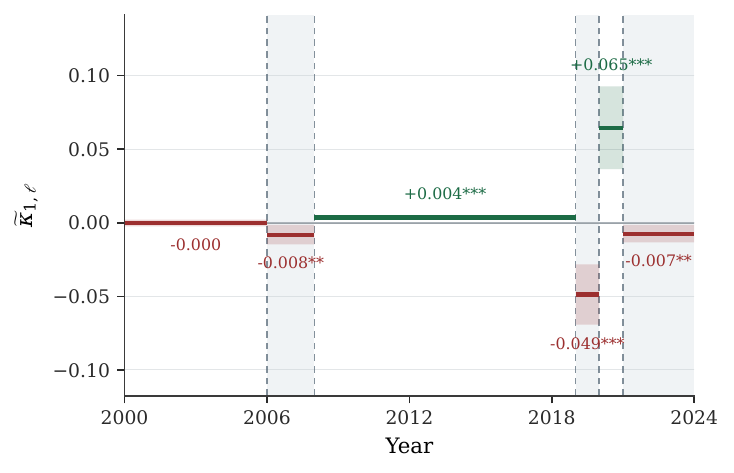}
    \end{minipage}
 
    \vspace{5pt}
    \begin{minipage}[t]{0.48\linewidth}\centering
        \includegraphics[width=\linewidth]{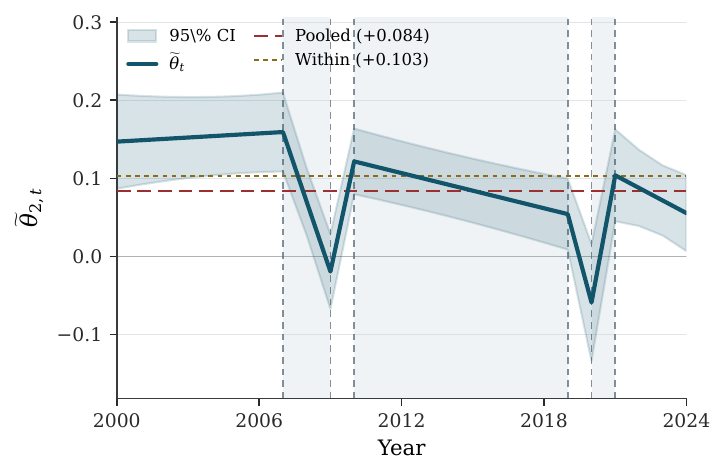}
    \end{minipage}\hfill
    \begin{minipage}[t]{0.48\linewidth}\centering
        \includegraphics[width=\linewidth]{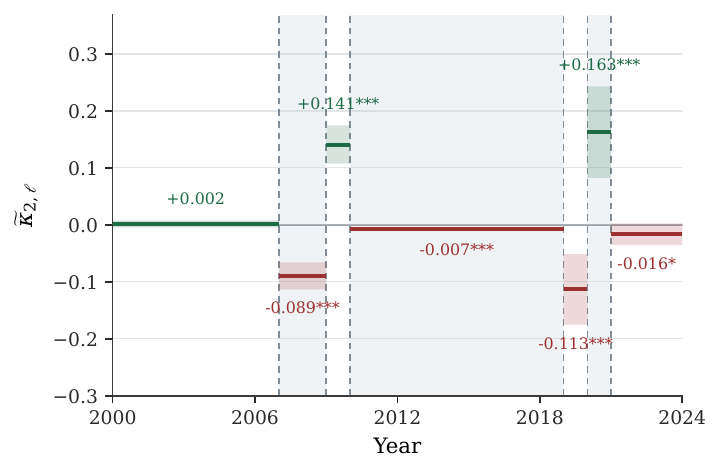}
    \end{minipage}
 
    \vspace{5pt}
    \begin{minipage}[t]{0.48\linewidth}\centering
        \includegraphics[width=\linewidth]{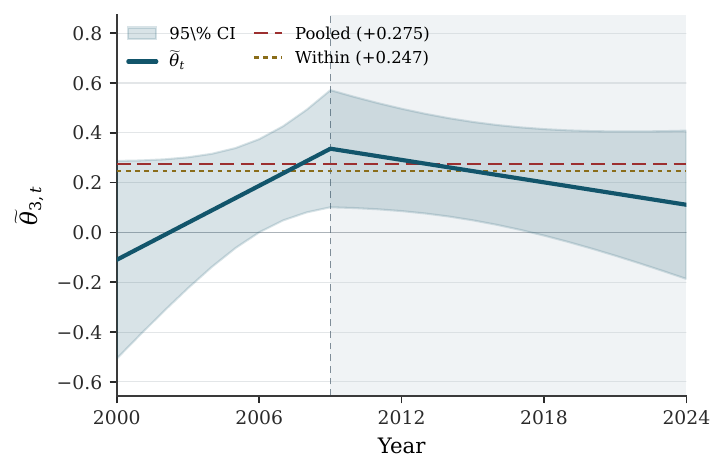}
    \end{minipage}\hfill
    \begin{minipage}[t]{0.48\linewidth}\centering
        \includegraphics[width=\linewidth]{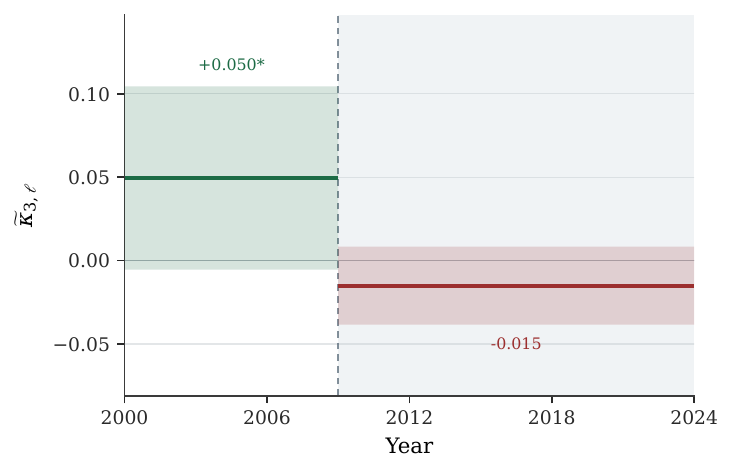}
    \end{minipage}
 
    \vspace{5pt}
    \begin{minipage}[t]{0.48\linewidth}\centering
        \includegraphics[width=\linewidth]{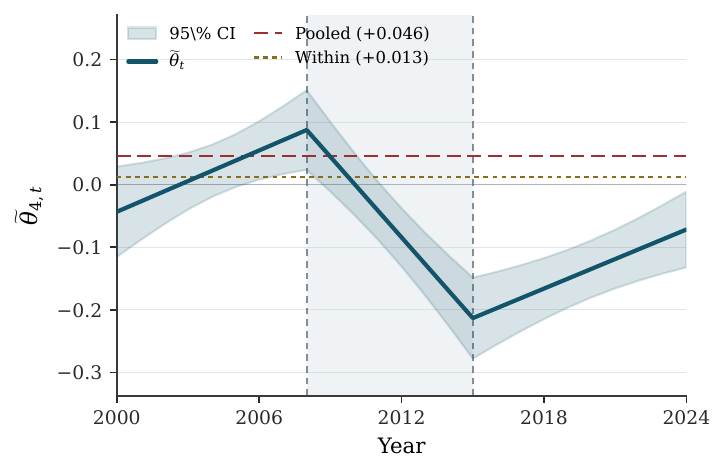}
    \end{minipage}\hfill
    \begin{minipage}[t]{0.48\linewidth}\centering
        \includegraphics[width=\linewidth]{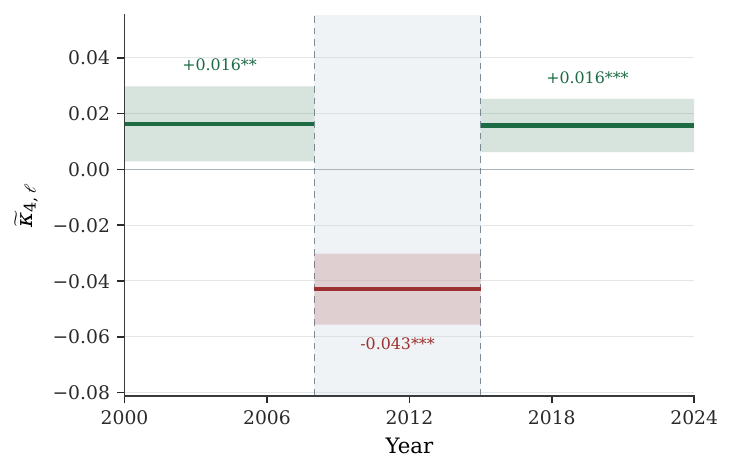}
    \end{minipage}
 
    \caption{\footnotesize{Component-wise estimates, 2000--2024. Rows are
    gross government debt, total investment, population growth and
    inflation. Left column: the coefficient path with its pointwise 95
    percent country-clustered band and the pooled and within benchmarks.
    Right column: the within-regime slopes with confidence intervals and
    printed values, starred at the one, five and ten percent levels. Shaded
    regions are the estimated regimes.}}
    \label{fig:dg_cw_panels}
\end{figure}
 
Figure~\ref{fig:dg_cw_panels} gives the four paths and their slopes. Read
row by row, they say that the crisis moved three of the four coefficients
and that only one of the three came back. Debt is flat at about $-0.019$ through the boom, deteriorates from 2006 at
$-0.008$ per year, and then recovers for eleven years at $+0.004$ per year
to a 2019 value at which $\theta_{1,2019}=0$ cannot be rejected
($\mathcal{W}=0.43$, chi-squared with one degree of freedom throughout).
The rate after 2008 differs from the rate before 2006 ($\mathcal{W}=6.01$)
and the 2006 level is not recovered ($\mathcal{W}=10.81$), so the crisis
put the coefficient on a new path rather than displacing it temporarily.
Around the pandemic the two rates again differ ($\mathcal{W}=10.44$) while
the level is restored ($\mathcal{W}=1.34$, not rejected). The 2006 turn
precedes the crisis, which is where the informative variation sits in the
account of \citet{SchularickTaylor2012}, though their leading indicator is
private credit growth rather than the public debt ratio used here.
 
Investment shows the same asymmetry more sharply. It is flat until 2007,
falls at $-0.089$ per year to 2009 and rebounds at $+0.141$ into 2010, but
the round trip does not close on either margin ($\mathcal{W}=6.86$ on the
rates and $\mathcal{W}=7.13$ on the level). What follows is a new regime
declining at $-0.008$ per year for nine years, the reduced form signature
of the persistent output losses documented by \citet{CerraSaxena2008} and
of the weak capital formation that \citet{Summers2015} places at the centre
of the post-crisis decade. Its pandemic movement, by contrast, closes on
both margins ($\mathcal{W}=0.75$ and $\mathcal{W}=0.00$), which is what
separates a year of suspended investment from a decade of foregone
investment.
 
Inflation supplies the clearest reversal. The coefficient rises at
$+0.016$ per year to 2008, so that before the crisis an economy with above
average inflation was typically an economy running hot, and then falls at
$-0.043$ for seven years to $-0.213$. A single kink covers those seven
years, which is the substantive point, since a sustained reorientation and
a sequence of annual shocks look nothing alike in this parameterisation.
Its duration is compositional. Advanced economies moved to the effective
lower bound with persistent disinflation and weak growth, over a period in
which \citet{BallMazumder2011} and \citet{CoibionGorodnichenko2015} find
the inflation and activity relation much looser than standard
specifications imply, while emerging and developing economies were
completing a long disinflation of their own in which exchange rate
movements pass through more strongly where policy frameworks are weaker
\citep{HaKoseOhnsorge2019}. Economies crossed between the two
configurations a few at a time, which is what produces a slope rather than
a jump. The 2015 kink falls inside the oil price collapse that began in
June 2014 and that \citet{BaumeisterKilian2016} trace largely to demand
and to market specific developments predating it. After 2015 the
coefficient rises at a rate indistinguishable from its pre-2008 rate
($\mathcal{W}=0.00$) and returns to its initial level
($\mathcal{W}=0.41$). The inflation of 2021 and 2022 falls inside that
final regime and generates no kink of its own.

\begin{figure}[ht]
    \centering
    \includegraphics[width=0.8\linewidth]{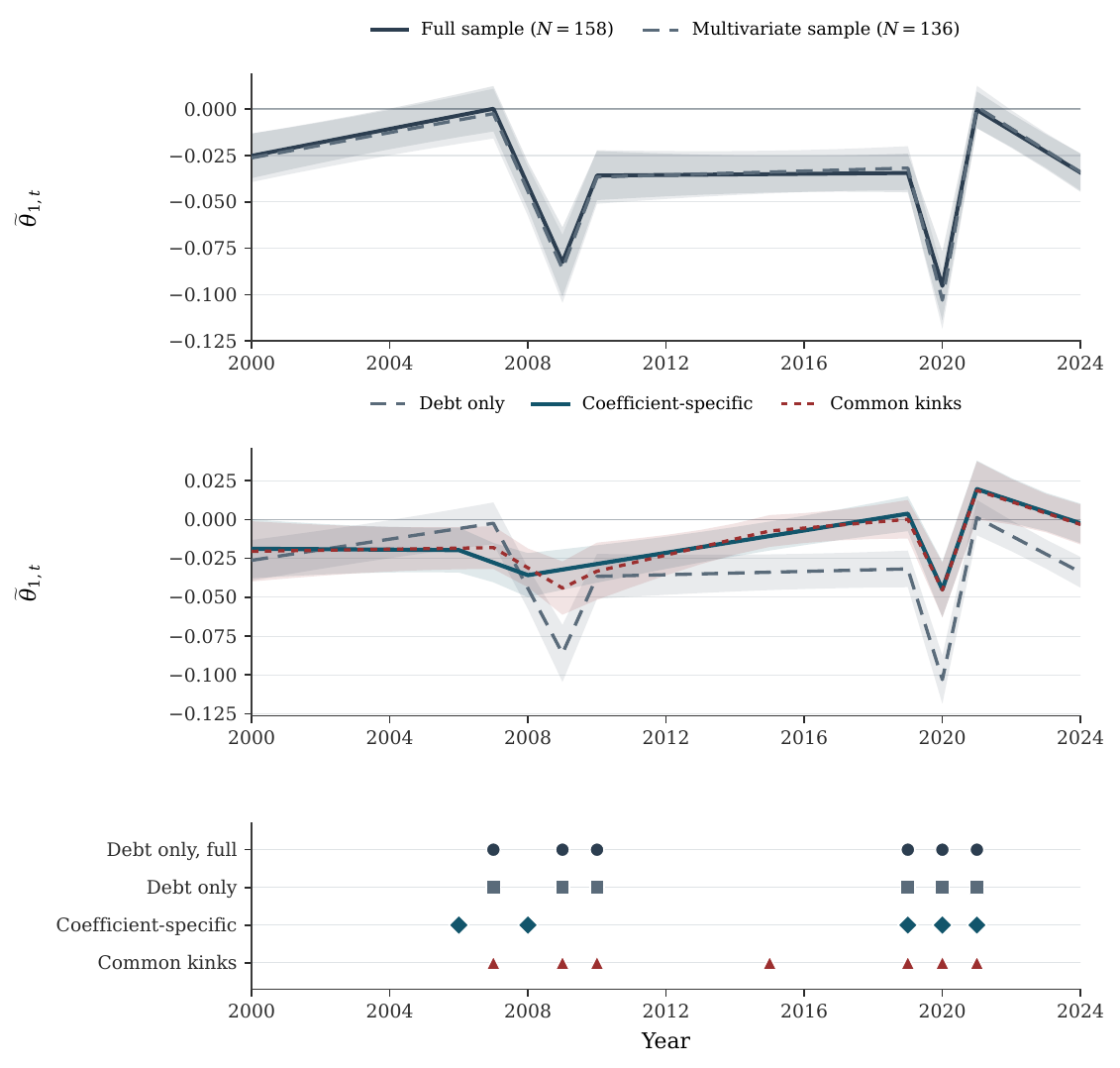}
    \caption{\footnotesize{The debt coefficient across specifications.
    Upper panel: the scalar specification on the full sample and on the
    economies with full covariate coverage. Middle panel: the scalar,
    coefficient-specific and common-kink estimates on the common sample.
    Lower panel: the kink dates each selects. Shaded regions are pointwise
    95 percent country-clustered bands.}}
    \label{fig:dg_cw_overview}
\end{figure}

Population growth carries one weak kink, at 2009. Neither of its two slopes
is significant at the five percent level, the change in rate is significant
only at ten percent ($\mathcal{W}=3.56$), and the initial level is restored
by 2024 ($\mathcal{W}=0.77$). It is the one coefficient here that neither
the crisis nor the pandemic moves.

 \begin{figure}[h!]
    \centering
    \includegraphics[width=0.94\linewidth]{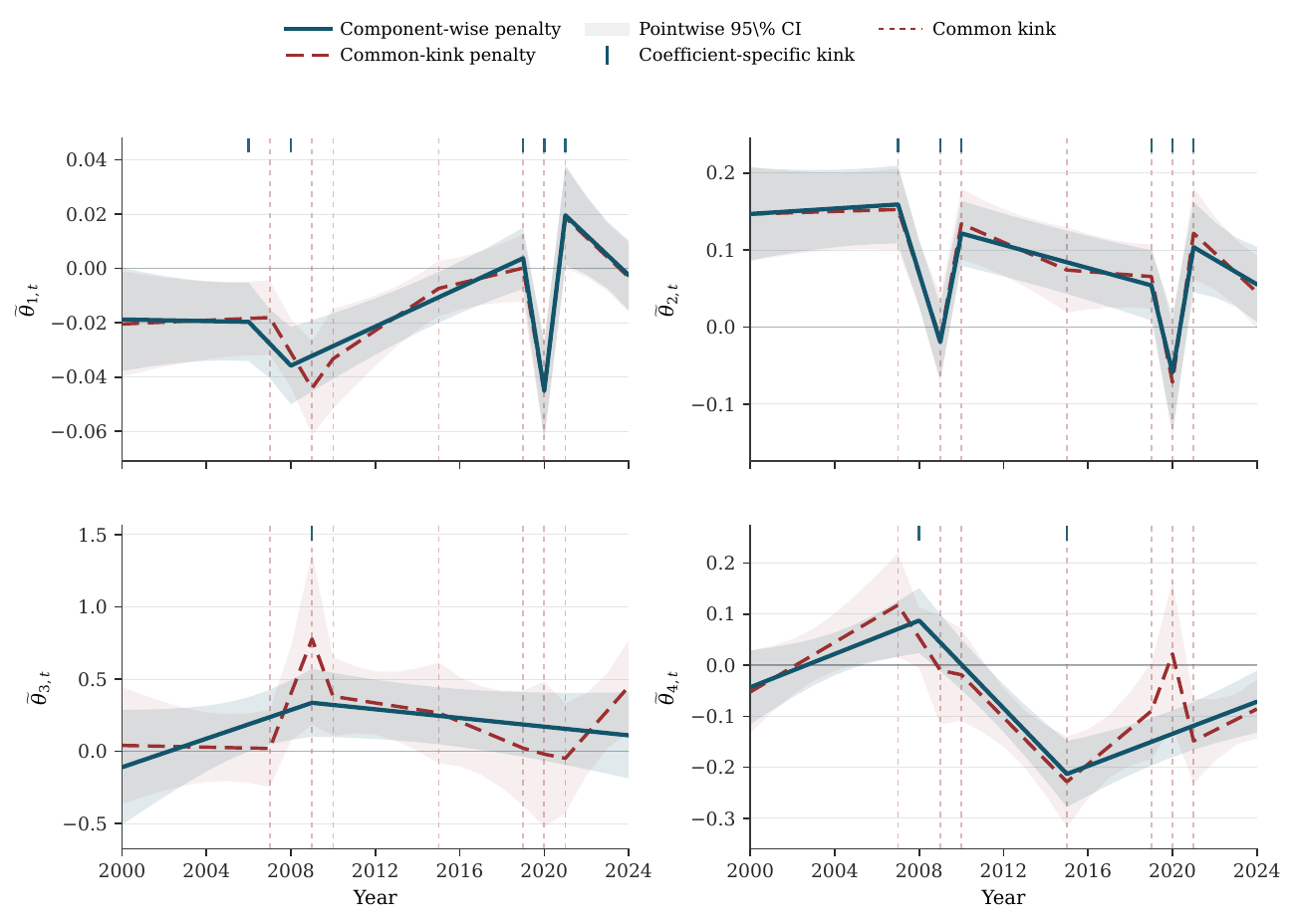}
    \caption{\footnotesize{Component-wise against common-kink estimates in
    the same four-regressor model. Panels are, in reading order, debt,
    investment, population growth and inflation. Solid paths use
    coefficient-specific kink sets, dashed paths the common dates.}}
    \label{fig:dg_cw_compare}
\end{figure}

Figure~\ref{fig:dg_cw_overview} puts the four specifications on one axis
and separates what conditioning does from what the sample does. The sample
restriction is immaterial, since the scalar model estimated on the 136
economies with full covariate coverage selects the same six dates as on the
full sample. Conditioning changes both the timing, as described above, and
the amplitude, halving the 2020 trough from $-0.103$ to $-0.045$. The common-kink penalty, which forces one date set on the whole coefficient
vector, is the more instructive comparison, since it is the restriction our
extension removes. It selects seven dates. Six of them are investment's and
the seventh is inflation's 2015, while debt's own 2006 and 2008 are lost.
The restriction overwrites the timing of precisely the coefficient of
interest, because among the four its signal is the weakest.
Figure~\ref{fig:dg_cw_compare} shows what that costs elsewhere. Population
growth is given six kinks it does not need, and inflation keeps the date
that ends its decline but loses the 2008 date that begins it, which is the
one that organises the whole path.
 
What survives every specification is the shape and the conclusion drawn
from it. The debt coefficient is kinked, it bends around both global
movements, and no constant coefficient describes it. Two caveats bound the
reading. The specification carries no year effects, so a common
macroeconomic movement loads onto whichever coefficient covaries with it,
and \citet{BlanchardLeigh2013} show how tightly output, fiscal policy and
forecast errors were bound together in exactly the years in which these
coefficients move most. And the estimates remain associations. What the
exercise establishes is when each association changed, how fast, and which
of the changes reversed.

\section{Conclusion}\label{sec:conclusion}

This paper studies panel data models in which the coefficients follow continuous piecewise-linear paths with an unknown number of kinks occurring at unknown dates. We estimate the kink dates using an adaptive group fused lasso applied to the second differences of the coefficient paths, show that the true dates are recovered exactly with probability approaching one, and establish the asymptotic normality of the post-kink within estimator. Continuity links adjacent regimes, so each parameter is informed by the regimes on either side of a kink and has its own convergence rate, without requiring balanced regime lengths. A component-wise extension also allows kink dates to differ across regressors. The usefulness of the design and the resulting estimator is demonstrated through an application in macro-finance, specifically by examining whether the relationship between debt and growth changes over calendar time.

\pagebreak

\section*{Declaration of generative AI and AI-assisted technologies in the manuscript preparation process}
During the preparation of this work the authors used Claude and ChatGPT in order to improve the language and presentation of the manuscript, to support the verification of mathematical arguments, and to accelerate the development of simulation code. After using this tool/service, the authors reviewed and edited the content as needed and take full responsibility for the content of the published article.

\pagebreak

\begin{table}[h]
\centering
\small
\caption{Selection performance, common-kink designs.}
\label{tab:mcsel}
\begin{tabular}{@{}cc ccc ccc ccc@{}}
\toprule
&& \multicolumn{3}{c}{DGP~1}
&  \multicolumn{3}{c}{DGP~2}
&  \multicolumn{3}{c}{DGP~3}\\
\cmidrule(lr){3-5}\cmidrule(lr){6-8}\cmidrule(lr){9-11}
$N$ & $T$
& Corr. & $\widehat K\!-\!K^0$ & Miscl.
& Corr. & $\widehat K\!-\!K^0$ & Miscl.
& Corr. & $\widehat K\!-\!K^0$ & Miscl.\\
\midrule
 50 & 20 & 99.3 & 0.01 & 0.4 & 97.8 & 0.01 & 0.9 & 80.0 & 0.07 & 4.3 \\
 50 & 40 & 99.4 & 0.01 & 0.3 & 99.0 & 0.01 & 0.5 & 83.9 & 0.15 & 9.0 \\
 50 & 80 & 99.3 & 0.01 & 0.6 & 98.8 & 0.01 & 0.5 & 82.0 & 0.18 & 8.1 \\
\addlinespace
100 & 20 & 99.4 & 0.01 & 0.3 & 99.5 & 0.00 & 0.3 & 93.9 & 0.05 & 3.0 \\
100 & 40 & 99.9 & 0.00 & 0.0 & 99.5 & 0.01 & 0.2 & 94.5 & 0.06 & 3.4 \\
100 & 80 & 99.8 & 0.00 & 0.1 & 99.1 & 0.01 & 0.6 & 95.6 & 0.04 & 1.7 \\
\addlinespace
200 & 20 & 99.8 & 0.00 & 0.2 & 99.8 & 0.00 & 0.1 & 98.7 & 0.01 & 0.7 \\
200 & 40 & 99.5 & 0.01 & 0.2 & 99.9 & 0.00 & 0.1 & 97.9 & 0.02 & 1.2 \\
200 & 80 & 99.9 & 0.00 & 0.1 & 99.8 & 0.00 & 0.1 & 98.3 & 0.02 & 0.6 \\
\midrule
&& \multicolumn{3}{c}{DGP~5}
&  \multicolumn{3}{c}{DGP~6}
&  \multicolumn{3}{c}{DGP~7}\\
\cmidrule(lr){3-5}\cmidrule(lr){6-8}\cmidrule(lr){9-11}
$N$ & $T$
& Corr. & $\widehat K\!-\!K^0$ & Miscl.
& Corr. & $\widehat K\!-\!K^0$ & Miscl.
& Corr. & $\widehat K\!-\!K^0$ & Miscl.\\
\midrule
 50 & 20 & 81.1 & 0.03 & 1.8 & 94.2 & 0.02 & 1.0 & 70.9 & 0.05 & 3.8 \\
 50 & 40 & 87.1 & 0.11 & 3.5 & 97.5 & 0.02 & 0.9 & 79.8 & 0.14 & 8.7 \\
 50 & 80 & 89.4 & 0.11 & 6.0 & 98.3 & 0.02 & 0.8 & 73.0 & 0.28 & 12.5 \\
\addlinespace
100 & 20 & 94.0 & 0.04 & 1.3 & 99.0 & 0.01 & 0.4 & 88.2 & 0.07 & 3.6 \\
100 & 40 & 93.8 & 0.06 & 1.9 & 99.1 & 0.01 & 0.4 & 85.0 & 0.15 & 9.3 \\
100 & 80 & 95.7 & 0.04 & 3.0 & 98.9 & 0.01 & 0.5 & 88.0 & 0.12 & 5.5 \\
\addlinespace
200 & 20 & 97.5 & 0.03 & 0.9 & 99.3 & 0.01 & 0.3 & 96.3 & 0.03 & 1.9 \\
200 & 40 & 99.2 & 0.01 & 0.2 & 99.6 & 0.00 & 0.2 & 94.2 & 0.06 & 3.7 \\
200 & 80 & 99.0 & 0.01 & 0.7 & 99.7 & 0.00 & 0.2 & 96.2 & 0.04 & 1.5 \\
\bottomrule
\end{tabular}
\par\smallskip
\begin{minipage}{0.92\textwidth}\footnotesize
\emph{Notes}: $1000$ replications. Corr.\ is the frequency, in
percent, of exact recovery of the true kink dates.
$\widehat K-K^0$ is the average kink-count bias. Miscl.\ is the
average percentage of dates assigned to the wrong regime, so that a
kink located one date off misclassifies $1/T$ of the sample while a
spurious interior kink misclassifies the share of dates beyond it.
DGP~5 has uneven regime lengths; DGPs~6 and~7 repeat DGPs~2 and~3
under $(\varpi_1,\varpi_2)=(0.5,0.2)$.
\end{minipage}
\end{table}

\pagebreak

\begin{table}[h]
\centering
\small
\caption{Selection performance of DGP~4 using component-wise estimator.}
\label{tab:mcselcw}
\begin{tabular}{@{}cc ccc ccc@{}}
\toprule
&& \multicolumn{3}{c}{Coordinate 1}
&  \multicolumn{3}{c}{Coordinate 2}\\
\cmidrule(lr){3-5}\cmidrule(lr){6-8}
$N$ & $T$
& Corr. & $\widehat K_1\!-\!K_1^0$ & Miscl.
& Corr. & $\widehat K_2\!-\!K_2^0$ & Miscl.\\
\midrule
 50 & 20 & 91.5 & 0.07 & 4.5 & 90.0 & 0.10 & 4.6 \\
 50 & 40 & 92.3 & 0.08 & 4.9 & 90.7 & 0.10 & 4.0 \\
 50 & 80 & 91.2 & 0.09 & 5.4 & 90.9 & 0.10 & 3.4 \\
\addlinespace
100 & 20 & 93.8 & 0.06 & 3.8 & 94.2 & 0.06 & 2.7 \\
100 & 40 & 93.3 & 0.07 & 4.0 & 93.8 & 0.06 & 2.6 \\
100 & 80 & 96.2 & 0.04 & 2.2 & 95.9 & 0.04 & 1.8 \\
\addlinespace
200 & 20 & 97.3 & 0.03 & 1.3 & 96.2 & 0.04 & 1.6 \\
200 & 40 & 97.1 & 0.03 & 1.8 & 97.3 & 0.03 & 1.2 \\
200 & 80 & 98.0 & 0.02 & 1.1 & 97.1 & 0.03 & 1.2 \\
\bottomrule
\end{tabular}
\par\smallskip
\begin{minipage}{0.85\textwidth}\footnotesize
\emph{Notes}: $1000$ replications. Statistics are as in
Table~\ref{tab:mcsel}, computed per coordinate.
\end{minipage}
\end{table}

\pagebreak

\begin{table}[h]
\centering
\small
\caption{Selection performance of DGP~8 using the component-wise
estimator.}
\label{tab:mcselcw8}
\begin{tabular}{@{}cc ccc ccc ccc@{}}
\toprule
&& \multicolumn{3}{c}{Coordinate 1 ($K_1^0=0$)}
&  \multicolumn{3}{c}{Coordinate 2 ($K_2^0=1$)}
&  \multicolumn{3}{c}{Coordinate 3 ($K_3^0=2$)}\\
\cmidrule(lr){3-5}\cmidrule(lr){6-8}\cmidrule(lr){9-11}
$N$ & $T$
& Corr. & $\widehat K_1\!-\!K_1^0$ & Miscl.
& Corr. & $\widehat K_2\!-\!K_2^0$ & Miscl.
& Corr. & $\widehat K_3\!-\!K_3^0$ & Miscl.\\
\midrule
 50 & 20 & 95.0 & 0.05 & 2.7 & 90.7 & 0.10 & 5.2 & 61.2 & 0.20 & 10.6 \\
 50 & 40 & 95.5 & 0.04 & 2.4 & 92.4 & 0.08 & 4.3 & 64.1 & 0.36 & 18.9 \\
 50 & 80 & 96.4 & 0.04 & 1.8 & 90.6 & 0.10 & 5.0 & 69.8 & 0.35 & 15.4 \\
\addlinespace
100 & 20 & 97.7 & 0.03 & 1.4 & 93.7 & 0.07 & 3.6 & 79.4 & 0.17 & 9.1 \\
100 & 40 & 97.3 & 0.03 & 1.2 & 95.1 & 0.05 & 2.4 & 81.7 & 0.20 & 10.3 \\
100 & 80 & 96.9 & 0.03 & 1.6 & 94.5 & 0.06 & 2.7 & 83.7 & 0.18 & 7.5 \\
\addlinespace
200 & 20 & 97.6 & 0.03 & 1.1 & 97.5 & 0.03 & 1.1 & 91.5 & 0.09 & 4.6 \\
200 & 40 & 98.3 & 0.02 & 0.7 & 97.2 & 0.03 & 1.5 & 93.2 & 0.07 & 3.7 \\
200 & 80 & 98.0 & 0.02 & 1.0 & 97.7 & 0.02 & 1.0 & 93.4 & 0.07 & 3.3 \\
\bottomrule
\end{tabular}
\par\smallskip
\begin{minipage}{0.90\textwidth}\footnotesize
\emph{Notes}: $1000$ replications. Statistics are as in
Table~\ref{tab:mcsel}, computed per coordinate. The three coordinates
carry a different number of kinks, so the design tests whether the
estimator recovers heterogeneous kink counts, including none in the
first coordinate.
\end{minipage}
\end{table}

\pagebreak

\begin{table}[h]
\centering
\small
\caption{Bias and RMSE of the slope and path estimates,
DGPs~1 to~3.}
\label{tab:mcest}
\begin{tabular}{@{}cc c ccc cccc@{}}
\toprule
&& DGP~1
&  \multicolumn{3}{c}{DGP~2}
&  \multicolumn{4}{c}{DGP~3}\\
\cmidrule(lr){3-3}\cmidrule(lr){4-6}\cmidrule(lr){7-10}
$N$ & $T$ & Path
& $\+\kappa_1$ & $\+\kappa_2$ & Path
& $\+\kappa_1$ & $\+\kappa_2$ & $\+\kappa_3$ & Path\\
\midrule
\multicolumn{10}{@{}l}{\emph{Panel A: Bias} ($\times1000$)}\\
\midrule
 50 & 20 & $-0.89$ & $-0.11$ & 0.19 & $-0.77$ & $-0.56$ & 0.62 & $-0.81$ & 0.11 \\
 50 & 40 & $-0.22$ & 0.07 & $-0.06$ & 0.57 & $-0.47$ & 0.13 & 0.12 & $-0.38$ \\
 50 & 80 & $-0.22$ & $-0.08$ & 0.08 & $-0.76$ & $-0.15$ & 0.00 & 0.07 & $-0.36$ \\
\addlinespace
100 & 20 & $-0.46$ & $-0.12$ & $-0.30$ & $-0.09$ & 0.23 & 0.36 & $-0.34$ & $-0.07$ \\
100 & 40 & 0.50 & 0.00 & $-0.09$ & $-0.15$ & $-0.29$ & 0.06 & 0.01 & $-0.02$ \\
100 & 80 & 0.23 & 0.01 & $-0.04$ & $-0.03$ & $-0.03$ & 0.04 & $-0.02$ & 0.00 \\
\addlinespace
200 & 20 & $-0.14$ & $-0.14$ & 0.01 & 0.15 & $-0.04$ & $-0.33$ & 0.46 & $-0.81$ \\
200 & 40 & $-0.55$ & 0.07 & $-0.07$ & $-0.44$ & $-0.02$ & $-0.02$ & 0.00 & $-0.47$ \\
200 & 80 & $-0.41$ & 0.00 & 0.00 & 0.11 & 0.00 & 0.02 & 0.00 & 0.09 \\
\midrule
\multicolumn{10}{@{}l}{\emph{Panel B: RMSE}}\\
\midrule
 50 & 20 & 0.049 & 0.014 & 0.012 & 0.062 & 0.030 & 0.019 & 0.021 & 0.082 \\
 50 & 40 & 0.035 & 0.005 & 0.005 & 0.042 & 0.009 & 0.007 & 0.007 & 0.054 \\
 50 & 80 & 0.024 & 0.002 & 0.002 & 0.029 & 0.003 & 0.002 & 0.003 & 0.038 \\
\addlinespace
100 & 20 & 0.035 & 0.010 & 0.009 & 0.041 & 0.021 & 0.013 & 0.014 & 0.051 \\
100 & 40 & 0.024 & 0.003 & 0.003 & 0.029 & 0.006 & 0.005 & 0.005 & 0.035 \\
100 & 80 & 0.017 & 0.001 & 0.001 & 0.021 & 0.002 & 0.002 & 0.002 & 0.025 \\
\addlinespace
200 & 20 & 0.024 & 0.007 & 0.006 & 0.030 & 0.015 & 0.009 & 0.010 & 0.034 \\
200 & 40 & 0.017 & 0.002 & 0.002 & 0.021 & 0.004 & 0.004 & 0.004 & 0.025 \\
200 & 80 & 0.012 & 0.001 & 0.001 & 0.015 & 0.002 & 0.001 & 0.001 & 0.017 \\
\bottomrule
\end{tabular}
\par\smallskip
\begin{minipage}{0.92\textwidth}\footnotesize
\emph{Notes}: $1000$ replications. Bias entries are multiplied by
$1000$. Slope statistics are per regime, pooled across the two
coordinates, and conditional on exact recovery of the kink set. Path
statistics are averaged over dates and coordinates and are
unconditional. In DGP~1 the fitted path under $\widehat K=0$ is a
single linear regime.
\end{minipage}
\end{table}

\pagebreak
\begin{table}[h]
\centering
\footnotesize
\caption{Bias and RMSE of the slope and path estimates, DGPs~5 to~7.}
\label{tab:mcestrob}
\begin{tabular}{@{}cc cccc ccc cccc@{}}
\toprule
&& \multicolumn{4}{c}{DGP~5}
&  \multicolumn{3}{c}{DGP~6}
&  \multicolumn{4}{c}{DGP~7}\\
\cmidrule(lr){3-6}\cmidrule(lr){7-9}\cmidrule(lr){10-13}
$N$ & $T$
& $\+\kappa_1$ & $\+\kappa_2$ & $\+\kappa_3$ & Path
& $\+\kappa_1$ & $\+\kappa_2$ & Path
& $\+\kappa_1$ & $\+\kappa_2$ & $\+\kappa_3$ & Path\\
\midrule
\multicolumn{13}{@{}l}{\emph{Panel A: Bias} ($\times1000$)}\\
\midrule
 50 & 20 & $-0.27$ & 0.15 & $-0.73$ & 0.32 & $-0.23$ & 0.82 & $-1.05$ & $-1.61$ & 1.01 & $-0.17$ & 1.03 \\
 50 & 40 & 0.58 & $-0.14$ & 0.63 & 1.05 & 0.04 & 0.04 & 0.31 & 0.27 & 0.00 & 0.14 & 0.47 \\
 50 & 80 & $-0.13$ & 0.07 & $-0.28$ & $-0.34$ & 0.03 & $-0.11$ & 0.55 & 0.04 & $-0.05$ & $-0.01$ & $-0.19$ \\
\addlinespace
100 & 20 & 0.24 & $-0.10$ & $-0.01$ & 0.27 & $-0.22$ & 0.60 & 0.11 & $-1.60$ & 0.44 & $-0.15$ & $-0.51$ \\
100 & 40 & $-0.09$ & 0.00 & $-0.15$ & 0.50 & 0.00 & $-0.18$ & $-0.28$ & $-0.25$ & $-0.12$ & 0.10 & 0.22 \\
100 & 80 & 0.03 & $-0.01$ & $-0.02$ & 0.25 & $-0.02$ & $-0.02$ & 0.20 & 0.12 & $-0.06$ & $-0.01$ & 0.13 \\
\addlinespace
200 & 20 & 0.48 & 0.13 & $-0.43$ & 0.22 & 0.35 & $-0.12$ & $-0.09$ & $-0.28$ & $-0.22$ & 0.49 & 0.67 \\
200 & 40 & $-0.08$ & 0.01 & $-0.11$ & $-0.17$ & 0.03 & $-0.08$ & $-0.60$ & 0.19 & $-0.04$ & 0.02 & 0.07 \\
200 & 80 & 0.08 & $-0.02$ & 0.07 & $-0.12$ & 0.02 & $-0.01$ & $-0.42$ & $-0.02$ & 0.05 & $-0.04$ & 0.21 \\
\midrule
\multicolumn{13}{@{}l}{\emph{Panel B: RMSE}}\\
\midrule
 50 & 20 & 0.051 & 0.010 & 0.037 & 0.080 & 0.017 & 0.016 & 0.079 & 0.036 & 0.022 & 0.024 & 0.100 \\
 50 & 40 & 0.016 & 0.004 & 0.014 & 0.054 & 0.006 & 0.006 & 0.054 & 0.011 & 0.009 & 0.010 & 0.070 \\
 50 & 80 & 0.005 & 0.001 & 0.005 & 0.036 & 0.002 & 0.002 & 0.038 & 0.004 & 0.003 & 0.004 & 0.051 \\
\addlinespace
100 & 20 & 0.036 & 0.007 & 0.026 & 0.051 & 0.012 & 0.011 & 0.051 & 0.026 & 0.016 & 0.018 & 0.067 \\
100 & 40 & 0.012 & 0.003 & 0.010 & 0.036 & 0.004 & 0.004 & 0.037 & 0.008 & 0.006 & 0.007 & 0.047 \\
100 & 80 & 0.004 & 0.001 & 0.004 & 0.025 & 0.002 & 0.001 & 0.027 & 0.003 & 0.002 & 0.002 & 0.033 \\
\addlinespace
200 & 20 & 0.026 & 0.005 & 0.019 & 0.035 & 0.008 & 0.008 & 0.037 & 0.018 & 0.011 & 0.013 & 0.044 \\
200 & 40 & 0.008 & 0.002 & 0.007 & 0.024 & 0.003 & 0.003 & 0.027 & 0.006 & 0.005 & 0.005 & 0.032 \\
200 & 80 & 0.003 & 0.001 & 0.003 & 0.017 & 0.001 & 0.001 & 0.019 & 0.002 & 0.002 & 0.002 & 0.023 \\
\bottomrule
\end{tabular}
\par\smallskip
\begin{minipage}{0.92\textwidth}\footnotesize
\emph{Notes}: $1000$ replications. Bias entries are multiplied by
$1000$. Slope statistics are per regime, pooled across the two
coordinates, and conditional on exact recovery of the kink set. Path
statistics are averaged over dates and coordinates and are
unconditional. In DGP~5 the outer regimes are short and the middle
regime long, so $\+\kappa_2$ is informed by two long adjacent regimes
while $\+\kappa_1$ and $\+\kappa_3$ are informed by a short own
regime. DGPs~6 and~7 repeat DGPs~2 and~3 under
$(\varpi_1,\varpi_2)=(0.5,0.2)$.
\end{minipage}
\end{table}

\pagebreak

\begin{table}[h]
\centering
\small
\caption{Bias and RMSE of DGP~4 using component-wise estimator.}
\label{tab:mccw}
\begin{tabular}{@{}cc ccc ccc@{}}
\toprule
&& \multicolumn{3}{c}{Coordinate 1}
&  \multicolumn{3}{c}{Coordinate 2}\\
\cmidrule(lr){3-5}\cmidrule(lr){6-8}
$N$ & $T$
& Regime 1 & Regime 2 & Path
& Regime 1 & Regime 2 & Path\\
\midrule
\multicolumn{8}{@{}l}{\emph{Panel A: Bias} ($\times1000$)}\\
\midrule
 50 & 20 & $-1.21$ & 0.07 & $-0.31$ & $-0.13$ & 0.34 & $-1.40$ \\
 50 & 40 & $-0.05$ & 0.12 & $-0.62$ & 0.09 & 0.05 & $-0.87$ \\
 50 & 80 & $-0.08$ & 0.05 & 0.33 & 0.01 & $-0.09$ & $-0.58$ \\
\addlinespace
100 & 20 & $-0.35$ & $-0.09$ & 0.74 & 0.17 & $-0.61$ & 0.04 \\
100 & 40 & $-0.09$ & $-0.02$ & 0.69 & $-0.05$ & $-0.13$ & 0.18 \\
100 & 80 & $-0.05$ & 0.02 & $-0.08$ & $-0.02$ & 0.09 & $-0.13$ \\
\addlinespace
200 & 20 & $-0.32$ & 0.07 & $-0.62$ & $-0.06$ & $-0.23$ & 0.24 \\
200 & 40 & $-0.03$ & 0.02 & $-0.64$ & $-0.09$ & 0.22 & $-0.41$ \\
200 & 80 & 0.02 & $-0.01$ & $-0.07$ & $-0.03$ & 0.02 & 0.01 \\
\midrule
\multicolumn{8}{@{}l}{\emph{Panel B: RMSE}}\\
\midrule
 50 & 20 & 0.027 & 0.009 & 0.068 & 0.010 & 0.019 & 0.068 \\
 50 & 40 & 0.008 & 0.003 & 0.046 & 0.003 & 0.007 & 0.048 \\
 50 & 80 & 0.003 & 0.001 & 0.033 & 0.001 & 0.003 & 0.033 \\
\addlinespace
100 & 20 & 0.020 & 0.006 & 0.047 & 0.007 & 0.013 & 0.045 \\
100 & 40 & 0.006 & 0.002 & 0.032 & 0.002 & 0.005 & 0.032 \\
100 & 80 & 0.002 & 0.001 & 0.022 & 0.001 & 0.002 & 0.022 \\
\addlinespace
200 & 20 & 0.014 & 0.004 & 0.031 & 0.005 & 0.009 & 0.031 \\
200 & 40 & 0.004 & 0.002 & 0.022 & 0.002 & 0.003 & 0.022 \\
200 & 80 & 0.001 & 0.001 & 0.015 & 0.001 & 0.001 & 0.015 \\
\bottomrule
\end{tabular}
\par\smallskip
\begin{minipage}{0.85\textwidth}\footnotesize
\emph{Notes}: $1000$ replications. Bias entries are multiplied by
$1000$. Slope statistics are conditional on exact recovery of that
coordinate's kink set. Path statistics are averaged over dates and
are unconditional.
\end{minipage}
\end{table}

\pagebreak

\begin{table}[htbp]
\centering
\caption{Bias and RMSE of DGP~8 using the component-wise estimator.}
\label{tab:mccw8}
\begingroup
\footnotesize
\setlength{\tabcolsep}{3.5pt}
\renewcommand{\arraystretch}{1.05}
\begin{adjustbox}{max width=\textwidth}
\begin{tabular}{@{}cc cc ccc cccc@{}}
\toprule
&& \multicolumn{2}{c}{Coordinate 1}
&  \multicolumn{3}{c}{Coordinate 2}
&  \multicolumn{4}{c}{Coordinate 3} \\
\cmidrule(lr){3-4}
\cmidrule(lr){5-7}
\cmidrule(lr){8-11}
$N$ & $T$
& Regime 1 & Path
& Regime 1 & Regime 2 & Path
& Regime 1 & Regime 2 & Regime 3 & Path \\
\midrule
\multicolumn{11}{@{}l}{\emph{Panel A: Bias} ($\times1000$)} \\
\midrule
 50 & 20 & $-0.11$ & $-0.74$ & 0.25 & 0.27 & $-0.04$ & $-0.15$ & 1.34 & $-1.08$ & $-0.52$ \\
 50 & 40 & 0.08 & 0.01 & 0.09 & $-0.11$ & 0.81 & $-0.18$ & 0.08 & $-0.20$ & 0.13 \\
 50 & 80 & 0.00 & 0.28 & $-0.04$ & $-0.08$ & $-0.39$ & $-0.10$ & 0.04 & $-0.17$ & $-1.16$ \\
\addlinespace
100 & 20 & 0.03 & 0.58 & $-0.09$ & $-0.25$ & $-0.38$ & $-1.82$ & 0.94 & $-1.30$ & $-0.83$ \\
100 & 40 & 0.00 & $-0.39$ & $-0.13$ & 0.06 & 0.83 & $-0.31$ & 0.31 & $-0.24$ & 0.32 \\
100 & 80 & 0.01 & $-0.13$ & 0.04 & $-0.05$ & $-0.25$ & $-0.15$ & 0.12 & $-0.14$ & $-0.03$ \\
\addlinespace
200 & 20 & $-0.12$ & $-0.39$ & 0.41 & 0.08 & 0.08 & 0.02 & 0.17 & 0.12 & 0.68 \\
200 & 40 & 0.05 & 0.10 & $-0.05$ & 0.05 & $-0.26$ & $-0.05$ & $-0.03$ & 0.15 & 0.23 \\
200 & 80 & $-0.01$ & 0.22 & $-0.01$ & 0.01 & 0.01 & $-0.01$ & $-0.01$ & $-0.01$ & $-0.14$ \\
\midrule
\multicolumn{11}{@{}l}{\emph{Panel B: RMSE}} \\
\midrule
 50 & 20 & 0.006 & 0.055 & 0.014 & 0.012 & 0.067 & 0.030 & 0.017 & 0.020 & 0.096 \\
 50 & 40 & 0.002 & 0.037 & 0.005 & 0.004 & 0.046 & 0.009 & 0.007 & 0.008 & 0.068 \\
 50 & 80 & 0.001 & 0.026 & 0.002 & 0.002 & 0.033 & 0.003 & 0.002 & 0.003 & 0.045 \\
\addlinespace
100 & 20 & 0.004 & 0.036 & 0.010 & 0.009 & 0.046 & 0.021 & 0.012 & 0.014 & 0.062 \\
100 & 40 & 0.001 & 0.025 & 0.003 & 0.003 & 0.032 & 0.006 & 0.005 & 0.005 & 0.041 \\
100 & 80 & 0.001 & 0.018 & 0.001 & 0.001 & 0.023 & 0.002 & 0.002 & 0.002 & 0.028 \\
\addlinespace
200 & 20 & 0.003 & 0.026 & 0.007 & 0.006 & 0.031 & 0.016 & 0.009 & 0.010 & 0.038 \\
200 & 40 & 0.001 & 0.018 & 0.002 & 0.002 & 0.022 & 0.005 & 0.003 & 0.004 & 0.027 \\
200 & 80 & 0.000 & 0.012 & 0.001 & 0.001 & 0.015 & 0.001 & 0.001 & 0.001 & 0.018 \\
\bottomrule
\end{tabular}
\end{adjustbox}
\endgroup
\par\smallskip
\begin{minipage}{\textwidth}
\footnotesize
\emph{Notes}: Based on $1{,}000$ replications. Bias entries are
multiplied by $1000$. Slope statistics are conditional on exact
recovery of the corresponding coordinate's kink set. Path statistics
are averaged over dates and are unconditional. Coordinate~1 is
globally linear and therefore has a single regime slope.
\end{minipage}
\end{table}

\pagebreak

\bibliography{kink_paper}

\appendix
\numberwithin{equation}{section}
\section{Proofs of the main results}\label{app:proofs}

\section*{Proof of Theorem~\ref{thm:consistency}.}

\noindent
\begin{proof}
Define $\*b_t=\sqrt N(\+\theta_t-\+\theta_t^0)$,
$\dot{\*b}_t=\sqrt N(\dot{\+\theta}_t-\+\theta_t^0)$,
$\widehat{\*b}_t=\sqrt N(\widehat{\+\theta}_t-\+\theta_t^0)$,
$\*b=[\*b_1',\ldots,\*b_T']'$, and
\begin{align}
    \bar b
    = \frac{1}{\sqrt T}
    \sqrt{(T-1)\|\*b_1\|^2+\sum_{t=2}^T\|\*b_t\|^2}.
    \label{eq:bbar}
\end{align}
Note that the scaling by $T^{-1/2}$ here is necessary to ensure that
$\bar b=O_p(1)$ if $\|\*b_t\|=O_p(1)$ for all $t$. The proof is in two
steps. Step 1 establishes the energy bound and the design matrix bounds
for the preliminary estimator $\dot{\+\theta}_t$ of \eqref{eq:initial},
which enters the penalised problem through the adaptive weights. Step 2
establishes the bound $\widehat{\bar b}=O_p(1)$ for the penalised
estimator and reads the three parts of the theorem off it.

We start by recalling that, by construction, for $i=1,\ldots,N$ and
$t=2,\ldots,T$,
\begin{align}
    \check y_{i,t} = \*x_{i,t}'\+\theta_t^0-\*x_{i,1}'\+\theta_1^0
    +\check\varepsilon_{i,t}.
\end{align}
This implies
\begin{align}
    \check y_{i,t}-\*x_{i,t}'\+\theta_t+\*x_{i,1}'\+\theta_1
    = \check\varepsilon_{i,t}
    -\frac{1}{\sqrt N}\bigl(\*x_{i,t}'\*b_t-\*x_{i,1}'\*b_1\bigr).
    \label{eq:resid}
\end{align}
Write $L_{NT}(\+\Theta_T)$ for the unpenalised objective in
\eqref{eq:initial}, normalised by $1/(NT)$. We now use \eqref{eq:resid}
to evaluate $N[L_{NT}(\+\Theta_T)-L_{NT}(\+\Theta_T^0)]$. Expanding the
square,
\begin{align}
    N\bigl[L_{NT}(\+\Theta_T)-L_{NT}(\+\Theta_T^0)\bigr]
    &= \frac1T\sum_{i=1}^N\sum_{t=2}^T
    \Bigl(\check\varepsilon_{i,t}
    -\frac{1}{\sqrt N}\bigl[\*x_{i,t}'\*b_t-\*x_{i,1}'\*b_1\bigr]\Bigr)^2
    \notag\\
    &\quad
    -\frac1T\sum_{i=1}^N\sum_{t=2}^T\check\varepsilon_{i,t}^2 \notag\\
    &= -\frac{2}{T\sqrt N}\sum_{i=1}^N\sum_{t=2}^T
    \check\varepsilon_{i,t}\bigl(\*x_{i,t}'\*b_t-\*x_{i,1}'\*b_1\bigr)
    \notag\\
    &\quad
    +\frac{1}{TN}\sum_{i=1}^N\sum_{t=2}^T
    \bigl(\*x_{i,t}'\*b_t-\*x_{i,1}'\*b_1\bigr)^2 \notag\\
    &=: -2L_1(\*b)+L_2(\*b),
    \label{eq:prelimdecomp}
\end{align}
with implicit definitions of $L_1(\*b)$ and $L_2(\*b)$. We bound the two
terms in turn. The same two bounds will be reused later on since the loss part of the penalised objective is identical.
Consider the term $L_1(\*b)$. Collecting the sum by parameter block, if
we introduce
\begin{align}
    \*E_N = \left[\begin{array}{c}
    -\dfrac{1}{\sqrt N}\displaystyle\sum_{t=2}^T\sum_{i=1}^N
    \check\varepsilon_{i,t}\*x_{i,1}\\[12pt]
    \dfrac{1}{\sqrt N}\displaystyle\sum_{i=1}^N
    \check\varepsilon_{i,2}\*x_{i,2}\\
    \vdots\\
    \dfrac{1}{\sqrt N}\displaystyle\sum_{i=1}^N
    \check\varepsilon_{i,T}\*x_{i,T}
    \end{array}\right],
\end{align}
then $L_1(\*b)=T^{-1}\*b'\*E_N$. With
$\mathbb D_0=\mathrm{diag}(\sqrt{T-1},1,\ldots,1)\otimes\*I_p$, we can
write $L_1(\*b)=T^{-1}(\mathbb D_0\*b)'(\mathbb D_0^{-1}\*E_N)$, where
\begin{align}
    \mathbb E\bigl\|\mathbb D_0^{-1}\*E_N\bigr\|^2
    &= \frac{1}{(T-1)N}\sum_{t=2}^T\sum_{s=2}^T
    \sum_{i=1}^N\sum_{j=1}^N
    \mathbb E\bigl(\check\varepsilon_{i,t}\*x_{i,1}'\*x_{j,1}
    \check\varepsilon_{j,s}\bigr) \notag\\
    &\quad+\sum_{t=2}^T\frac1N\sum_{i=1}^N\sum_{j=1}^N
    \mathbb E\bigl(\check\varepsilon_{i,t}\*x_{i,t}'\*x_{j,t}
    \check\varepsilon_{j,t}\bigr) \notag\\
    &= \frac{O(T^2)}{T-1}+(T-1)\,C \notag\\
    &= O(T),
    \label{eq:scoremoment}
\end{align}
which holds by Assumptions 4.1(b) and 4.1(c), where Assumption 4.1(b) is used
in its uniform over $t$ form,
\begin{align}
    \sup_{2\le t\le T}\frac1N\sum_{i=1}^N\sum_{j=1}^N
    \bigl|\mathbb E\bigl(\check\varepsilon_{i,t}\*x_{i,t}'\*x_{j,t}
    \check\varepsilon_{j,t}\bigr)\bigr|\le C,
    \label{eq:supform}
\end{align}
so that each of the $T-1$ per date blocks contributes at most $C$. We
convert \eqref{eq:scoremoment} into a stochastic order by Markov's
inequality. For any $\delta>0$, choosing
$M_\delta=\sqrt{C'/\delta}$ with $C'$ the constant in
\eqref{eq:scoremoment},
\begin{align}
    \mathbb P\Bigl(\bigl\|\mathbb D_0^{-1}\*E_N\bigr\|
    >M_\delta\sqrt T\Bigr)
   & \le \frac{\mathbb E\|\mathbb D_0^{-1}\*E_N\|^2}{M_\delta^2\,T}
    \notag\\ &\le \frac{C'}{M_\delta^2}  \notag\\ &= \delta,
\end{align}
so that $\|\mathbb D_0^{-1}\*E_N\|=O_p(\sqrt T)$. Hence, by the
Cauchy--Schwarz inequality and
$\|\mathbb D_0\*b\|=\sqrt T\,\bar b$ from \eqref{eq:bbar},
\begin{align}
    |L_1(\*b)|
    &\le \frac1T\bigl\|\mathbb D_0\*b\bigr\|
    \bigl\|\mathbb D_0^{-1}\*E_N\bigr\| \notag\\
    &= \frac{\sqrt T\,\bar b}{T}\,O_p\bigl(\sqrt T\bigr) \notag\\
    &= O_p(\bar b).
    \label{eq:L1}
\end{align}

 Moving to the term $L_2(\*b)$. For each $t$, we have the
identity
\begin{align}
    \*x_{i,1}'(\*b_t-\*b_1)+(\*x_{i,t}-\*x_{i,1})'\*b_t
    &= \*x_{i,1}'\*b_t-\*x_{i,1}'\*b_1
    +\*x_{i,t}'\*b_t-\*x_{i,1}'\*b_t \notag\\
    &= \*x_{i,t}'\*b_t-\*x_{i,1}'\*b_1,
    \label{eq:identity}
\end{align}
so that, setting $\*u_t=[(\*b_t-\*b_1)',\*b_t']'\in\mathbb R^{2p}$, each
summand of $L_2(\*b)$ is a quadratic form in the joint second moment
matrix appearing in Assumption 4.2(a). By the rules for quadratic forms,
$\*x'\*A\*x\ge\mu_{\min}(\*A)\|\*x\|^2$ (see Abadir and Magnus, 2005,
Exercise 7.53), and hence, by Assumption 4.2(a),
\begin{align}
    \frac1N\sum_{i=1}^N
    \bigl(\*x_{i,t}'\*b_t-\*x_{i,1}'\*b_1\bigr)^2
    \ge \underline c\bigl(\|\*b_t-\*b_1\|^2+\|\*b_t\|^2\bigr)
    \quad\text{w.p.1}.
    \label{eq:curvraw}
\end{align}
To extract the marginal norms from \eqref{eq:curvraw}, note that for any
$\*u,\*v\in\mathbb R^p$,
\begin{align}
    \|\*u-\*v\|^2+\|\*v\|^2
    = \begin{bmatrix}\*u'&\*v'\end{bmatrix}
    \left(\begin{bmatrix}1&-1\\-1&2\end{bmatrix}\otimes\*I_p\right)
    \begin{bmatrix}\*u\\\*v\end{bmatrix},
\end{align}
and that the eigenvalues of the Kronecker product $\*C\otimes\*I_p$ are
the eigenvalues of $\*C$, each repeated $p$ times. The characteristic
polynomial of the displayed $2\times2$ matrix is
$(1-\mu)(2-\mu)-1=\mu^2-3\mu+1=0$, with roots $\mu=(3\pm\sqrt5)/2$, both
strictly positive, so that
\begin{align}
    \|\*u-\*v\|^2+\|\*v\|^2
    \ge \alpha\bigl(\|\*u\|^2+\|\*v\|^2\bigr),
    \label{eq:alpha}
\end{align}
where $ \alpha=\frac{3-\sqrt5}{2}>0$. 
Applying \eqref{eq:alpha} with $\*u=\*b_1$ and $\*v=\*b_t$ inside
\eqref{eq:curvraw}, and summing over $t=2,\ldots,T$,
\begin{align}
    L_2(\*b)
    &\ge \frac{\underline c\,\alpha}{T}\sum_{t=2}^T
    \bigl(\|\*b_1\|^2+\|\*b_t\|^2\bigr) \notag\\
    &= \frac{\underline c\,\alpha}{T}
    \Bigl[(T-1)\|\*b_1\|^2+\sum_{t=2}^T\|\*b_t\|^2\Bigr] \notag\\
    &= \underline c\,\alpha\,\bar b^2 > 0
    \quad\text{w.p.1}.
    \label{eq:L2}
\end{align}

\medskip
\noindent Now, from the definition of $\dot{\+\Theta}_T$ as the
minimiser of $L_{NT}(\+\Theta_T)$, we have
$N[L_{NT}(\dot{\+\Theta}_T)-L_{NT}(\+\Theta_T^0)]\le0$. Evaluating
\eqref{eq:prelimdecomp} at $\dot{\+\Theta}_T$ and combining with
\eqref{eq:L1} and \eqref{eq:L2},
$\underline c\,\alpha\,\dot{\bar b}^2\le2|L_1(\dot{\*b})|
\le {\mathfrak{G}}_{NT}\,\dot{\bar b}$ with ${\mathfrak{G}}_{NT}=O_p(1)$, so that, on the event
$\{\dot{\bar b}>0\}$, the case $\dot{\bar b}=0$ being trivial,
$\dot{\bar b}\le {\mathfrak{G}}_{NT}/(\underline c\,\alpha)=O_p(1)$, that is,
\begin{align}
    (T-1)\bigl\|\dot{\*b}_1\bigr\|^2
    +\sum_{t=2}^T\bigl\|\dot{\*b}_t\bigr\|^2
    = O_p(T).
    \label{eq:prelimenergy}
\end{align}
The first term in \eqref{eq:prelimenergy} gives
$\|\dot{\*b}_1\|=O_p(1)$, that is,
$\dot{\+\theta}_1-\+\theta_1^0=O_p(N^{-1/2})$, where the factor $T-1$
reflects that $\+\theta_1$ accumulates identifying variation from all
$T-1$ differenced equations. Moreover, by the same argument that gives
part (c) below applied to \eqref{eq:prelimenergy},
$\max_{1\le t\le T}\|\dot{\+\theta}_t-\+\theta_t^0\|=O_p(\sqrt{T/N})$.
Now, denote
\begin{align}
    \*A_t=\frac1N\sum_{i=1}^N\*x_{i,t}\*x_{i,t}',
    \qquad
    \*C_t=\frac1N\sum_{i=1}^N\*x_{i,t}\*x_{i,1}',
    \qquad
    \+\zeta_t=\frac1N\sum_{i=1}^N\*x_{i,t}\check\varepsilon_{i,t}.
    \label{eq:ACzeta}
\end{align}
First, since $\*x_{i,t}=\*x_{i,1}+(\*x_{i,t}-\*x_{i,1})$, for any
$\*v\in\mathbb R^p$,
\begin{align}
    \*v'\*A_t\*v
   & = \frac1N\sum_{i=1}^N\bigl(\*x_{i,t}'\*v\bigr)^2
   \notag\\& = \begin{bmatrix}\*v'&\*v'\end{bmatrix}
    \Biggl(\frac1N\sum_{i=1}^N
    \begin{bmatrix}\*x_{i,1}\\\*x_{i,t}-\*x_{i,1}\end{bmatrix}
    \begin{bmatrix}\*x_{i,1}\\\*x_{i,t}-\*x_{i,1}\end{bmatrix}'\Biggr)
    \begin{bmatrix}\*v\\\*v\end{bmatrix},
\end{align}
which, by Assumption 4.2(a) and $\|[\*v',\*v']'\|^2=2\|\*v\|^2$, lies in
$[2\underline c\,\|\*v\|^2,\,2\overline c\,\|\*v\|^2]$ w.p.1. Hence
$\mu_{\min}(\*A_t)\ge2\underline c$,
$\mu_{\max}(\*A_t)\le2\overline c$, and
$\|\*A_t^{-1}\|\le(2\underline c)^{-1}$ w.p.1, uniformly in $t$. Second,
writing $\*G_t:=  \frac{1}{N}\sum_{i=1}^N
    \begin{bmatrix}\*x_{i,1}\\ \*x_{i,t}-\*x_{i,1}\end{bmatrix}
    \begin{bmatrix}\*x_{i,1}\\ \*x_{i,t}-\*x_{i,1}\end{bmatrix}'$ for the joint moment matrix in Assumption 4.2(a),
\begin{align}
    \*C_t
  &  = \frac1N\sum_{i=1}^N\*x_{i,1}\*x_{i,1}'
    +\frac1N\sum_{i=1}^N(\*x_{i,t}-\*x_{i,1})\*x_{i,1}'
   \notag\\& = \*G_{t,11}+\*G_{t,21},
\end{align}
the upper left and lower left $p\times p$ blocks of $\*G_t$. The
spectral norm of a diagonal block of a positive semidefinite matrix is
bounded by its largest eigenvalue, and for the off diagonal block, the
Cauchy--Schwarz inequality in the inner product induced by $\*G_t$
gives, for any unit vectors $\*u,\*v\in\mathbb R^p$,
$|\*u'\*G_{t,21}\*v|
\le(\*u'\*G_{t,22}\*u)^{1/2}(\*v'\*G_{t,11}\*v)^{1/2}\le\overline c$,
so that
\begin{align}
    \|\*C_t\|\le\|\*G_{t,11}\|+\|\*G_{t,21}\|\le2\overline c
    \quad\text{w.p.1, uniformly in }t.
    \label{eq:Cbound}
\end{align}
Third, by \eqref{eq:supform},
$\mathbb E\|\sqrt N\+\zeta_t\|^2\le C$ uniformly in $t$, so that, by
Markov's inequality applied date by date, for each fixed $t$ and any
$\delta>0$,
\begin{align}
    \mathbb P\Bigl(\bigl\|\sqrt N\+\zeta_t\bigr\|
    >\sqrt{C/\delta}\Bigr)
    \le \frac{\mathbb E\|\sqrt N\+\zeta_t\|^2}{C/\delta}
    \le \delta,
\end{align}
that is, $\+\zeta_t=O_p(N^{-1/2})$ for each fixed $t$. The first order
condition of \eqref{eq:initial} with respect to $\+\theta_t$, $t\ge2$,
is $N^{-1}\sum_{i=1}^N\*x_{i,t}(\check y_{i,t}
-\*x_{i,t}'\dot{\+\theta}_t+\*x_{i,1}'\dot{\+\theta}_1)
=\*0_{p\times1}$, and substituting \eqref{eq:resid} evaluated at
$\dot{\+\Theta}_T$,
\begin{align}
    \*A_t\bigl(\dot{\+\theta}_t-\+\theta_t^0\bigr)
    = \+\zeta_t+\*C_t\bigl(\dot{\+\theta}_1-\+\theta_1^0\bigr).
    \label{eq:prelimfoc}
\end{align}
There is no penalty contribution in \eqref{eq:prelimfoc} because the
preliminary objective is unpenalised. Inverting \eqref{eq:prelimfoc}
and applying the bounds just established together with
$\dot{\+\theta}_1-\+\theta_1^0=O_p(N^{-1/2})$,
\begin{align}
    \bigl\|\dot{\+\theta}_t-\+\theta_t^0\bigr\|
    &\le \bigl\|\*A_t^{-1}\bigr\|
    \Bigl(\|\+\zeta_t\|
    +\|\*C_t\|\,\bigl\|\dot{\+\theta}_1-\+\theta_1^0\bigr\|\Bigr)
    \notag\\
    &\le \frac{1}{2\underline c}
    \Bigl(O_p\bigl(N^{-1/2}\bigr)
    +2\overline c\,O_p\bigl(N^{-1/2}\bigr)\Bigr)
    \notag\\
    &= O_p\bigl(N^{-1/2}\bigr)
    \label{eq:prelimrate}
\end{align}
for each fixed $t\ge2$, which together with the result for $t=1$ gives
the per date rate of the preliminary estimator.

We now use this to record a bound on the adaptive weights
over the true kink set. It is convenient to separate the case with no
true kinks from the case with at least one. If $K^0=0$, then
$\mathcal T^0_{K^0}=\varnothing$, no second difference is active, and the
weight bound below is vacuous, so we assume $K^0\ge1$ in what follows.
For $K^0\ge1$, recall
\begin{align}
    J_{\min}
    := \min_{1\le\ell\le K^0}
    \bigl\|\+\kappa_{\ell+1}^0-\+\kappa_\ell^0\bigr\|,
\end{align}
and note that, since a kink at $T_\ell^0$ is the nonzero second
difference
$\Delta^2\+\theta_{T_\ell^0}^0=\+\kappa_{\ell+1}^0-\+\kappa_\ell^0$, we
have
$\min_{t\in\mathcal T^0_{K^0}}\|\Delta^2\+\theta_t^0\|=J_{\min}$.

We first establish the uniform preliminary rate over the kink set, which
requires uniformity over the at most $3K^0$ dates entering the second
differences there. Let
$\mathcal T^{0\pm}_{K^0}
=\bigcup_{t\in\mathcal T^0_{K^0}}\{t-1,t,t+1\}$ denote the dilated kink
set, of cardinality at most $3K^0$. By \eqref{eq:supform}, the union
bound and Markov's inequality, for any $M>0$,
\begin{align}
    \mathbb P\Bigl(\max_{s\in\mathcal T^{0\pm}_{K^0}}
    \bigl\|\sqrt N\+\zeta_s\bigr\|>M\sqrt{K^0}\Bigr)
   & \le \sum_{s\in\mathcal T^{0\pm}_{K^0}}
    \frac{\mathbb E\|\sqrt N\+\zeta_s\|^2}{M^2K^0}
   \notag\\& \le \frac{3K^0\,C}{M^2K^0}
    \notag\\&= \frac{3C}{M^2},
\end{align}
so that
$\max_{s\in\mathcal T^{0\pm}_{K^0}}\|\+\zeta_s\|=O_p(\sqrt{K^0/N})$.
Combining this with \eqref{eq:prelimfoc}, the matrix bounds, and
$\dot{\+\theta}_1-\+\theta_1^0=O_p(N^{-1/2})$, exactly as in
\eqref{eq:prelimrate},
\begin{align}
    \max_{s\in\mathcal T^{0\pm}_{K^0},\,s\ge2}
    \bigl\|\dot{\+\theta}_s-\+\theta_s^0\bigr\|
    = O_p\bigl(\sqrt{K^0/N}\bigr),
\end{align}
and since $\Delta^2(\dot{\+\theta}_t-\+\theta_t^0)$ involves only dates
in $\mathcal T^{0\pm}_{K^0}$ when $t\in\mathcal T^0_{K^0}$, with the
date $1$ covered by $\|\dot{\*b}_1\|=O_p(1)$ directly,
\begin{align}
    \max_{t\in\mathcal T^0_{K^0}}
    \bigl\|\Delta^2\dot{\+\theta}_t-\Delta^2\+\theta_t^0\bigr\|
    = O_p\bigl(\sqrt{K^0/N}\bigr).
    \label{eq:unifprelim}
\end{align}
To make the order in \eqref{eq:unifprelim} explicit, define the bounded
sequence
\begin{align}
    {\mathfrak{G}}_{NT}'
    := \frac{\max_{t\in\mathcal T^0_{K^0}}
    \|\Delta^2\dot{\+\theta}_t-\Delta^2\+\theta_t^0\|}
    {\sqrt{K^0/N}},
    \label{eq:Gdef}
\end{align}
so that ${\mathfrak{G}}_{NT}'=O_p(1)$ and
$\max_{t\in\mathcal T^0_{K^0}}
\|\Delta^2\dot{\+\theta}_t-\Delta^2\+\theta_t^0\|
={\mathfrak{G}}_{NT}'\sqrt{K^0/N}$. Define the event
\begin{align}
    \mathcal J_{NT}
    = \Bigl\{\max_{t\in\mathcal T^0_{K^0}}
    \bigl\|\Delta^2\dot{\+\theta}_t-\Delta^2\+\theta_t^0\bigr\|
    \le J_{\min}/2\Bigr\},
\end{align}
the event on which the preliminary second difference is within
$J_{\min}/2$ of the truth at every true kink. By \eqref{eq:Gdef}, its
complement is
\begin{align}
    \mathbb P(\mathcal J_{NT}^c)
    &= \mathbb P\bigl({\mathfrak{G}}_{NT}'\sqrt{K^0/N}>J_{\min}/2\bigr) \notag\\
    &= \mathbb P\bigl({\mathfrak{G}}_{NT}'>a_N\bigr),
    \label{eq:ANevent}
\end{align}
where $a_N:=\tfrac12\sqrt{N/K^0}\,J_{\min}$, which diverges by the
second clause of Assumption~\ref{ass:signal}(a).  Because
${\mathfrak{G}}_{NT}'=O_p(1)$, for every $\epsilon>0$ there is $M_\epsilon$ with
$\sup_N\mathbb P({\mathfrak{G}}_{NT}'>M_\epsilon)<\epsilon$, and as $a_N\to\infty$ we
have $a_N>M_\epsilon$ for all $N$ large, so that
$\mathbb P({\mathfrak{G}}_{NT}'>a_N)\le\mathbb P({\mathfrak{G}}_{NT}'>M_\epsilon)<\epsilon$
eventually. As $\epsilon$ was arbitrary,
$\mathbb P(\mathcal J_{NT}^c)\to0$, that is $\mathbb P(\mathcal J_{NT})\to1$.
On $\mathcal J_{NT}$, the reverse triangle inequality and
$\|\Delta^2\+\theta_t^0\|\ge J_{\min}$ for $t\in\mathcal T^0_{K^0}$ give
\begin{align}
    \bigl\|\Delta^2\dot{\+\theta}_t\bigr\|
    &\ge \bigl\|\Delta^2\+\theta_t^0\bigr\|
    -\bigl\|\Delta^2\dot{\+\theta}_t-\Delta^2\+\theta_t^0\bigr\| \notag\\
    &\ge J_{\min}-\frac{J_{\min}}{2}
    = \frac{J_{\min}}{2},
\end{align}
and therefore, on $\mathcal J_{NT}$,
\begin{align}
    \max_{t\in\mathcal T^0_{K^0}}\dot\omega_t
  &  = \max_{t\in\mathcal T^0_{K^0}}
    \bigl\|\Delta^2\dot{\+\theta}_t\bigr\|^{-\zeta_1}
   \notag\\& \le \Bigl(\frac{J_{\min}}{2}\Bigr)^{-\zeta_1}
    \notag\\&= 2^{\zeta_1}J_{\min}^{-\zeta_1}.
    \label{eq:wmax}
\end{align}
Since $\mathbb P(\mathcal J_{NT})\to1$, this gives
$\max_{t\in\mathcal T^0_{K^0}}\dot\omega_t
=O_p(J_{\min}^{-\zeta_1})$. This is the key active set weight bound. It
keeps the adaptive penalty asymptotically weak at the true kinks, so
that in Step 2 the penalty cannot shrink the genuine slope changes
toward zero.

From the definition of $\widehat{\+\Theta}_T$ as the minimiser of
$\mathcal L_{\vartheta_1}(\+\Theta_T)$, we have
$N[\mathcal L_{\vartheta_1}(\widehat{\+\Theta}_T)
-\mathcal L_{\vartheta_1}(\+\Theta_T^0)]\le0$. We will show that
$N[\mathcal L_{\vartheta_1}(\+\Theta_T)
-\mathcal L_{\vartheta_1}(\+\Theta_T^0)]$ is bounded below by a quadratic
in $\bar b$ whose leading coefficient is positive, which forces the
minimiser to have $\widehat{\bar b}=O_p(1)$. Using \eqref{eq:resid} and
splitting the penalty at the true kink set,
\begin{align}
    & N\bigl[\mathcal L_{\vartheta_1}(\+\Theta_T)
    -\mathcal L_{\vartheta_1}(\+\Theta_T^0)\bigr] \notag\\
    &\quad= -\frac{2}{T\sqrt N}\sum_{i=1}^N\sum_{t=2}^T
    \check\varepsilon_{i,t}\bigl(\*x_{i,t}'\*b_t-\*x_{i,1}'\*b_1\bigr)
    \notag\\
    &\qquad
    +\frac{1}{TN}\sum_{i=1}^N\sum_{t=2}^T
    \bigl(\*x_{i,t}'\*b_t-\*x_{i,1}'\*b_1\bigr)^2 \notag\\
    &\qquad
    +N\vartheta_1\sum_{t\in\mathcal T^0_{K^0}}
    \dot\omega_t\Bigl[\Bigl\|\Delta^2\+\theta_t^0
    +\frac{1}{\sqrt N}\Delta^2\*b_t\Bigr\|
    -\bigl\|\Delta^2\+\theta_t^0\bigr\|\Bigr] \notag\\
    &\qquad
    +N\vartheta_1\sum_{t\in\mathcal T^{0c}_{K^0}}
    \dot\omega_t\Bigl\|\frac{1}{\sqrt N}\Delta^2\*b_t\Bigr\| \notag\\
    &\quad=: -2L_1(\*b)+L_2(\*b)+L_3(\*b)+L_4(\*b),
\end{align}
where $\mathcal T^{0c}_{K^0}
=\{2,\ldots,T-1\}\setminus\mathcal T^0_{K^0}$ and implicit definitions
of $L_3(\*b)$ and $L_4(\*b)$. Note the split of the penalty into its
kink and non-kink parts. This split is necessary as the order of
$\dot\omega_t$ depends on whether or not $t$ is a true kink date. The
terms $L_1(\*b)$ and $L_2(\*b)$ are identical to those in
\eqref{eq:prelimdecomp}, and therefore satisfy \eqref{eq:L1} and
\eqref{eq:L2}. Since $\dot\omega_t\ge0$, we have $L_4(\*b)\ge0$ term by
term. It remains to bound $L_3(\*b)$.

If $K^0=0$, then $\mathcal T^0_{K^0}=\varnothing$ and $L_3(\*b)=0$, so
take $K^0\ge1$. By the triangle inequality,
$\|\Delta^2\+\theta_t^0+N^{-1/2}\Delta^2\*b_t\|
-\|\Delta^2\+\theta_t^0\|\ge-N^{-1/2}\|\Delta^2\*b_t\|$, so that, by
$\sum_{i=1}^n a_ib_i\le(\max_{i=1,\ldots,n}a_i)\sum_{i=1}^n b_i$
whenever $a_i\ge0$ and $b_i\ge0$,
\begin{align}
    L_3(\*b)
    \ge -\sqrt N\vartheta_1
    \Bigl(\max_{t\in\mathcal T^0_{K^0}}\dot\omega_t\Bigr)
    \sum_{t\in\mathcal T^0_{K^0}}\bigl\|\Delta^2\*b_t\bigr\|.
    \label{eq:L3start}
\end{align}
For the sum, by the triangle inequality and the Cauchy--Schwarz
inequality applied to the three term sum with coefficients $(1,2,1)$,
\begin{align}
    \bigl\|\Delta^2\*b_t\bigr\|
    &\le \|\*b_{t+1}\|+2\|\*b_t\|+\|\*b_{t-1}\| \notag\\
    &\le \sqrt{1^2+2^2+1^2}\,
    \bigl(\|\*b_{t+1}\|^2+\|\*b_t\|^2+\|\*b_{t-1}\|^2\bigr)^{1/2},
\end{align}
so that, squaring and summing over the kink set, and noting that each
index $s\in\{1,\ldots,T\}$ belongs to at most three of the windows
$\{t-1,t,t+1\}$ with $t\in\mathcal T^0_{K^0}$,
\begin{align}
    \sum_{t\in\mathcal T^0_{K^0}}\bigl\|\Delta^2\*b_t\bigr\|^2
   & \le 6\sum_{t\in\mathcal T^0_{K^0}}
    \bigl(\|\*b_{t+1}\|^2+\|\*b_t\|^2+\|\*b_{t-1}\|^2\bigr)
    \notag\\ & \le 18\,\|\*b\|^2.
\end{align}
Hence, by the Cauchy--Schwarz inequality over the $K^0$ kink dates and
$\|\*b\|\le\|\mathbb D_0\*b\|=\sqrt T\,\bar b$, since every diagonal
entry of $\mathbb D_0$ is at least one,
\begin{align}
    \sum_{t\in\mathcal T^0_{K^0}}\bigl\|\Delta^2\*b_t\bigr\|
    &\le \sqrt{K^0}\Bigl(\sum_{t\in\mathcal T^0_{K^0}}
    \bigl\|\Delta^2\*b_t\bigr\|^2\Bigr)^{1/2} \notag\\
    &\le \sqrt{18K^0T}\,\bar b.
    \label{eq:CSsum}
\end{align}
Inserting \eqref{eq:wmax} and \eqref{eq:CSsum} into \eqref{eq:L3start},
it follows that, by the first clause of Assumption~\ref{ass:signal}(a),
\begin{align}
    L_3(\*b)
    &\ge -\sqrt N\vartheta_1\,
    2^{\zeta_1}J_{\min}^{-\zeta_1}\,\sqrt{18K^0T}\,\bar b\,
    \bigl(1+o_p(1)\bigr) \notag\\
    &= -O_p\Bigl(\sqrt{NTK^0}\,\vartheta_1
    J_{\min}^{-\zeta_1}\Bigr)\bar b \notag\\
    &= -O_p(1)\,\bar b.
    \label{eq:L3}
\end{align}
By putting everything together, and using the fact that
$L_4(\*b)\ge0$, we obtain
\begin{align}
    N\bigl[\mathcal L_{\vartheta_1}(\+\Theta_T)
    -\mathcal L_{\vartheta_1}(\+\Theta_T^0)\bigr]
    &= -2L_1(\*b)+L_2(\*b)+L_3(\*b)+L_4(\*b) \notag\\
    &\ge \underline c\,\alpha\,\bar b^2-{\mathfrak{H}}_{NT}\,\bar b,
    \label{eq:quadbound}
\end{align}
where ${\mathfrak{H}}_{NT}=O_p(1)$ collects the $O_p(\bar b)$ contributions of
$2|L_1(\*b)|$ and $|L_3(\*b)|$. The right-hand side of
\eqref{eq:quadbound} is a quadratic in $\bar b$ whose leading
coefficient $\underline c\,\alpha>0$ is deterministic, so it is strictly
positive once $\bar b>{\mathfrak{H}}_{NT}/(\underline c\,\alpha)$. Fix $\epsilon>0$
and choose $M$ such that
$\mathbb P({\mathfrak{H}}_{NT}>\underline c\,\alpha\,M)<\epsilon$ for all $N$, which
is possible since ${\mathfrak{H}}_{NT}=O_p(1)$. On the event
$\{{\mathfrak{H}}_{NT}\le\underline c\,\alpha\,M\}$, the right-hand side of
\eqref{eq:quadbound} is strictly positive whenever $\bar b>M$, so no
$\+\Theta_T$ with $\bar b>M$ can satisfy
$N[\mathcal L_{\vartheta_1}(\+\Theta_T)
-\mathcal L_{\vartheta_1}(\+\Theta_T^0)]\le0$. Since
$\widehat{\+\Theta}_T$ minimises $\mathcal L_{\vartheta_1}$ it does
satisfy that inequality, when
$\mathbb P(\widehat{\bar b}>M)\le
\mathbb P({\mathfrak{H}}_{NT}>\underline c\,\alpha\,M)<\epsilon$. As $\epsilon$ was
arbitrary, $\widehat{\bar b}=O_p(1)$, that is,
\begin{align}
    (T-1)\bigl\|\widehat{\*b}_1\bigr\|^2
    +\sum_{t=2}^T\bigl\|\widehat{\*b}_t\bigr\|^2
    = O_p(T).
    \label{eq:energy}
\end{align}
The three parts of the theorem now follow from \eqref{eq:energy}. The
first term gives $\|\widehat{\*b}_1\|=O_p(1)$, that is,
$\widehat{\+\theta}_1-\+\theta_1^0=O_p(N^{-1/2})$, which is part (a),
where the factor $T-1$ records that $\+\theta_1$ accumulates identifying
variation from all $T-1$ differenced equations. Dividing
\eqref{eq:energy} by $NT$,
\begin{align}
    \frac1T\sum_{t=1}^T\bigl\|\widehat{\+\theta}_t-\+\theta_t^0\bigr\|^2
   & = \frac{1}{NT}\sum_{t=1}^T\bigl\|\widehat{\*b}_t\bigr\|^2
   \notag\\& \le \frac{O_p(T)}{NT}
    \notag\\&= O_p(N^{-1}),
\end{align}
which is part (b). Finally, since every diagonal entry of $\mathbb D_0$
is at least one,
$\max_{1\le t\le T}\|\widehat{\*b}_t\|\le\|\mathbb D_0\widehat{\*b}\|
=\sqrt T\,\widehat{\bar b}=O_p(\sqrt T)$, so that
\begin{align}
    \max_{1\le t\le T}
    \bigl\|\widehat{\+\theta}_t-\+\theta_t^0\bigr\|
   & = \frac{1}{\sqrt N}\max_{1\le t\le T}\bigl\|\widehat{\*b}_t\bigr\|
   \notag\\& = O_p\bigl(\sqrt{T/N}\bigr),
\end{align}
which is part (c). The required results are implied by this.
\end{proof}

\section*{Proof of Theorem~\ref{thm:sign}.}

\begin{proof}
We work with the second differences directly. Define
$\+\gamma=\+\theta_2-\+\theta_1$ and $\+\nu_s=\Delta^2\+\theta_s$ for
$s=2,\ldots,T-1$. Since $\+\theta_t=\+\theta_1+\sum_{r=2}^t\Delta\+\theta_r$
and telescoping $\Delta\+\theta_r-\Delta\+\theta_{r-1}=\+\nu_{r-1}$ gives
$\Delta\+\theta_r=\+\gamma+\sum_{s=2}^{r-1}\+\nu_s$ for $r\ge2$, we have
\begin{align}
    \+\theta_t
    &= \+\theta_1+\sum_{r=2}^t\Bigl(\+\gamma+\sum_{s=2}^{r-1}\+\nu_s\Bigr)
    \notag\\
    &= \+\theta_1+(t-1)\+\gamma+\sum_{s=2}^{t-1}(t-s)\+\nu_s
    \label{eq:reparam}
\end{align}
for $t=1,\ldots,T$, where the inner double sum counts each $\+\nu_s$
exactly $t-s$ times. The parameter count is $p+p+(T-2)p=Tp$ and the map is
triangular, so $(\+\theta_1,\+\gamma,\+\nu_2,\ldots,\+\nu_{T-1})\mapsto
\+\Theta_T$ is a linear bijection of $\mathbb R^{Tp}$. The objective
$\mathcal L_{\vartheta_1}$ is convex, being the sum of a convex quadratic
and a nonnegative combination of norms, and convexity is preserved under
the linear reparametrisation \eqref{eq:reparam}, in which the penalty
$\vartheta_1\sum_{s=2}^{T-1}\dot\omega_s\|\+\nu_s\|$ is separable across
the $\+\nu_s$. For a convex objective the inclusion of $\*0$ in the
subdifferential characterises a global minimiser, so it suffices to read
the first order condition with respect to each $\+\nu_s$ at
$\widehat{\+\Theta}_T$.

Write
\begin{align}
    \widehat e_{i,t}
    := \check y_{i,t}-\*x_{i,t}'\widehat{\+\theta}_t
    +\*x_{i,1}'\widehat{\+\theta}_1
\end{align}
for the fitted residuals. From \eqref{eq:reparam} the loss depends on
$\+\nu_s$ only through the levels $\+\theta_t$ with $t>s$, with
$\partial\+\theta_t/\partial\+\nu_s'=(t-s)\*I_p$ for $t>s$ and
$\*0_{p\times p}$ otherwise, while
$\partial\mathcal L_{\vartheta_1}^{\mathrm{loss}}/\partial\+\theta_t
=-\tfrac{2}{NT}\sum_{i=1}^N\*x_{i,t}\widehat e_{i,t}$. The chain rule
therefore gives, for the loss part,
\begin{align}
    \frac{\partial\mathcal L_{\vartheta_1}^{\mathrm{loss}}}
    {\partial\+\nu_s}
   & = \sum_{t=s+1}^T(t-s)\,
    \frac{\partial\mathcal L_{\vartheta_1}^{\mathrm{loss}}}
    {\partial\+\theta_t}
  \notag \\ &  = -\frac{2}{NT}\sum_{t=s+1}^T(t-s)\sum_{i=1}^N\*x_{i,t}\widehat e_{i,t},
\end{align}
and adding the subdifferential of the penalty, the first order condition
with respect to $\+\nu_s$ reads
\begin{align}
    \vartheta_1\,\dot\omega_s\,\widehat{\*g}_s
    = \frac{2}{NT}\sum_{t=s+1}^{T}(t-s)\sum_{i=1}^N
    \*x_{i,t}\,\widehat e_{i,t},
    \label{eq:kkt}
\end{align}
where $\widehat{\*g}_s=\widehat{\+\nu}_s/\|\widehat{\+\nu}_s\|$ if
$\widehat{\+\nu}_s\neq\*0_{p\times1}$, by the subdifferential of the norm
at a nonzero point, and $\|\widehat{\*g}_s\|\le1$ otherwise. In particular,
if $\widehat{\+\nu}_s\neq\*0_{p\times1}$ for some $s$, then taking norms on
both sides of \eqref{eq:kkt} and using $\|\widehat{\*g}_s\|=1$,
\begin{align}
    \vartheta_1\,\dot\omega_s
    &= \frac{2}{NT}\Bigl\|\sum_{t=s+1}^{T}(t-s)\sum_{i=1}^N
    \*x_{i,t}\,\widehat e_{i,t}\Bigr\|
    \notag\\
    &= \frac{2}{T}\Bigl\|\sum_{t=s+1}^T(t-s)\,\mathcal{E}_t\Bigr\|,
    \label{eq:kktnorm}
\end{align}
where $\mathcal{E}_t:=\frac1N\sum_{i=1}^N\*x_{i,t}\,\widehat e_{i,t}$. We
will show that, w.p.a.1, the left side of \eqref{eq:kktnorm} exceeds the
right side simultaneously at every $s\in\mathcal T^{0c}_{K^0}$, so that
\eqref{eq:kktnorm} cannot hold at any non-kink date and therefore
$\widehat{\+\nu}_s=\*0_{p\times1}$ at every one of them. If
$\|\Delta^2\dot{\+\theta}_s\|=0$ for some $s$, then $\dot\omega_s=+\infty$
and any candidate with $\+\nu_s\neq\*0_{p\times1}$ has infinite objective
value, so the minimiser sets $\widehat{\+\nu}_s=\*0_{p\times1}$ at such
dates and we may assume $\dot\omega_s<\infty$ throughout.

We first bound the left side of \eqref{eq:kktnorm} from below,
uniformly over $s\in\mathcal T^{0c}_{K^0}$. This requires the uniform rate
of the preliminary estimator over the whole path, which we record
explicitly. The preliminary energy bound \eqref{eq:prelimenergy} states
$(T-1)\|\dot{\*b}_1\|^2+\sum_{t=2}^T\|\dot{\*b}_t\|^2=O_p(T)$ with
$\dot{\*b}_t=\sqrt N(\dot{\+\theta}_t-\+\theta_t^0)$, and since every term is
nonnegative and the factor $T-1$ is at least one,
\begin{align}
    \max_{1\le t\le T}\|\dot{\*b}_t\|^2
    &\le (T-1)\|\dot{\*b}_1\|^2+\sum_{t=2}^T\|\dot{\*b}_t\|^2
  \notag\\&  = O_p(T),
\end{align}
so that $\max_{1\le t\le T}\|\dot{\+\theta}_t-\+\theta_t^0\|
=N^{-1/2}\max_{1\le t\le T}\|\dot{\*b}_t\|=O_p(\sqrt{T/N})$. To make this
order explicit, define the bounded sequence
\begin{align}
    \widetilde{\mathfrak{G}}_{NT}
    := \sqrt{N/T}\,\max_{1\le t\le T}
    \bigl\|\dot{\+\theta}_t-\+\theta_t^0\bigr\|,
    \label{eq:Gtildedef}
\end{align}
where $\widetilde{\mathfrak{G}}_{NT}=O_p(1)$ so that $\max_{1\le t\le T}\|\dot{\+\theta}_t-\+\theta_t^0\|
=\widetilde{\mathfrak{G}}_{NT}\sqrt{T/N}$. For $s\in\mathcal T^{0c}_{K^0}$ we
have $\Delta^2\+\theta_s^0=\*0_{p\times1}$, so that, by the triangle
inequality and \eqref{eq:Gtildedef},
\begin{align}
    \bigl\|\Delta^2\dot{\+\theta}_s\bigr\|
    &= \bigl\|\Delta^2\bigl(\dot{\+\theta}_s-\+\theta_s^0\bigr)\bigr\|
    \notag\\
    &\le \|\dot{\+\theta}_{s+1}-\+\theta_{s+1}^0\|
    +2\|\dot{\+\theta}_s-\+\theta_s^0\|
    +\|\dot{\+\theta}_{s-1}-\+\theta_{s-1}^0\|
    \notag\\
    &\le 4\widetilde{\mathfrak{G}}_{NT}\sqrt{T/N},
    \label{eq:floorprep}
\end{align}
where \eqref{eq:floorprep} holds for all $s\in\mathcal T^{0c}_{K^0}$
simultaneously since the right side does not depend on $s$. This uniformity
is why the preliminary uniform rate $\sqrt{T/N}$, rather than the pointwise
rate $N^{-1/2}$, is the rate that enters here. Hence
\begin{align}
    \min_{s\in\mathcal T^{0c}_{K^0}}\dot\omega_s
    &= \min_{s\in\mathcal T^{0c}_{K^0}}
    \bigl\|\Delta^2\dot{\+\theta}_s\bigr\|^{-\zeta_1}
    \notag\\
    &\ge \bigl(4\widetilde{\mathfrak{G}}_{NT}\bigr)^{-\zeta_1}
    \Bigl(\frac NT\Bigr)^{\zeta_1/2}.
    \label{eq:floor}
\end{align}

\medskip
\noindent We now bound the right side of \eqref{eq:kktnorm} from above,
uniformly over $s$. By the construction of $\check y_{i,t}$ we have
$\check y_{i,t}=\*x_{i,t}'\+\theta_t^0-\*x_{i,1}'\+\theta_1^0
+\check\varepsilon_{i,t}$, so that with
$\widehat{\*u}_t:=\widehat{\+\theta}_t-\+\theta_t^0$,
\begin{align}
    \widehat e_{i,t}
    = \check\varepsilon_{i,t}
    -\*x_{i,t}'\widehat{\*u}_t+\*x_{i,1}'\widehat{\*u}_1,
\end{align}
and therefore, averaging over $i$ and using the notation of
\eqref{eq:ACzeta},
\begin{align}
    \mathcal{E}_t
   & = \frac1N\sum_{i=1}^N\*x_{i,t}\widehat e_{i,t}
   \notag \\ & = \+\zeta_t-\*A_t\widehat{\*u}_t+\*C_t\widehat{\*u}_1.
    \label{eq:pdecomp}
\end{align}
From the proof of Theorem~\ref{thm:consistency} we have
$\|\*A_t\|\le2\overline c$ and $\|\*C_t\|\le2\overline c$ w.p.1 uniformly
in $t$, so that, by the elementary inequality
$(a+b+c)^2\le3(a^2+b^2+c^2)$,
\begin{align}
    \sum_{t=2}^T\|\mathcal{E}_t\|^2
    &\le 3\sum_{t=2}^T\|\+\zeta_t\|^2
    +12\,\overline c^{\,2}\sum_{t=2}^T\|\widehat{\*u}_t\|^2
    \notag\\
    &\quad
    +12\,\overline c^{\,2}\,(T-1)\,\|\widehat{\*u}_1\|^2
    \quad\text{w.p.1}.
    \label{eq:psum}
\end{align}
We bound the three terms in \eqref{eq:psum}. First, by
Assumption~\ref{ass:errors}(b), $\mathbb E\|\sqrt N\+\zeta_t\|^2\le C$
uniformly in $t$, so that $\mathbb E\sum_{t=2}^T\|\+\zeta_t\|^2\le(T-1)C/N$,
and Markov's inequality gives $\sum_{t=2}^T\|\+\zeta_t\|^2=O_p(T/N)$.
Second, $\sum_{t=2}^T\|\widehat{\*u}_t\|^2=O_p(T/N)$ by
Theorem~\ref{thm:consistency}(b). Third,
$(T-1)\|\widehat{\*u}_1\|^2=O_p(T/N)$ by
Theorem~\ref{thm:consistency}(a). Substituting into \eqref{eq:psum} and
defining the bounded sequence
\begin{align}
    \mathfrak{B}_{NT}
    := \Bigl(\frac NT\sum_{t=2}^T\|\mathcal{E}_t\|^2\Bigr)^{1/2},
    \label{eq:Bdef}
\end{align}
where $\mathfrak{B}_{NT}=O_p(1)$ so that 
we have $\sum_{t=2}^T\|\mathcal{E}_t\|^2=\mathfrak{B}_{NT}^2\,T/N$. Next, for
any $s$,
\begin{align}
    \sum_{t=s+1}^T(t-s)^2
    &= \sum_{h=1}^{T-s}h^2
    \notag \\ &\le \sum_{h=1}^{T}h^2
    \notag \\ & = \frac{T(T+1)(2T+1)}{6}
    \notag \\ &\le T^3,
    \label{eq:rampbound}
\end{align}
using $(T+1)(2T+1)=2T^2+3T+1\le6T^2$ for $T\ge1$. Hence, by the
Cauchy--Schwarz inequality, \eqref{eq:Bdef} and \eqref{eq:rampbound},
\begin{align}
    \Bigl\|\sum_{t=s+1}^T(t-s)\,\mathcal{E}_t\Bigr\|
    &\le \Bigl(\sum_{t=s+1}^T(t-s)^2\Bigr)^{1/2}
    \Bigl(\sum_{t=2}^T\|\mathcal{E}_t\|^2\Bigr)^{1/2}
    \notag\\
    &\le T^{3/2}\,\mathfrak{B}_{NT}\sqrt{\frac TN}
    \notag\\
    &= \mathfrak{B}_{NT}\,\frac{T^2}{\sqrt N},
    \label{eq:ceiling}
\end{align}
simultaneously over all $s$, since the right side does not depend on $s$.

 We now argue by contradiction. Suppose that
$\widehat{\+\nu}_s\neq\*0_{p\times1}$ for some $s\in\mathcal T^{0c}_{K^0}$
with probability not tending to zero. On this event \eqref{eq:kktnorm}
holds at that $s$, and combining the floor \eqref{eq:floor} with the
ceiling \eqref{eq:ceiling},
\begin{align}
    \vartheta_1\bigl(4\widetilde{\mathfrak{G}}_{NT}\bigr)^{-\zeta_1}
    \Bigl(\frac NT\Bigr)^{\zeta_1/2}
    &\le \vartheta_1\,\dot\omega_s
  \notag \\ &  = \frac{2}{T}\Bigl\|\sum_{t=s+1}^T(t-s)\,\mathcal{E}_t\Bigr\|
 \notag \\ &   \le \frac{2\,\mathfrak{B}_{NT}\,T}{\sqrt N},
    \label{eq:floorceiling}
\end{align}
where the bound is uniform in $s$ and so holds on the whole event.
Multiplying both sides of \eqref{eq:floorceiling} by
$(4\widetilde{\mathfrak{G}}_{NT})^{\zeta_1}\sqrt N/T$ and collecting powers
of $N$ and $T$,
\begin{align}
    \frac{N^{(\zeta_1+1)/2}\,\vartheta_1}{T^{1+\zeta_1/2}}
  &  \le 2\bigl(4\widetilde{\mathfrak{G}}_{NT}\bigr)^{\zeta_1}\,
    \mathfrak{B}_{NT}
    \notag\\&=: \mathfrak{F}_{NT},
    \label{eq:contra}
\end{align}
where $\mathfrak{F}_{NT}=O_p(1)$ as a product of powers of the $O_p(1)$
sequences $\widetilde{\mathfrak{G}}_{NT}$ and $\mathfrak{B}_{NT}$.But Assumption~\ref{ass:signal}(b) states that
$N^{(\zeta_1+1)/2}\vartheta_1/T^{1+\zeta_1/2}\to_p\infty$, and a sequence
that diverges in probability cannot lie below an $O_p(1)$ sequence on an
event of probability bounded away from zero. This is the required
contradiction. Hence $\widehat{\+\nu}_s=\*0_{p\times1}$ for all
$s\in\mathcal T^{0c}_{K^0}$ w.p.a.1, and since
$\Delta^2\widehat{\+\theta}_s=\widehat{\+\nu}_s$ by definition, the proof
is complete.
\end{proof}

\section*{Proof of Corollary~\ref{cor:kinks}}

\begin{proof}
We first show that Assumption~\ref{ass:signal} implies
$\sqrt{N/T}\,J_{\min}\to\infty$. By the first clause of part~(a),
w.p.a.1 $\vartheta_1\le(c_1+1)J_{\min}^{\zeta_1}(NTK^0)^{-1/2}$.
Substituting this bound into the ratio from part~(b) yields, w.p.a.1,
\begin{align}
\frac{N^{(\zeta_1+1)/2}\vartheta_1}{T^{1+\zeta_1/2}}
&\le (c_1+1)\,
\frac{N^{(\zeta_1+1)/2}J_{\min}^{\zeta_1}}
{T^{1+\zeta_1/2}\sqrt{NTK^0}}
\notag\\ &= (c_1+1)\,
\frac{J_{\min}^{\zeta_1}N^{\zeta_1/2}}
{T^{(3+\zeta_1)/2}\sqrt{K^0}}
\notag\\& \le (c_1+1)
\left(\sqrt{\frac{N}{T}}\,J_{\min}\right)^{\zeta_1},
\end{align}
where the final inequality follows from $T^{3/2}\sqrt{K^0}\ge1$.
Since the left-hand side diverges in probability by part~(b),
the deterministic right-hand side must diverge as well, and hence
$\sqrt{N/T}\,J_{\min}\to\infty$.

Next, let $\mathcal{S}_{NT}$ denote the event of
Theorem~\ref{thm:sign}, on which $\|\Delta^2\widehat{\+\theta}_t\|=0$
for all $t\in\mathcal{T}^{0c}_{K^0}$, so that
$\widehat{\mathcal{T}}_{\widehat K}\subseteq\mathcal{T}^0_{K^0}$ on
$\mathcal{S}_{NT}$, and $\mathbb{P}(\mathcal{S}_{NT})\to1$.

It remains to show that every true kink date survives. Fix
$t\in\mathcal{T}^0_{K^0}$. Since
\(
    \Delta^2\widehat{\+\theta}_t-\Delta^2\+\theta_t^0
    =(\widehat{\+\theta}_{t+1}-\+\theta_{t+1}^0)
    -2(\widehat{\+\theta}_t-\+\theta_t^0)
    +(\widehat{\+\theta}_{t-1}-\+\theta_{t-1}^0),
\)
the triangle inequality and
$\|\Delta^2\+\theta_t^0\|\ge J_{\min}$ at a true kink give
\(
    \|\Delta^2\widehat{\+\theta}_t\|
    \ge J_{\min}
    -4\max_{1\le s\le T}\|\widehat{\+\theta}_s-\+\theta_s^0\|,
\)
the constant $4$ being the absolute sum of the coefficients $(1,-2,1)$.
By Theorem~\ref{thm:consistency}(c), the sequence
\(
    \mathfrak{M}_{NT}
    :=\sqrt{N/T}\,
    \max_{1\le s\le T}\|\widehat{\+\theta}_s-\+\theta_s^0\|
\)
satisfies $\mathfrak{M}_{NT}=O_p(1)$. Since the lower bound above is
free of $t$,
\(
    \min_{t\in\mathcal{T}^0_{K^0}}
    \|\Delta^2\widehat{\+\theta}_t\|
    \ge J_{\min}-4\mathfrak{M}_{NT}\sqrt{T/N}
    =\sqrt{T/N}\,\bigl(a_{NT}-4\mathfrak{M}_{NT}\bigr),
    \qquad
    a_{NT}:=\sqrt{N/T}\,J_{\min},
\)
so that
\(
    \mathbb{P}\Bigl(\min_{t\in\mathcal{T}^0_{K^0}}
    \|\Delta^2\widehat{\+\theta}_t\|=0\Bigr)
    \le\mathbb{P}\bigl(\mathfrak{M}_{NT}\ge a_{NT}/4\bigr).
\)
Fix $\epsilon>0$. Since $\mathfrak{M}_{NT}=O_p(1)$, there exists
$M_\epsilon<\infty$ such that
$\sup_{N,T}\mathbb{P}(\mathfrak{M}_{NT}>M_\epsilon)<\epsilon$, and
since $a_{NT}\to\infty$ by the first part, $a_{NT}/4>M_\epsilon$ for
all $N,T$ large enough, whence
\(
    \mathbb{P}\bigl(\mathfrak{M}_{NT}\ge a_{NT}/4\bigr)
    \le\mathbb{P}\bigl(\mathfrak{M}_{NT}>M_\epsilon\bigr)
    <\epsilon
\)
eventually. As $\epsilon$ was arbitrary,
$\mathbb{P}(\min_{t\in\mathcal{T}^0_{K^0}}
\|\Delta^2\widehat{\+\theta}_t\|>0)\to1$, and every true kink date is
estimated as a kink w.p.a.1.

Combining the two parts, w.p.a.1 the estimated kink set contains no
date outside $\mathcal{T}^0_{K^0}$ and every date inside it, so
$\mathbb{P}(\widehat{\mathcal{T}}_{\widehat K}
=\mathcal{T}^0_{K^0})\to1$. Both claims of the corollary follow.
\end{proof}



\section*{Proofs of the post-selection results}\label{app:postproofs}

Throughout, $C$ denotes a finite positive constant whose
value may change between occurrences, and we retain the conventions
$I_0^0=0$, $I_{K^0+2}^0=0$ and
$\mathcal K_0^0=\mathcal K_{K^0+2}^0:=\varnothing$. Because the true
kink dates $\mathcal T^0_{K^0}$ are nonrandom, so is every quantity
constructed from them alone, and in particular $I_\ell^0$,
$u_{\ell,t}$, $h_{j,t}$, $r_j$, $s_j$, $w_t$, $\*M_{K^0}$ and
$\mathbb D_{K^0+1}$. We write $\+\Phi:=\+\Phi(\mathcal T^0_{K^0})$,
$\+\Phi^\eta:=\+\Phi^\eta(\mathcal T^0_{K^0})$, $\*M:=\*M_{K^0}$ and
$\mathbb D:=\mathbb D_{K^0+1}$ whenever no confusion arises, and we
let $\*y$ stack the observations conformably
with~\eqref{eq:pk_expanded}. Recall that
\begin{align}
    r_j^2:=\sum_{t=1}^{T}h_{j,t}^2,
    \qquad j=0,\ldots,K^0+1,
    \label{eq:rjdef}
\end{align}
so that $\mathbb D=\mathrm{diag}(r_0,\ldots,r_{K^0+1})\otimes\*I_p$,
and that
\begin{align}
    \+\Psi_{NT}
    :=\frac1{\sqrt N}\+\Phi^{\eta\prime}\*C\+\varepsilon.
    \label{eq:psidef}
\end{align}

The basis norms in~\eqref{eq:rjdef} are sums of squares of the
loadings~\eqref{eq:hdef}, which are linear in the elapsed fraction of
a regime. All the constants below therefore descend from three
elementary sums, which we record first. For an integer $L\ge1$, let
\begin{align}
    A(L):=\sum_{q=0}^{L-1}\Bigl(\frac qL\Bigr)^2,
    \qquad
    B(L):=\sum_{q=0}^{L-1}\Bigl(1-\frac qL\Bigr)^2,
    \qquad
    F(L):=\sum_{q=0}^{L-1}\Bigl(1-\frac qL\Bigr)\Bigl(\frac qL\Bigr),
    \label{eq:ABFdef}
\end{align}
with $A(0):=B(0):=F(0):=0$. Here $A(L)$ and $B(L)$ are the squared
norms of an increasing and of a decreasing loading over a regime of
length $L$, while $F(L)$ is the inner product of the two, which is the
only source of dependence between adjacent boundary levels.

\begin{lemma}[Closed forms and bounds for the elementary sums]
\label{lem:ABF}
For every integer $L\ge1$,
\begin{align}
    A(L)=\frac{(L-1)(2L-1)}{6L}=\frac L3-\frac12+\frac1{6L},
    \qquad
    B(L)=\frac{(L+1)(2L+1)}{6L}=\frac L3+\frac12+\frac1{6L},
    \label{eq:ABclosed}
\end{align}
and $F(L)=(L^2-1)/(6L)$. Moreover,
\begin{align}
    \frac L3\le B(L)\le L,
    \qquad
    A(L)\le\frac L3,
    \qquad
    F(L)^2\le\frac14A(L)B(L)
    \label{eq:ABFbounds}
\end{align}
for every integer $L\ge1$, while $A(L)\ge L/8$ for every integer
$L\ge2$ and $A(1)=0$.
\end{lemma}

\begin{proof}
Since $\sum_{q=0}^{L-1}q^2=(L-1)L(2L-1)/6$, we have
\begin{align}
    A(L)
    &=\frac1{L^2}\sum_{q=0}^{L-1}q^2
    \notag \\
    &=\frac{(L-1)L(2L-1)}{6L^2}
    \notag \\
    &=\frac{(L-1)(2L-1)}{6L}
    \notag \\
    &=\frac{2L^2-3L+1}{6L}
    \notag \\
    &=\frac L3-\frac12+\frac1{6L}.
\end{align}
Substituting $m=L-q$ in the definition of $B(L)$, so that $m$ runs
over $1,\ldots,L$ as $q$ runs over $0,\ldots,L-1$, and using
$\sum_{m=1}^{L}m^2=L(L+1)(2L+1)/6$,
\begin{align}
    B(L)
    &=\sum_{m=1}^{L}\Bigl(\frac mL\Bigr)^2
    \notag \\
    &=\frac{L(L+1)(2L+1)}{6L^2}
    \notag \\
    &=\frac{(L+1)(2L+1)}{6L}
    \notag \\
    &=\frac{2L^2+3L+1}{6L}
    \notag \\
    &=\frac L3+\frac12+\frac1{6L}.
\end{align}
For the third sum, expanding the product and using
$\sum_{q=0}^{L-1}q=(L-1)L/2$,
\begin{align}
    F(L)
    &=\frac1L\sum_{q=0}^{L-1}q-\frac1{L^2}\sum_{q=0}^{L-1}q^2
    \notag \\
    &=\frac{L-1}{2}-\frac{(L-1)(2L-1)}{6L}
    \notag \\
    &=(L-1)\,\frac{3L-(2L-1)}{6L}
    \notag \\
    &=\frac{(L-1)(L+1)}{6L}
    \notag \\
    &=\frac{L^2-1}{6L}.
\end{align}

Consider~\eqref{eq:ABFbounds}. The bound $B(L)>L/3$ is immediate
from~\eqref{eq:ABclosed}, since $1/2+1/(6L)>0$. The bound $B(L)\le L$
is equivalent to $1/2+1/(6L)\le2L/3$, whose left side is nonincreasing
and whose right side is increasing in $L$, and at $L=1$ both sides
equal $2/3$; hence it holds for every $L\ge1$. Likewise $A(L)\le L/3$
is equivalent to $1/(6L)\le1/2$, which holds for every $L\ge1$. For
the third bound, $A(1)=F(1)=0$ renders it trivial at $L=1$, while for
$L\ge2$ the closed forms give
\begin{align}
    \frac{F(L)^2}{A(L)B(L)}
    &=\frac{(L-1)^2(L+1)^2/(6L)^2}
    {\bigl[(L-1)(2L-1)/(6L)\bigr]\bigl[(L+1)(2L+1)/(6L)\bigr]}
    \notag \\
    &=\frac{(L-1)(L+1)}{(2L-1)(2L+1)}
    \notag \\
    &=\frac{L^2-1}{4L^2-1}
    \notag \\
    &<\frac14,
    \label{eq:Fratio}
\end{align}
the last inequality because $4(L^2-1)=4L^2-4<4L^2-1$. Finally, writing
$g(L):=A(L)-L/8=5L/24-1/2+1/(6L)$ by~\eqref{eq:ABclosed}, we have
$g(2)=5/12-1/2+1/12=0$ and $g'(L)=5/24-1/(6L^2)>0$ for every $L\ge1$,
since then $1/(6L^2)\le1/6<5/24$. Hence $g(L)\ge0$ for every $L\ge2$,
and $A(1)=0$ follows from the closed form.
\end{proof}

The loading on boundary level $j$ is supported on the two regimes
adjacent to it, so its squared norm decomposes into one increasing and
one decreasing contribution, and Lemma~\ref{lem:ABF} places it between
absolute multiples of the combined length of those two regimes. The
following lemma records this, and is used at every point below at
which a scale is converted into a rate.

\begin{lemma}[Closed form and order of the basis norms]
\label{lem:rjorder}
For every $j=0,\ldots,K^0+1$,
\begin{align}
    r_j^2=A(I_j^0)+B(I_{j+1}^0),
    \label{eq:rjclosed}
\end{align}
and, uniformly over $K^0$ and over every admissible configuration
$\mathcal T^0_{K^0}$,
\begin{align}
    \frac18\bigl(I_j^0+I_{j+1}^0\bigr)
    \le r_j^2
    \le I_j^0+I_{j+1}^0.
    \label{eq:rjbound}
\end{align}
In particular $r_j^2\ge1/8>0$ for every $j$, so $\mathbb D$ is
nonsingular. Moreover, for every $j=1,\ldots,K^0+1$,
\begin{align}
    \frac18\Bigl[\bigl(I_{j-1}^0+I_j^0\bigr)
    \wedge\bigl(I_j^0+I_{j+1}^0\bigr)\Bigr]
    \le r_{j-1}^2\wedge r_j^2
    \le\bigl(I_{j-1}^0+I_j^0\bigr)
    \wedge\bigl(I_j^0+I_{j+1}^0\bigr).
    \label{eq:rminorder}
\end{align}
\end{lemma}

\begin{proof}
Fix $j\in\{0,\ldots,K^0+1\}$. By~\eqref{eq:hdef}, $h_{j,t}=0$ unless
$t\in\mathcal K_j^0\cup\mathcal K_{j+1}^0$, and these two sets are
disjoint, so~\eqref{eq:rjdef} gives
\begin{align}
    r_j^2
    &=\sum_{t=1}^{T}h_{j,t}^2
    \notag \\
    &=\sum_{t\in\mathcal K_j^0}h_{j,t}^2
    +\sum_{t\in\mathcal K_{j+1}^0}h_{j,t}^2.
    \label{eq:rjsplit}
\end{align}
Since $\mathcal K_j^0=\{T_{j-1}^0,\ldots,T_j^0-1\}$, the change of
index $q=t-T_{j-1}^0$ maps $\mathcal K_j^0$ onto
$\{0,1,\ldots,I_j^0-1\}$, and on $\mathcal K_j^0$ we have
$h_{j,t}=u_{j,t}=q/I_j^0$ by~\eqref{eq:hdef}, whence
\begin{align}
    \sum_{t\in\mathcal K_j^0}h_{j,t}^2
    &=\sum_{q=0}^{I_j^0-1}\Bigl(\frac{q}{I_j^0}\Bigr)^2
    \notag \\
    &=A(I_j^0)
    \label{eq:firstsum}
\end{align}
by~\eqref{eq:ABFdef}. Similarly $q=t-T_j^0$ maps
$\mathcal K_{j+1}^0$ onto $\{0,1,\ldots,I_{j+1}^0-1\}$, and there
$h_{j,t}=1-u_{j+1,t}=1-q/I_{j+1}^0$, so
\begin{align}
    \sum_{t\in\mathcal K_{j+1}^0}h_{j,t}^2
    &=\sum_{q=0}^{I_{j+1}^0-1}
    \Bigl(1-\frac{q}{I_{j+1}^0}\Bigr)^2
    \notag \\
    &=B(I_{j+1}^0).
    \label{eq:secondsum}
\end{align}
Both identities remain valid at the endpoints under our conventions,
since $\mathcal K_0^0=\varnothing$ with $A(0)=0$ and
$\mathcal K_{K^0+2}^0=\varnothing$ with $B(0)=0$.
Substituting~\eqref{eq:firstsum} and~\eqref{eq:secondsum}
into~\eqref{eq:rjsplit} gives~\eqref{eq:rjclosed}.

We establish~\eqref{eq:rjbound} for the three ranges of $j$
separately. For $j=0$, \eqref{eq:rjclosed} and $A(0)=0$ give
$r_0^2=B(I_1^0)$, while $I_0^0+I_1^0=I_1^0$; Lemma~\ref{lem:ABF} then
yields $I_1^0/8\le I_1^0/3\le B(I_1^0)\le I_1^0$, as required. For
$1\le j\le K^0$ we have $I_j^0\ge1$ and $I_{j+1}^0\ge1$, so the closed
forms in~\eqref{eq:ABclosed} apply to both terms
of~\eqref{eq:rjclosed} and the constants $\mp1/2$ cancel:
\begin{align}
    r_j^2-\frac{I_j^0+I_{j+1}^0}{8}
    &=\Bigl[\frac{I_j^0}{3}-\frac12+\frac1{6I_j^0}\Bigr]
    +\Bigl[\frac{I_{j+1}^0}{3}+\frac12+\frac1{6I_{j+1}^0}\Bigr]
    -\frac{I_j^0}{8}-\frac{I_{j+1}^0}{8}
    \notag \\
    &=\frac{5I_j^0}{24}+\frac{5I_{j+1}^0}{24}
    +\frac1{6I_j^0}+\frac1{6I_{j+1}^0}
    \notag \\
    &>0,
    \label{eq:interiorlower}
\end{align}
which gives the lower bound, while the upper bound follows from
Lemma~\ref{lem:ABF} because $r_j^2\le I_j^0/3+I_{j+1}^0\le
I_j^0+I_{j+1}^0$. For $j=K^0+1$, \eqref{eq:rjclosed} and $B(0)=0$ give
$r_{K^0+1}^2=A(I_{K^0+1}^0)$, while
$I_{K^0+1}^0+I_{K^0+2}^0=I_{K^0+1}^0$. Because $T_{K^0}^0<T$ and
$T_{K^0+1}^0=T+1$, we have $I_{K^0+1}^0=T+1-T_{K^0}^0\ge2$, so
Lemma~\ref{lem:ABF} yields
\begin{align}
    \frac{I_{K^0+1}^0}{8}
   & \le A(I_{K^0+1}^0)
\notag\\&    \le\frac{I_{K^0+1}^0}{3}
  \notag\\&    \  \le I_{K^0+1}^0.
\end{align}
The restriction $I_{K^0+1}^0\ge2$ is needed here and nowhere else:
$A(1)=0$, so the lower bound would fail for a final regime of unit
length. Since $I_j^0+I_{j+1}^0\ge1$ for every $j$,
\eqref{eq:rjbound} implies $r_j^2\ge1/8>0$, so $\mathbb D$ is
nonsingular.

Finally, \eqref{eq:rminorder} follows on applying~\eqref{eq:rjbound}
at $j-1$ and at $j$ and using that the minimum of two quantities is
monotone in each of them.
\end{proof}

We may now prove Lemma~\ref{lem:basis}. The argument establishes an
exact expansion of the normalised quadratic form: the normalisation by
$r_j$ reproduces the squared norm of the coefficient vector exactly,
and the only discrepancy is the overlap of adjacent loadings on their
common regime, which Lemma~\ref{lem:ABF} bounds.

\begin{proof}[Proof of Lemma~\ref{lem:basis}]
The lemma asserts an ordering between symmetric matrices, and we
first reduce it to a scalar statement. Write
\begin{align}
    \*M:=(\*D^0)^{-1}\*H^{0\prime}\*H^0(\*D^0)^{-1},
    \label{eq:Mdef}
\end{align}
which is well defined because $r_j>0$ for every $j$ by
Lemma~\ref{lem:rjorder}, so $\*D^0$ is nonsingular. By the definition
of the ordering $\preceq$, the claim
$\tfrac12\*I_{K^0+2}\preceq\*M\preceq\tfrac32\*I_{K^0+2}$ holds if and
only if
\begin{align}
    \tfrac12\|\*c\|^2\le\*c'\*M\*c\le\tfrac32\|\*c\|^2
    \qquad\text{for every }\*c\in\mathbb R^{K^0+2},
    \label{eq:scalarclaim}
\end{align}
and, since
$\*c'\*M\*c=\bigl\|\*H^0(\*D^0)^{-1}\*c\bigr\|^2$, it suffices to
bound the squared length of $\*H^0(\*D^0)^{-1}\*c$ between $\tfrac12$
and $\tfrac32$ times the squared length of $\*c$. This is what we do.

Fix an arbitrary $\*c=[c_0,\ldots,c_{K^0+1}]'\in\mathbb R^{K^0+2}$,
which plays the role of a candidate vector of boundary levels, and
set
\begin{align}
    \gamma_j:=\frac{c_j}{r_j},
    \qquad j=0,\ldots,K^0+1,
    \label{eq:gammadef}
\end{align}
for its entries after normalisation by the basis scales. Put
$\*v:=\*H^0(\*D^0)^{-1}\*c$, so that
\begin{align}
    v_t=\sum_{j=0}^{K^0+1}h_{j,t}\,\gamma_j,
    \qquad t=1,\ldots,T.
    \label{eq:vdef}
\end{align}
By~\eqref{eq:knotrep}, $v_t$ is the value at date $t$ of the
piecewise linear path whose boundary levels are $\gamma_0,\ldots,
\gamma_{K^0+1}$, so $\|\*v\|^2$ is the energy of that path over the
sample and $\|\*c\|^2$ is the squared length of the coefficients that
generate it. The lemma states that the two are equivalent up to the
factors $\tfrac12$ and $\tfrac32$, uniformly in the configuration.

The regimes $\mathcal K_1^0,\ldots,\mathcal K_{K^0+1}^0$ partition
$\{1,\ldots,T\}$, so
\begin{align}
    \|\*v\|^2
  &  =\sum_{t=1}^{T}v_t^2
     \notag \\
    &=\sum_{\ell=1}^{K^0+1}\sum_{t\in\mathcal K_\ell^0}v_t^2.
    \label{eq:vsplit}
\end{align}
Fix a regime $\ell$ and a date $t\in\mathcal K_\ell^0$.
By~\eqref{eq:hdef}, $h_{j,t}\neq0$ requires
$t\in\mathcal K_j^0\cup\mathcal K_{j+1}^0$, which for
$t\in\mathcal K_\ell^0$ forces $j=\ell$ or $j=\ell-1$. Exactly two
terms therefore survive in~\eqref{eq:vdef}, and abbreviating
\begin{align}
    a_\ell:=\gamma_{\ell-1},
    \qquad
    b_\ell:=\gamma_\ell,
    \label{eq:abdef}
\end{align}
we obtain $v_t=h_{\ell-1,t}a_\ell+h_{\ell,t}b_\ell$. Note for later
that $b_\ell=a_{\ell+1}$, so each normalised coefficient appears
twice, once as the right endpoint of its own regime and once as the
left endpoint of the next.

On $\mathcal K_\ell^0$ the change of index $q=t-T_{\ell-1}^0$ runs
over $\{0,\ldots,I_\ell^0-1\}$ and gives
$h_{\ell-1,t}=1-q/I_\ell^0$ and $h_{\ell,t}=q/I_\ell^0$
by~\eqref{eq:hdef}, so by the definitions in~\eqref{eq:ABFdef},
\begin{align}
    \sum_{t\in\mathcal K_\ell^0}h_{\ell-1,t}^2=B(I_\ell^0),
    \qquad
    \sum_{t\in\mathcal K_\ell^0}h_{\ell,t}^2=A(I_\ell^0),
    \qquad
    \sum_{t\in\mathcal K_\ell^0}h_{\ell-1,t}h_{\ell,t}=F(I_\ell^0).
    \label{eq:threesums}
\end{align}
Expanding the square of $v_t$ and summing over
$t\in\mathcal K_\ell^0$ therefore yields
\begin{align}
    \sum_{t\in\mathcal K_\ell^0}v_t^2
    =a_\ell^2B(I_\ell^0)+2a_\ell b_\ell F(I_\ell^0)
    +b_\ell^2A(I_\ell^0),
    \label{eq:regimeexpand}
\end{align}
and substituting~\eqref{eq:regimeexpand} into~\eqref{eq:vsplit},
\begin{align}
    \|\*v\|^2
    =\sum_{\ell=1}^{K^0+1}
    \Bigl[a_\ell^2B(I_\ell^0)+2a_\ell b_\ell F(I_\ell^0)
    +b_\ell^2A(I_\ell^0)\Bigr].
    \label{eq:vquad}
\end{align}

The normalisation is chosen so that $\|\*c\|^2$ admits a
decomposition indexed by the same regimes. By~\eqref{eq:gammadef} and
the closed form $r_j^2=A(I_j^0)+B(I_{j+1}^0)$
of~\eqref{eq:rjclosed},
\begin{align}
    \|\*c\|^2
  &  =\sum_{j=0}^{K^0+1}c_j^2
     \notag \\
    &=\sum_{j=0}^{K^0+1}\gamma_j^2\,r_j^2
     \notag \\
    &=\sum_{j=0}^{K^0+1}\gamma_j^2A(I_j^0)
    +\sum_{j=0}^{K^0+1}\gamma_j^2B(I_{j+1}^0).
    \label{eq:csplit}
\end{align}
The two sums are reindexed by regime. In the first, the term $j=0$
vanishes because $I_0^0=0$ and $A(0)=0$, so setting $\ell=j$ leaves
$\sum_{\ell=1}^{K^0+1}\gamma_\ell^2A(I_\ell^0)$, which is
$\sum_{\ell=1}^{K^0+1}b_\ell^2A(I_\ell^0)$ by~\eqref{eq:abdef}. In
the second, the term $j=K^0+1$ vanishes because $I_{K^0+2}^0=0$ and
$B(0)=0$, so setting $\ell=j+1$ leaves
$\sum_{\ell=1}^{K^0+1}\gamma_{\ell-1}^2B(I_\ell^0)$, which is
$\sum_{\ell=1}^{K^0+1}a_\ell^2B(I_\ell^0)$. This is where the
identity $b_\ell=a_{\ell+1}$ is used. Hence
\begin{align}
    \|\*c\|^2
    =\sum_{\ell=1}^{K^0+1}
    \Bigl[a_\ell^2B(I_\ell^0)+b_\ell^2A(I_\ell^0)\Bigr].
    \label{eq:cquad}
\end{align}

Comparing~\eqref{eq:vquad} with~\eqref{eq:cquad}, the squared and the
mixed terms separate and subtraction gives
\begin{align}
    \|\*v\|^2-\|\*c\|^2
    =2\sum_{\ell=1}^{K^0+1}a_\ell b_\ell F(I_\ell^0).
    \label{eq:excess}
\end{align}
The normalisation by $r_j$ therefore reproduces $\|\*c\|^2$ exactly,
and the entire discrepancy between the path energy and the
coefficient length is carried by the inner products $F(I_\ell^0)$ of
the two hat functions that overlap on regime $\ell$. Bounding the
right side of~\eqref{eq:excess} by half of $\|\*c\|^2$ will therefore
deliver~\eqref{eq:scalarclaim}.

Fix $\ell$ and abbreviate $L:=I_\ell^0$. If $L=1$ then $F(1)=0$ by
Lemma~\ref{lem:ABF} and the corresponding term
in~\eqref{eq:excess} vanishes, so the bound below holds trivially and
we may assume $L\ge2$. Then $A(L)>0$, again by
Lemma~\ref{lem:ABF}, so $\lambda:=\sqrt{B(L)/A(L)}$ is well defined
and strictly positive. The weighted arithmetic-geometric mean  
inequality $2|a_\ell b_\ell|\le\lambda a_\ell^2
+\lambda^{-1}b_\ell^2$ (a special case of Young's Inequality) gives
\begin{align}
    2\bigl|a_\ell b_\ell\bigr|F(L)
    &\le F(L)\sqrt{\frac{B(L)}{A(L)}}\,a_\ell^2
    +F(L)\sqrt{\frac{A(L)}{B(L)}}\,b_\ell^2
    \notag \\
    &=\frac{F(L)}{\sqrt{A(L)B(L)}}
    \Bigl[B(L)a_\ell^2+A(L)b_\ell^2\Bigr]
    \notag \\
    &\le\frac12\Bigl[B(L)a_\ell^2+A(L)b_\ell^2\Bigr],
    \label{eq:amgm}
\end{align}
the last inequality by the third bound in~\eqref{eq:ABFbounds}, which
states $F(L)^2\le\tfrac14A(L)B(L)$ and hence
$F(L)/\sqrt{A(L)B(L)}\le\tfrac12$. The choice of $\lambda$ is what
makes the two weights combine into the single factor
$F(L)/\sqrt{A(L)B(L)}$, and it is exactly this factor that
Lemma~\ref{lem:ABF} controls. Summing~\eqref{eq:amgm} over
$\ell=1,\ldots,K^0+1$ and applying the triangle inequality
to~\eqref{eq:excess},
\begin{align}
    \Bigl|\,\|\*v\|^2-\|\*c\|^2\Bigr|
    &\le2\sum_{\ell=1}^{K^0+1}\bigl|a_\ell b_\ell\bigr|F(I_\ell^0)
     \notag \\
    &\le\frac12\sum_{\ell=1}^{K^0+1}
    \Bigl[B(I_\ell^0)a_\ell^2+A(I_\ell^0)b_\ell^2\Bigr]
   \notag \\
    &  =\frac12\|\*c\|^2,
    \label{eq:excessbound}
\end{align}
the final equality by~\eqref{eq:cquad}.

Inequality~\eqref{eq:excessbound} states that $\|\*v\|^2$ differs
from $\|\*c\|^2$ by at most $\tfrac12\|\*c\|^2$, that is
\begin{align}
    \frac12\|\*c\|^2
   & \le\|\*v\|^2
     \notag \\&=\bigl\|\*H^0(\*D^0)^{-1}\*c\bigr\|^2
     \notag \\
    &=\*c'\*M\*c
     \notag \\
    &\le\frac32\|\*c\|^2,
    \label{eq:sandwichc}
\end{align}
which is~\eqref{eq:scalarclaim}. Since $\*c$ was arbitrary, the left
inequality in~\eqref{eq:sandwichc} says
$\*c'\bigl(\*M-\tfrac12\*I_{K^0+2}\bigr)\*c\ge0$ for every $\*c$, so
$\*M-\tfrac12\*I_{K^0+2}$ is positive semidefinite, and the right
inequality says
$\*c'\bigl(\tfrac32\*I_{K^0+2}-\*M\bigr)\*c\ge0$ for every $\*c$, so
$\tfrac32\*I_{K^0+2}-\*M$ is positive semidefinite. By the definition
of $\preceq$ these two statements are
\begin{align}
    \frac12\*I_{K^0+2}
    \preceq(\*D^0)^{-1}\*H^{0\prime}\*H^0(\*D^0)^{-1}
    \preceq\frac32\*I_{K^0+2},
    \label{eq:loewner}
\end{align}
equivalently every eigenvalue of the normalised basis Gram lies in
$[\tfrac12,\tfrac32]$, which is the second assertion of the lemma.
The first assertion is~\eqref{eq:rjbound} of
Lemma~\ref{lem:rjorder}. The constants $\tfrac12$ and $\tfrac32$
descend only from the bound $F(L)/\sqrt{A(L)B(L)}\le\tfrac12$ of
Lemma~\ref{lem:ABF}, which holds for every integer $L\ge1$ and
involves neither $K^0$ nor the regime lengths, so~\eqref{eq:loewner}
holds uniformly over $K^0$ and over every admissible configuration
$\mathcal T^0_{K^0}$.
\end{proof}
The remaining proofs rest on three facts, which we record here rather
than inside the proofs that use them. The first is the algebra of the
block selectors. For $j=0,\ldots,K^0+1$, let
$\*P_j:=\*e_{j+1}'\otimes\*I_p$, where $\*e_{j+1}$ is the $(j+1)$th
column of $\*I_{K^0+2}$, so that $\*P_j$ extracts the $j$th boundary
block from any conformable stacked vector. Since
$\mathbb D=\mathrm{diag}(r_0,\ldots,r_{K^0+1})\otimes\*I_p$, the mixed
product rule for Kronecker products gives
\begin{align}
    \*P_j\+\eta=\+\eta_j,
    \qquad
    \*P_j\mathbb D=r_j\*P_j,
    \qquad
    \*P_j\*P_k'
    =\bigl(\*e_{j+1}'\*e_{k+1}\bigr)\otimes\*I_p
    =\delta_{jk}\*I_p
    \label{eq:selectorids}
\end{align}
for all $j,k\in\{0,\ldots,K^0+1\}$, where $\delta_{jk}$ is the
Kronecker delta. Moreover, if $\*S$ is a symmetric positive definite
$p(K^0+2)\times p(K^0+2)$ matrix and $\*H$ is an $l\times p(K^0+2)$
matrix with $\*H\*H'=\*I_l$, then, putting $\*u:=\*H'\*v$ for any
$\*v\in\mathbb R^l$, so that $\|\*u\|^2=\*v'\*H\*H'\*v=\|\*v\|^2$, the
variational characterisation of the extreme eigenvalues of a symmetric
matrix gives
$\mu_{\min}(\*S)\|\*v\|^2\le\*u'\*S\*u=\*v'\*H\*S\*H'\*v
\le\mu_{\max}(\*S)\|\*v\|^2$, that is,
\begin{align}
    \mu_{\min}(\*S)\*I_l\preceq\*H\*S\*H'\preceq\mu_{\max}(\*S)\*I_l.
    \label{eq:sandwich}
\end{align}
Two consequences of~\eqref{eq:selectorids} and~\eqref{eq:sandwich} are
used repeatedly. First, $\*H\*H'=\*I_l$ implies
$\|\*H\|_{op}=\mu_{\max}(\*H\*H')^{1/2}=1$, so the boundedness
requirement of Assumption~\ref{ass:postclt} is met automatically by
every contrast constructed below. Second, taking
$\*S=\*Q_0^{-1}\+\Sigma_0\*Q_0^{-1}$, which is symmetric and positive
definite because $\*Q_0$ and $\+\Sigma_0$ are, any limit of
$\*H_{NT}\*S\*H_{NT}'$ along a sequence with
$\*H_{NT}\*H_{NT}'=\*I_l$ inherits the bounds~\eqref{eq:sandwich} and
is therefore positive definite. In Corollaries~\ref{cor:kappa}
and~\ref{cor:thetat} only the existence of the limiting variance need
be supposed; its nonsingularity is automatic.

The second fact converts Assumption~\ref{ass:postrank}, which
concerns $\widehat{\*Q}_N$, into the statement about
$\widehat{\*Q}_N^{-1}$ that the least squares algebra actually uses.
Both $\widehat{\*Q}_N$ and $\*Q_0$ are symmetric, so Weyl's
inequality gives
$|\mu_{\min}(\widehat{\*Q}_N)-\mu_{\min}(\*Q_0)|
\le\|\widehat{\*Q}_N-\*Q_0\|_{op}$. Since $\mu_{\min}(\*Q_0)\ge c>0$
and $\|\widehat{\*Q}_N-\*Q_0\|_{op}=o_p(1)$ by
Assumption~\ref{ass:postrank}, the event
$\mathcal A_{NT}:=\{\mu_{\min}(\widehat{\*Q}_N)\ge c/2\}$ satisfies
\begin{align}
    \mathbb P\bigl(\mathcal A_{NT}\bigr)
    \ge\mathbb P\Bigl(\bigl\|\widehat{\*Q}_N-\*Q_0\bigr\|_{op}
    \le\tfrac c2\Bigr)\to1.
    \label{eq:AevProb}
\end{align}
On $\mathcal A_{NT}$ the matrix $\widehat{\*Q}_N$ is nonsingular with
$\|\widehat{\*Q}_N^{-1}\|_{op}=\mu_{\min}(\widehat{\*Q}_N)^{-1}
\le2/c$, and the resolvent identity
$\widehat{\*Q}_N^{-1}-\*Q_0^{-1}
=\widehat{\*Q}_N^{-1}(\*Q_0-\widehat{\*Q}_N)\*Q_0^{-1}$ together with
submultiplicativity of the operator norm and
$\|\*Q_0^{-1}\|_{op}=\mu_{\min}(\*Q_0)^{-1}\le1/c$ gives
$\|\widehat{\*Q}_N^{-1}-\*Q_0^{-1}\|_{op}
\le(2/c)\cdot o_p(1)\cdot(1/c)=o_p(1)$. As
$\mathbb P(\mathcal A_{NT})\to1$ by~\eqref{eq:AevProb}, we conclude
that $\widehat{\*Q}_N$ is nonsingular w.p.a.1 and that
\begin{align}
    \bigl\|\widehat{\*Q}_N^{-1}\bigr\|_{op}=O_p(1),
    \qquad
    \bigl\|\widehat{\*Q}_N^{-1}-\*Q_0^{-1}\bigr\|_{op}=o_p(1).
    \label{eq:Qinv}
\end{align}

The third fact controls $\+\Psi_{NT}$ of~\eqref{eq:psidef}, which
enters the remainder term of every limit below. Write
$\+\Psi_{NT}=N^{-1/2}\sum_{i=1}^N\+\Phi_i^{\eta\prime}\*C_T
\+\varepsilon_i$, where $\+\Phi_i^\eta$ is the $T\times p(K^0+2)$
block of $\+\Phi^\eta$ belonging to unit $i$ and
$\+\varepsilon_i:=[\varepsilon_{i,1},\ldots,\varepsilon_{i,T}]'$. The
$j$th $p\times1$ block of $\+\Phi_i^{\eta\prime}\*C_T\+\varepsilon_i$
is $\sum_{t=1}^{T}h_{j,t}\*x_{i,t}
(\varepsilon_{i,t}-\overline\varepsilon_i)$, where
$\overline\varepsilon_i:=T^{-1}\sum_{s=1}^{T}\varepsilon_{i,s}$. By
Assumption~\ref{ass:errors}(a) we have
$\mathbb E(\varepsilon_{i,s}\*x_{i,t})=\*0_{p\times1}$ for all $s$ and
$t$, whence
\begin{align}
    \mathbb E\bigl(\*x_{i,t}\overline\varepsilon_i\bigr)
    &=\frac1T\sum_{s=1}^{T}
    \mathbb E\bigl(\*x_{i,t}\varepsilon_{i,s}\bigr)
    \notag \\
    &=\*0_{p\times1},
\end{align}
so each block has mean zero, and, as $h_{j,t}$ and $\mathbb D$ are
nonrandom,
$\mathbb E(\mathbb D^{-1}\+\Psi_{NT})=\*0_{p(K^0+2)\times1}$. Mean
zero gives
$\mathbb E\|\mathbb D^{-1}\+\Psi_{NT}\|^2
=\mathrm{tr}\,\mathrm{Var}(\mathbb D^{-1}\+\Psi_{NT})$, which
converges to $\mathrm{tr}\,\+\Sigma_0<\infty$ by
Assumption~\ref{ass:postclt}(a) and, being convergent, is bounded by
some $C<\infty$ for all $N$ and $T$. Markov's inequality then gives,
for every $M>0$,
$\mathbb P(\|\mathbb D^{-1}\+\Psi_{NT}\|>M)\le C/M^2$ uniformly in
$N$ and $T$, that is,
\begin{align}
    \bigl\|\mathbb D^{-1}\+\Psi_{NT}\bigr\|=O_p(1).
    \label{eq:Psinorm}
\end{align}

We now prove Theorem~\ref{thm:oracle}.

\begin{proof}[Proof of Theorem~\ref{thm:oracle}]
We have $\widehat{\*Q}_N=\mathbb D^{-1}N^{-1}
\+\Phi^{\eta\prime}\*C\+\Phi^\eta\mathbb D^{-1}$, and $\mathbb D$ is
nonsingular by Lemma~\ref{lem:rjorder}, so multiplying on either side
by $\mathbb D$ and by $N$ gives the factorisation
\begin{align}
    \+\Phi^{\eta\prime}\*C\+\Phi^\eta
    =N\,\mathbb D\widehat{\*Q}_N\mathbb D.
    \label{eq:gramfact}
\end{align}
By~\eqref{eq:Qinv}, $\widehat{\*Q}_N$ is nonsingular w.p.a.1, and
on that event $\+\Phi^{\eta\prime}\*C\+\Phi^\eta$ is nonsingular as a
product of nonsingular matrices, with
\begin{align}
    \bigl[\+\Phi^{\eta\prime}\*C\+\Phi^\eta\bigr]^{-1}
    =\frac1N\,\mathbb D^{-1}\widehat{\*Q}_N^{-1}\mathbb D^{-1}.
    \label{eq:graminv}
\end{align}
All statements below are understood to hold on this event, whose
probability tends to one.

We first show that the oracle estimator of the boundary levels is the
least squares estimator of the centered boundary-level regression. By
definition $\widetilde{\+\eta}^{\,o}=\*M\widetilde{\+\psi}^{\,o}$ with
$\widetilde{\+\psi}^{\,o}=[\+\Phi'\*C\+\Phi]^{-1}\+\Phi'\*C\*y$, and
by the row-by-row identity~\eqref{eq:designequiv} we have
$\+\Phi=\+\Phi^\eta\*M$, whence
\begin{align}
    \+\Phi'\*C\+\Phi=\*M'\+\Phi^{\eta\prime}\*C\+\Phi^\eta\*M,
    \qquad
    \+\Phi'\*C\*y=\*M'\+\Phi^{\eta\prime}\*C\*y.
    \label{eq:transformed}
\end{align}
The matrix $\*M$ is nonsingular by~\eqref{eq:Minverse}, so the first
identity in~\eqref{eq:transformed} inverts to
$[\+\Phi'\*C\+\Phi]^{-1}
=\*M^{-1}[\+\Phi^{\eta\prime}\*C\+\Phi^\eta]^{-1}\*M^{-1\prime}$, and
therefore
\begin{align}
    \widetilde{\+\psi}^{\,o}
    &=\*M^{-1}\bigl[\+\Phi^{\eta\prime}\*C\+\Phi^\eta\bigr]^{-1}
    \*M^{-1\prime}\*M'\+\Phi^{\eta\prime}\*C\*y
    \notag \\
    &=\*M^{-1}\bigl[\+\Phi^{\eta\prime}\*C\+\Phi^\eta\bigr]^{-1}
    \+\Phi^{\eta\prime}\*C\*y.
\end{align}
Multiplying by $\*M$ gives
\begin{align}
    \widetilde{\+\eta}^{\,o}
    =\bigl[\+\Phi^{\eta\prime}\*C\+\Phi^\eta\bigr]^{-1}
    \+\Phi^{\eta\prime}\*C\*y.
    \label{eq:etaols}
\end{align}

We next derive the estimation error. Stacking~\eqref{eq:DGP} in the
representation $y_{i,t}=\mu_i+\+\phi_{i,t}^{\eta\prime}\+\eta^0
+\varepsilon_{i,t}$ gives
$\*y=\+\mu+\+\Phi^\eta\+\eta^0+\+\varepsilon$, where $\+\mu$ stacks
$\mu_i\+\iota_T$ over $i$. Since $\*C=\*I_N\otimes\*C_T$ with
$\*C_T=\*I_T-T^{-1}\+\iota_T\+\iota_T'$, we have
\begin{align}
    \*C_T\+\iota_T
    &=\+\iota_T-\frac1T\+\iota_T
    \bigl(\+\iota_T'\+\iota_T\bigr)
    \notag \\
    &=\+\iota_T-\+\iota_T
    \notag \\
    &=\*0_{T\times1},
\end{align}
and hence $\*C\+\mu=\*0$. Therefore
$\*C\*y=\*C\+\Phi^\eta\+\eta^0+\*C\+\varepsilon$, and substituting
into~\eqref{eq:etaols},
\begin{align}
    \widetilde{\+\eta}^{\,o}
    &=\bigl[\+\Phi^{\eta\prime}\*C\+\Phi^\eta\bigr]^{-1}
    \+\Phi^{\eta\prime}\*C\+\Phi^\eta\+\eta^0
    +\bigl[\+\Phi^{\eta\prime}\*C\+\Phi^\eta\bigr]^{-1}
    \+\Phi^{\eta\prime}\*C\+\varepsilon
    \notag \\
    &=\+\eta^0
    +\bigl[\+\Phi^{\eta\prime}\*C\+\Phi^\eta\bigr]^{-1}
    \+\Phi^{\eta\prime}\*C\+\varepsilon,
\end{align}
which is~\eqref{eq:oracleerr}. Inserting~\eqref{eq:graminv} together
with $\+\Phi^{\eta\prime}\*C\+\varepsilon=\sqrt N\+\Psi_{NT}$
from~\eqref{eq:psidef},
\begin{align}
    \widetilde{\+\eta}^{\,o}-\+\eta^0
    &=\frac1N\,\mathbb D^{-1}\widehat{\*Q}_N^{-1}\mathbb D^{-1}
    \cdot\sqrt N\+\Psi_{NT}
    \notag \\
    &=\frac1{\sqrt N}\,\mathbb D^{-1}\widehat{\*Q}_N^{-1}
    \mathbb D^{-1}\+\Psi_{NT},
\end{align}
and multiplying on the left by $\sqrt N\,\mathbb D$ gives
\begin{align}
    \sqrt N\,\mathbb D
    \bigl(\widetilde{\+\eta}^{\,o}-\+\eta^0\bigr)
    =\widehat{\*Q}_N^{-1}\mathbb D^{-1}\+\Psi_{NT},
    \label{eq:oracleident}
\end{align}
which holds identically in finite samples. The normalisation is
therefore not imposed; it is what the least squares algebra returns
once~\eqref{eq:gramfact} is substituted.

Let $\{\*H_{NT}\}$ satisfy the conditions of
Assumption~\ref{ass:postclt}, and suppose $\+\Upsilon$ exists and is
positive definite. Multiplying~\eqref{eq:oracleident} by $\*H_{NT}$
and adding and subtracting
$\*H_{NT}\*Q_0^{-1}\mathbb D^{-1}\+\Psi_{NT}$,
\begin{align}
    \sqrt N\,\*H_{NT}\mathbb D
    \bigl(\widetilde{\+\eta}^{\,o}-\+\eta^0\bigr)
    =\underbrace{\*H_{NT}\*Q_0^{-1}\mathbb D^{-1}\+\Psi_{NT}}_{=:\,
    \*Z_{NT}}
    +\underbrace{\*H_{NT}\bigl(\widehat{\*Q}_N^{-1}-\*Q_0^{-1}\bigr)
    \mathbb D^{-1}\+\Psi_{NT}}_{=:\,\*R_{NT}}.
    \label{eq:decomp}
\end{align}
Assumption~\ref{ass:postclt}(b) gives
$\*Z_{NT}\to_d\mathcal N(\*0_{l\times1},\+\Upsilon)$, while
submultiplicativity of the operator norm gives
\begin{align}
    \|\*R_{NT}\|
    &\le\|\*H_{NT}\|_{op}\,
    \bigl\|\widehat{\*Q}_N^{-1}-\*Q_0^{-1}\bigr\|_{op}\,
    \bigl\|\mathbb D^{-1}\+\Psi_{NT}\bigr\|
    \notag \\
    &=O(1)\cdot o_p(1)\cdot O_p(1)
    \notag \\
    &=o_p(1),
\end{align}
by Assumption~\ref{ass:postclt}, \eqref{eq:Qinv}
and~\eqref{eq:Psinorm} respectively. Slutsky's theorem applied
to~\eqref{eq:decomp} completes the proof.
\end{proof}

Theorem~\ref{thm:normality} follows by transferring the oracle limit
across the recovery event of Corollary~\ref{cor:kinks}.

\begin{proof}[Proof of Theorem~\ref{thm:normality}]
Define $\mathcal E_{NT}:=\{\widehat{\mathcal T}_{\widehat K}
=\mathcal T^0_{K^0}\}$. Since $\mathcal E_{NT}
=\{\widehat K=K^0\}\cap\{\widehat T_1=T_1^0,\ldots,
\widehat T_{K^0}=T_{K^0}^0\}$, Corollary~\ref{cor:kinks} gives
\begin{align}
    \mathbb P\bigl(\mathcal E_{NT}\bigr)
    &=\mathbb P\Bigl(\widehat T_1=T_1^0,\ldots,
    \widehat T_{K^0}=T_{K^0}^0\;\Big|\;\widehat K=K^0\Bigr)
    \,\mathbb P\bigl(\widehat K=K^0\bigr)
    \notag \\
    &\to1
    \label{eq:recovprob}
\end{align}
as the product of two factors each tending to one. On
$\mathcal E_{NT}$ the estimated and true kink sets coincide, hence so
do the estimated and true regime lengths, so
$\+\Phi(\widehat{\mathcal T}_{\widehat K})=\+\Phi$ and
$\*M_{\widehat K}=\*M$. The post-kink estimator~\eqref{eq:postest} is
a fixed function of the design and the response, so
$\widetilde{\+\psi}_{\widehat K}=\widetilde{\+\psi}^{\,o}$ on
$\mathcal E_{NT}$, and therefore
\begin{align}
    \widetilde{\+\eta}_{\widehat K}
    &=\*M_{\widehat K}\widetilde{\+\psi}_{\widehat K}
    \notag \\
    &=\*M\widetilde{\+\psi}^{\,o}
    \notag \\
    &=\widetilde{\+\eta}^{\,o}
    \qquad\text{on }\mathcal E_{NT}.
    \label{eq:coincide}
\end{align}
Put $\*Z_{NT}:=\sqrt N\,\*H_{NT}\mathbb D
(\widetilde{\+\eta}_{\widehat K}-\+\eta^0)$ and
$\*Z^{\,o}_{NT}:=\sqrt N\,\*H_{NT}\mathbb D
(\widetilde{\+\eta}^{\,o}-\+\eta^0)$. By~\eqref{eq:coincide} we have
$\{\*Z_{NT}\ne\*Z^{\,o}_{NT}\}\subseteq\mathcal E_{NT}^c$, so for
every $\delta>0$,
\begin{align}
    \mathbb P\bigl(\|\*Z_{NT}-\*Z^{\,o}_{NT}\|>\delta\bigr)
    \le\mathbb P\bigl(\mathcal E_{NT}^c\bigr)\to0
\end{align}
by~\eqref{eq:recovprob}, that is,
$\*Z_{NT}-\*Z^{\,o}_{NT}=o_p(1)$. Since
$\*Z^{\,o}_{NT}\to_d\mathcal N(\*0_{l\times1},\+\Upsilon)$ by
Theorem~\ref{thm:oracle}, Slutsky's theorem gives
$\*Z_{NT}\to_d\mathcal N(\*0_{l\times1},\+\Upsilon)$.
\end{proof}

Every rate in Section~\ref{sec:postselection} is obtained from
Theorem~\ref{thm:normality} by an appropriate choice of $\*H_{NT}$,
and we begin with the boundary levels themselves, for
which~\eqref{eq:etarate} is claimed.

\begin{proof}[Proof of~\eqref{eq:etarate}]
Fix $j\in\{0,\ldots,K^0+1\}$ and take $\*H_{NT}=\*P_j$, which is
nonrandom and satisfies $\*P_j\*P_j'=\*I_p$
by~\eqref{eq:selectorids}, hence $\|\*P_j\|_{op}=1$. As $\*P_j$,
$\*Q_0$ and $\+\Sigma_0$ do not depend on $N$ or $T$, the limit
\begin{align}
    \+\Upsilon_j^\eta
    &:=\lim\,\*P_j\*Q_0^{-1}\+\Sigma_0\*Q_0^{-1}\*P_j'
    \notag \\
    &=\*P_j\*Q_0^{-1}\+\Sigma_0\*Q_0^{-1}\*P_j'
\end{align}
exists trivially and equals the $(j+1)$th diagonal $p\times p$ block
of $\*Q_0^{-1}\+\Sigma_0\*Q_0^{-1}$, which is positive definite
by~\eqref{eq:sandwich}. Theorem~\ref{thm:normality} therefore applies
and, using $\*P_j\mathbb D=r_j\*P_j$ and $\*P_j\+\eta=\+\eta_j$
from~\eqref{eq:selectorids},
\begin{align}
    \sqrt N\,\*P_j\mathbb D
    \bigl(\widetilde{\+\eta}_{\widehat K}-\+\eta^0\bigr)
    &=\sqrt N\,r_j\*P_j
    \bigl(\widetilde{\+\eta}_{\widehat K}-\+\eta^0\bigr)
    \notag \\
    &=\sqrt N\,r_j\bigl(\widetilde{\+\eta}_j-\+\eta_j^0\bigr)
    \notag \\
    &\to_d\mathcal N\bigl(\*0_{p\times1},\+\Upsilon_j^\eta\bigr).
    \label{eq:etaclt}
\end{align}
A sequence converging in distribution is bounded in probability, so
$\sqrt N\,r_j(\widetilde{\+\eta}_j-\+\eta_j^0)=O_p(1)$ and hence
$\widetilde{\+\eta}_j-\+\eta_j^0=O_p([Nr_j^2]^{-1/2})$.
By~\eqref{eq:rjbound} we have $\tfrac18(I_j^0+I_{j+1}^0)\le r_j^2
\le I_j^0+I_{j+1}^0$, so $[Nr_j^2]^{-1/2}$ and
$[N(I_j^0+I_{j+1}^0)]^{-1/2}$ differ by a factor lying in
$[1,2\sqrt2]$, and~\eqref{eq:etarate} follows. Since
$\+\Upsilon_j^\eta$ is positive definite the limit
in~\eqref{eq:etaclt} is nondegenerate, so the rate is exact: no
normalisation of larger order leaves the limit bounded, and none of
smaller order leaves it nondegenerate.
\end{proof}

The slopes are scaled contrasts of adjacent boundary levels
by~\eqref{eq:slopefromknots}, and Corollary~\ref{cor:kappa} follows
from Theorem~\ref{thm:normality} once the corresponding contrast is
shown to be orthonormal.

\begin{proof}[Proof of Corollary~\ref{cor:kappa}]
Fix $j\in\{1,\ldots,K^0+1\}$. By~\eqref{eq:sj},
$s_j=I_j^0(r_{j-1}^{-2}+r_j^{-2})^{-1/2}$, which is well defined and
strictly positive because $r_{j-1},r_j>0$ by
Lemma~\ref{lem:rjorder}, and recall
\begin{align}
    \*H_{j,NT}^{\kappa}
    :=\frac{s_j}{I_j^0}
    \bigl(r_j^{-1}\*P_j-r_{j-1}^{-1}\*P_{j-1}\bigr).
\end{align}
Since $\*P_j\mathbb D=r_j\*P_j$ and
$\*P_{j-1}\mathbb D=r_{j-1}\*P_{j-1}$ by~\eqref{eq:selectorids},
the two scale factors cancel and
\begin{align}
    \*H_{j,NT}^{\kappa}\mathbb D
    &=\frac{s_j}{I_j^0}
    \bigl(r_j^{-1}r_j\*P_j-r_{j-1}^{-1}r_{j-1}\*P_{j-1}\bigr)
    \notag \\
    &=\frac{s_j}{I_j^0}\bigl(\*P_j-\*P_{j-1}\bigr).
    \label{eq:kappacontrast}
\end{align}
Applying~\eqref{eq:kappacontrast} to
$\widetilde{\+\eta}_{\widehat K}-\+\eta^0$ and using
$\*P_j\+\eta=\+\eta_j$,
\begin{align}
    \*H_{j,NT}^{\kappa}\mathbb D
    \bigl(\widetilde{\+\eta}_{\widehat K}-\+\eta^0\bigr)
    &=\frac{s_j}{I_j^0}
    \Bigl[\bigl(\widetilde{\+\eta}_j-\+\eta_j^0\bigr)
    -\bigl(\widetilde{\+\eta}_{j-1}-\+\eta_{j-1}^0\bigr)\Bigr]
    \notag \\
    &=s_j\Bigl[\frac{\widetilde{\+\eta}_j
    -\widetilde{\+\eta}_{j-1}}{I_j^0}
    -\frac{\+\eta_j^0-\+\eta_{j-1}^0}{I_j^0}\Bigr].
    \label{eq:kappastep}
\end{align}
The second term in brackets of~\eqref{eq:kappastep} is $\+\kappa_j^0$
by~\eqref{eq:slopefromknots}. For the first, the identity
$\widetilde{\+\eta}_{\widehat K}
=\*M_{\widehat K}\widetilde{\+\psi}_{\widehat K}$ holds in finite
samples by construction, so $\widetilde{\+\psi}_{\widehat K}
=\*M_{\widehat K}^{-1}\widetilde{\+\eta}_{\widehat K}$, and on
$\mathcal E_{NT}$ the block row of $\*M_{\widehat K}^{-1}$
in~\eqref{eq:Minverse} corresponding to $\+\kappa_j$ carries
$-(I_j^0)^{-1}\*I_p$ in block column $j-1$ and $(I_j^0)^{-1}\*I_p$ in
block column $j$, whence
$\widetilde{\+\kappa}_j
=(\widetilde{\+\eta}_j-\widetilde{\+\eta}_{j-1})/I_j^0$. Therefore
\begin{align}
    \*H_{j,NT}^{\kappa}\mathbb D
    \bigl(\widetilde{\+\eta}_{\widehat K}-\+\eta^0\bigr)
    =s_j\bigl(\widetilde{\+\kappa}_j-\+\kappa_j^0\bigr)
    \qquad\text{on }\mathcal E_{NT}.
    \label{eq:kappaequals}
\end{align}
Moreover, using $\*P_j\*P_k'=\delta_{jk}\*I_p$
from~\eqref{eq:selectorids}, the two cross terms vanish and
\begin{align}
    \*H_{j,NT}^{\kappa}\*H_{j,NT}^{\kappa\prime}
    &=\frac{s_j^2}{(I_j^0)^2}
    \bigl(r_j^{-1}\*P_j-r_{j-1}^{-1}\*P_{j-1}\bigr)
    \bigl(r_j^{-1}\*P_j'-r_{j-1}^{-1}\*P_{j-1}'\bigr)
    \notag \\
    &=\frac{s_j^2}{(I_j^0)^2}
    \Bigl[r_j^{-2}\*P_j\*P_j'
    -r_j^{-1}r_{j-1}^{-1}\*P_j\*P_{j-1}'
    -r_{j-1}^{-1}r_j^{-1}\*P_{j-1}\*P_j'
    +r_{j-1}^{-2}\*P_{j-1}\*P_{j-1}'\Bigr]
    \notag \\
    &=\frac{s_j^2}{(I_j^0)^2}
    \bigl(r_{j-1}^{-2}+r_j^{-2}\bigr)\*I_p
    \notag \\
    &=\*I_p,
    \label{eq:kappaortho}
\end{align}
the last equality by~\eqref{eq:sj}, which gives
$s_j^2/(I_j^0)^2=(r_{j-1}^{-2}+r_j^{-2})^{-1}$. Hence
$\|\*H_{j,NT}^{\kappa}\|_{op}=1$, so
Assumption~\ref{ass:postclt} applies to this sequence, and by the
discussion following~\eqref{eq:sandwich} the limit
$\+\Upsilon_j^{\kappa}$ is positive definite whenever it exists.
Theorem~\ref{thm:normality} with $\*H_{NT}=\*H_{j,NT}^{\kappa}$ now
gives
\begin{align}
    \sqrt N\,\*H_{j,NT}^{\kappa}\mathbb D
    \bigl(\widetilde{\+\eta}_{\widehat K}-\+\eta^0\bigr)
    \to_d\mathcal N\bigl(\*0_{p\times1},\+\Upsilon_j^{\kappa}\bigr),
\end{align}
and by~\eqref{eq:kappaequals} the left side equals
$\sqrt N\,s_j(\widetilde{\+\kappa}_j-\+\kappa_j^0)$ on
$\mathcal E_{NT}$, whose complement has probability tending to zero
by~\eqref{eq:recovprob}. The two therefore differ by $o_p(1)$, and
Slutsky's theorem gives
\begin{align}
    \sqrt N\,s_j
    \bigl(\widetilde{\+\kappa}_j-\+\kappa_j^0\bigr)
    \to_d\mathcal N\bigl(\*0_{p\times1},\+\Upsilon_j^{\kappa}\bigr).
    \label{eq:kappaclt}
\end{align}

It remains to establish the order of $s_j^2$. For any $u,v>0$,
\begin{align}
    \frac12\bigl(u\wedge v\bigr)
    &\le\frac1{u^{-1}+v^{-1}}\notag\\&=\frac{uv}{u+v}
    \notag\\&\le u\wedge v,
    \label{eq:harmonic}
\end{align}
since, taking $u\le v$ without loss of generality, the upper bound
follows from $uv/(u+v)\le uv/v=u$ and the lower bound from
$uv/(u+v)\ge uv/(2v)=u/2$. Applying~\eqref{eq:harmonic} with
$u=r_{j-1}^2$ and $v=r_j^2$,
\begin{align}
    \frac12\bigl(r_{j-1}^2\wedge r_j^2\bigr)
    \le\frac{s_j^2}{(I_j^0)^2}
    =\frac1{r_{j-1}^{-2}+r_j^{-2}}
    \le r_{j-1}^2\wedge r_j^2,
    \label{eq:sjharmonic}
\end{align}
and combining~\eqref{eq:sjharmonic} with~\eqref{eq:rminorder},
\begin{align}
    \frac1{16}\,(I_j^0)^2
    \Bigl[\bigl(I_{j-1}^0+I_j^0\bigr)
    \wedge\bigl(I_j^0+I_{j+1}^0\bigr)\Bigr]
    \le s_j^2
    \le(I_j^0)^2
    \Bigl[\bigl(I_{j-1}^0+I_j^0\bigr)
    \wedge\bigl(I_j^0+I_{j+1}^0\bigr)\Bigr].
    \label{eq:sjorder}
\end{align}
Hence $s_j^2\asymp(I_j^0)^2[(I_{j-1}^0+I_j^0)\wedge
(I_j^0+I_{j+1}^0)]$ with the absolute constants $1/16$ and $1$,
uniformly over every admissible configuration. As \eqref{eq:kappaclt}
gives $\sqrt N\,s_j(\widetilde{\+\kappa}_j-\+\kappa_j^0)=O_p(1)$, we
obtain
\begin{align}
    \widetilde{\+\kappa}_j-\+\kappa_j^0
    &=O_p\bigl([Ns_j^2]^{-1/2}\bigr)
    \notag \\
    &=O_p\Bigl(\Bigl[N(I_j^0)^2
    \bigl\{(I_{j-1}^0+I_j^0)\wedge(I_j^0+I_{j+1}^0)\bigr\}
    \Bigr]^{-1/2}\Bigr).
\end{align}
The limit in~\eqref{eq:kappaclt} being nondegenerate, the rate is
exact and not merely an upper bound.
\end{proof}

The coefficient path is a convex combination of two adjacent boundary
levels by~\eqref{eq:knotrep}, and Corollary~\ref{cor:thetat} follows
by the same route. The rate claimed there holds uniformly over the
dates of the regime, however, and does not follow from the pointwise
limit; we therefore return to the underlying identity for that part of
the argument.

\begin{proof}[Proof of Corollary~\ref{cor:thetat}]
Fix $\ell\in\{1,\ldots,K^0+1\}$ and $t\in\mathcal K_\ell^0$, and
abbreviate $u:=u_{\ell,t}\in[0,1)$. Identity~\eqref{eq:knotrep} was
obtained by substituting
$\+\theta_{T_{\ell-1}^0}^0=\+\eta_{\ell-1}^0$
and~\eqref{eq:slopefromknots} into~\eqref{eq:Kink}, both of which are
identities in the parameters rather than statements about their true
values. The linear map $\*L_t(\mathcal T^0_{K^0})$
of~\eqref{eq:pk_path} therefore satisfies
$\*L_t(\mathcal T^0_{K^0})=(1-u)\*P_{\ell-1}\*M+u\*P_\ell\*M$, so
applying both sides to $\widetilde{\+\psi}_{\widehat K}$ on
$\mathcal E_{NT}$ and using $\widetilde{\+\eta}_{\widehat K}
=\*M\widetilde{\+\psi}_{\widehat K}$ gives
$\widetilde{\+\theta}_t=(1-u)\widetilde{\+\eta}_{\ell-1}
+u\widetilde{\+\eta}_\ell$. Subtracting~\eqref{eq:knotrep},
\begin{align}
    \widetilde{\+\theta}_t-\+\theta_t^0
    =(1-u)\bigl(\widetilde{\+\eta}_{\ell-1}
    -\+\eta_{\ell-1}^0\bigr)
    +u\bigl(\widetilde{\+\eta}_\ell-\+\eta_\ell^0\bigr).
    \label{eq:thetaerr}
\end{align}
With
$w_t=[(1-u)^2r_{\ell-1}^{-2}+u^2r_\ell^{-2}]^{-1/2}$ and
$\*H_{t,NT}^{\theta}:=w_t[(1-u)r_{\ell-1}^{-1}\*P_{\ell-1}
+u\,r_\ell^{-1}\*P_\ell]$, \eqref{eq:selectorids} gives, exactly as
in~\eqref{eq:kappacontrast},
\begin{align}
    \*H_{t,NT}^{\theta}\mathbb D
    \bigl(\widetilde{\+\eta}_{\widehat K}-\+\eta^0\bigr)
    &=w_t\Bigl[(1-u)\bigl(\widetilde{\+\eta}_{\ell-1}
    -\+\eta_{\ell-1}^0\bigr)
    +u\bigl(\widetilde{\+\eta}_\ell-\+\eta_\ell^0\bigr)\Bigr]
    \notag \\
    &=w_t\bigl(\widetilde{\+\theta}_t-\+\theta_t^0\bigr)
\end{align}
on $\mathcal E_{NT}$, by~\eqref{eq:thetaerr}. The cross terms again
vanish by $\*P_{\ell-1}\*P_\ell'=\*0$, so
\begin{align}
    \*H_{t,NT}^{\theta}\*H_{t,NT}^{\theta\prime}
    &=w_t^2\bigl[(1-u)^2r_{\ell-1}^{-2}+u^2r_\ell^{-2}\bigr]\*I_p
    \notag \\
    &=\*I_p,
\end{align}
hence $\|\*H_{t,NT}^{\theta}\|_{op}=1$, so
Assumption~\ref{ass:postclt} applies and $\+\Upsilon_t^{\theta}$ is
positive definite whenever it exists. Theorem~\ref{thm:normality} and
Slutsky's theorem, applied exactly as in the proof of
Corollary~\ref{cor:kappa}, give
\begin{align}
    \sqrt N\,w_t\bigl(\widetilde{\+\theta}_t-\+\theta_t^0\bigr)
    \to_d\mathcal N\bigl(\*0_{p\times1},\+\Upsilon_t^{\theta}\bigr).
\end{align}

For the rate we return to~\eqref{eq:thetaerr}. Since $u\in[0,1)$, both
weights are nonnegative and sum to one, so the triangle inequality
gives, for every $t\in\mathcal K_\ell^0$,
\begin{align}
    \bigl\|\widetilde{\+\theta}_t-\+\theta_t^0\bigr\|
    &\le(1-u)\bigl\|\widetilde{\+\eta}_{\ell-1}
    -\+\eta_{\ell-1}^0\bigr\|
    +u\bigl\|\widetilde{\+\eta}_\ell-\+\eta_\ell^0\bigr\|
    \notag \\
    &\le\bigl[(1-u)+u\bigr]
    \max\Bigl\{\bigl\|\widetilde{\+\eta}_{\ell-1}
    -\+\eta_{\ell-1}^0\bigr\|,
    \bigl\|\widetilde{\+\eta}_\ell-\+\eta_\ell^0\bigr\|\Bigr\}
    \notag \\
    &=\max\Bigl\{\bigl\|\widetilde{\+\eta}_{\ell-1}
    -\+\eta_{\ell-1}^0\bigr\|,
    \bigl\|\widetilde{\+\eta}_\ell-\+\eta_\ell^0\bigr\|\Bigr\}.
    \label{eq:thetatriangle}
\end{align}
The right side of~\eqref{eq:thetatriangle} does not depend on $t$, so
the same bound holds for $\max_{t\in\mathcal K_\ell^0}
\|\widetilde{\+\theta}_t-\+\theta_t^0\|$. By~\eqref{eq:etarate}
applied at $\ell-1$ and at $\ell$, the two norms are
$O_p([Nr_{\ell-1}^2]^{-1/2})$ and $O_p([Nr_\ell^2]^{-1/2})$
respectively, and the maximum of two terms that are $O_p(a_N)$ and
$O_p(b_N)$ is $O_p(a_N\vee b_N)$. Hence
\begin{align}
    \max_{t\in\mathcal K_\ell^0}
    \bigl\|\widetilde{\+\theta}_t-\+\theta_t^0\bigr\|
    &=O_p\Bigl(\bigl[N\bigl(r_{\ell-1}^2\wedge r_\ell^2\bigr)
    \bigr]^{-1/2}\Bigr)
    \notag \\
    &=O_p\Bigl(\Bigl[N\bigl\{(I_{\ell-1}^0+I_\ell^0)
    \wedge(I_\ell^0+I_{\ell+1}^0)\bigr\}\Bigr]^{-1/2}\Bigr),
    \label{eq:thetaunifrate}
\end{align}
the last equality by~\eqref{eq:rminorder}, which is the second claim
of Corollary~\ref{cor:thetat}.

Finally, the two statements are consistent, in that the normalisation
$w_t$ is of the order claimed. Since $u\in[0,1)$ we have
$(1-u)^2+u^2\le1$, the maximum over $[0,1]$ being attained at the
endpoints, so
\begin{align}
    (1-u)^2r_{\ell-1}^{-2}+u^2r_\ell^{-2}
    &\le\bigl[(1-u)^2+u^2\bigr]
    \max\bigl\{r_{\ell-1}^{-2},r_\ell^{-2}\bigr\}
    \notag \\
    &\le\frac1{r_{\ell-1}^2\wedge r_\ell^2},
    \label{eq:wtfloor}
\end{align}
whence $w_t^2\ge r_{\ell-1}^2\wedge r_\ell^2$, which
by~\eqref{eq:rminorder} is of the order of
$(I_{\ell-1}^0+I_\ell^0)\wedge(I_\ell^0+I_{\ell+1}^0)$.
\end{proof}

\section*{Proof of Theorem~\ref{thm:cbc}}

\begin{lemma}[Consistency of the component-wise penalised estimator]
\label{lem:cwcons}
Suppose that Assumptions~\ref{ass:errors} and~\ref{ass:rank}(a) hold,
and that Assumption~\ref{ass:signal}(a) holds with $J_{\min}$,
$\vartheta_1$, $\zeta_1$, and $K^0$ replaced by $J^{\mathrm{cw}}_{\min}$,
$\vartheta_2$, $\zeta_2$, and $M^0$, respectively, and with part (a)
replaced by $\vartheta_2\to_p0$ when $M^0=0$. Let
$\widehat{\+\theta}{}^{\mathrm{cw}}_t
:=[\widehat\theta^{\mathrm{cw}}_{1,t},\ldots,
\widehat\theta^{\mathrm{cw}}_{p,t}]'$ collect the minimiser
of~\eqref{eq:objective_componentwise}. Then, as $N,T\to\infty$,
\begin{enumerate}[label=\textnormal{(\alph*)}]
\item $\widehat{\+\theta}{}^{\mathrm{cw}}_1-\+\theta_1^0
      =O_p\!\left(N^{-1/2}\right)$;
\item $\dfrac1T\displaystyle\sum_{t=1}^T
      \bigl\|\widehat{\+\theta}{}^{\mathrm{cw}}_t-\+\theta_t^0\bigr\|^2
      =O_p\!\left(N^{-1}\right)$;
\item $\displaystyle\max_{1\le t\le T}
      \bigl\|\widehat{\+\theta}{}^{\mathrm{cw}}_t-\+\theta_t^0\bigr\|
      =O_p\!\left(\sqrt{T/N}\right)$.
\end{enumerate}
\end{lemma}

\begin{proof}[Proof of Lemma~\ref{lem:cwcons}]
Throughout, $\*b_t=\sqrt N(\+\theta_t-\+\theta_t^0)$ with scalar
components $b_{j,t}$, $\*b=[\*b_1',\ldots,\*b_T']'$, and $\bar b$ is as
in~\eqref{eq:bbar}. Define the active pair set
\(
    \mathcal S^0
    :=\bigl\{(j,t):1\le j\le p,\;
    t\in\mathcal T^0_{j,K_j^0}\bigr\}\) with \(
    |\mathcal S^0|=\sum_{j=1}^pK_j^0=M^0,
\)
so that, by the component-wise analog of
Proposition~\ref{prop:FD}, $\Delta^2\theta_{j,t}^0\neq0$ if and only if
$(j,t)\in\mathcal S^0$, with
$\Delta^2\theta^0_{j,T^0_{j,\ell}}
=\kappa^0_{j,\ell+1}-\kappa^0_{j,\ell}$ and hence
\begin{align}
    \min_{(j,t)\in\mathcal S^0}
    \bigl|\Delta^2\theta_{j,t}^0\bigr|
    =J^{\mathrm{cw}}_{\min}.
    \label{eq:cwJmin}
\end{align}
If $M^0=0$ the penalty in~\eqref{eq:objective_componentwise} vanishes at
the truth, the term $L_3^{\mathrm{cw}}(\*b)$ below is identically zero,
and the argument goes through with the obvious simplifications, so we
take $M^0\ge1$. The proof is in four steps. Step 1 collects the
properties of the preliminary estimator. Step 2 bounds the adaptive
weights over the active pairs. Step 3 decomposes the objective and
bounds each term. Step 4 completes the quadratic argument.

\medskip
\noindent\emph{Step 1: the preliminary estimator.}\;
The preliminary estimator $\dot{\+\theta}_t$
of~\eqref{eq:initial} is the same unpenalised regression as in the
baseline model, so Step 1 of the proof of
Theorem~\ref{thm:consistency}, which nowhere involves the penalty,
applies verbatim. In particular the energy
bound~\eqref{eq:prelimenergy} holds, giving
$\dot{\+\theta}_1-\+\theta_1^0=O_p(N^{-1/2})$, the first order
condition~\eqref{eq:prelimfoc} holds with the design
bounds $\mu_{\min}(\*A_t)\ge2\underline c$,
$\|\*A_t^{-1}\|\le(2\underline c)^{-1}$, and
$\|\*C_t\|\le2\overline c$ w.p.1 uniformly in $t$, and
$\mathbb E\|\sqrt N\+\zeta_t\|^2\le C$ uniformly in $t$
by~\eqref{eq:supform}.

\medskip
\noindent\emph{Step 2: the weight bound over the active pairs.}\;
Let
\(
    \mathcal D^0
    :=\bigcup_{(j,t)\in\mathcal S^0}\{t-1,t,t+1\},
    \) with \(    |\mathcal D^0|\le3M^0
\)
be  the  set of dates entering a second difference at some
active pair. By the union bound and Markov's inequality, for any $M>0$,
\begin{align}
    \mathbb P\Bigl(\max_{s\in\mathcal D^0}
    \bigl\|\sqrt N\+\zeta_s\bigr\|>M\sqrt{M^0}\Bigr)
 &   \le\sum_{s\in\mathcal D^0}
    \frac{\mathbb E\|\sqrt N\+\zeta_s\|^2}{M^2M^0}
   \notag\\
    &  \le\frac{3M^0C}{M^2M^0}
    \notag\\ &=\frac{3C}{M^2},
\end{align}
so that $\max_{s\in\mathcal D^0}\|\+\zeta_s\|=O_p(\sqrt{M^0/N})$.
Combining this with the first order
condition~\eqref{eq:prelimfoc}, the design bounds of Step 1, and
$\dot{\+\theta}_1-\+\theta_1^0=O_p(N^{-1/2})$,
\begin{align}
    \max_{s\in\mathcal D^0,\,s\ge2}
    \bigl\|\dot{\+\theta}_s-\+\theta_s^0\bigr\|
  &  \le\frac{1}{2\underline c}
    \Bigl(\max_{s\in\mathcal D^0}\|\+\zeta_s\|
    +2\overline c\,
    \bigl\|\dot{\+\theta}_1-\+\theta_1^0\bigr\|\Bigr)
   \notag\\
    &  =O_p\bigl(\sqrt{M^0/N}\bigr),
    \label{eq:cwprelimmax}
\end{align}
where $M^0\ge1$ absorbs the $O_p(N^{-1/2})$ contribution, and the date
$s=1$, which enters only when some coordinate kinks at $t=2$, is covered
directly by $\dot{\+\theta}_1-\+\theta_1^0=O_p(N^{-1/2})$. Since a
scalar component is bounded by the Euclidean norm of its vector, for
every $(j,t)\in\mathcal S^0$,
\begin{align}
    \bigl|\Delta^2\dot\theta_{j,t}-\Delta^2\theta_{j,t}^0\bigr|
    &\le\bigl\|\Delta^2\bigl(\dot{\+\theta}_t-\+\theta_t^0\bigr)\bigr\|
    \notag\\&\le4\max_{s\in\mathcal D^0\cup\{1\}}
    \bigl\|\dot{\+\theta}_s-\+\theta_s^0\bigr\|,
    \label{eq:cwscalardev}
\end{align}
the constant $4$ being the absolute sum of the coefficients $(1,-2,1)$,
so that we have 
$\max_{(j,t)\in\mathcal S^0}
|\Delta^2\dot\theta_{j,t}-\Delta^2\theta_{j,t}^0|
=O_p(\sqrt{M^0/N})$. Define
\begin{align}
    \mathcal J^{\mathrm{cw}}_{NT}
    :=\Bigl\{\max_{(j,t)\in\mathcal S^0}
    \bigl|\Delta^2\dot\theta_{j,t}-\Delta^2\theta_{j,t}^0\bigr|
    \le J^{\mathrm{cw}}_{\min}/2\Bigr\}.
    \label{eq:cwevent}
\end{align}
Writing the maximum as ${\mathfrak G}^{\mathrm{cw}}_{NT}\sqrt{M^0/N}$
with ${\mathfrak G}^{\mathrm{cw}}_{NT}=O_p(1)$, the complement of
$\mathcal J^{\mathrm{cw}}_{NT}$ is the event
$\{{\mathfrak G}^{\mathrm{cw}}_{NT}
>\tfrac12\sqrt{N/M^0}\,J^{\mathrm{cw}}_{\min}\}$, and the threshold
diverges by the second clause of the modified
Assumption~\ref{ass:signal}(a). An $O_p(1)$ sequence exceeds a
divergent threshold with vanishing probability, so
$\mathbb P(\mathcal J^{\mathrm{cw}}_{NT})\to1$. On
$\mathcal J^{\mathrm{cw}}_{NT}$, the reverse triangle inequality
and~\eqref{eq:cwJmin} give, for every $(j,t)\in\mathcal S^0$,
$|\Delta^2\dot\theta_{j,t}|\ge J^{\mathrm{cw}}_{\min}/2$, and therefore
\begin{align}
    \max_{(j,t)\in\mathcal S^0}\dot\omega^{\mathrm{cw}}_{j,t}
    &=\max_{(j,t)\in\mathcal S^0}
    \bigl|\Delta^2\dot\theta_{j,t}\bigr|^{-\zeta_2}
   \notag\\  &\le2^{\zeta_2}\,
    \bigl(J^{\mathrm{cw}}_{\min}\bigr)^{-\zeta_2}.
    \label{eq:cwwmax}
\end{align}

\medskip
\noindent\emph{Step 3: decomposition and bounds.}\;
Using~\eqref{eq:resid} and splitting the penalty at $\mathcal S^0$,
\begin{align}
    N\bigl[\mathcal L^{\mathrm{cw}}_{\vartheta_2}(\+\Theta_T)
    -\mathcal L^{\mathrm{cw}}_{\vartheta_2}(\+\Theta_T^0)\bigr]
    =-2L_1(\*b)+L_2(\*b)
    +L_3^{\mathrm{cw}}(\*b)+L_4^{\mathrm{cw}}(\*b),
    \label{eq:cwdecomp}
\end{align}
where $L_1(\*b)$ and $L_2(\*b)$ are exactly as
in~\eqref{eq:prelimdecomp}, since the loss
in~\eqref{eq:objective_componentwise} is identical to that
in~\eqref{eq:objective}, and
\begin{align}
    L_3^{\mathrm{cw}}(\*b)
    &:=N\vartheta_2\sum_{(j,t)\in\mathcal S^0}
    \dot\omega^{\mathrm{cw}}_{j,t}
    \Bigl[\Bigl|\Delta^2\theta_{j,t}^0
    +\frac{1}{\sqrt N}\Delta^2 b_{j,t}\Bigr|
    -\bigl|\Delta^2\theta_{j,t}^0\bigr|\Bigr],
    \notag\\
    L_4^{\mathrm{cw}}(\*b)
    &:=N\vartheta_2\sum_{(j,t)\notin\mathcal S^0}
    \dot\omega^{\mathrm{cw}}_{j,t}\,
    \frac{1}{\sqrt N}\bigl|\Delta^2 b_{j,t}\bigr|
    \;\ge\;0,
    \label{eq:cwL34}
\end{align}
the second display using $\Delta^2\theta_{j,t}^0=0$ off the active set.
The bounds~\eqref{eq:L1} and~\eqref{eq:L2} therefore apply unchanged,
$|L_1(\*b)|=O_p(\bar b)$ and
$L_2(\*b)\ge\underline c\,\alpha\,\bar b^2$ w.p.1 with
$\alpha=(3-\sqrt5)/2$. For $L_3^{\mathrm{cw}}(\*b)$, the triangle
inequality gives
$|\Delta^2\theta_{j,t}^0+N^{-1/2}\Delta^2 b_{j,t}|
-|\Delta^2\theta_{j,t}^0|\ge-N^{-1/2}|\Delta^2 b_{j,t}|$, so that,
by~\eqref{eq:cwwmax},
\begin{align}
    L_3^{\mathrm{cw}}(\*b)
    \ge-\sqrt N\vartheta_2\,
    2^{\zeta_2}\bigl(J^{\mathrm{cw}}_{\min}\bigr)^{-\zeta_2}
    \sum_{(j,t)\in\mathcal S^0}\bigl|\Delta^2 b_{j,t}\bigr|
    \label{eq:cwL3start}
\end{align}
on $\mathcal J^{\mathrm{cw}}_{NT}$. For the sum, the scalar analog of
the window bound gives
$|\Delta^2 b_{j,t}|\le|b_{j,t+1}|+2|b_{j,t}|+|b_{j,t-1}|$ and hence, by
the Cauchy--Schwarz inequality with coefficients $(1,2,1)$,
$|\Delta^2 b_{j,t}|^2\le6(b_{j,t+1}^2+b_{j,t}^2+b_{j,t-1}^2)$. For each
fixed $j$ the dates $\mathcal T^0_{j,K_j^0}$ are distinct, so each index
$s$ belongs to at most three windows $\{t-1,t,t+1\}$ with
$t\in\mathcal T^0_{j,K_j^0}$, and summing first over
$t\in\mathcal T^0_{j,K_j^0}$ and then over $j$,
\begin{align}
    \sum_{(j,t)\in\mathcal S^0}\bigl|\Delta^2 b_{j,t}\bigr|^2
  & \le18\sum_{j=1}^p\sum_{s=1}^Tb_{j,s}^2
    \notag\\
    & =18\,\|\*b\|^2.
    \label{eq:cwwindow}
\end{align}
By the Cauchy--Schwarz inequality over the $M^0$ active pairs and
$\|\*b\|\le\|\mathbb D_0\*b\|=\sqrt T\,\bar b$,
\begin{align}
    \sum_{(j,t)\in\mathcal S^0}\bigl|\Delta^2 b_{j,t}\bigr|
 &   \le\sqrt{M^0}
    \Bigl(\sum_{(j,t)\in\mathcal S^0}
    \bigl|\Delta^2 b_{j,t}\bigr|^2\Bigr)^{1/2}
   \notag\\
    &  \le\sqrt{18M^0T}\,\bar b.
    \label{eq:cwCSsum}
\end{align}
Inserting~\eqref{eq:cwCSsum} into~\eqref{eq:cwL3start} and applying the
first clause of the modified Assumption~\ref{ass:signal}(a), by which
$\sqrt{NTM^0}\,\vartheta_2
(J^{\mathrm{cw}}_{\min})^{-\zeta_2}=O_p(1)$,
\begin{align}
    L_3^{\mathrm{cw}}(\*b)
 &   \ge-O_p\Bigl(\sqrt{NTM^0}\,\vartheta_2
    \bigl(J^{\mathrm{cw}}_{\min}\bigr)^{-\zeta_2}\Bigr)\bar b
  \notag\\
    &   =-O_p(1)\,\bar b.
    \label{eq:cwL3}
\end{align}

\medskip
\noindent\emph{Step 4: the quadratic argument.}\;
Combining~\eqref{eq:cwdecomp} with the bounds of Step 3 and
$L_4^{\mathrm{cw}}(\*b)\ge0$,
\begin{align}
    N\bigl[\mathcal L^{\mathrm{cw}}_{\vartheta_2}(\+\Theta_T)
    -\mathcal L^{\mathrm{cw}}_{\vartheta_2}(\+\Theta_T^0)\bigr]
    \ge\underline c\,\alpha\,\bar b^2
    -{\mathfrak H}^{\mathrm{cw}}_{NT}\,\bar b,
    \label{eq:cwquad}
\end{align}
where ${\mathfrak H}^{\mathrm{cw}}_{NT}=O_p(1)$ collects the
contributions of $2|L_1(\*b)|$ and $|L_3^{\mathrm{cw}}(\*b)|$. The
minimiser satisfies
$N[\mathcal L^{\mathrm{cw}}_{\vartheta_2}
(\widehat{\+\Theta}{}^{\mathrm{cw}}_T)
-\mathcal L^{\mathrm{cw}}_{\vartheta_2}(\+\Theta_T^0)]\le0$, and the
right side of~\eqref{eq:cwquad} is strictly positive once
$\bar b>{\mathfrak H}^{\mathrm{cw}}_{NT}/(\underline c\,\alpha)$, so the
argument closing the proof of Theorem~\ref{thm:consistency} applies
word for word and yields $\widehat{\bar b}{}^{\mathrm{cw}}=O_p(1)$, that
is,
\begin{align}
    (T-1)\bigl\|\widehat{\*b}{}^{\mathrm{cw}}_1\bigr\|^2
    +\sum_{t=2}^T\bigl\|\widehat{\*b}{}^{\mathrm{cw}}_t\bigr\|^2
    =O_p(T).
    \label{eq:cwenergy}
\end{align}
Parts (a), (b), and (c) are read off~\eqref{eq:cwenergy} exactly as in
the proof of Theorem~\ref{thm:consistency}, the first term giving
part (a), division by $NT$ giving part (b), and
$\max_t\|\widehat{\*b}{}^{\mathrm{cw}}_t\|
\le\sqrt T\,\widehat{\bar b}{}^{\mathrm{cw}}=O_p(\sqrt T)$ giving
part (c).
\end{proof}

\begin{proof}
The proof is in four steps. Step 1 establishes part (i), that every
inactive scalar second difference is set exactly to zero w.p.a.1.
Step 2 records the component-wise detection condition. Step 3 shows
that every active scalar second difference survives. Step 4 combines
the steps into parts (ii) and (iii). If $M^0=0$ there are no active
pairs, Steps 2 and 3 are vacuous, and the result reduces to Step 1, so
we take $M^0\ge1$ where the distinction matters.

\medskip
\noindent\emph{Step 1: no spurious kinks.}\;
Reparametrise coordinate by coordinate. For each $j$, set
$\gamma_j:=\theta_{j,2}-\theta_{j,1}$ and
$\nu_{j,s}:=\Delta^2\theta_{j,s}$ for $s=2,\ldots,T-1$, so that, by the
scalar version of the telescoping argument
behind~\eqref{eq:reparam},
\begin{align}
    \theta_{j,t}
    =\theta_{j,1}+(t-1)\gamma_j
    +\sum_{s=2}^{t-1}(t-s)\,\nu_{j,s},
    \label{eq:cwreparam2}
\end{align}
for $t=1,\ldots,T,$ where each $\nu_{j,s}$ being counted exactly $t-s$ times. Stacking over $j$,
the map $(\+\theta_1,\+\gamma,\{\+\nu_s\})\mapsto\+\Theta_T$ is the
same linear bijection of $\mathbb R^{Tp}$ as in the proof of
Theorem~\ref{thm:sign}, now read one coordinate at a time. The
objective~\eqref{eq:objective_componentwise} is convex, being the sum
of a convex quadratic and a nonnegative combination of absolute values,
convexity is preserved under the linear
reparametrisation~\eqref{eq:cwreparam2}, and the penalty
$\vartheta_2\sum_{s}\sum_j\dot\omega^{\mathrm{cw}}_{j,s}|\nu_{j,s}|$ is
separable across the pairs $(j,s)$. The inclusion of zero in the
subdifferential therefore characterises the global minimiser, and it
suffices to read the first order condition with respect to each scalar
$\nu_{j,s}$.

Write $\widehat e_{i,t}
:=\check y_{i,t}-\*x_{i,t}'\widehat{\+\theta}{}^{\mathrm{cw}}_t
+\*x_{i,1}'\widehat{\+\theta}{}^{\mathrm{cw}}_1$ for the fitted
residuals. From~\eqref{eq:cwreparam2}, the loss depends on
$\nu_{j,s}$ only through the scalars $\theta_{j,t}$ with $t>s$, with
$\partial\theta_{j,t}/\partial\nu_{j,s}=(t-s)$ for $t>s$ and zero
otherwise, while
$\partial\mathcal L^{\mathrm{cw},\mathrm{loss}}_{\vartheta_2}
/\partial\theta_{j,t}
=-\tfrac{2}{NT}\sum_{i=1}^Nx_{i,t,j}\,\widehat e_{i,t}$ for $t\ge2$.
The chain rule and the subdifferential of the absolute value therefore
give, for each pair $(j,s)$,
\begin{align}
    \vartheta_2\,\dot\omega^{\mathrm{cw}}_{j,s}\,\widehat g_{j,s}
    =\frac{2}{NT}\sum_{t=s+1}^T(t-s)\sum_{i=1}^N
    x_{i,t,j}\,\widehat e_{i,t},
    \label{eq:cwkkt}
\end{align}
where $\widehat g_{j,s}=\mathrm{sgn}(\widehat\nu_{j,s})$ if
$\widehat\nu_{j,s}\neq0$ and $|\widehat g_{j,s}|\le1$ otherwise. If
$|\Delta^2\dot\theta_{j,s}|=0$ for some pair, then
$\dot\omega^{\mathrm{cw}}_{j,s}=+\infty$, any candidate with
$\nu_{j,s}\neq0$ has infinite objective value, the minimiser sets
$\widehat\nu_{j,s}=0$ there, and we may take
$\dot\omega^{\mathrm{cw}}_{j,s}<\infty$ throughout. If
$\widehat\nu_{j,s}\neq0$, taking absolute values
in~\eqref{eq:cwkkt} and using $|\widehat g_{j,s}|=1$,
\begin{align}
    \vartheta_2\,\dot\omega^{\mathrm{cw}}_{j,s}
    =\frac{2}{T}\Bigl|\sum_{t=s+1}^T(t-s)\,
    \mathcal E_{j,t}\Bigr|,
    \label{eq:cwkktnorm}
\end{align}
where $   \mathcal E_{j,t}
    :=\frac1N\sum_{i=1}^Nx_{i,t,j}\,\widehat e_{i,t}$
and $\mathcal E_{j,t}$ is the $j$th component of the vector
$\mathcal E_t$ of the proof of Theorem~\ref{thm:sign}, now formed from
the component-wise fit.

\medskip
\noindent\emph{The floor.}\;
Fix an inactive pair, $(j,s)\notin\mathcal S^0$ with
$s\in\{2,\ldots,T-1\}$, so that $\Delta^2\theta_{j,s}^0=0$, while other
coordinates may well kink at $s$. Since a scalar component is bounded
by the norm of its vector, and
using~\eqref{eq:Gtildedef} for the uniform preliminary rate,
\begin{align}
    \bigl|\Delta^2\dot\theta_{j,s}\bigr|
    &=\bigl|\Delta^2\dot\theta_{j,s}-\Delta^2\theta_{j,s}^0\bigr|
     \notag\\
    &\le\bigl\|\Delta^2\bigl(\dot{\+\theta}_s
    -\+\theta_s^0\bigr)\bigr\|
     \notag\\
    &\le4\widetilde{\mathfrak G}_{NT}\sqrt{T/N},
    \label{eq:cwfloorprep}
\end{align}
uniformly over all inactive pairs since the right side depends on
neither $j$ nor $s$. Hence
\begin{align}
    \min_{(j,s)\notin\mathcal S^0}
    \dot\omega^{\mathrm{cw}}_{j,s}
    \ge\bigl(4\widetilde{\mathfrak G}_{NT}\bigr)^{-\zeta_2}
    \Bigl(\frac NT\Bigr)^{\zeta_2/2}.
    \label{eq:cwfloor}
\end{align}

\medskip
\noindent\emph{The ceiling.}\;
With $\widehat{\*u}_t
:=\widehat{\+\theta}{}^{\mathrm{cw}}_t-\+\theta_t^0$, the residual
decomposition~\eqref{eq:pdecomp} holds for the component-wise fit,
$\mathcal E_t=\+\zeta_t-\*A_t\widehat{\*u}_t+\*C_t\widehat{\*u}_1$, and
the elementary bound~\eqref{eq:psum} applies with the design bounds of
Step 1 of the proof of Theorem~\ref{thm:consistency}. The three terms
are $O_p(T/N)$ by, respectively, Assumption~\ref{ass:errors}(b) with
Markov's inequality, Lemma~\ref{lem:cwcons}(b), and
Lemma~\ref{lem:cwcons}(a). Defining
\begin{align}
    \mathfrak B^{\mathrm{cw}}_{NT}
    :=\Bigl(\frac NT\sum_{t=2}^T
    \bigl\|\mathcal E_t\bigr\|^2\Bigr)^{1/2},
    \label{eq:cwBdef}
\end{align}
where $ \mathfrak B^{\mathrm{cw}}_{NT}=O_p(1)$, so 
we have $\sum_{t=2}^T\mathcal E_{j,t}^2
\le\sum_{t=2}^T\|\mathcal E_t\|^2
=(\mathfrak B^{\mathrm{cw}}_{NT})^2\,T/N$ for every $j$. By the
Cauchy--Schwarz inequality and the ramp bound~\eqref{eq:rampbound},
\begin{align}
    \Bigl|\sum_{t=s+1}^T(t-s)\,\mathcal E_{j,t}\Bigr|
 &   \le\Bigl(\sum_{t=s+1}^T(t-s)^2\Bigr)^{1/2}
    \Bigl(\sum_{t=2}^T\mathcal E_{j,t}^2\Bigr)^{1/2}
   \notag\\
    &  \le T^{3/2}\,\mathfrak B^{\mathrm{cw}}_{NT}\sqrt{\frac TN}
     \notag\\
    &=\mathfrak B^{\mathrm{cw}}_{NT}\,\frac{T^2}{\sqrt N},
    \label{eq:cwceiling}
\end{align}
simultaneously over all pairs $(j,s)$, since the right side depends on
neither.

\medskip
\noindent\emph{The contradiction.}\;
Suppose $\widehat\nu_{j,s}\neq0$ at some inactive pair with
probability not tending to zero. On this
event~\eqref{eq:cwkktnorm} holds at that pair, and combining the
floor~\eqref{eq:cwfloor} with the ceiling~\eqref{eq:cwceiling},
\begin{align}
    \vartheta_2\bigl(4\widetilde{\mathfrak G}_{NT}\bigr)^{-\zeta_2}
    \Bigl(\frac NT\Bigr)^{\zeta_2/2}
    \le\frac{2\,\mathfrak B^{\mathrm{cw}}_{NT}\,T}{\sqrt N}.
    \label{eq:cwfloorceiling}
\end{align}
Multiplying both sides
of~\eqref{eq:cwfloorceiling} by
$(4\widetilde{\mathfrak G}_{NT})^{\zeta_2}\sqrt N/T$ and collecting
powers of $N$ and $T$,
\begin{align}
    \frac{N^{(\zeta_2+1)/2}\,\vartheta_2}{T^{1+\zeta_2/2}}
  &  \le2\bigl(4\widetilde{\mathfrak G}_{NT}\bigr)^{\zeta_2}
    \mathfrak B^{\mathrm{cw}}_{NT}
   \notag\\ & =O_p(1),
    \label{eq:cwcontra}
\end{align}
which contradicts the modified Assumption~\ref{ass:signal}(b), by which
the left side diverges in probability. Since the floor and the ceiling
are uniform over the inactive pairs, the contradiction excludes all of
them simultaneously, so
$\widehat\nu_{j,s}=0$ for every $(j,s)\notin\mathcal S^0$ w.p.a.1, and
since $\Delta^2\widehat\theta^{\mathrm{cw}}_{j,s}
=\widehat\nu_{j,s}$ by definition, part (i) follows.

\medskip
\noindent\emph{Step 2: the detection condition.}\;
By the first clause of the modified Assumption~\ref{ass:signal}(a),
w.p.a.1
$\vartheta_2\le(c_1+1)
(J^{\mathrm{cw}}_{\min})^{\zeta_2}(NTM^0)^{-1/2}$. Inserting this cap
into the modified Assumption~\ref{ass:signal}(b), a tuning
parameter can exist only if
\(
    (J^{\mathrm{cw}}_{\min})^{\zeta_2}N^{\zeta_2/2}
    T^{-(3+\zeta_2)/2}(M^0)^{-1/2}\to\infty,
\)
and multiplying by $T^{3/2}\sqrt{M^0}\ge1$ gives
$(\sqrt{N/T}\,J^{\mathrm{cw}}_{\min})^{\zeta_2}\to\infty$, hence
\begin{align}
    \sqrt{N/T}\,J^{\mathrm{cw}}_{\min}\to\infty.
    \label{eq:cwdetectapp}
\end{align}

\medskip
\noindent\emph{Step 3: every true kink is detected.}\;
Fix $(j,t)\in\mathcal S^0$. By the scalar-versus-vector bound and the
triangle inequality,
\begin{align}
    \bigl|\Delta^2\widehat\theta^{\mathrm{cw}}_{j,t}\bigr|
    &\ge\bigl|\Delta^2\theta_{j,t}^0\bigr|
    -\bigl|\Delta^2\bigl(\widehat\theta^{\mathrm{cw}}_{j,t}
    -\theta_{j,t}^0\bigr)\bigr|
   \notag\\& \ge J^{\mathrm{cw}}_{\min}
    -4\max_{1\le s\le T}
    \bigl\|\widehat{\+\theta}{}^{\mathrm{cw}}_s
    -\+\theta_s^0\bigr\|,
    \label{eq:cwkinkfloor}
\end{align}
using~\eqref{eq:cwJmin} in the second inequality. By
Lemma~\ref{lem:cwcons}(c) the maximum
in~\eqref{eq:cwkinkfloor} is $O_p(\sqrt{T/N})$, so that, uniformly over
$(j,t)\in\mathcal S^0$,
\begin{align}
    \min_{(j,t)\in\mathcal S^0}
    \bigl|\Delta^2\widehat\theta^{\mathrm{cw}}_{j,t}\bigr|
    \ge J^{\mathrm{cw}}_{\min}
    \Bigl(1-O_p\bigl((J^{\mathrm{cw}}_{\min})^{-1}
    \sqrt{T/N}\bigr)\Bigr),
    \label{eq:cwminfloor}
\end{align}
and the $O_p$ term is $o_p(1)$
by~\eqref{eq:cwdetectapp}. Hence every active pair has a strictly
positive estimated second difference w.p.a.1.

\medskip
\noindent\emph{Step 4: combination.}\;
By Steps 1 and 3, w.p.a.1 the estimated scalar kink set
$\{(j,t):|\Delta^2\widehat\theta^{\mathrm{cw}}_{j,t}|>0\}$ contains no
pair outside $\mathcal S^0$ and every pair inside it, and therefore
equals $\mathcal S^0$. Reading this coordinate by coordinate,
$\widehat{\mathcal T}_{j,\widehat K_j}
=\mathcal T^0_{j,K_j^0}$ for every $j=1,\ldots,p$ jointly w.p.a.1,
which gives $\widehat K_j=K_j^0$ for all $j$, part (ii), and, on that
event, exact coincidence of all estimated dates with the true ones,
part (iii).
\end{proof}

\section*{Proof of Proposition~\ref{prop:cwnorm}}

\begin{proof}
The proof is in five steps. Step 1 collects the per-coordinate basis
facts. Step 2 reduces to the true design. Step 3 derives the exact
scaling identity. Step 4 establishes part (i). Step 5 extracts
parts (ii) and (iii).

\medskip

Fix $j\in\{1,\ldots,p\}$ and $m\in\{0,\ldots,K_j^0+1\}$. The loading
$h_{j,m,t}$ of~\eqref{eq:cwhdef} vanishes off
$\mathcal K_{j,m}^0\cup\mathcal K_{j,m+1}^0$, and these two sets are
disjoint, so
\begin{align}
    r_{j,m}^2
    &=\sum_{t\in\mathcal K_{j,m}^0}u_{j,m,t}^2
    +\sum_{t\in\mathcal K_{j,m+1}^0}
    \bigl(1-u_{j,m+1,t}\bigr)^2
    \notag \\
    &=A\bigl(I_{j,m}^0\bigr)+B\bigl(I_{j,m+1}^0\bigr),
    \label{eq:cwrjclosed}
\end{align}
the second equality because the change of index $q=t-T_{j,m-1}^0$
maps $\mathcal K_{j,m}^0$ onto $\{0,\ldots,I_{j,m}^0-1\}$ with
$u_{j,m,t}=q/I_{j,m}^0$, and $q=t-T_{j,m}^0$ maps
$\mathcal K_{j,m+1}^0$ onto $\{0,\ldots,I_{j,m+1}^0-1\}$ with
$1-u_{j,m+1,t}=1-q/I_{j,m+1}^0$, so the two sums are those
of~\eqref{eq:ABFdef}, and both identities remain valid at the
endpoints under the conventions
$\mathcal K_{j,0}^0=\mathcal K_{j,K_j^0+2}^0=\varnothing$ with
$A(0)=B(0)=0$. The argument establishing~\eqref{eq:rjbound} in the
proof of Lemma~\ref{lem:rjorder} depends on the knot sequence only
through~\eqref{eq:cwrjclosed} and applies verbatim with
$I_{j,m}^0$ in place of $I_m^0$, the endpoint case using
$I_{j,K_j^0+1}^0=T+1-T_{j,K_j^0}^0\ge2$, which holds because
$T_{j,K_j^0}^0<T$. Hence, uniformly over every admissible
configuration,
\begin{align}
    \frac18\bigl(I_{j,m}^0+I_{j,m+1}^0\bigr)
    \le r_{j,m}^2
    \le I_{j,m}^0+I_{j,m+1}^0,
    \label{eq:cwrjbound}
\end{align}
so that $r_{j,m}^2\ge1/8>0$ for every pair $(j,m)$ and
$\mathbb D^{\mathrm{cw}}$ is nonsingular. Next, let
$\*P_{j,m}$ denote the $1\times q^0$ unit row whose nonzero entry
sits in the position that $\eta_{j,m}^0$ occupies in the stacked
vector $\+\eta^{0,\mathrm{cw}}$ of Section~\ref{sec:component}.
Since $\mathbb D^{\mathrm{cw}}$ is diagonal with $r_{j,m}$ in that
same position,
\begin{align}
    \*P_{j,m}\+\eta^{0,\mathrm{cw}}=\eta_{j,m}^0,
    \qquad
    \*P_{j,m}\mathbb D^{\mathrm{cw}}=r_{j,m}\*P_{j,m},
    \qquad
    \*P_{j,m}\*P_{j',m'}'=\delta_{(j,m)(j',m')},
    \label{eq:cwselectorids}
\end{align}
where $\delta_{(j,m)(j',m')}$ equals one if $(j,m)=(j',m')$ and zero
otherwise. Further, recall from~\eqref{eq:cwknotdef} and the
interpolation identity of Section~\ref{sec:component} that
\begin{align}
    \kappa_{j,\ell}^0
    =\frac{\eta_{j,\ell}^0-\eta_{j,\ell-1}^0}{I_{j,\ell}^0},
    \qquad
    \theta_{j,t}^0
    =(1-u_{j,\ell,t})\eta_{j,\ell-1}^0
    +u_{j,\ell,t}\eta_{j,\ell}^0
    \label{eq:cwmaps}
\end{align}
for $t\in\mathcal K_{j,\ell}^0$ and $\ell=1,\ldots,K_j^0+1$, and
that $\widetilde\kappa^{\mathrm{cw}}_{j,\ell}$ and
$\widetilde\theta^{\mathrm{cw}}_{j,t}$ are defined by the same two
maps applied to $\widetilde{\+\eta}^{\mathrm{cw}}$ at the estimated
kink sets. Finally, recall from Section~\ref{sec:component} that
\begin{align}
    \widehat{\*Q}_N^{\mathrm{cw}}
    :=\bigl(\mathbb D^{\mathrm{cw}}\bigr)^{-1}\,\frac1N\,
    \+\Phi^{\eta,\mathrm{cw}\prime}\*C\,\+\Phi^{\eta,\mathrm{cw}}\,
    \bigl(\mathbb D^{\mathrm{cw}}\bigr)^{-1},
    \qquad
    \+\Psi_{NT}^{\mathrm{cw}}
    :=\frac1{\sqrt N}\,
    \+\Phi^{\eta,\mathrm{cw}\prime}\*C\,\+\varepsilon,
    \label{eq:cwQhatapp}
\end{align}
with $\*C=\*I_N\otimes\*C_T$ the centering matrix.

\medskip
Moving on, define
\(
    \mathcal A^{\mathrm{cw}}_{NT}
    :=\bigl\{\widehat{\mathcal T}_{j,\widehat K_j}
    =\mathcal T^0_{j,K_j^0}\;\forall\;j=1,\ldots,p\bigr\}.
\)
By Proposition~\ref{thm:cbc}(ii) and (iii),
\begin{align}
    \mathbb P\bigl(\mathcal A^{\mathrm{cw}}_{NT}\bigr)
    &=\mathbb P\Bigl(\widehat T_{j,\ell}=T_{j,\ell}^0,
    \;\ell=1,\ldots,K_j^0,\;\forall\;j
    \;\Big|\;\widehat K_j=K_j^0\;\forall\;j\Bigr)
    \,\mathbb P\bigl(\widehat K_j=K_j^0\;\forall\;j\bigr)
    \notag \\
    &\to1
    \label{eq:cwrecovprob}
\end{align}
as the product of two factors each tending to one. On
$\mathcal A^{\mathrm{cw}}_{NT}$ the estimated kink sets coincide
with the true ones, hence so do the estimated regime lengths and
loadings, so the design built from the estimated sets equals
$\+\Phi^{\eta,\mathrm{cw}}$, the estimator~\eqref{eq:cwols} equals
its oracle version evaluated at the true coordinate-specific dates,
and the maps defining $\widetilde\kappa^{\mathrm{cw}}_{j,\ell}$ and
$\widetilde\theta^{\mathrm{cw}}_{j,t}$ coincide
with~\eqref{eq:cwmaps}. Exactly as in the proof of
Theorem~\ref{thm:normality}, the feasible and oracle quantities
differ on an event whose probability vanishes
by~\eqref{eq:cwrecovprob}, so every limit and every stochastic order
established below for the oracle quantities transfers to the
feasible ones by Slutsky's theorem, and we work on
$\mathcal A^{\mathrm{cw}}_{NT}$ for the remainder of the proof.

\medskip
Now, 
by~\eqref{eq:cwQhatapp} and the nonsingularity of
$\mathbb D^{\mathrm{cw}}$,
\begin{align}
    \+\Phi^{\eta,\mathrm{cw}\prime}\*C\+\Phi^{\eta,\mathrm{cw}}
    =N\,\mathbb D^{\mathrm{cw}}\widehat{\*Q}_N^{\mathrm{cw}}
    \mathbb D^{\mathrm{cw}}.
    \label{eq:cwgramfact}
\end{align}
Both $\widehat{\*Q}_N^{\mathrm{cw}}$ and $\*Q_0^{\mathrm{cw}}$ are
symmetric, so Weyl's inequality and Assumption~\ref{ass:cwrank}(a)
give
\begin{align}
    \bigl|\mu_{\min}\bigl(\widehat{\*Q}_N^{\mathrm{cw}}\bigr)
    -\mu_{\min}\bigl(\*Q_0^{\mathrm{cw}}\bigr)\bigr|
    &\le\bigl\|\widehat{\*Q}_N^{\mathrm{cw}}
    -\*Q_0^{\mathrm{cw}}\bigr\|_{op}
    \notag \\
    &=o_p(1),
\end{align}
so the event
$\{\mu_{\min}(\widehat{\*Q}_N^{\mathrm{cw}})\ge c/2\}$ has
probability tending to one. On this event
$\widehat{\*Q}_N^{\mathrm{cw}}$ is nonsingular with
$\|(\widehat{\*Q}_N^{\mathrm{cw}})^{-1}\|_{op}
=\mu_{\min}(\widehat{\*Q}_N^{\mathrm{cw}})^{-1}\le2/c$, and the
resolvent identity
$(\widehat{\*Q}_N^{\mathrm{cw}})^{-1}-(\*Q_0^{\mathrm{cw}})^{-1}
=(\widehat{\*Q}_N^{\mathrm{cw}})^{-1}
(\*Q_0^{\mathrm{cw}}-\widehat{\*Q}_N^{\mathrm{cw}})
(\*Q_0^{\mathrm{cw}})^{-1}$ together with submultiplicativity of the
operator norm and
$\|(\*Q_0^{\mathrm{cw}})^{-1}\|_{op}
=\mu_{\min}(\*Q_0^{\mathrm{cw}})^{-1}\le1/c$ gives
\begin{align}
    \bigl\|\bigl(\widehat{\*Q}_N^{\mathrm{cw}}\bigr)^{-1}
    -\bigl(\*Q_0^{\mathrm{cw}}\bigr)^{-1}\bigr\|_{op}
    &\le\frac2c\cdot o_p(1)\cdot\frac1c
    \notag \\
    &=o_p(1),
    \label{eq:cwQinv}
\end{align}
and in particular
$\|(\widehat{\*Q}_N^{\mathrm{cw}})^{-1}\|_{op}=O_p(1)$. On the same
event, $\+\Phi^{\eta,\mathrm{cw}\prime}\*C\+\Phi^{\eta,\mathrm{cw}}$
is nonsingular by~\eqref{eq:cwgramfact} as a product of nonsingular
matrices, with
\begin{align}
    \bigl[\+\Phi^{\eta,\mathrm{cw}\prime}\*C
    \+\Phi^{\eta,\mathrm{cw}}\bigr]^{-1}
    =\frac1N\,\bigl(\mathbb D^{\mathrm{cw}}\bigr)^{-1}
    \bigl(\widehat{\*Q}_N^{\mathrm{cw}}\bigr)^{-1}
    \bigl(\mathbb D^{\mathrm{cw}}\bigr)^{-1}.
    \label{eq:cwgraminv}
\end{align}
By~\eqref{eq:cwdesign}, the DGP~\eqref{eq:DGP} stacks as
$\*y=(\*I_N\otimes\+{\iota}_T)\+\mu
+\+\Phi^{\eta,\mathrm{cw}}\+\eta^{0,\mathrm{cw}}+\+\varepsilon$,
where $\+\mu:=[\mu_1,\ldots,\mu_N]'$, and
$\*C(\*I_N\otimes\+{\iota}_T)=\*0$, so
premultiplying by $\*C$ annihilates the fixed effects,
\begin{align}
    \*C\*y
    =\*C\+\Phi^{\eta,\mathrm{cw}}\+\eta^{0,\mathrm{cw}}
    +\*C\+\varepsilon.
\end{align}
Substituting into~\eqref{eq:cwols} and using symmetry and
idempotency of $\*C$,
\begin{align}
    \widetilde{\+\eta}^{\mathrm{cw}}
    &=\bigl[\+\Phi^{\eta,\mathrm{cw}\prime}\*C
    \+\Phi^{\eta,\mathrm{cw}}\bigr]^{-1}
    \+\Phi^{\eta,\mathrm{cw}\prime}\*C
    \+\Phi^{\eta,\mathrm{cw}}\+\eta^{0,\mathrm{cw}}
    +\bigl[\+\Phi^{\eta,\mathrm{cw}\prime}\*C
    \+\Phi^{\eta,\mathrm{cw}}\bigr]^{-1}
    \+\Phi^{\eta,\mathrm{cw}\prime}\*C\+\varepsilon
    \notag \\
    &=\+\eta^{0,\mathrm{cw}}
    +\bigl[\+\Phi^{\eta,\mathrm{cw}\prime}\*C
    \+\Phi^{\eta,\mathrm{cw}}\bigr]^{-1}
    \+\Phi^{\eta,\mathrm{cw}\prime}\*C\+\varepsilon.
    \label{eq:cwerrorrep}
\end{align}
Inserting~\eqref{eq:cwgraminv} together with
$\+\Phi^{\eta,\mathrm{cw}\prime}\*C\+\varepsilon
=\sqrt N\+\Psi_{NT}^{\mathrm{cw}}$ from~\eqref{eq:cwQhatapp},
\begin{align}
    \widetilde{\+\eta}^{\mathrm{cw}}-\+\eta^{0,\mathrm{cw}}
    &=\frac1N\,\bigl(\mathbb D^{\mathrm{cw}}\bigr)^{-1}
    \bigl(\widehat{\*Q}_N^{\mathrm{cw}}\bigr)^{-1}
    \bigl(\mathbb D^{\mathrm{cw}}\bigr)^{-1}
    \cdot\sqrt N\+\Psi_{NT}^{\mathrm{cw}}
    \notag \\
    &=\frac1{\sqrt N}\,\bigl(\mathbb D^{\mathrm{cw}}\bigr)^{-1}
    \bigl(\widehat{\*Q}_N^{\mathrm{cw}}\bigr)^{-1}
    \bigl(\mathbb D^{\mathrm{cw}}\bigr)^{-1}
    \+\Psi_{NT}^{\mathrm{cw}},
\end{align}
and multiplying on the left by $\sqrt N\,\mathbb D^{\mathrm{cw}}$
gives the exact identity
\begin{align}
    \sqrt N\,\mathbb D^{\mathrm{cw}}
    \bigl(\widetilde{\+\eta}^{\mathrm{cw}}
    -\+\eta^{0,\mathrm{cw}}\bigr)
    =\bigl(\widehat{\*Q}_N^{\mathrm{cw}}\bigr)^{-1}
    \bigl(\mathbb D^{\mathrm{cw}}\bigr)^{-1}
    \+\Psi_{NT}^{\mathrm{cw}}.
    \label{eq:cwident}
\end{align}

\medskip
Further, we first control the normalised score. Every entry of
$\+\Phi^{\eta,\mathrm{cw}}$ is $h_{j,m,t}x_{i,t,j}$ with
deterministic $h_{j,m,t}$, and $\*C$ is deterministic, so each entry
of $\mathbb E(\+\Phi^{\eta,\mathrm{cw}\prime}\*C\+\varepsilon)$ is a
finite linear combination of terms
$\mathbb E(x_{i,s,j}\varepsilon_{i,t})=0$ by
Assumption~\ref{ass:errors}(a), whence
$\mathbb E[(\mathbb D^{\mathrm{cw}})^{-1}
\+\Psi_{NT}^{\mathrm{cw}}]=\*0_{q^0\times1}$. Mean zero gives
\begin{align}
    \mathbb E\bigl\|\bigl(\mathbb D^{\mathrm{cw}}\bigr)^{-1}
    \+\Psi_{NT}^{\mathrm{cw}}\bigr\|^2
    &=\mathrm{tr}\,\operatorname{Var}\!\left(
    \bigl(\mathbb D^{\mathrm{cw}}\bigr)^{-1}
    \+\Psi_{NT}^{\mathrm{cw}}\right)
    \notag \\
    &\to\mathrm{tr}\,\+\Sigma_0^{\mathrm{cw}}
    \notag \\
    &<\infty
\end{align}
by Assumption~\ref{ass:cwrank}(b), and, being convergent, the
second moment is bounded by some $C<\infty$ for all $N$ and $T$.
Markov's inequality then gives, for every $M>0$,
\begin{align}
    \mathbb P\Bigl(\bigl\|\bigl(\mathbb D^{\mathrm{cw}}\bigr)^{-1}
    \+\Psi_{NT}^{\mathrm{cw}}\bigr\|>M\Bigr)
    &\le\frac{\mathbb E\bigl\|
    \bigl(\mathbb D^{\mathrm{cw}}\bigr)^{-1}
    \+\Psi_{NT}^{\mathrm{cw}}\bigr\|^2}{M^2}
    \notag \\
    &\le\frac{C}{M^2}
\end{align}
uniformly in $N$ and $T$, that is,
\begin{align}
    \bigl\|\bigl(\mathbb D^{\mathrm{cw}}\bigr)^{-1}
    \+\Psi_{NT}^{\mathrm{cw}}\bigr\|=O_p(1).
    \label{eq:cwPsinorm}
\end{align}
Now let $\{\*H_{NT}^{\mathrm{cw}}\}$ be as in
Assumption~\ref{ass:cwrank}(b), with $\+\Upsilon^{\mathrm{cw}}$
existing and positive definite.
Multiplying~\eqref{eq:cwident} by $\*H_{NT}^{\mathrm{cw}}$ and
adding and subtracting
$\*H_{NT}^{\mathrm{cw}}(\*Q_0^{\mathrm{cw}})^{-1}
(\mathbb D^{\mathrm{cw}})^{-1}\+\Psi_{NT}^{\mathrm{cw}}$,
\begin{align}
    \sqrt N\,\*H_{NT}^{\mathrm{cw}}\mathbb D^{\mathrm{cw}}
    \bigl(\widetilde{\+\eta}^{\mathrm{cw}}
    -\+\eta^{0,\mathrm{cw}}\bigr)
    &=\*H_{NT}^{\mathrm{cw}}\bigl(\*Q_0^{\mathrm{cw}}\bigr)^{-1}
    \bigl(\mathbb D^{\mathrm{cw}}\bigr)^{-1}
    \+\Psi_{NT}^{\mathrm{cw}}
    \notag \\
    &\quad
    +\*H_{NT}^{\mathrm{cw}}
    \Bigl[\bigl(\widehat{\*Q}_N^{\mathrm{cw}}\bigr)^{-1}
    -\bigl(\*Q_0^{\mathrm{cw}}\bigr)^{-1}\Bigr]
    \bigl(\mathbb D^{\mathrm{cw}}\bigr)^{-1}
    \+\Psi_{NT}^{\mathrm{cw}}.
    \label{eq:cwdecompose}
\end{align}
The first term converges in distribution to
$\mathcal N(\*0_{l\times1},\+\Upsilon^{\mathrm{cw}})$ by
Assumption~\ref{ass:cwrank}(b), while the norm of the second is
bounded by
\begin{align}
    \bigl\|\*H_{NT}^{\mathrm{cw}}\bigr\|_{op}
    \bigl\|\bigl(\widehat{\*Q}_N^{\mathrm{cw}}\bigr)^{-1}
    -\bigl(\*Q_0^{\mathrm{cw}}\bigr)^{-1}\bigr\|_{op}
    \bigl\|\bigl(\mathbb D^{\mathrm{cw}}\bigr)^{-1}
    \+\Psi_{NT}^{\mathrm{cw}}\bigr\|
    &=O(1)\cdot o_p(1)\cdot O_p(1)
    \notag \\
    &=o_p(1),
\end{align}
by~\eqref{eq:cwQinv} and~\eqref{eq:cwPsinorm}. Slutsky's theorem
applied to~\eqref{eq:cwdecompose}, together with this and the above, gives part (i).

An immediate consequence is the rate of each boundary level. Fix
$(j,m)$ and apply $\*P_{j,m}$ to~\eqref{eq:cwident}.
Using~\eqref{eq:cwselectorids}, \eqref{eq:cwQinv}
and~\eqref{eq:cwPsinorm},
\begin{align}
    \sqrt N\,r_{j,m}
    \bigl(\widetilde\eta^{\mathrm{cw}}_{j,m}-\eta_{j,m}^0\bigr)
    &=\*P_{j,m}
    \bigl(\widehat{\*Q}_N^{\mathrm{cw}}\bigr)^{-1}
    \bigl(\mathbb D^{\mathrm{cw}}\bigr)^{-1}
    \+\Psi_{NT}^{\mathrm{cw}}
    \notag \\
    &=O_p(1),
    \label{eq:cwetaOp}
\end{align}
so that, by~\eqref{eq:cwrjbound},
\begin{align}
    \widetilde\eta^{\mathrm{cw}}_{j,m}-\eta_{j,m}^0
    &=O_p\bigl(\bigl[Nr_{j,m}^2\bigr]^{-1/2}\bigr)
    \notag \\
    &=O_p\Bigl(\bigl[N\bigl(I_{j,m}^0+I_{j,m+1}^0\bigr)
    \bigr]^{-1/2}\Bigr).
    \label{eq:cwetarate}
\end{align}

\medskip

Fix $j$ and $\ell\in\{1,\ldots,K_j^0+1\}$.
By~\eqref{eq:cwrjbound}, $r_{j,\ell-1},r_{j,\ell}>0$, so
$s_{j,\ell}=I_{j,\ell}^0
(r_{j,\ell-1}^{-2}+r_{j,\ell}^{-2})^{-1/2}$ is well defined and
strictly positive. By~\eqref{eq:cwmaps} and the above, we have
\begin{align}
    \widetilde\kappa^{\mathrm{cw}}_{j,\ell}-\kappa_{j,\ell}^0
    &=\frac{\widetilde\eta^{\mathrm{cw}}_{j,\ell}
    -\widetilde\eta^{\mathrm{cw}}_{j,\ell-1}}{I_{j,\ell}^0}
    -\frac{\eta_{j,\ell}^0-\eta_{j,\ell-1}^0}{I_{j,\ell}^0}
    \notag \\
    &=\frac{\bigl(\widetilde\eta^{\mathrm{cw}}_{j,\ell}
    -\eta_{j,\ell}^0\bigr)
    -\bigl(\widetilde\eta^{\mathrm{cw}}_{j,\ell-1}
    -\eta_{j,\ell-1}^0\bigr)}{I_{j,\ell}^0}
    \label{eq:cwkappaerr}
\end{align}
$\text{on }\mathcal A^{\mathrm{cw}}_{NT},$
so that, by the triangle inequality and~\eqref{eq:cwetaOp} at
$\ell-1$ and $\ell$,
\begin{align}
    \bigl|\widetilde\kappa^{\mathrm{cw}}_{j,\ell}
    -\kappa_{j,\ell}^0\bigr|
    &\le\frac{\bigl|\widetilde\eta^{\mathrm{cw}}_{j,\ell}
    -\eta_{j,\ell}^0\bigr|
    +\bigl|\widetilde\eta^{\mathrm{cw}}_{j,\ell-1}
    -\eta_{j,\ell-1}^0\bigr|}{I_{j,\ell}^0}
    \notag \\
    &=\frac{O_p\bigl([Nr_{j,\ell}^2]^{-1/2}\bigr)
    +O_p\bigl([Nr_{j,\ell-1}^2]^{-1/2}\bigr)}{I_{j,\ell}^0}
    \notag \\
    &=O_p\Bigl(\bigl[N\bigl(I_{j,\ell}^0\bigr)^2
    \bigl(r_{j,\ell-1}^2\wedge r_{j,\ell}^2\bigr)\bigr]^{-1/2}
    \Bigr),
    \label{eq:cwkappaOp}
\end{align}
the last equality because
$[Nr_{j,\ell}^2]^{-1/2}+[Nr_{j,\ell-1}^2]^{-1/2}
\le2[N(r_{j,\ell-1}^2\wedge r_{j,\ell}^2)]^{-1/2}$.
Applying~\eqref{eq:cwrjbound} at $\ell-1$ and $\ell$, together with
the monotonicity of the minimum in each argument,
\begin{align}
    \widetilde\kappa^{\mathrm{cw}}_{j,\ell}-\kappa_{j,\ell}^0
    =O_p\Bigl(\Bigl[N\bigl(I_{j,\ell}^0\bigr)^2
    \bigl\{\bigl(I_{j,\ell-1}^0+I_{j,\ell}^0\bigr)
    \wedge\bigl(I_{j,\ell}^0+I_{j,\ell+1}^0\bigr)\bigr\}
    \Bigr]^{-1/2}\Bigr),
\end{align}
which is part (ii).

For part (iii), fix $j$ and $\ell$, let
$t\in\mathcal K_{j,\ell}^0$, and abbreviate
$u:=u_{j,\ell,t}\in[0,1)$. By~\eqref{eq:cwmaps} and Step 2,
\begin{align}
    \widetilde\theta^{\mathrm{cw}}_{j,t}-\theta_{j,t}^0
    =(1-u)\bigl(\widetilde\eta^{\mathrm{cw}}_{j,\ell-1}
    -\eta_{j,\ell-1}^0\bigr)
    +u\bigl(\widetilde\eta^{\mathrm{cw}}_{j,\ell}
    -\eta_{j,\ell}^0\bigr)
    \label{eq:cwthetaerr}
\end{align}
$\text{on }\mathcal A^{\mathrm{cw}}_{NT}$. 
Since $u\in[0,1)$, both weights are nonnegative and sum to one, so
the triangle inequality gives, for every
$t\in\mathcal K_{j,\ell}^0$,
\begin{align}
    \bigl|\widetilde\theta^{\mathrm{cw}}_{j,t}
    -\theta_{j,t}^0\bigr|
    &\le(1-u)\bigl|\widetilde\eta^{\mathrm{cw}}_{j,\ell-1}
    -\eta_{j,\ell-1}^0\bigr|
    +u\bigl|\widetilde\eta^{\mathrm{cw}}_{j,\ell}
    -\eta_{j,\ell}^0\bigr|
    \notag \\
    &\le\max\Bigl\{
    \bigl|\widetilde\eta^{\mathrm{cw}}_{j,\ell-1}
    -\eta_{j,\ell-1}^0\bigr|,
    \bigl|\widetilde\eta^{\mathrm{cw}}_{j,\ell}
    -\eta_{j,\ell}^0\bigr|\Bigr\}.
    \label{eq:cwthetatriangle}
\end{align}
The right side of~\eqref{eq:cwthetatriangle} does not depend on
$t$, so the same bound holds for
$\max_{t\in\mathcal K_{j,\ell}^0}
|\widetilde\theta^{\mathrm{cw}}_{j,t}-\theta_{j,t}^0|$.
By~\eqref{eq:cwetarate} applied at $\ell-1$ and $\ell$, and the
maximum of two terms that are $O_p(a_N)$ and $O_p(b_N)$ being
$O_p(a_N\vee b_N)$,
\begin{align}
    \max_{t\in\mathcal K_{j,\ell}^0}
    \bigl|\widetilde\theta^{\mathrm{cw}}_{j,t}
    -\theta_{j,t}^0\bigr|
    &=O_p\Bigl(\bigl[N\bigl(r_{j,\ell-1}^2\wedge
    r_{j,\ell}^2\bigr)\bigr]^{-1/2}\Bigr)
    \notag \\
    &=O_p\Bigl(\Bigl[N\bigl\{
    \bigl(I_{j,\ell-1}^0+I_{j,\ell}^0\bigr)
    \wedge\bigl(I_{j,\ell}^0+I_{j,\ell+1}^0\bigr)\bigr\}
    \Bigr]^{-1/2}\Bigr),
    \label{eq:cwthetarate}
\end{align}
the last equality by~\eqref{eq:cwrjbound} applied at $\ell-1$ and
$\ell$, which is the uniform claim of part (iii). The initial level
is the case $t=1$, $\ell=1$, $u_{j,1,1}=0$
of~\eqref{eq:cwthetaerr}, in which
$\widetilde\theta^{\mathrm{cw}}_{j,1}-\theta_{j,1}^0
=\widetilde\eta^{\mathrm{cw}}_{j,0}-\eta_{j,0}^0$,
and~\eqref{eq:cwetarate} at $m=0$ with $I_{j,0}^0=0$ gives
$\widetilde\theta^{\mathrm{cw}}_{j,1}-\theta_{j,1}^0
=O_p\bigl((NI_{j,1}^0)^{-1/2}\bigr)$. This completes the proof.
\end{proof}

\section*{Proof of Proposition~\ref{prop:breaks}}

\begin{proof}
Throughout, $\Lambda_j:=I_j^0+I_{j+1}^0$ denotes the adjacent span of
boundary $j$ of the induced kink configuration, that is the combined
length of the two regimes informing it. We first compute the induced
configuration, then verify that it is admissible and exactly recovered,
then record once and for all how a boundary-level rate is read off, and
finally derive the three claims.

The induced kink dates are $T_{2q-1}^0=T_q^b-1$ and $T_{2q}^0=T_q^b$ for
$q=1,\ldots,m$, so that $K^0=2m$ is fixed, and the endpoint conventions
$T_0^0=1=T_0^b$ and $T_{2m+1}^0=T+1=T_{m+1}^b$ are inherited from the
break model. Differencing consecutive dates gives, for $q=1,\ldots,m$,
\begin{align}
    I_{2q-1}^0
    &=\bigl(T_q^b-1\bigr)-T_{q-1}^b
    \notag \\
    &=I_q^b-1,
    \label{eq:brodd}
\end{align}
and, for the intervening regime of the same pair,
\begin{align}
    I_{2q}^0
    &=T_q^b-\bigl(T_q^b-1\bigr)
    \notag \\
    &=1,
    \label{eq:breven}
\end{align}
while the final regime has
\begin{align}
    I_{2m+1}^0
    &=(T+1)-T_m^b
    \notag \\
    &=I_{m+1}^b.
    \label{eq:brlast}
\end{align}
The configuration is admissible. The dates are strictly increasing, since
$T_{2q}^0<T_{2q+1}^0$ is $I_{q+1}^b\ge2$ by \eqref{eq:brodd} and
$T_1^0>T_0^0$ is $I_1^b\ge2$; the last kink lies strictly before the
sample end, since $T_{2m}^0=T+1-I_{m+1}^b<T$ again by $I_{m+1}^b\ge2$;
and $I_{2m+1}^0=I_{m+1}^b\ge2$, which is exactly the requirement under
which Lemma~\ref{lem:rjorder} delivers its lower bound at the right
endpoint, the sum $A(L)$ vanishing at $L=1$. This is the only role played
by the spacing condition at $q=m+1$; at $q\le m$ it serves instead to
keep $I_{2q-1}^0\ge1$, so that consecutive breaks do not share a second
difference. Substituting \eqref{eq:brodd}--\eqref{eq:brlast} into the
definition of the adjacent span, and using $I_0^0=I_{2m+2}^0=0$,
\begin{align}
    \Lambda_0
    &=I_1^b-1,
    \notag \\
    \Lambda_{2q-1}
    &=\bigl(I_q^b-1\bigr)+1
    =I_q^b,
    \notag \\
    \Lambda_{2q}
    &=1+\bigl(I_{q+1}^b-1\bigr)
    =I_{q+1}^b,
    \notag \\
    \Lambda_{2m}
    &=1+I_{m+1}^b,
    \notag \\
    \Lambda_{2m+1}
    &=I_{m+1}^b,
    \label{eq:brspan}
\end{align}
the second line holding for $q=1,\ldots,m$ and the third for
$q=1,\ldots,m-1$. Since $I_q^b\ge2$ implies $I_q^b-1\ge I_q^b/2$, every
entry of \eqref{eq:brspan} is bounded between $I^b/2$ and $2I^b$, where
$I^b$ is the break-regime length carrying its index, so that
$\Lambda_{2q-2}\asymp\Lambda_{2q-1}\asymp I_q^b$ for $q=1,\ldots,m+1$,
with absolute constants depending on neither $m$ nor the configuration.

We next identify the boundary levels. Since $\+\eta_0^0=\+\theta_1^0$ and
$\+\eta_j^0=\+\theta_{T_j^0}^0$ for $j=1,\ldots,K^0$ by
\eqref{eq:etainduction}, and since $\+\theta_t^0=\+\alpha_q^0$ throughout
break regime $q$, evaluating the piecewise-constant path at the two dates
of the $q$th pair gives
\begin{align}
    \+\eta_{2q-1}^0
    &=\+\theta_{T_q^b-1}^0
    \notag \\
    &=\+\alpha_q^0,
    \label{eq:brknotodd}
\end{align}
because $T_q^b-1$ is the last date of break regime $q$, and
\begin{align}
    \+\eta_{2q}^0
    &=\+\theta_{T_q^b}^0
    \notag \\
    &=\+\alpha_{q+1}^0,
    \label{eq:brknoteven}
\end{align}
because $T_q^b$ is the first date of break regime $q+1$, both for
$q=1,\ldots,m$. At the left endpoint $\+\eta_0^0=\+\alpha_1^0$, so
reindexing \eqref{eq:brknoteven} yields $\+\eta_{2q-2}^0=\+\alpha_q^0$
for every $q=1,\ldots,m+1$. The slopes follow from
\eqref{eq:slopefromknots}. For the flat regimes,
\begin{align}
    \+\kappa_{2q-1}^0
    &=\frac{\+\eta_{2q-1}^0-\+\eta_{2q-2}^0}{I_{2q-1}^0}
    \notag \\
    &=\frac{\+\alpha_q^0-\+\alpha_q^0}{I_q^b-1}
    \notag \\
    &=\*0_{p\times1},
    \label{eq:brslopeodd}
\end{align}
the division being legitimate because $I_q^b-1\ge1$, and the same
computation at $j=2m+1$, using
$\+\eta_{2m+1}^0=\+\eta_{2m}^0=\+\alpha_{m+1}^0$ from
\eqref{eq:etaendpoint} and the flatness of the last break regime, gives
$\+\kappa_{2m+1}^0=\*0_{p\times1}$. For the intervening regimes,
\begin{align}
    \+\kappa_{2q}^0
    &=\frac{\+\alpha_{q+1}^0-\+\alpha_q^0}{1}
    \notag \\
    &=\+\delta_q^0,
    \label{eq:brslopeeven}
\end{align}
which is \eqref{eq:breakslope}. The second differences at the two dates
of the $q$th pair are therefore
$\+\kappa_{2q}^0-\+\kappa_{2q-1}^0=\+\delta_q^0$ and
$\+\kappa_{2q+1}^0-\+\kappa_{2q}^0=-\+\delta_q^0$, both of norm
$\|\+\delta_q^0\|$, and since every induced kink date belongs to exactly
one pair,
\begin{align}
    J_{\min}
    &=\min_{1\le\ell\le K^0}
    \bigl\|\+\kappa_{\ell+1}^0-\+\kappa_\ell^0\bigr\|
    \notag \\
    &=\min_{1\le q\le m}\bigl\|\+\delta_q^0\bigr\|,
    \label{eq:brJmin}
\end{align}
which is positive because $\+\delta_q^0\neq\*0_{p\times1}$ and $m$ is
fixed. Corollary~\ref{cor:kinks} therefore gives
$\mathbb P(\widehat{\mathcal T}_{\widehat K}=\mathcal T^0_{K^0})\to1$,
that is, exact recovery of all $2m$ dates and hence of the $m$ adjacent
pairs, so that $\widehat T_q^b=\widehat T_{2q}=T_q^b$ on that event. Two
sequences coinciding on an event of probability tending to one differ by
$o_p(1)$, so every order established below transfers to the feasible
estimator by Slutsky's theorem, and we work on that event throughout.

The rates all descend from a single limit. Fix $j$ and take
$\*H_{NT}=\*P_j$ in Theorem~\ref{thm:normality}(i), which is admissible
because $\*P_j\*P_j'=\*I_p$ by \eqref{eq:selectorids}, and whose limiting
variance is the $(j+1)$th diagonal block of
$\*Q_0^{-1}\+\Sigma_0\*Q_0^{-1}$, positive definite by the argument
following \eqref{eq:sandwich}. Using $\*P_j\mathbb D_{K^0+1}=r_j\*P_j$
and $\*P_j\+\eta=\+\eta_j$,
\begin{align}
    \sqrt N\,r_j
    \bigl(\widetilde{\+\eta}_j-\+\eta_j^0\bigr)
    &=\sqrt N\,\*P_j\,\mathbb D_{K^0+1}
    \bigl(\widetilde{\+\eta}_{\widehat K}-\+\eta^0\bigr)
    \notag \\
    &\to_d\mathcal N\bigl(\*0_{p\times1},\+\Upsilon_j^\eta\bigr),
    \label{eq:brclt}
\end{align}
so that $\widetilde{\+\eta}_j-\+\eta_j^0=O_p([Nr_j^2]^{-1/2})$, the
finitely many indices $j=0,\ldots,2m+1$ being handled together because
$m$ is fixed. Lemma~\ref{lem:rjorder} gives $r_j^2\asymp\Lambda_j$, and
\eqref{eq:brspan} gives $\Lambda_j\asymp I^b$ for the break-regime length
carrying the index, so
\begin{align}
    \widetilde{\+\eta}_{2q-2}-\+\eta_{2q-2}^0
    &=O_p\Bigl(\bigl[N\Lambda_{2q-2}\bigr]^{-1/2}\Bigr)
    \notag \\
    &=O_p\Bigl(\bigl[NI_q^b\bigr]^{-1/2}\Bigr),
    \label{eq:bretarate}
\end{align}
and identically for $\widetilde{\+\eta}_{2q-1}$, both for
$q=1,\ldots,m+1$. Every bound below is obtained from
\eqref{eq:bretarate}.

Consider part (i). By Remark~\ref{rem:breakindex} the regime level is
reported as $\widetilde{\+\alpha}_q=\widetilde{\+\eta}_{2q-2}$, whose
true counterpart is $\+\alpha_q^0$, so \eqref{eq:bretarate} is already
the claim,
\begin{align}
    \widetilde{\+\alpha}_q-\+\alpha_q^0
    &=O_p\Bigl(\bigl[NI_q^b\bigr]^{-1/2}\Bigr).
    \label{eq:bralpharate}
\end{align}
For the path, break regime $q\le m$ is not a kink regime but the union of
two of them,
\begin{align}
    \bigl\{T_{q-1}^b,\ldots,T_q^b-1\bigr\}
    &=\mathcal K_{2q-1}^0\cup\mathcal K_{2q}^0,
    \label{eq:brsplit}
\end{align}
the second set being the singleton $\{T_q^b-1\}$ because $I_{2q}^0=1$.
Applied to that singleton, the generic bound of
Corollary~\ref{cor:thetat}(ii) would return
$[N\{\Lambda_{2q-1}\wedge\Lambda_{2q}\}]^{-1/2}$, which by
\eqref{eq:brspan} is the slower $[N\{I_q^b\wedge I_{q+1}^b\}]^{-1/2}$, so
the claim does not follow regime by regime and we work from the
interpolation identity. Let $t$ lie in break regime $q$ and let $\ell$ be
the kink regime containing it, so that $\ell\in\{2q-1,2q\}$. Subtracting
\eqref{eq:knotrep} from its estimated counterpart, the two carrying the
same weights,
\begin{align}
    \widetilde{\+\theta}_t-\+\theta_t^0
    &=\bigl(1-u_{\ell,t}\bigr)
    \bigl(\widetilde{\+\eta}_{\ell-1}-\+\eta_{\ell-1}^0\bigr)
    +u_{\ell,t}
    \bigl(\widetilde{\+\eta}_\ell-\+\eta_\ell^0\bigr),
    \label{eq:brthetaerr}
\end{align}
and, the weights being nonnegative and summing to one,
\begin{align}
    \bigl\|\widetilde{\+\theta}_t-\+\theta_t^0\bigr\|
    &\le\max_{j\in\{\ell-1,\ell\}}
    \bigl\|\widetilde{\+\eta}_j-\+\eta_j^0\bigr\|.
    \label{eq:brthetamax}
\end{align}
If $t\in\mathcal K_{2q-1}^0$ then $\ell=2q-1$ and the two boundaries
entering \eqref{eq:brthetamax} are $2q-2$ and $2q-1$, both of span
$\asymp I_q^b$. If instead $t=T_{2q-1}^0$, the unique date of
$\mathcal K_{2q}^0$, then $\ell=2q$ and the elapsed fraction is
\begin{align}
    u_{2q,t}
    &=\frac{t-T_{2q-1}^0}{I_{2q}^0}
    \notag \\
    &=0,
    \label{eq:brufrac}
\end{align}
so \eqref{eq:brthetaerr} collapses to
$\widetilde{\+\theta}_t-\+\theta_t^0
=\widetilde{\+\eta}_{2q-1}-\+\eta_{2q-1}^0$ and only the boundary
$2q-1$, again of span $I_q^b$, is involved. In either case
\eqref{eq:bretarate} and \eqref{eq:brthetamax} give a bound free of $t$,
so that, uniformly over the dates of break regime $q$,
\begin{align}
    \widetilde{\+\theta}_t-\+\alpha_q^0
    &=\widetilde{\+\theta}_t-\+\theta_t^0
    \notag \\
    &=O_p\Bigl(\bigl[NI_q^b\bigr]^{-1/2}\Bigr),
    \label{eq:brpathrate}
\end{align}
the first equality because $\+\theta_t^0=\+\alpha_q^0$ throughout the
regime. For $q=m+1$ the break regime coincides with
$\mathcal K_{2m+1}^0$, whose two boundaries have spans
$\Lambda_{2m}=1+I_{m+1}^b$ and $\Lambda_{2m+1}=I_{m+1}^b$, so the same
argument applies with $I_{m+1}^b$ in place of $I_q^b$.

Part (ii) is a contrast of two adjacent boundary levels. By
\eqref{eq:brknotodd} and \eqref{eq:brknoteven} we have
$\+\eta_{2q}^0-\+\eta_{2q-1}^0=\+\delta_q^0$, and the reported magnitude
of Remark~\ref{rem:breakindex} is $\widetilde{\+\delta}_q
=\widetilde{\+\eta}_{2q}-\widetilde{\+\eta}_{2q-1}$, which coincides with
$\widetilde{\+\kappa}_{2q}$ because $I_{2q}^0=1$ and the block row of
$\*M_{K^0}^{-1}$ in \eqref{eq:Minverse} belonging to $\+\kappa_{2q}$
carries $-(I_{2q}^0)^{-1}\*I_p$ and $(I_{2q}^0)^{-1}\*I_p$ in block
columns $2q-1$ and $2q$. Subtracting and applying the triangle
inequality,
\begin{align}
    \bigl\|\widetilde{\+\delta}_q-\+\delta_q^0\bigr\|
    &\le\bigl\|\widetilde{\+\eta}_{2q}-\+\eta_{2q}^0\bigr\|
    +\bigl\|\widetilde{\+\eta}_{2q-1}-\+\eta_{2q-1}^0\bigr\|
    \notag \\
    &=O_p\Bigl(\bigl[N\Lambda_{2q}\bigr]^{-1/2}\Bigr)
    +O_p\Bigl(\bigl[N\Lambda_{2q-1}\bigr]^{-1/2}\Bigr)
    \notag \\
    &=O_p\Bigl(\bigl[N\bigl(\Lambda_{2q-1}\wedge\Lambda_{2q}\bigr)
    \bigr]^{-1/2}\Bigr)
    \notag \\
    &=O_p\Bigl(\bigl[N\bigl\{I_q^b\wedge I_{q+1}^b\bigr\}
    \bigr]^{-1/2}\Bigr),
    \label{eq:brdeltarate}
\end{align}
the third equality because $a^{-1/2}+b^{-1/2}\le2(a\wedge b)^{-1/2}$ for
positive $a$ and $b$, and the fourth by \eqref{eq:brspan}. The rate is
governed by the shorter of the two regimes flanking the break, as it must
be for a contrast of their levels, and the unit length of the intervening
regime means that no division by a diverging quantity takes place.

Part (iii) concerns the slopes that the break specification does not
carry. Fix $q\in\{0,\ldots,m\}$. The same block row of $\*M_{K^0}^{-1}$,
now at index $2q+1$, gives $\widetilde{\+\kappa}_{2q+1}
=(\widetilde{\+\eta}_{2q+1}-\widetilde{\+\eta}_{2q})/I_{2q+1}^0$, while
$\+\kappa_{2q+1}^0=\*0_{p\times1}$ by \eqref{eq:brslopeodd} and
correspondingly $\+\eta_{2q+1}^0=\+\eta_{2q}^0$. Hence, exactly as in
\eqref{eq:brdeltarate} but with the contrast divided by $I_{2q+1}^0$,
\begin{align}
    \widetilde{\+\kappa}_{2q+1}-\+\kappa_{2q+1}^0
    &=\frac{\bigl(\widetilde{\+\eta}_{2q+1}-\+\eta_{2q+1}^0\bigr)
    -\bigl(\widetilde{\+\eta}_{2q}-\+\eta_{2q}^0\bigr)}
    {I_{2q+1}^0}
    \notag \\
    &=O_p\Bigl(\bigl[N\bigl(I_{2q+1}^0\bigr)^2
    \bigl(\Lambda_{2q}\wedge\Lambda_{2q+1}\bigr)\bigr]^{-1/2}\Bigr).
    \label{eq:brkappastep}
\end{align}
By \eqref{eq:brodd} and \eqref{eq:brlast}, $I_{2q+1}^0\asymp I_{q+1}^b$,
and by \eqref{eq:brspan} both $\Lambda_{2q}$ and $\Lambda_{2q+1}$ are
$\asymp I_{q+1}^b$, the endpoint cases being $\Lambda_0=I_1^b-1$ at $q=0$
and $\Lambda_{2m+1}=I_{m+1}^b$ at $q=m$. Substituting into
\eqref{eq:brkappastep},
\begin{align}
    \widetilde{\+\kappa}_{2q+1}-\+\kappa_{2q+1}^0
    &=O_p\Bigl(\bigl[N\bigl(I_{q+1}^b\bigr)^3\bigr]^{-1/2}\Bigr),
    \label{eq:brkapparate}
\end{align}
which is the trend rate of Remark~\ref{rem:trend} on the flat regime
itself. The cube is visible in \eqref{eq:brkappastep}: both boundary
levels enclosing regime $2q+1$ are informed by that regime and by an
adjacent regime of unit length, so their spans are of the order of
$I_{q+1}^b$ and no strength is borrowed across the break, while dividing
their contrast by the regime length contributes the remaining two powers.
This completes the proof.
\end{proof}

\end{document}